\documentclass[11pt]{article}
\usepackage{mathtools,amssymb,amsfonts,graphicx}

\DeclareFontFamily{U}{bbold}{}
\DeclareFontShape{U}{bbold}{m}{n}
 {  <-5.5> s*[1.04] bbold5
    <5.5-6.5> s*[1.04] bbold6
    <6.5-7.5> s*[1.04] bbold7
    <7.5-8.5> s*[1.04] bbold8
    <8.5-9.5> s*[1.04] bbold9
    <9.5-11.5> s*[1.04] bbold10
    <11.5-16> s*[1.04] bbold12
    <16-> s*[1.04] bbold17
 }{}

\usepackage[bbgreekl]{mathbbol}
\DeclareSymbolFontAlphabet{\mathbbm}{bbold}
\DeclareSymbolFontAlphabet{\mathbb}{AMSb}%

\usepackage{hyperref}
\usepackage[nosort]{cite}
\usepackage{verbatim}
\usepackage{multicol,color,xcolor,longtable}

\DeclareFontFamily{U}{matha}{\hyphenchar\font45}
\DeclareFontShape{U}{matha}{m}{n}{
      <5> <6> <7> <8> <9> <10> gen * matha
      <10.95> matha10 <12> <14.4> <17.28> <20.74> <24.88> matha12
      }{}
\DeclareSymbolFont{matha}{U}{matha}{m}{n}

\DeclareMathSymbol{\oleft}{2}{matha}{"68}
\DeclareMathSymbol{\oright}{2}{matha}{"69}

\usepackage[small,bf,hang]{caption}
\usepackage{slashed}
\usepackage{latexsym,epsfig}

\definecolor{darkred}{rgb}{0.65,0.15,0}
\definecolor{darkgreen}{rgb}{.05,.5,.05}
\hypersetup{pdfborder={0 0 0},colorlinks=true,urlcolor=darkred,citecolor=blue,linkcolor=darkred,linktocpage=true}

\usepackage{mathrsfs}
\usepackage{dsfont}
\usepackage[Symbolsmallscale]{upgreek}

\makeatletter
\g@addto@macro\bfseries{\boldmath}
\makeatother

\newcommand{\vir}{\mathfrak{vir}}
\newcommand{\hevir}{\hat{\mathfrak{e}}_{8}\oleft\mathfrak{vir}}
\newcommand{\Vir}{\mathrm{Vir}}

\newcommand{\virm}{\mathfrak{vir}^{-}}
\newcommand{\hevirm}{\hat{\mathfrak{e}}_{8}\oleft\mathfrak{vir}^-}
\newcommand{\Virm}{\mathrm{Vir}^{-}}

\newcommand{\nn}{\nonumber}

\newcommand{\ints}{\mathds{Z}}

\newcommand{\cx}{\mathds{C}}

\def\bea{\begin{eqnarray}}\def\eea{\end{eqnarray}}
\newcommand{\be}{\begin{equation}}
\newcommand{\ee}{\end{equation}}
\newcommand{\CR}{\nonumber \\*}

\def\cL{{\cal L}}
\def\cD{{\cal D}}
\def\cI{{\mathcal{I}_1}}
\def\cE{{\mathcal{E}}}
\def\cF{{\mathcal{F}}}
\def\cG{{\mathcal{G}}}

\def\cP{{\mathcal{P}}}
\def\cQ{{\mathcal{Q}}}

\newcommand{\ord}[1]{{\scriptscriptstyle (#1)}}

\newcommand{\cM}{{\mathcal{M}}}
\newcommand{\cV}{{\mathcal{V}}}
\newcommand{\cJ}{{\mathcal{J}}}
\newcommand{\cS}{\mathcal{S}}

\newcommand{\dL}{{\mathsf{d}}}
\newcommand{\dK}{{\mathsf{K}}}
\newcommand{\dd}{{\mathrm{d}}}

\newcommand{\lb}{\left[}
\newcommand{\rb}{\right]}

\newcommand{\mf}[1]{{\mathfrak{#1}}}

\makeatletter

\@addtoreset{equation}{section}
\makeatother

\newcommand{\Tr}{{\mathrm{Tr}}}

\newcommand{\ket}[1]{ |#1 \rangle}
\newcommand{\bra}[1]{ \langle #1 |}
\newcommand{\braket}[2]{ \langle #1 | #2 \rangle}

\newcommand{\ul}[1]{{\underline{#1}}}

\def\hE8{\widehat{E}_{8}}
\def\he8{\hat{\mathfrak{e}}_{8}}

\def\T{\mathsf{T}}

\makeatletter
\def\sbm{\@ifstar\sbm@starred\sbm@unstarred}
\newcommand{\sbm@unstarred}[1]{%
\raisebox{1.1pt}{\scalebox{1.}[.88]{\ensuremath{[}}}%
\mathbbm{#1}\raisebox{1.1pt}{\scalebox{1.}[.88]{\ensuremath{]}}}%
}
\newcommand{\sbm@starred}[1]{%
\raisebox{1.1pt}{\scalebox{1.}[.88]{\ensuremath{[}}}%
{#1}\raisebox{1.1pt}{\scalebox{1.}[.88]{\ensuremath{]}}}%
}
\makeatother

\newcommand{\hbm}[1]{\widehat{\mathbbm{#1}}}
\newcommand{\mhbm}[1]{\raisebox{1.5pt}{$^{\scriptscriptstyle\cM}$}\hspace*{-1pt}\hbm{#1}}
\makeatletter
\DeclareRobustCommand\fJ{\@ifnextchar({\@shiftfJ}{\mathfrak{J}}}
\def\@shiftfJ(#1){\mathfrak{J}^{(#1)}}
\DeclareRobustCommand\fP{\@ifnextchar({\@shiftfP}{\mathcal{P}}}
\def\@shiftfP(#1){\mathcal{P}^{(#1)}}
\makeatother
\def\fQ{\mathcal{Q}}

\def\tJ{\mathsf{J}}
\def\tP{\mathsf{P}}
\def\tQ{\mathsf{Q}}

\begin{document}

{\flushright {CPHT-RR016.032021}\\
{QMUL-PH-21-15}\\[5mm]}

\begin{center}
  {\LARGE \sc  $E_9$ exceptional field theory\\[4mm] II. The complete dynamics}
    \\[13mm]

{\large
Guillaume Bossard${}^{1}$, Franz Ciceri${}^2$, Gianluca Inverso${}^{3,4}$,\\[1ex] Axel Kleinschmidt${}^{2,5}$ and  Henning Samtleben${}^6$}

\vspace{8mm}
${}^1${\it Centre de Physique Th\'eorique, CNRS, Institut Polytechnique de Paris, \\ FR-91128 Palaiseau cedex, France}
\vskip 1.2 ex
${}^2${\it Max-Planck-Institut f\"{u}r Gravitationsphysik (Albert-Einstein-Institut)\\
Am M\"{u}hlenberg 1, DE-14476 Potsdam, Germany}
\vskip 1.2 ex
${}^3${\it Centre for Research in String Theory, School of Physics and Astronomy, \\
Queen Mary University of London, 327 Mile End Road, London, E1 4NS, United Kingdom}
\vskip 1.2 ex
${}^4${\it INFN, Sezione di Padova \\
		Via Marzolo 8, 35131 Padova, Italy}
\vskip 1.2 ex
${}^5${\it International Solvay Institutes\\
ULB-Campus Plaine CP231, BE-1050 Brussels, Belgium}
\vskip 1.2 ex
${}^6${\it Univ Lyon, Ens de Lyon, Univ Claude Bernard, CNRS,\\
Laboratoire de Physique, FR-69342 Lyon, France}

\end{center}

\vspace{5mm}

\begin{center} 
\hrule

\vspace{6mm}

\begin{tabular}{p{14cm}}
{\small%
We construct the first complete exceptional field theory that is based on an infinite-dimensional symmetry group. $E_9$ exceptional field theory provides a unified description of eleven-dimensional and type IIB supergravity covariant under the affine Kac--Moody symmetry of two–dimensional maximal supergravity. We present two equivalent formulations of the dynamics, which both rely on a pseudo-Lagrangian supplemented by a twisted self-duality equation. One formulation involves a minimal set of fields and gauge symmetries, which uniquely determine the entire dynamics. The other formulation extends $\mathfrak{e}_9$ by half of the Virasoro algebra and makes direct contact with the integrable structure of two-dimensional supergravity. Our results apply directly to other affine Kac--Moody groups, such as the Geroch group of general relativity.
}
\end{tabular}
\vspace{5mm}
\hrule
\end{center}

\thispagestyle{empty}

\newpage
\setcounter{page}{1}


\tableofcontents

\section{Introduction}
\label{sec:intro}

This paper completes the work started in \cite{Bossard:2018utw}, constructing the exceptional field theory
for the Kac--Moody group $E_{9}$.
Exceptional field theories (ExFT) are the duality covariant formulations of maximal supergravity
built on the exceptional hidden symmetry groups $E_n$ that arise in dimensional reduction of eleven-dimensional (or type IIB) supergravity~\cite{Cremmer:1978ds,Julia:1982gx,Cremmer:1997ct,Cremmer:1998px}. 
$E_n$ ExFT is based on a split of the coordinates of eleven-dimensional supergravity into $D$ external
and $n=11-D$ internal coordinates. 
The latter are embedded into an extended space-time
with coordinates transforming in a representation of the exceptional group $E_n$.
The original physical coordinates are recovered as the solution of an $E_n$ covariant section constraint
that constrains the coordinate dependence of all fields and gauge parameters. 
On the extended space-time, the original geometric diffeomorphisms and gauge symmetries are unified
into generalised diffeomorphisms~\cite{Hull:2007zu,Pacheco:2008ps,Hillmann:2009pp,Berman:2010is,Coimbra:2011ky,Berman:2012vc,Coimbra:2012af,Cederwall:2013naa,Cederwall:2013oaa,Aldazabal:2013mya,Hohm:2013jma,Bossard:2017aae,Cederwall:2017fjm}
which provide the central symmetry structure organising and defining the theories. 

Exceptional field theories have been constructed for all finite-dimensional duality groups,
i.e.\ for $n\le 8$, and $D\ge3$ external dimensions
\cite{Hohm:2013pua,Hohm:2013vpa,Hohm:2013uia,Hohm:2014fxa,Hohm:2015xna,Abzalov:2015ega,Musaev:2015ces,Berman:2015rcc}.
The resulting actions are modelled after the structure of $D$-dimensional maximal (gauged) supergravities,
with all fields living on the full extended space-time (subject to the section contraint) and the non-abelian
gauge structure of $D$-dimensional supergravity replaced by the infinite-dimensional algebraic structure
of generalised diffeomorphisms. In particular, the scalar fields parametrise the coset space
$E_{n}/K(E_{n})$ and couple via a (gauged) sigma-model on this space, where $K(E_n)$ is the maximal compact subgroup of $E_n$.

For $n=9$, the group  $E_9$ is the infinite-dimensional affine group that appears as the global 
symmetry of maximal two-dimensional supergravity~\cite{Julia:1981wc,Nicolai:1987kz},
generalising the Geroch group in the reduction of four-dimensional Einstein gravity~\cite{Geroch:1970nt,Belinsky:1971nt,Maison:1978es,Breitenlohner:1986um}.
The affine Lie algebra $\mf{e}_9$ of $E_9$ is by definition the sum of the centrally extended loop algebra $\hat{\mf{e}}_8$ over $\mf{e}_8$ and a one-dimensional derivation algebra.\footnote{We denote the derivation of $\mf{e}_9$ by $\dL$ and it is related to the Virasoro generator $L_0$ by $\dL = L_0 +h$ in a module of weight $h\in \mathds{C}$.}
The study of the internal sector of ExFT based on this affine algebra was initiated in \cite{Bossard:2017aae}
with the construction of generalised diffeomorphisms acting on fields whose internal coordinates $Y^M$ transform
in the basic lowest weight representation $R(\Lambda_0)_{-1}$ of $E_9$,
and a section constraint of the generic form
\begin{align}
\label{eq:SCintro}
Y^{MN}{}_{PQ} \,\partial_M \otimes \partial_N = 0
\end{align}
where $Y^{MN}{}_{PQ}$ is the  $E_9$ invariant tensor~\eqref{eq:Y-def}. 
In $E_9$ ExFT, all fields live on the full extended space-time with two external coordinates $x^\mu$ and internal coordinates $Y^M$, subject to the section constraint (\ref{eq:SCintro}).

In a first paper \cite{Bossard:2018utw}, we have constructed the internal part of the action of $E_9$ ExFT, usually
referred to as the potential, as a scalar field Lagrangian invariant under generalised diffeomorphisms.
More precisely, this is the truncation of ExFT obtained after dropping all dependence
on the two external coordinates $x^\mu$ and truncating the gauge fields.
The goal of this paper is to extend this construction to the full $E_9$
exceptional field theory.

With respect to its lower-rank cousins, $E_9$ ExFT comes with a number of additional technical challenges. First of all,
the affine nature of the duality group requires all bosonic objects to appear in infinite-dimensional representations. This is reflected
by the scalar fields parametrising the infinite-dimensional coset space
\begin{equation}
\frac{ \widehat{E}_{8}\rtimes\bigl( \mathds{R}^+_{L_0}\ltimes\mathds{R}_{L_{-1}} \bigr) }
 {K(E_9)}\,,
 \label{intro:coset}
\end{equation}
where $\widehat{E}_8$ denotes the centrally extended loop group over $E_8$,
and the factor $\mathds{R}^+_{L_0}\ltimes\mathds{R}_{L_{-1}}$ is obtained by
exponentiating the Virasoro generators $L_0$ and $L_{-1}$.

Secondly, while the ExFT field content in generic dimensions is largely based on $D$-dimen\-sional maximal supergravity as imposed by the tensor hierarchy~\cite{deWit:2005hv,deWit:2008ta,Palmkvist:2013vya}, closure of the gauge algebra in general requires the introduction 
of additional $p$-forms of rank $p\ge D-2$. 
These additional fields are covariantly constrained by algebraic conditions analogous to the section contraint (\ref{eq:SCintro}). For sufficiently large $D$, the constrained  fields may not be visible at
the level of the Lagrangian, but starting from $D\le4$, they become an inevitable part of the dynamical equations of exceptional field theory. In particular, for $D=2$, such constrained fields appear in all sectors of the theory. 
Indeed, the results of \cite{Bossard:2018utw} show that the construction of an invariant scalar potential
requires the coupling of an additional scalar field $\chi_M$ obeying the algebraic constraints
\begin{align}
Y^{MN}{}_{PQ} \,\chi_M \otimes \partial_N  = 0 = Y^{MN}{}_{PQ} \,\chi_M \otimes \chi_N
\;,
\label{intro:chiM}
\end{align}
analogous to \eqref{eq:SCintro}. 
Similarly, the gauge fields
\begin{align}
\mathbbm{A}_\mu= (A_\mu^M, B_\mu{}^M{}_N)
\,,
\label{intro:gauge}
\end{align}
combine vectors $A_\mu^M$ transforming in the
 basic lowest weight (coordinate) representation $R(\Lambda_0)_{-1}$ of $E_9$
with vector fields $B_\mu{}^M{}_N$ constrained in their last index analogously to \eqref{intro:chiM}.
The set of vector fields (\ref{intro:gauge}) reflects the structure of 
gauge parameters of $E_9$ generalised diffeomorphisms~\cite{Bossard:2017aae}.

Thirdly, as known from the lower-rank exceptional field theories, their construction in an even number
of external dimensions $D$ is typically hampered by the fact that in $D$-dimensional supergravity 
only a subgroup of the duality group $E_n$ is realised off-shell while the full duality group is only 
realised on the equations of motion. The most straightforward construction of the ExFT dynamical field equations
thus draws on a manifestly duality invariant pseudo-Lagrangian which must be supplemented by a set of
first order (self-)duality equations, in the spirit of the so-called democratic formulation of
supergravity theories~\cite{Bergshoeff:2001pv}.
In two-dimensional maximal supergravity, the Lagrangian only depends on the 128 propagating scalar fields parametrising $E_8/Spin(16)$, the dilaton $\rho$ associated to $L_0$, and the conformal factor $\sigma$ of the external metric, associated to the central charge $\dK$.
The infinitely many fields parametrising the coset space (\ref{intro:coset}) are defined on-shell by virtue
of an infinite set of duality relations typically formulated in terms of a linear system~\cite{Belinsky:1971nt,Maison:1978es,Breitenlohner:1986um,Nicolai:1987kz}
and its expansion in terms of a spectral parameter. 
In order to reflect these structures, $E_9$ ExFT will be based on a manifestly duality invariant pseudo-Lagrangian which features all the scalar fields
parametrising (\ref{intro:coset}), together with  duality equations relating them to the physical ones. However, the spectral parameter of the linear system introduced in \cite{Maison:1978es,Breitenlohner:1986um} depends on the $D=2$ space-time coordinates, a property that is a priori incompatible with the definition of fields in $E_9$ lowest weight modules like $A_\mu^M$ in  ExFT.  This was not yet an issue in the construction of the  ExFT scalar potential \cite{Bossard:2018utw}
which corresponds to a truncation of the theory in which all duality equations are consistently projected out.
It will, however, be relevant in the construction of the full dynamics, and one of the main achievements of this paper will be to show that the ExFT duality equations reproduce the Breitenlohner--Maison linear system.  

From the above discussion and the algebra of generalised diffeomorphisms, one might expect that the fields of $E_{9}$ ExFT comprise the coset scalar fields in \eqref{intro:coset}, the unimodular metric $\tilde{g}_{\mu\nu}$, the constrained field $\chi_M$ and  the vector fields \eqref{intro:gauge}.
We will indeed define a minimal formulation of the theory featuring  these fields (in the gauge in which the scalar fields lie in $E_9 / K(E_9)$, i.e.  with the $L_{-1}$ scalar $\tilde{\rho}=0$), as well as a single additional one-form field $\chi_\mu$. 
However, the relation of the $E_{9}$ ExFT duality equation to the Breitenlohner--Maison linear system cannot be unraveled in this minimal formulation. 
This can be resolved by adopting a formulation of the Breitenlohner--Maison linear system that circumvents the problem of the space-time dependent spectral parameter through the introduction of scalar fields in the negative Virasoro group~\cite{Julia:1996nu,Paulot:2004hh}, enlarging the scalar target coset space~\eqref{intro:coset} to 
\begin{equation}\label{VirE9coset}
\frac{\widehat{E}_{8} \rtimes \Virm }{K(E_9)}
\, ,
\end{equation}
where  $ \Virm$ is the group associated to the algebra $\virm=\langle L_{n} | n\leq 0\rangle$. We shall define a Virasoro-extended formulation of $E_{9}$ ExFT, in which the algebra of generalised diffeomorphisms is based on $\widehat{E}_{8} \rtimes \Virm$. The set of vector fields is enlarged accordingly, with one vector field  $B_\mu^{\ord{n}}{}^M{}_N$ associated to each negative Virasoro generator $L_{-n}$.  The truncation of the Virasoro-extended formulation of ExFT to $D=2$ supergravity (via the trivial solution of the section constraint) will be shown to reproduce the Breitenlohner--Maison linear system in the formulation of~\cite{Julia:1996nu,Paulot:2004hh}.

The scalar fields in \eqref{VirE9coset} are parametrised by the coset representative
\begin{equation}
\cV = \Gamma \hat{V} 
= \,\rho^{-L_0} \, e^{-\phi_1 L_{-1}} \, e^{-\phi_2 L_{-2}} \,  \cdots
  \mathring{V}\,e^{Y_{1A} T^A_{-1}}\,e^{Y_{2A} T^A_{-2}} \, \cdots\,e^{-\sigma\dK}
\label{intro:full coset repr}
\end{equation}
upon introducing an additional infinite tower of scalar fields $\phi_n$
associated with the negative half of the Virasoro algebra. The fields $\rho$ and $\phi_n$
combine into the $\Gamma$-factor in (\ref{intro:full coset repr})
whose presence replaces the space-time dependent spectral parameter of the standard linear system on (\ref{intro:coset})
\cite{Maison:1978es,Breitenlohner:1986um}.
The $E_8$-valued matrix $\mathring{V}$ in (\ref{intro:full coset repr}) carries the propagating scalar fields of $D=2$ maximal supergravity
while the negative mode generators $T^A_{-n}$ of the loop algebra are associated with the
infinite tower of dual potentials defined by the linear system.

The first-order duality relations encoding the ExFT scalar field dynamics are most compactly formulated 
in terms of the covariant currents 
\begin{align}
\fJ_\mu\ =\ \cM^{-1} \cD_\mu \cM\ =\ \fJ_{\mu\,\alpha} T^\alpha =
 \sum_{m\in\mathds{Z}}( \, \tJ_\mu{}^m_{A}\,  T^A_m + \tJ_{\mu\,m}\,  L_{m}\,  ) + \tJ_{\mu\,\dK} \,\dK\,.
 \label{intro:J}
\end{align}
obtained from the generalised metric $\cM = \cV^\dagger \cV$,
and explicitly expanded in terms of the generators of  $\hE8$ and the full Virasoro algebra.
As standard in ExFT,
the covariant derivative $\cD_\mu$ employs the vector fields (\ref{intro:gauge}) 
(and a tower of Virasoro descendants $B_\mu^{\ord{n}}{}^M{}_N$ of the constrained vectors) 
to gauge the action of Virasoro-extended $E_9$ generalised diffeomorphisms.
Its explicit connection is defined in (\ref{covariant J definition}) in the main text.
From (\ref{intro:J}), one may define a tower of shifted currents
\begin{equation}\label{intro:shifted cov current}
\fJ(m)_\mu = \cS_0\big(\, \Gamma^{-1} \cS_m(\Gamma \fJ_\mu \Gamma^{-1})\,\Gamma \,\big) 
+ \chi^\upgamma_{\mu,m} \dK\,,
\end{equation}
where the operators $\cS_m$ shift the mode number of the $\widehat{E}_8$ loop generators
and the Virasoro generators by $m$ units and project out the component along the central charge $\dK$, while  $\Gamma$ is the $\Virm$ part of the coset representative (\ref{intro:full coset repr}). 
Covariance of the shifted currents (\ref{intro:shifted cov current}) under rigid $\hE8\rtimes\Virm$ requires the introduction
of a tower of new (external) one-forms $\chi^\upgamma_{\mu,m}$ as their $\dK$ components.\footnote{The superscript $\upgamma$ is not an index, but rather a label related to the $\Gamma$ conjugation in \eqref{intro:shifted cov current}, see Section~\ref{sec:extE9exft}.} This is in complete analogy to the appearance of the constrained scalar field $\chi_M$~(\ref{intro:chiM}) in the shifted current along internal derivatives which was introduced in \cite{Bossard:2018utw} to define the ExFT scalar potential. 
As a key result of this paper, we obtain the ExFT scalar twisted self-duality equation in the form 
\begin{equation}
\star\!\fJ = \fJ(1)\,,
\label{intro:twsd cov J}
\end{equation}
where $\star$ denotes Hodge duality with respect to the unimodular two-dimensional metric 
in conformal gauge $\tilde{g}_{\mu\nu}=\eta_{\mu\nu}$.
Upon dropping internal derivatives and gauge fields, the duality equation (\ref{intro:twsd cov J})
reduces to a linear system for the field equations of $D=2$ supergravity. 
The standard form of the linear system \cite{Maison:1978es,Breitenlohner:1986um} is recovered
after integrating the $\virm$ part of the duality equation in order to obtain polynomial expressions for
the additional scalar fields $\phi_m$ in (\ref{intro:full coset repr}) 
in terms of the fields $\rho$ and $\tilde\rho\sim \phi_1$ parametrising the $\mathds{R}^+_{L_0}\ltimes\mathds{R}_{L_{-1}}$ 
factor in (\ref{intro:coset}).

Next, we construct in this paper a pseudo-Lagrangian which together with (\ref{intro:twsd cov J}) determines the full
dynamics of $E_9$ ExFT. According to the general structure of gauged supergravity in $D=2$ dimensions \cite{Samtleben:2007an},
and more generally to the structure of two-dimensional gauged sigma models with WZW terms~\cite{Hull:1990ms},
the full Virasoro-extended $E_9$ ExFT pseudo-Lagrangian is expected to extend the potential term of \cite{Bossard:2018utw} by
a topological term coupling the currents of the coset space (\ref{VirE9coset}) to the gauge fields,
as well as by a suitably gauged version of the linear system of duality equations. The pseudo-Lagrangian consists accordingly of the potential term of \cite{Bossard:2018utw} and a topological term written as a top-form:
\begin{align}
\label{intro:final cov Lagrangian J}
\rho^{-1}\cL_{\text{top}}\ &=\ 
  \cD\chi^\upgamma_1 
- \frac12\omega^{\alpha\beta}\fJ_\alpha \wedge \fJ(1)_\beta 
-  \frac{4}{3}\sum_{n=2}^\infty(n^3-n)\mathsf{P}_n \wedge (\tP_{n+1}+\tP_{n-1}) 
\CR&\hphantom{=\ }
+  \sum_{n=1}^\infty \tP_n \wedge (\chi^\upgamma_{1+n} - \chi^\upgamma_{1-n})
+ \chi^\upgamma_{M,1}\,{\cF}^M
+ \hbm{F}^\upgamma_1 
+  \mhbm{F}^\upgamma_1\ .
\end{align}
The various ingredients are defined and detailed in Section~\ref{sec:extE9exft}.
In particular, the covariant derivative of $\chi^\upgamma_1 = \chi^\upgamma_{\mu,1} \dd x^\mu$ is defined in \eqref{Dchigamma}. The second term in (\ref{intro:final cov Lagrangian J}) is the gauged WZW-like term and carries 
the algebra cocycle $\omega^{\alpha\beta}$  corresponding to the central charge component of the commutator
\begin{equation}
\omega^{\alpha\beta} \fJ_{\mu\,\alpha} \, \fJ(1)_{\nu\,\beta} = -[\fJ_\mu,\,\fJ(1)_\nu]\big|_\dK
\,.
\label{intro:algebra cocycle}
\end{equation}
The currents $\tP_{\hspace{-1pt}\mu\,n}=\frac12 (\cD_\mu\Gamma\Gamma^{-1})_{-n}$ for $n\ge 1$ are defined as the $\virm$ components of the 
Maurer--Cartan form from (\ref{intro:full coset repr}). 
The field $\chi^\upgamma_{M,1}$ in (\ref{intro:final cov Lagrangian J}) is related by the field redefinition \eqref{eq:chiredef} to the field $\chi_M$ appearing in the potential in \cite{Bossard:2018utw}.
Finally, the last three terms in (\ref{intro:final cov Lagrangian J}) are proportional to the 
non-abelian field strengths
\begin{equation}
\mathbbm{F}_{\mu\nu}=(\mathcal{F}_{\mu\nu}^M,\mathcal{G}_{\mu\nu}^{\ord{k}}{}^M{}_N)\,, 
\qquad
k\in \mathds{N}^+
\,,
\end{equation}
of the gauge fields (\ref{intro:gauge}) (and their Virasoro-extensions). Their precise expressions
are given in (\ref{hatF gamma definition}) and (\ref{mhbm F def}) below.
Remarkably, the full ExFT pseudo-Lagrangian, after gauge-fixing $\phi_n \to 0$ for the $\virm$ scalar fields at $n\ge2$,
is simply given by the combination
\begin{equation}\label{intro:Ltop-pot}
\cL_{\text{ext}} = \cL_{\text{top}} -\star V
\,,
\end{equation}
of the topological term (\ref{intro:final cov Lagrangian J}) with the
potential $V$ constructed in \cite{Bossard:2018utw}.
In their most explicit form, these two terms are given in
(\ref{final cov Lagrangian J GAUGEFIX}) and (\ref{scalpot}) below.
In particular, this pseudo-Lagrangian does not carry a traditional kinetic term 
for the scalar fields.

In order to make contact with the lower-rank exceptional field theories
and to match the field equations with those of eleven-dimensional (and type IIB) supergravity,
it turns out to be convenient to pass eventually from the Virasoro-extended to the minimal formulation of $E_9$ ExFT
in which not only the $\virm$ scalar fields $\phi_n$ are gauge fixed to zero, but 
also the infinite tower of constrained and auxiliary one-forms 
$\{B_\mu^{\ord{n}}{}^M{}_N, \chi_{\mu,n}^\upgamma\}$ featuring in (\ref{intro:final cov Lagrangian J}) is integrated out 
in favour of the original gauge fields (\ref{intro:gauge}) and a single auxiliary one-form $\chi_\mu$. Although we shall prove that the minimal and Virasoro-extended formulations of the theory describe the same dynamics, these two formulations differ in structure and will be discussed separately. In particular, the current $\cJ$ in the minimal formulation is valued in the Lie algebra of the group $\widehat{E}_{8}\rtimes( \mathds{R}^+_{L_0}\ltimes\mathds{R}_{L_{-1}} )$ 
which is not closed under Hermitian conjugation unlike the current~\eqref{intro:J}.
In this formulation, we relax the conformal gauge and consider an arbitrary unimodular metric $\tilde{g}_{\mu\nu}$, thereby allowing for the definition of external diffeomorphisms. To distinguish the currents in the two formulations we will use a different notation, with the minimal formulation current 
\begin{align}
\cJ_\mu\ =\  \cJ_{\mu\,\alpha} T^\alpha =
 \sum_{m\in\mathds{Z}}  \, J_\mu{}^m_{A}\,  T^A_m +  J_{\mu\,0}\,  L_{0} +  J_{\mu\,-1}\,  L_{-1}  + J_{\mu\,\dK} \,\dK\; .
 \label{intro:Jm}
\end{align}
In particular, $J_{\mu\,\dK}$ depends explicitly on the unimodular metric $\tilde{g}_{\mu\nu}$, see~\eqref{Jdef}, and $J_{\mu\,-1} = B_{\mu}{}^M{}_M$. The duality equation takes the form of a twisted self-duality 
 \be 
 \cJ  = \rho^{-1}  \star  \cM^{-1}  ( \mathcal{S}_1(  \cJ  ) + \chi   \dK )^\dagger  \cM\; , \label{MindualInttro}  
 \ee
 associated to an $E_9$ invariant symmetric bilinear form. The resulting pseudo-Lagrangian, now written as a density, takes the form
\begin{align}
 \label{intro:PseudoLagrangian} 
 \cL_{\rm min} = \cL_1 + \cL_2 +\frac14 \rho \, \varepsilon^{\mu\nu} \varepsilon^{\sigma\rho}  \tilde{g}^{\kappa\lambda} \cD_\mu \tilde{g}_{\sigma\kappa} \cD_\nu  \tilde{g}_{\rho\lambda} + \frac{\rho^{-1}}{4} \cM^{MN} \partial_M  \tilde{g}^{\mu\nu} \partial_N  \tilde{g}_{\mu\nu} -V\; , 
 \end{align}
where $\cL_2$ is a topological term similar in structure to \eqref{intro:final cov Lagrangian J}, whereas $\cL_1$ depends explicitly on the metric $\tilde{g}_{\mu\nu}$ and plays the role of a kinetic term. They are defined below in  (\ref{eq:kinmin}) and
(\ref{eq:topmin}) and their  variation with respect to  $B_\mu{}^M{}_N$ is the contraction of the duality equation \eqref{MindualInttro} with $\delta B_\mu{}^M{}_N$. 
The pseudo-Lagrangian \eqref{intro:PseudoLagrangian} is a sum of terms separately invariant under internal diffeomorphisms, whose relative coefficients are determined by external diffeomorphism invariance, as is usually the case in ExFT.
The invariance of the system of equations under external diffeomorphisms, including the duality equation \eqref{MindualInttro}, requires us to consider the additional equation 
\begin{equation}  
\cF^M =  \star  \rho^{-1} T^{\alpha M}{}_N \cM^{NP}  \cJ^-_{P\; \alpha}\; ,  
\end{equation}
in its entirety while it only appears as an Euler--Lagrange equation contracted with the constrained variation $\delta \chi_M$. 
The shifted internal current $\cJ^-_{M\; \alpha}$ 
is defined in~\eqref{InternalJminus} as in \cite{Bossard:2018utw}.

Upon partially solving the section constraint (\ref{eq:SCintro}), most of the components of the duality equation \eqref{MindualInttro} simply determine the non-vanishing components of the vector field $B_{\mu}{}^M{}_N$. All the solutions to the constraint (\ref{eq:SCintro}) can be mapped under $E_9$ to a form compatible with the parabolic gauge \eqref{intro:full coset repr}, such that the fields only depend on the $D=3$ external coordinates and the 248 internal coordinates of $E_8$ ExFT.  Using this solution and integrating out the unconstrained components of $B_{\mu}{}^M{}_N$ one obtains that the pseudo-Lagrangian \eqref{intro:PseudoLagrangian} becomes equivalent to the one of $E_8$ ExFT \cite{Hohm:2014fxa}. This proves that the dynamics of the theory reproduces the one of eleven-dimensional supergravity or type IIB supergravity depending on the choice of solution to the section constraint, confirming the
dynamical content of the pseudo-Lagrangian~\eqref{intro:PseudoLagrangian}.

The rest of this paper is organised as follows:
In Section~\ref{sec:prelim}, we review the algebraic structures and how they appear in two-dimensional ungauged supergravity through a (Virasoro-extended) linear system. Section~\ref{sec:E9gauge} describes the structure of $E_9$ generalised diffeomorphisms and the associated gauge fields, including their Virasoro-extensions.  In Section~\ref{sec:extE9exft}, we construct the Virasoro-extended $E_9$ ExFT by first determining a gauge-invariant topological term. The Virasoro-extended topological term can be gauge-fixed and combined with the potential term.
The minimal formulation is presented in Section~\ref{sec:E9EFT} where the gauge-fixing is used to reduce the formulation of $E_9$ ExFT to a finite set of fields.  In this formulation, we moreover study external diffeomorphisms and show consistency with $E_8$ ExFT.  Because the two formulations of $E_9$ ExFT are  different in structure, we have exposed them in a way that can be read independently, so that in particular  Section~\ref{sec:extE9exft} is not a prerequisite to Section~\ref{sec:E9EFT} from \ref{InternalDifMin} onwards. The equivalence of the two formulations is proved in Section~\ref{FromVirtoMin}. 
In Section~\ref{sec:conclusion}, we discuss possible applications and generalisations of our results.
Several appendices contain additional technical details.

\section{\texorpdfstring{Preliminaries on $D=2$ supergravity and algebraic structures}{Preliminaries on D=2 supergravity and algebraic structures}}
\label{sec:prelim}

In this section, we fix our notation for $E_9$ and its extension by Virasoro generators. We also review the so-called linear system~\cite{Geroch:1970nt,Belinsky:1971nt,Maison:1978es,Breitenlohner:1986um,Nicolai:1987kz} of the bosonic part of $D=2$ maximal supergravity and how the affine symmetry arises. In Section~\ref{sec:extcoset}, we  discuss a less well-known extension~\cite{Julia:1996nu,Paulot:2004hh} of the linear system that also features the Virasoro algebra.

\subsection{\texorpdfstring{$E_9$, its Lie algebra and Virasoro extension}{E9, its Lie algebra and Virasoro extension}}

The split real affine Kac--Moody algebra $\mf{e}_9$ has the basis $T_m^A$, $\dK$ and $\dL$ with non-trivial commutation relations
\begin{align}
\label{eq:e9}
\lb T^A_m, T^B_n \rb = f^{AB}{}_C T^C_{m+n} +m \, \eta^{AB} \delta_{m,-n} \dK\,,\quad\quad \lb \dL , T^A_m \rb = -m T^A_m\,.
\end{align}
The element $\dK$ is central in the Lie algebra and the indices $A,B,C=1,\ldots,248$ parametrise the adjoint of the underlying exceptional $\mf{e}_8$ Lie algebra where $\mf{e}_8$ is realised as a subalgebra by only considering the elements $T^A_0$ and has the Killing form $\eta^{AB}$. The mode number index $m\in\ints$ arises from the loop algebra construction of affine Lie algebras~\cite{Kac,Goddard:1986bp}. The element $\dL$ is called the derivation. We shall also encounter the centrally extended loop algebra 
\begin{align}
    \hat{\mf{e}}_8=\langle T^A_m, \dK\rangle
\end{align}
that differs from $\mf{e}_9$ by the omission of the derivation $\dL$. 

Lowest weight representations, denoted by $R(\Lambda)_h$, are determined by giving a weight $\Lambda$ of $\hat{\mf{e}}_8$ and a conformal weight $h$. The latter corresponds to the eigenvalue under $\dL$ while $\Lambda$ summarises the eigenvalues under the Cartan subalgebra of $\hat{\mf{e}}_8$, that consists of the Cartan subalgebra of $\mf{e}_8$ and $\dK$. The most relevant instance for us is the so-called basic representation $R(\Lambda_0)_h$ that is constructed in Fock space notation from an $\mf{e}_8$-invariant ground state $|0\rangle$ 
\begin{align}
\label{eq:Fock}
T^A_0 |0\rangle =0 \,,\quad \dK |0\rangle = |0\rangle \,,\quad \dL |0\rangle = h |0\rangle\,.
\end{align}
The lowest weight condition means that $T^A_n|0\rangle =0$  for $n>0$ and the module is the unique irreducible quotient obtained by acting with $T^A_n$ for $n<0$ on the ground state. 

The Kac--Moody group associated with $\mf{e}_9$ will be denoted by $E_9$ and we refer to Appendix~\ref{app:kacmoody} for a discussion of some of the subtleties arising when defining this infinite-dimensional group. The subgroup that is generated by only $T_m^A$ and $\dK$ is the centrally extended loop group and denoted by $\widehat{E}_8$ in this paper.

The Sugawara construction~\cite{Sugawara:1967rw,Goddard:1986bp} provides an infinite set of additional Virasoro generators $L_m$ ($m\in\ints$) acting on any lowest weight module $R(\Lambda)_h$. These generators satisfy the Virasoro algebra
\begin{align}
\lb L_m, L_n \rb = (m-n) L_{m+n}  + \frac{c_{\mf{vir}}}{12} m (m^2-1) \delta_{m,-n} \dK
\end{align}
where the Virasoro central charge is determined by $\Lambda$. For the basic representation we have $c_{\mf{\vir}}=8$.  The commutation relations with the $\hat{\mf{e}}_8$ generators in this representation are
\begin{align}
\label{eq:vire9}
\lb L_m , T^A_n \rb = - n T^A_{m+n}\,,\quad \lb L_m, \dK \rb =0 \,, 
\end{align}
and we see that the action of $L_0$ agrees with that of $\dL$ and they can be related in the lowest weight representation as
$\dL = L_0 +h$.

Denoting the Virasoro algebra by $\vir=\langle L_m\,|\, m\in\ints\rangle$, we can form the extended algebra
\begin{align}
\hevir\,,
\end{align}
where the symbol $\oleft$ indicates a semi-direct sum of Lie algebras since the Virasoro algebra acts on $\hat{\mf{e}}_8$ according to~\eqref{eq:vire9}.
We shall denote the generators of $\hevir$ collectively as $T^\alpha$, so that this index runs over $\hat{\mf{e}}_8$ and $\vir$. For any fixed $m\in \ints$ we define an $\hat{\mf{e}}_8$-invariant bilinear form by
\begin{align}
\eta_{m\,\alpha\beta} T^\alpha \otimes T^\beta = \sum_{n\in\ints} \eta^{AB} T^A_n \otimes T^B_{m-n} - L_m \otimes \dK - \dK \otimes L_m\,.
\label{eta def}
\end{align}
For $m=0$, this form coincides with the standard non-degenerate bilinear form $\mf{e}_9$ when identifying $L_0$ with $\dL$. In this case we simply write $\eta_{\alpha\beta} \equiv \eta_{0\,\alpha\beta}$.

We shall also use the so-called shift operators $\cS_m$ for $m\in\ints$ that are defined on $\hevir$ by
\begin{align}
\cS_m( \dK) = 0 
\,,\quad \cS_m(L_n) = L_{m+n} \,,\quad \cS_m(T^A_n) = T^A_{m+n}\,,
\label{eq:shift op def}
\end{align}
which implies 
that
\begin{align}
\eta_{(n+m)\,\alpha\beta} T^\alpha \otimes T^\beta = \eta_{n\,\alpha\beta} T^\alpha \otimes \cS_m(T^\beta) - L_{n+m}\otimes \dK \,.
\label{eta and shift ops}
\end{align}
The shift operators are not invariant under $\hat{\mf{e}}_8$ and their properties are discussed in more detail in~\cite{Bossard:2018utw} and in Appendix~\ref{app:shift}.
Therefore, when we define a shifted object we may want to introduce its $\dK$ completion, such that the completed expression transforms as an algebra-valued object. 
This will be explained in Section~\ref{sec:extE9exft}.
Notice that the definition for $\cS_0$ differs from that in~\cite{Bossard:2018utw}. 
There, we defined $\cS_0(\dK) = \dK$ while in this paper we find it convenient to use $\cS_0(\dK) = 0$.

Finally, we define the Hermitian conjugates of the generators of $\hevir$ in the basic representation $R(\Lambda_0)_h$ by
\begin{align}
L_n^\dagger  = L_{-n}^\dagger\,,\quad \dK^\dagger = \dK \,,\quad T^{A\dagger}_n = \eta_{AB} T^B_{-n}\,.
\label{eq:hermconj}
\end{align}
Writing the Kac--Moody group as $E_9$, this defines a unitary subgroup $K(E_9)$ consisting of the elements $k$ that satisfy $kk^\dagger = k^\dagger k= 1$ when acting  on $R(\Lambda_0)_h$.
Under Hermitian conjugation the shift operators transform as
\begin{equation}
\cS_{m}(X^\dagger) = \big(\cS_{-m}(X) \big)^\dagger \,,\qquad X\in\hevir\,.
\label{eq:shiftophermconj}
\end{equation}

\subsection{\texorpdfstring{$D=2$ supergravity and the linear system}{D=2 supergravity and the linear system}}

In this section we review how the infinite-dimensional algebras presented above appear as symmetries of two-dimensional supergravity \cite{Geroch:1970nt,Belinsky:1971nt,Maison:1978es,Breitenlohner:1986um,Nicolai:1987kz}.
The entire discussion applies to two-dimensional gravity coupled to a dilaton and scalars in a non-linear sigma model on a coset space $G/H$ based on a simple Lie group $G$.
This includes the dimensional reduction to two dimensions of pure $D=4$ General Relativity, with $G/H=SL(2)/SO(2)$, as well as reductions of many supergravity theories.
For definiteness, we will take $G/H=E_8/Spin(16)$ as it appears in maximal supergravity, but all the results are easily generalised to the other cases.
We thus begin with (ungauged) maximal supergravity in three dimensions, whose bosonic field content comprises a metric and scalar fields parametrising the symmetric space $E_{8}/Spin(16)$.
Dimensionally reducing to two dimensions, the metric decomposes as\footnote{Notice that there is no notion of Einstein frame in two-dimensional gravity. The metric we present is thus just the dimensional split of the three-dimensional Einstein frame metric.}
\begin{equation} \label{3Dmetric}
\dd s^2_{\scalebox{0.5}{3D}} = 
e^{2\sigma}\tilde{g}_{\mu\nu}\dd x^\mu\dd x^\nu + \rho^2( \dd\varphi + w_\mu \dd x^\mu)^2
\end{equation}
where $\tilde{g}_{\mu\nu}$ is the two-dimensional unimodular metric with $\det\tilde{g}=-1$, $e^{2\sigma}$ is the conformal factor, $\varphi$ is the Kaluza--Klein coordinate, and all fields only depend on the two-dimensional space-time coordinates $x^\mu$.
The vector field $w_\mu$ is non-dynamical and we set it to zero in this section.
The field $\rho$ is a scalar in two dimensions and referred to as the dilaton.
The $E_{8}/Spin(16)$ scalar fields are encoded in a coset representative $\mathring{V}(x)$ which transforms as
\begin{equation}
\mathring{V}(x) \to \mathring{h}(x) \mathring{V}(x) \mathring{g}\,, \quad \mathring{g}\in E_{8}\,,\ \ \mathring{h}(x)\in Spin(16)\,.
\end{equation}
The equations of motion are phrased in terms of the coset and $Spin(16)$ components of the Maurer--Cartan form
\begin{equation}
\dd \mathring{V} \mathring{V}^{-1}  = \mathring{P} + \mathring{Q}  \,,\qquad 
\mathring{P} = (\mathring{P})^\T  \,,\ \ 
\mathring{Q} = -(\mathring{Q})^\T  \,.
\end{equation}
The symbol $^\T$ denotes transposition in $\mf{e}_{8}$, i.e. the anti-involution that singles out the maximal compact subalgebra $\mf{so}(16)$ and it agrees with restriction of the Hermitian conjugation defined in~\eqref{eq:hermconj} to $T_0^A$.
The scalar field $\rho$ is free:
\begin{equation}
\dd\star\dd\rho=0\,,
\label{rho ungauged eq}
\end{equation}
where $\star$ denotes Hodge duality with respect to the unimodular metric $\tilde{g}_{\mu\nu}$.%
\footnote{We use conventions such that $\star1=\dd x^0\wedge\dd x^1$. 
For one-forms we have $(\star\omega)_\mu = \tilde g_{\mu\nu}\varepsilon^{\nu\rho} \omega_\rho$ with $\varepsilon^{01}=1=-\varepsilon_{01}$.
}
The equations of motion for the $E_{8}$ scalars are
\begin{equation}
\dd\star(\rho\, \mathring{V}^{-1} \mathring{P}\, \mathring{V}) = 0
\label{e8 scalars ungauged equation}
\end{equation}
and finally the conformal factor is entirely specified by the Virasoro constraint which is most easily written in conformal gauge $\tilde{g}_{\mu\nu}=\eta_{\mu\nu}$ and light-cone coordinates $x^\pm$ such that $\eta_{+-}=\eta_{-+}=1$ and $\eta_{\pm\pm}=0$:
\begin{equation}
\partial_\pm\sigma\,\partial_\pm\rho 
- \frac12 \partial_\pm\partial_\pm\rho 
- \frac12 \rho\, \eta_{AB} \, P_\pm^A P_\pm^B = 0\,.
\label{ungauged Vir constraint}
\end{equation}
There is also a second-order equation for $\sigma$, which is implied by the other ones
\begin{equation}\label{sigma 2nd order ungauged}
\partial_+\partial_- \sigma + \frac12 \eta_{AB} P^A_+\,P^B_- = 0 \,.
\end{equation}

Equation \eqref{rho ungauged eq} implies that we can define a dual scalar field $\tilde\rho$ such that
\begin{equation}
\dd\tilde\rho = \star\dd\rho\,.
\label{tilde rho duality ungauged}
\end{equation}
Combining this relation with \eqref{e8 scalars ungauged equation} we can construct infinitely many scalar fields dual to the currents $\mathring{P}$.
The first of these duality relations reads 
\begin{equation}
\dd Y_1 = 2\rho \   \mathring{V}^{-1} \star\!\mathring{P}\, \mathring{V}\,,
\label{Y1 duality ungauged}
\end{equation}
and the whole tower is best encoded into a linear system which we now describe in the form given in~\cite{Breitenlohner:1986um,Nicolai:1987kz}.

We introduce a (constant) spectral parameter $w\in\mathds{C}$ 
and define an $E_{8}$ valued function $\hat{V}(w)$ with the requirement that it reduces to $\mathring{V}$ as $w\to\infty$:\footnote{Dependence on the space-time coordinates $x^\mu$ is understood.}
\begin{equation}
\hat{V}(w) \in E_{8}\,,\qquad \lim_{w\to\infty}\hat{V}(w)=\mathring{V}\,.
\label{Vhat regularity}
\end{equation}
Then, equations \eqref{rho ungauged eq} and \eqref{e8 scalars ungauged equation} imply integrability of the linear system\footnote{The negative signs in \eqref{BMsys} may appear unusual compared to the expression in \cite{Breitenlohner:1986um}. The explanation is that the $\upgamma$ parameter we use here corresponds to $1/t$ there.}
\begin{equation}
\dd\hat{V}(w)\hat{V}^{-1}(w) = 
\mathring{Q} - \frac{1+\upgamma^2}{1-\upgamma^2} \mathring{P}
    - \frac{2\upgamma}{1-\upgamma^2} \star\! \mathring{P}\,.
\label{BMsys}
\end{equation}
In order to reproduce the $\rho$ factor in \eqref{e8 scalars ungauged equation}, $\upgamma$ must be a space-time dependent function of $\rho$ and $\tilde\rho$ and satisfy
\begin{equation}
\frac{\dd\upgamma}{\upgamma} =
\frac{1+\upgamma^2}{1-\upgamma^2}\, \rho^{-1}\dd\rho
+ \frac{2\upgamma}{1-\upgamma^2}\, \rho^{-1} \dd\tilde\rho\,,
\label{BMgamma sys}
\end{equation}
which is solved by
\begin{equation}
\upgamma_\pm = 
\frac{1}{\rho}\left( s(w)-\tilde\rho \pm \sqrt{(s(w)-\tilde\rho)^2-\rho}\right)\,,\qquad
\upgamma=\upgamma_+ = \frac{1}{\,\upgamma_-\!}\,.
\label{BMgamma}
\end{equation}
Here, $s(w)$ is an integration constant through which $\upgamma$ depends on the spectral parameter.
The simplest choice compatible with the asymptotics described below is 
\begin{equation}
s(w)=w\,,
\end{equation}
which we will use in the remainder of this section.
More general choices are possible and we will make use of this fact later.\footnote{The meaning of the symbols $s$ and $w$ here is changed with respect to \cite{Breitenlohner:1986um}. Also notice that \cite{Breitenlohner:1986um} is in Euclidean signature, which causes some further sign flips with respect to our discussion here.}
Because of the square root, the function $\upgamma$ then defines a double covering of the $w$ plane.
The two sheets correspond to two solutions of \eqref{BMsys} that are analytic continuations of each other and are captured by writing
\begin{equation}
\hat{V}(w) = V\big(\upgamma(w)\big)\,.
\label{hatV = V of gamma}
\end{equation}
It is then straightforward to see that if $V\big(\upgamma(w)\big)$ satisfies \eqref{BMsys}, so does $V^{-\T}\!\big(1/\upgamma(w)\big)$, where $V^{-\T}=(V^\T)^{-1}$.
We then have that the monodromy matrix
\begin{equation}
\mathfrak{M}(w) = V^\T\big(1/\upgamma(w)\big) V\big(\upgamma(w)\big)
\label{monodromy matrix}
\end{equation}
is symmetric, single-valued in $w$ and constant, i.e. independent of the space-time coordinates $x^\mu$ as a consequence of \eqref{BMsys}.
The monodromy matrix entirely specifies a solution of the linear system.%
\footnote{Solutions to the physical equations of motion are extracted from $\mf M(w)$ by solving a Riemann--Hilbert factorisation problem. See \cite{Aniceto:2019rhg} and references therein for recent developments in the context of general relativity, and \cite{Katsimpouri:2013wka} for a generalisation to coset spaces other than $SL(2)/SO(2)$.}
It can be acted upon by constant elements of the loop group over $E_{8}$, namely $E_{8}$-valued functions of the spectral parameter:
\begin{equation}
\mathfrak{M}(w) \to g^\T\!(w) \, \mathfrak{M}(w) \,  g(w)\,,\qquad
g(w) \in E_{8}\,.
\end{equation}
This defines the global $\widehat{E}_{8}$ symmetry of the model.
For compatibility with \eqref{Vhat regularity}, we require $V\big(\upgamma(w)\big)$ to be expandable as a series in $\upgamma$ and reduce to $\mathring{V}$ as $\upgamma\to\infty$. 
Then, the right $\widehat{E}_{8}$ action on $V\big(\upgamma(w)\big)$ must be compensated on the left by a local transformation that preserves its regularity properties:
\begin{equation}
V\big(\upgamma(w),\,x\big) \to 
h\big(\upgamma(w),\,x\big) \, V\big(\upgamma(w),\,x\big) \, g(w)\,,\qquad
h\big(\upgamma(w),\,x\big) = h^{-\T}\!\big(1/\upgamma(w),\,x\big)\,.
\label{BM coset trf}
\end{equation}
where we have denoted explicitly which objects are space-time dependent (and $\upgamma$ itself is space-time dependent through $\rho$ and $\tilde\rho$).
The compensating transformation belongs to \emph{one} $K(E_9)=K(\widehat{E}_{8})$ subgroup of $\widehat{E}_{8}$ and leaves the monodromy matrix invariant.
It is defined in terms of a \emph{field-dependent} anti-involution which acts by inversion of $\upgamma(w)$ and differs from the Hermitian conjugation we introduced in \eqref{eq:hermconj} which, in the spectral parameter representation, acts by inversion of $w$.
This distinction will become relevant in the next section.
We see that $V\big(\upgamma(w)\big)$ is a coset representative for $\widehat{E}_{8} / K(E_9)$.
The conformal factor $e^\sigma$ also transforms under $\widehat{E}_{8}$, compatibly with \eqref{ungauged Vir constraint}, in terms of a group cocycle that defines the central extension of $\widehat{E}_{8}$.

An infinite tower of dual fields and the associated duality equations generalising \eqref{Y1 duality ungauged} are obtained by expanding \eqref{BMsys} around $w\to\infty$ with%
\footnote{We provide more details on the definitions of Kac--Moody groups associated to a give Kac--Moody algebra in Appendix~\ref{app:kacmoody}.}
\begin{equation}
V\big(\upgamma(w)\big) = 
\mathring{V}\, e^{Y_{1A}T_{-1}^A}\, e^{Y_{2A}T_{-2}^A}\, e^{Y_{3A}T_{-3}^A}\cdots\,.
\label{Vhat Y expansion}
\end{equation}
Where we have defined
\begin{equation}
T_m^A = w^m T^A
\end{equation}
with $T^A$ a basis of generators for $\mathfrak{e}_{8}$.
We then see that the generators $T^A_m$ with $m<0$ correspond to shifts of the dual $Y_m$ potentials that do not affect the physical fields in $\mathring{V}$. 
Positive loop level generators $T^A_m$ with $m>0$, instead, correspond to hidden symmetries that mix $\mathring{V}$ with the dual potentials.

There are three more global symmetry generators manifest in the linear system:
a rescaling of $\rho$ combined with appropriate rescalings of the $Y_m$ fields, captured by the action of the $L_0$ generator
\begin{equation}
L_0 = -w \frac{\partial}{\partial w}\,,
\end{equation}
a shift of $\tilde\rho$ corresponding to the choice of integration constant in \eqref{tilde rho duality ungauged} combined with a redefinition of the $Y_m$ fields captured by the Virasoro generator
\begin{equation}
L_{-1} = -\frac{\partial}{\partial w}\,,
\end{equation}
and finally a constant shift of $\sigma$ corresponding to the choice of integration constant in \eqref{ungauged Vir constraint}, corresponding to $\dK$.
Notice that the generators $L_0,\,L_{-1}$ do not commute with the loop algebra but normalise it.

\subsection{Extended coset and twisted self-duality}
\label{sec:extcoset}

In order to make contact with exceptional field theory, it is desirable to rewrite the linear system in a form that is manifestly covariant under all the global symmetries described above and  independent of the $K(E_9)$ gauge. 
Such a form should also incorporate the duality equation between $\rho$ and $\tilde\rho$ on the same footing as the ones for the $Y_m$ fields.
Furthermore, we will need to write our coset representatives and currents in a lowest weight representation of $E_9$. 
The natural definition of the maximal unitary subgroup $K(E_9)$ will then be in terms of the Hermitian conjugation \eqref{eq:hermconj} --- corresponding here to the inversion $w\to1/w$ combined with $E_{8}$ transposition, rather than the field-dependent involution defined in \eqref{BM coset trf}.
These issues are addressed by extending the group theoretical structures found so far and rephrasing the linear system as a twisted self-duality constraint on an enlarged set of dual fields.

This approach follows \cite{Julia:1996nu,Paulot:2004hh} with several adaptations.
It is based on the similarity between \eqref{BMgamma sys} and \eqref{BMsys} and the association of $\rho$ and $\tilde\rho$ with the Virasoro generators $L_0$ and $L_{-1}$.
The basic idea is to regard $\upgamma(w)$ as a diffeomorphism on the $w$ plane.
Infinitesimally, such changes of variables are generated by the Virasoro elements
\begin{equation}
L_m = -w^{m+1}\frac{\partial}{\partial w}\,.
\end{equation}
There are some issues with such an interpretation, because the naive exponentiation of the Virasoro algebra does not form a group~\cite{Milnor:1985xz}.
This is reflected for instance in the fact that \eqref{BMgamma} with $s=w$ defines a double covering of the $w$ plane and hence while $\upgamma^{-1}(w)$ is well defined, the double inverse is not unique.
To circumvent these issues, we shall focus only on the behaviour of fields and group elements around $w\to+\infty$.
We shall consider redefinitions of the spectral parameter that preserve the asymptotics at $+\infty$, namely we consider only transformations of the form
\begin{equation}
w \to f(w) = f_{-1} w + f_0 + f_{1} w^{-1} + f_2 w^{-2} + \ldots
\label{w formal power series}
\end{equation}
with real coefficient and with $f_{-1}>0$. 
It is not necessary for the power series to converge (i.e., we accept formal power series).
Group multiplication is given by composition of two such transformations, and the coefficients of the inverse of \eqref{w formal power series} are finite expressions, uniquely determined order by order in terms of the $f_i$.
There is a one-to-one correspondence between \eqref{w formal power series} and (formal) products of exponentials of the non-positive part of the Virasoro algebra:
\begin{equation}
F^{-1} w = f(w)\,,\qquad
F^{-1} =   \cdots \, e^{-\frac{f_1}{f_{-1}} L_{-2}}\,e^{-\frac{f_0}{f_{-1}} L_{-1}} \,(f_{-1})^{-L_0} \,,
\label{Vir minus element}
\end{equation}
where each exponent is a rational function of finitely many $f_i$, whose explicit expression we will not need.
We shall denote $\mathfrak{vir}^-$ the algebra generated by $L_m$ with $m\le0$ and by $\Virm$ the group we just described.
The function \eqref{BMgamma} belongs to this set when expanded around $w\to+\infty$, so that we can write close to $w\to+\infty$
\begin{equation}
\frac{1}{\rho}\left( s(w)-\tilde\rho \pm \sqrt{(s(w)-\tilde\rho)^2-\rho}\right) = \frac{2s(w)}{\rho} -\frac{2\tilde\rho}{\rho} -\frac{\rho}{2s(w)}
-\frac{\rho\,\tilde\rho}{2 s(w)^2} + \ldots
\label{sqrt expansion}
\end{equation}
where we now also allow the integration constant $s(w)$ to take a more general form than just $s(w)=w$, compatibly with the regularity properties of $\hat{V}(w)$ stated above, so that
\begin{equation}
s(w) = s_{-1} w + s_0 + s_1 w^{-1} +s_2 w^{-2} + \ldots
\label{s expansion}
\end{equation}
with $s_{-1}>0$.%
\footnote{In the linear system, the choice of $s(w)$ amounts to a redefinition of the $Y_m$ fields with $m\ge2$.}
Namely, $s(w)$ is itself a (constant) element of $\Virm$.

The key point of this reformulation is to regard $\upgamma(w)$ as an \textit{arbitrary} space-time-dependent element of $\Virm$, which we shall write as
\begin{align}
\Gamma^{-1} = \cdots \, e^{\phi_3 L_{-3}}\,e^{\phi_2 L_{-2}}\,e^{\phi_1 L_{-1}}\,\rho^{L_0}\,,\qquad
\upgamma(w) = \Gamma^{-1}w = 
\frac{w}{\rho}-\frac{\phi_1}{\rho}-\frac{\phi_2}{\rho\,w}-\frac{\phi_3}{\rho\,w^2} 
+\ldots
\label{upgamma as series}
\end{align}
where $\phi_m(x)$, $m>0$ are infinitely many scalar fields generalising the dual potential $\tilde\rho(x)$.
The natural action of $\Virm$ elements on $\hat{V}(w)$ can be used, in particular, to rephrase \eqref{hatV = V of gamma} as
\begin{equation}\label{Vhat V relation}
\hat{V}(w) = V\big(\upgamma(w)\big) = \Gamma^{-1} V(w) \Gamma\,,
\end{equation}
where expansion at $w\to+\infty$ is understood and we must now regard $\upgamma(w)$ as the arbitrary series \eqref{upgamma as series} rather than the expression that appears in the linear system.
We therefore see that we may define an extended coset representative which includes the $\Virm$ group element above:
\begin{equation}
\cV = \Gamma V\big(\upgamma(w)\big) = V(w) \Gamma \,.
\end{equation}
We can then combine this expression with the transformation property of $V\big(\upgamma(w)\big)$ in \eqref{BM coset trf} to deduce how $\cV$ transforms under the loop group.
One finds that the compensating transformation in \eqref{BM coset trf} is brought to the left of $\cV$ after conjugation by $\Gamma$ and therefore it depends on $w$ directly, rather than through $\upgamma(w)$:
\begin{equation}
\cV = \Gamma V\big(\upgamma(w)\big) 
\ \to\ 
h(w)\, \cV\, g(w)\,,\qquad
h(w) = h^{-\mathsf{T}}(1/w) = \big(h^{-1}(w)\big)^\dagger\,.
\label{cV loop transformation}
\end{equation}
The compensating element $h(w)$ belongs again to a $K(E_9)$ subgroup of the loop group but now, crucially, it is defined by a field-independent involution that acts by inversion of $w$ rather than $\upgamma$.
Indeed this is the spectral parameter representation of the Hermitian conjugation \eqref{eq:hermconj} that we were seeking.
The extended coset representative also transforms under rigid $\Virm$ transformations such as \eqref{Vir minus element}:
\begin{equation}
\cV = \Gamma V\big(\upgamma(w)\big) 
\ \to\ 
\Gamma  V\big(\upgamma(w)\big) F
=
\Gamma F\, V\Big( \upgamma\big(f(w)\big) \Big)\,.
\label{cV Virm transformation}
\end{equation}
In this case no compensating transformations are required.
We conclude that we have found an extended set of fields parameterising the coset space
\begin{equation}
\frac{\widehat{E}_{8} \rtimes \Virm }{K(E_9)}
\end{equation}
where $K(E_9)$ is defined in terms of the field independent Hermitian conjugation \eqref{eq:hermconj}.
This allows us to define the coset representative $\cV$ in arbitrary representations of $\widehat{E}_{8}$.
At this point we remind the reader that in the spectral parameter representation $\dK$ is represented trivially.
The field associated to the central charge is $\sigma$ and we will include it in $\cV$ later when we work in a faithful representation of the algebra.

We can now introduce the twisted self-duality constraint equivalent to \eqref{BMsys} and \eqref{BMgamma sys}.
We introduce the Hermitian and anti-Hermitian parts of the Maurer--Cartan form
\begin{equation}
\label{eq:PQungauged}
(\dd \cV \cV^{-1})(w) = P(w) + Q(w)\,,\qquad 
P^\T(1/w) = P(w)\,,\quad 
Q^\T(1/w) =-Q(w)\,.
\end{equation}
Notice that while $\dd \cV \cV^{-1} \in \hat{\mf{e}}_{8}\oleft\virm$, $P$ and $Q$ take values also along the positive part of $\vir$.
The twisted self-duality constraint takes the form%
\footnote{We stress that we have so far defined $P(w)$ in the spectral parameter representation so that $P(w)|_\dK = 0$ by construction.
The twisted self-duality constraint introduced here must be modified when written in a faithful representation of $\hat{\mf{e}}_{8}\oleft\vir$ to take into account the central charge sector. This will be done in Section~\ref{sec:extE9exft}.}
\begin{equation}
\star\!P(w) = \cS_{1}\big(P(w)\big) = w P(w)\,.
\label{twsd ungauged, P}
\end{equation}
Under the symmetries of the extended coset space, $P(w)$ only transforms by conjugation with the local $K(E_9)$ transformation defined in \eqref{cV loop transformation}.
The shift operator does not commute with $K(E_9)$, but one can see that the commutator is again proportional to \eqref{twsd ungauged, P} and hence the twisted self-duality constraint is invariant.
In Lorentzian signature $\star^2P(w)=P(w)$.
The right-hand side of \eqref{twsd ungauged, P} does not apparently square to $P$, but taking into account \eqref{eq:shiftophermconj} and Hermiticity of $P$ it can be easily shown to define a $\mathds{Z}_2$ action.
On the other hand, \eqref{twsd ungauged, P} also \textit{implies} a cascade of duality relations
\begin{equation}
\star^{|m|}\!P(w) = \cS_{m}\big(P(w)\big)\,,\qquad m\in\mathds{Z}\,.
\label{cascade ungauged, P}
\end{equation}

We now show that \eqref{twsd ungauged, P} is equivalent to the linear system in the triangular $K(E_9)$ gauge of \eqref{Vhat regularity} and \eqref{Vhat Y expansion}.
Let us first focus on the loop components of the extended Maurer--Cartan form.
We define the components of $P(w)$:
\begin{equation}
P(w) = \sum_{m\in\mathds{Z}} P_A^m T^A_m + \sum_{m\in\mathds{Z}} P_m L_m\,,
\end{equation}
where $P^0_A T^A_0 = \mathring{P}$ is the $\mf{e}_{8}$ coset element and does not depend on $w$.
Equation \eqref{cascade ungauged, P} implies
\begin{equation}
P_A^m = \star^{|m|} P_A^0\,,
\end{equation}
which in turn gives
\begin{equation}
\big(\dd\cV\cV^{-1}\big)(w)\Big|_{\text{loop}} 
\ =\ \Gamma \,\big(\dd\hat{V}\hat{V}^{-1}\big)(w)\, \Gamma^{-1} 
\ =\ \mathring{Q} +\frac{1+w^{-2}}{1-w^{-2}}\mathring{P} +\frac{2w^{-1}}{1-w^{-2}}\star\!\mathring{P}
\end{equation}
where it is understood that the denominators are expanded in a geometric series for $|w|>1$.
Because $\mathring{P}$ and $\mathring{Q}$ are $w$ independent, we see that by conjugating this expression with $\Gamma^{-1}$, \eqref{BMsys} is reproduced in a geometric expansion for $|\upgamma(w)|>1$.
We now only have to show that the $\vir$ components of \eqref{twsd ungauged, P} restrict the a priori arbitrary $\upgamma(w)$ to solve \eqref{BMgamma sys}.
Following the same steps as above we find for the $\vir$ components of the Maurer--Cartan form
\begin{align}
\label{dGamma as Vir series}
\dd\Gamma\Gamma^{-1} 
&=
-\rho^{-1}\dd\rho \Big(L_0 + 2\sum_{n=1}^\infty L_{-2n} \Big)
-\rho^{-1}\star\!\dd\rho \Big( 2\sum_{n=1}^\infty L_{-2n+1} \Big) \\[1ex]
&= 
\rho^{-1} \dd\rho \ \ \frac{1+w^{-2}}{1-w^{-2}} w \frac{\partial}{\partial w}
+ \rho^{-1} \star\!\dd\rho \ \ \frac{2w^{-1}}{1-w^{-2}} w  \frac{\partial}{\partial w}\ .
\end{align}
Observing that $\dd\big(\Gamma^{-1}w\big) = \dd\upgamma(w) = -\Gamma^{-1} \dd\Gamma \Gamma^{-1} w$, we see that it is sufficient to apply the expression above to $w$, multiply by $\Gamma^{-1}$ from the left and substitute \eqref{tilde rho duality ungauged} to reproduce \eqref{BMgamma sys}, which concludes our proof.
This relation can also be reinterpeted as defining duality equations for all the $\phi_m$ fields.
Expanding $\dd\Gamma\Gamma^{-1}$ in the first few orders and using twisted self-duality we find
\begin{equation}
\dd\phi_1 = 2\star\!\dd\rho\,,\qquad
\dd\phi_2 = \dd(\rho^2)\,,\qquad
\dd\phi_3 -\phi_1\dd\phi_2 = 2\rho^2 \star\!\dd\rho\,,\qquad\ldots
\label{vir dualities}
\end{equation}
We see that $\phi_1$ is the same as $\tilde\rho$ up to a factor of 2 and in fact, contrary to the loop case where all dual potentials are non-locally related to each other, the whole tower of duality relations can be integrated to algebraic expressions in $\rho$ and $\tilde\rho$ (or $\phi_1$) exclusively, so that the only truly non-local relation is \eqref{tilde rho duality ungauged}:
\begin{equation}
\phi_1 = 2\tilde\rho -2s_0\,,\qquad
\phi_2 = \rho^2 -2s_1\,,\qquad
\phi_3 = 2\rho^2(\tilde\rho-s_0)-2s_2\,,\quad\ldots
\end{equation}
The integration constants $s_i$ are those appearing in \eqref{s expansion} and we have set $s_{-1}=1/2$ using the rigid $L_0$ symmetry included in \eqref{cV Virm transformation}, as was already implied by the parameterisation \eqref{upgamma as series}.
We stress that there is no solution of twisted self-duality such that the $\virm$ scalar fields vanish unless $\dd\rho=0$.

We conclude this section by rewriting \eqref{twsd ungauged, P} in terms of the $K(E_9)$ invariant and $E_{8}\rtimes\Virm$ covariant current
\begin{equation}
J = 2 \cV^{-1} P \cV\,.
\label{ungauged J}
\end{equation}
Dressing \eqref{twsd ungauged, P} with $\cV$, we find
\begin{equation}
\star\!J(w) = \upgamma(w)J(w) = \cS^\upgamma_1(J(w))\,.
\label{twsd ungauged J}
\end{equation}
The operators $\cS^\upgamma_m$ act as multiplication by $\upgamma(w)^m$. Their definition in a lowest weight representation will be presented in Section~\ref{sec:extE9exft} and Appendix~\ref{app:shift}.

\section{\texorpdfstring{Gauge structure of $E_9$ exceptional field theory}{Gauge structure of E9 exceptional field theory}}
\label{sec:E9gauge}

In this section, we exhibit the $E_9$ ExFT gauge structures, beginning with a review of $E_9$ generalised diffeomorphisms and defining a Dorfman structure for $E_9$.
In Section~\ref{sec:gauges} we then introduce a minimal set of vector fields that are needed to covariantise external derivatives, and we define their field strengths.
We also define a natural set of transformation properties for these fields in terms of the Dorfman product.
However, some modifications to these structures will be required in order to construct the dynamical theory.
As a necessary step to introduce the Virasoro-extended formulation of the theory, in Section~\ref{sec:extgauge} we extend the definition of generalised diffeomorphisms so that they gauge the larger algebra $\hevirm$. The vector field contents and their transformations are extended as well. 
In Section~\ref{sec:E9EFT}, when considering the minimal formulation of $E_9$ exceptional field theory, we will amend the (non-extended) gauge transformations of the vector fields and of their field strengths by a different choice of trivial parameters compared to this section, and also introduce extra terms dependent on the external metric $\tilde g_{\mu\nu}$.

\subsection{Generalised diffeomorphisms and Dorfman structure}
\label{sec:gendiffeo}

The internal space of $E_9$ exceptional field theory has coordinates $Y^M$ transforming according to the $R(\Lambda_0)_{-1}$ and the generalised Lie derivative was introduced in~\cite{Bossard:2017aae}. It has a gauge parameter $\Lambda^M$ that lies in the $R(\Lambda_0)_{-1}$ representation of $E_9$ and an ancillary parameter $\Sigma^M{}_N$ lying in the representation $R(\Lambda_0)_0\otimes \overline{R(\Lambda_0)_{-1}}$. Due to the Fock space structure of the $R(\Lambda_0)_h$ representation introduced in~\eqref{eq:Fock} we write the parameter $\Lambda^M$ as a ket vector $|\Lambda\rangle$ and the ancillary parameter $\Sigma^M{}_N$ as an operator $\Sigma$ in this space that also changes the $\dL$ weight. The two notations are related by expanding out the ket vector in a basis and writing $|\Lambda\rangle = \Lambda^M |e_M\rangle$. The coordinates $Y^M$ arise in this way and the derivatives $\langle \partial| = \langle e^M| \partial_M$ transform in the conjugate representation $\overline{R(\Lambda_0)_{-1}}$.

The generalised Lie derivative acting on a vector $|V\rangle$ belonging to $R(\Lambda_0)_{-1}$ is given by
\begin{align}
\label{eq:GL}
\mathcal{L}_{(\Lambda,\Sigma)} |V\rangle = \langle \partial_V | \Lambda \rangle |V\rangle -\eta_{\alpha\beta} \langle \partial_\Lambda | T^\alpha | \Lambda\rangle T^\beta |V\rangle -\langle \partial_\Lambda | \Lambda\rangle |V\rangle - \eta_{-1\,\alpha\beta} \Tr(T^\alpha \Sigma)T^\beta |V\rangle\,.
\end{align}
Here, $\langle \partial|$ is the derivative with respect to the internal coordinates written as a bra vector and the subscript indicates which object the derivative is acting on. 
Thus, the first term in~\eqref{eq:GL} is the transport term, the second the rotation term and the third a weight term. The final term realises the ancillary transformations. In index notation,~\eqref{eq:GL} reads 
\be \cL_{(\Lambda,\Sigma)} V^M = \Lambda^N \partial_N V^M - \partial_N \Lambda^M V^N + Y^{MN}{}_{PQ} \partial_N \Lambda^P V^Q - \eta_{-1\, \alpha\beta} T^{\alpha P}{}_Q \Sigma^Q{}_P \, T^{\beta M}{}_N V^N \; , \ee
where the $Y$ tensor defining the section constraint in \eqref{eq:SCintro} is defined as
\be Y^{MN}{}_{PQ} = \delta^M_N \delta^N_Q - \delta^M_Q \delta^N_P - \eta_{\alpha\beta} T^{\alpha M}{}_Q T^{\beta N}{}_P \; . \label{eq:Y-def}\ee %
More generally, the action of the generalised Lie derivative on a generic field $\Phi$, with respect to the pair of parameters $\mathbbm{\Lambda}=(\ket{\Lambda},\Sigma)$, takes the form 
\begin{equation}
\mathcal {L}_{\mathbbm{\Lambda}}\Phi= \langle \partial_\Phi | \Lambda \rangle \Phi+\sbm{\Lambda}_\alpha\,\delta^\alpha \Phi\,,\label{eq:GL2}%
\end{equation}
where we defined the linear combination $\sbm{\,\cdot\,}_\alpha$ for any pair of parameters
\begin{equation}
\sbm\Lambda_\alpha\equiv \eta_{\alpha\beta} \langle \partial_\Lambda | T^\beta | \Lambda\rangle+\eta_{-1\,\alpha\beta} \Tr(T^\beta \Sigma)\,,\label{eq:balpha}
\end{equation}
and where $\delta^\alpha$ denotes an infinitesimal rigid $\mathfrak{e}_9\oleft L_{-1}$ variation. We insist that this includes the variation with respect to $\dL$ and not $L_0$. For a vector $\ket{V}$ in the $R(\Lambda_0)_{-1}$ representation, as in \eqref{eq:GL}, the weight term that appears then follows from the infinitesimal scaling associated with the action $\dL\ket{V}=(L_0-1)\ket{V}$. 

Covariance of the generalised Lie derivative under rigid $\mathfrak{e}_9\oleft L_{-1}$ transformations requires the combination $\sbm\Lambda_\alpha$ to be a projection onto the adjoint representation. This implies that the parameters $\ket{\Lambda}$ and $\Sigma$ must transform separately as
\begin{align}
X_\alpha\, \delta^\alpha \ket{\Lambda}&=-X_\alpha T^\alpha\ket{\Lambda}+X_0\ket{\Lambda},\nonumber\\
X_\alpha\, \delta^\alpha \Sigma&= -X_\alpha[T^\alpha, \Sigma]-X_0\Sigma+X_{-1}\ket{\Lambda}\bra{\partial_\Lambda}\,,\label{eq:tpara}
\end{align}
where here by definition 
\begin{equation}
X_\alpha T^\alpha=X_0L_0+X_{\dK} \dK+\sum\limits_{n\in\mathds{Z}} X^A_n T^n_A+X_{-1} L_{-1}\,.
\end{equation}
The last term in \eqref{eq:tpara} indicates that the parameters $\ket{\Lambda}$ and $\Sigma$ transform together in an indecomposable representation under $L_{-1}$. In the following, the `doubled notation' $\mathbbm{\Lambda}=(\ket{\Lambda},\Sigma)$ will be consistently used for other indecomposable pairs of fields.

As is usual in ExFT, the algebra of generalised Lie derivatives closes only when an appropriate (strong) section constraint is fulfilled. The section constraint \eqref{eq:SCintro} is written explicitly in terms of $\mf{e}_9$ generators~\cite{Bossard:2017aae}
\begin{align}
\label{eq:SC}
\eta_{\alpha\beta} \langle \partial_1 | T^\alpha \otimes \langle \partial_2| T^\beta + \langle\partial_1| \otimes \langle \partial_2 | -  \langle\partial_2| \otimes \langle \partial_1 | =0\,,
\end{align}
where $\langle\partial_1|$ and $\langle \partial_2|$ denote two partial derivatives acting on any objects in the theory. It moreover implies that
\begin{align}
\label{eq:SC2}
\eta_{-k\,\alpha\beta} \langle \partial_1| T^\alpha \otimes \langle \partial_2| T^\beta &= 0\,, \;\;\text{for all} \,\,k\in \mathds{N}^+ \,,\nn\\
\eta_{+1\,\alpha\beta} \left(\langle \partial_1| T^\alpha \otimes \langle \partial_2| T^\beta +\langle \partial_2| T^\alpha \otimes \langle \partial_1| T^\beta\right)  &= 0\,.
\end{align}
The ancillary parameter $\Sigma$ is also section constrained~\cite{Bossard:2017aae}, in the same way as happens for $E_8$~\cite{Hohm:2014fxa,Cederwall:2015ica}. This can be expressed by writing it out in bases as $\Sigma= \Sigma^M{}_N |e_M\rangle \langle e^N| = |\Sigma\rangle \langle \pi_\Sigma|$, where we have introduced a suggestive notation involving the bra vector $\langle \pi_\Sigma|$.\footnote{$\Sigma$ is not factorised, but the notation $|\Sigma\rangle \langle \pi_\Sigma|$ helps to describe the section constraint for $\Sigma$.} The constrained nature of $\Sigma$ corresponds to replacing either $\langle \partial_1|$ or $\langle \partial_2|$ in~\eqref{eq:SC} by it. The constraint on $\Sigma$ also ensures that the trace operation in~\eqref{eq:GL} is well-defined even though we are acting on an infinite-dimensional space. As an operator $\Sigma$ has finite rank. 

Any solution of the section constraint can be brought to the following form by an $E_9$ transformation:
\begin{equation}\label{eq:secsol}
\bra\partial 
=
\bra0\,\partial_\varphi + \bra0 T^A_{+1} \partial_A \ ,
\end{equation}
where $\partial_A$ must satisfy the section constraint of $E_8$ exceptional field theory~\cite{Hohm:2014fxa}.

The generalised Lie derivative closes according to
\begin{align}
\label{eq:dfffclos}
\left[ \mathcal{L}_{\mathbbm{\Lambda}_1},\mathcal{L}_{\mathbbm{\Lambda}_2} \right] =  \mathcal{L}_{[\mathbbm{\Lambda}_1,\mathbbm{\Lambda}_2]_E}\,,
\end{align}
with $\mathbbm{\Lambda}_1=(\ket{\Lambda_1},\Sigma_1)$ and $\mathbbm{\Lambda}_2=(\ket{\Lambda_2},\Sigma_2)$, and where the exceptional Lie bracket is given by
\begin{align}
[\mathbbm{\Lambda}_1,\mathbbm{\Lambda}_2]_E=(\ket{\Lambda_{12}},\Sigma_{12})
\end{align}
with~\cite{Bossard:2017aae}
\begin{align}
\label{eq:Ebracket}
| \Lambda_{12} \rangle &= \frac12 \left( \mathcal{L}_{(\Lambda_1,0)} |\Lambda_2\rangle-\mathcal{L}_{(\Lambda_2,0)} |\Lambda_1\rangle \right)\,,\nn\\
\Sigma_{12} &= \mathcal{L}_{(\Lambda_1,0)} \Sigma_2 - \frac12 \eta_{-1\, \alpha \beta} \Tr(T^\alpha \Sigma_1)T^\beta \Sigma_2\\
&\hspace{4.5mm}  -\frac14 \eta_{1\,\alpha\beta}\Bigl(  \langle \partial_{\Lambda_2}| T^\alpha |\Lambda_2\rangle  T^\beta |\Lambda_1\rangle - \langle \partial_{\Lambda_2}|  T^\alpha |\Lambda_1\rangle  T^\beta |\Lambda_2\rangle \Bigr) \langle \partial_{\Lambda_2} | - (1\leftrightarrow 2)\,,\nn
\end{align}

The action of the generalised Lie derivative on a pair of parameters is given by
\begin{align}
\label{eq:GLpair}
\mathcal{L}_{\mathbbm{\Lambda}_1} \mathbbm{\Lambda}_2 &= \Big( \mathcal{L}_{\mathbbm{\Lambda}_1} |\Lambda_2\rangle\; ,\;  
\mathcal{L}_{(\Lambda_1,0)} \Sigma_2
-\eta_{-1\,\alpha\beta} \Tr(T^\alpha \Sigma_1) \left( T^\beta \Sigma_2 -\Sigma_2  T^\beta \right) - \Tr (\Sigma_1) |\Lambda_2\rangle \langle \partial_{\Lambda_2}| \Big)\nn\\
&= \Big( \mathcal{L}_{\mathbbm{\Lambda}_1} |\Lambda_2\rangle \; ,\;  \langle \partial_{\Sigma_2} | \Lambda_1 \rangle \Sigma_2 
- \eta_{\alpha\beta}  \langle \partial_{\Lambda_1} | T^\alpha | \Lambda_1\rangle T^\beta \Sigma_2 + \Sigma_2 |\Lambda_1 \rangle \langle \partial_{\Lambda_1}|\nn\\
&\hspace {60mm} -\eta_{-1\,\alpha\beta} \Tr(T^\alpha \Sigma_1)  T^\beta \Sigma_2  - \Tr (\Sigma_1) |\Lambda_2\rangle \langle \partial_{\Lambda_2}| \Big)\,,
\end{align}
where the term $\Tr (\Sigma_1) |\Lambda_2\rangle \langle \partial_{\Lambda_2}|$ is due to the indecomposable structure of the pair $\mathbbm{\Lambda}_2$ under $L_{-1}$ transformations. In the second step of~\eqref{eq:GLpair} we have used the section constraint to simplify a few terms.

\medskip

In order to bring out the covariance properties of the gauge fields and field strengths in the next section, it is convenient to introduce a generalised Dorfman structure~\cite{Hohm:2017wtr}. This is defined via the following (non-commutative and non-associative) product which involves exclusively pairs of parameters 
\begin{align} 
\label{eq:DorfmanProduct} 
\mathbbm{\Lambda}_1\! \circ\! \mathbbm{\Lambda}_2 = 
\bigg( \mathcal{L}_{\mathbbm{\Lambda}_1} \Lambda_2 \ , \ \ &
\mathcal{L}_{\mathbbm{\Lambda}_1} \Sigma_2{}
+ \eta_{1\,\alpha\beta} \langle \partial_{\Lambda_1} | 
  T^\alpha | \Lambda_1\rangle T^\beta |\Lambda_2\rangle \langle \partial_{\Lambda_1}|
\CR&\ 
+ \eta_{\alpha\beta} \Tr( T^\alpha \Sigma_1) T^\beta |\Lambda_2\rangle \langle \partial_{\Sigma_1}| 
- |\Lambda_2\rangle \langle \partial_{\Sigma_1} | \Sigma_1
\bigg)\,.
\end{align}
A key property of the Dorfman product, which is not satisfied by the generalised Lie derivative, is that it obeys the Leibniz property
\begin{align}
\mathbbm{\Lambda}_1\circ\big(\mathbbm{\Lambda}_2\circ\mathbbm{\Lambda}_3\big)= \big(\mathbbm{\Lambda}_1\circ \mathbbm{\Lambda}_2\big)\circ\mathbbm{\Lambda}_3+\mathbbm{\Lambda}_2\circ\big(\mathbbm{\Lambda}_1\circ\mathbbm{\Lambda}_3\big) \; .\label{eq:Leibniz}
\end{align} 
This relation can be proved by showing that the Dorfman product closes according to the antisymmetric Dorfman bracket  $\left[ \mathbbm{\Lambda}_1, \mathbbm{\Lambda}_2 \right]_D \equiv  \frac12\left(\mathbbm{\Lambda}_1\circ \mathbbm{\Lambda}_2 - \mathbbm{\Lambda}_2\circ \mathbbm{\Lambda}_1\right) $, and that the symmetric bracket $\{ \mathbbm{\Lambda}_1, \mathbbm{\Lambda}_2 \} \equiv  \frac12\left(\mathbbm{\Lambda}_1\circ \mathbbm{\Lambda}_2 + \mathbbm{\Lambda}_2\circ \mathbbm{\Lambda}_1\right) $ is trivial (with respect to $\circ$ itself). These properties respectively correspond to the projections of~\eqref{eq:Leibniz} onto its antisymmetric and symmetric parts in $\mathbbm{\Lambda}_1$ and $\mathbbm{\Lambda}_2$, and are discussed in more detail in Appendix~\ref{app:triv}. Let us simply note here that, on the first entry of a pair of parameters, they imply that under the generalised Lie derivative
\begin{align}
\label{eq:Dorf1}
\mathcal{L}_{[\mathbbm{\Lambda}_1,\mathbbm{\Lambda}_2]_D} = \mathcal{L}_{\left[ \mathbbm{\Lambda}_1, \mathbbm{\Lambda}_2\right]_E}\,,\;\;\;\;\;\;\;\;\mathcal{L}_{\{\mathbbm{\Lambda}_1,\mathbbm{\Lambda}_2\}} = 0\,,
\end{align}
where the E-bracket was defined in \eqref{eq:Ebracket}.

\subsection{Covariant derivatives and field strengths}
\label{sec:gauges}
Using the Dorfman structure we can conveniently deduce the form and properties of the covariant field strengths associated to a pair of `vector fields' 
\begin{align}
\mathbbm{A}_\mu= (A_\mu^M, B_\mu{}^M{}_N)=(\ket{A_\mu}, B_\mu)\,.
\end{align}
The first component $\ket{A_\mu}$ is the usual vector field in two external dimensions, and it gauges the $\ket{\Lambda}$-diffeomorphisms. It is thus valued in the basic representation of $E_9$ and it carries weight $-1$ under $\dL$. Its second component partner $B_\mu$ is associated to $\Sigma$-diffeomorphisms and is therefore constrained on its lower index and carries weight $+1$. This means the two components of $\mathbbm{A}_\mu$ behave as their respective parameters under an infinitesimal $\mathfrak{e}_9\oleft L_{-1}$ variation~\eqref{eq:tpara} and, in particular, also transform in an indecomposable representation under $L_{-1}$.

We define a covariant derivative as
\begin{align}
\label{eq:covD}
\mathcal{D}_\mu = \partial_\mu - \mathcal{L}_{\mathbbm{A}_\mu}\,.
\end{align}
As usual, its covariance under generalised diffeomorphisms is ensured if, acting on any vector $V^M$, one has
\begin{align}
\label{eq:deltacond}
\delta_{\mathbbm{\Lambda}} \mathcal{D}_\mu V^M = \mathcal{L}_{\mathbbm{\Lambda}} \mathcal{D}_\mu V^M\,,
\end{align}
where $\mathbbm{\Lambda}=(\ket{\Lambda},\Sigma)$. Using \eqref{eq:dfffclos}, we find that this can be achieved by the transformation
\begin{align}
\label{eq:deltaA}
\delta_{\mathbbm{\Lambda}} \mathbbm{A}_\mu &=  \partial_\mu \mathbbm{\Lambda} - [\mathbbm{A}_\mu, \mathbbm{\Lambda}]_E &&\hspace{-35mm}+\text{trivial parameters} \nonumber\\
&= \partial_\mu \mathbbm{\Lambda} - \mathbbm{A}_\mu \circ \mathbbm{\Lambda} &&\hspace{-35mm}+\text{trivial parameters} \nonumber\\
& = \partial_\mu \mathbbm{\Lambda} +  \mathbbm{\Lambda} \circ \mathbbm{A}_\mu &&\hspace{-35mm}+ \text{trivial parameters} \,,
\end{align}
in terms of the E-bracket \eqref{eq:Ebracket} and the Dorfman product~\eqref{eq:DorfmanProduct}. We have given three equivalent ways of writing the transformation and they differ by trivial parameters according to~\eqref{eq:Dorf1}. This is consistent since the transformation of $\mathbbm{A}_\mu$ is only defined up to trivial parameters, as $\mathbbm{A}_\mu$ only appears as the parameter of a generalised Lie derivative in ~\eqref{eq:deltacond}. 

In exceptional field theory, one typically chooses to work with the second transformation in~\eqref{eq:deltaA} (without trivial parameters), which can be written in more compact form as
\begin{align}
\delta_{\mathbbm{\Lambda}} \mathbbm{A}_\mu = \mathbbm{D}_\mu \mathbbm{\Lambda}\,,\label{eq:DorfdeltaA}
\end{align}
in terms of the Dorfman covariant derivative
\begin{align}
\mathbbm{D}_\mu = \partial_\mu - \mathbbm{A}_\mu \circ\,.\label{eq:Dorfcovdev}
\end{align}
However, for the purpose of Section~\ref{sec:E9EFT}, we will consider in this paper the third version of the transformation in~\eqref{eq:deltaA} (without trivial parameters). For the components of $\mathbbm{A}_{\mu}$ we then get explicitly
\begin{align}
\delta_{\mathbbm{\Lambda}} |A_\mu\rangle = \partial_\mu |\Lambda \rangle+\mathcal{L}_{\mathbbm{\Lambda}}\ket{A_\mu}\,, \label{GaugeAused}
\end{align}
and
\begin{align}
\delta_{\mathbbm{\Lambda}} B_\mu{} = &\,\partial_\mu \Sigma+\mathcal{L}_{\mathbbm{\Lambda}}B_\mu \nn\\
&+\eta_{1\,\alpha\beta} \langle \partial_\Lambda | T^\alpha |\Lambda \rangle T^\beta |A_\mu \rangle \langle \partial_\Lambda | +\eta_{\alpha\beta} \Tr(T^\alpha \Sigma) T^\beta | A_\mu\rangle \langle \partial_\Sigma| - |A_\mu\rangle \langle \partial_\Sigma| \Sigma\,, \label{GaugeBused}
\end{align}
that we have written without displaying internal space indices. We also recall that the second term contains a term of the form $\Tr (\Sigma) |A_\mu\rangle \langle \partial_{A}|$ which is due to the indecomposable structure of the pair of gauge fields $\mathbbm{A_\mu}$.

Using the covariant derivative~\eqref{eq:covD} or its Dorfman version~\eqref{eq:Dorfcovdev}, we can introduce a pair of field strengths 
\begin{align}
    \mathbbm{F}_{\mu\nu} = (\mathcal{F}_{\mu\nu}^M, \mathcal{G}_{\mu\nu}^{\hspace{2.5mm} M\hspace{-0.5mm}}{}_N) =(\ket{\mathcal{F}_{\mu\nu}}, \mathcal{G}_{\mu\nu})
\end{align}
by
\begin{align}
\left[ \mathcal{D}_\mu, \mathcal{D}_\nu\right] = -\mathcal{L}_{\mathbbm{F}_{\mu\nu}}\,,\;\;\;\;\;\;\;\;\;\;\left[ \mathbbm{D}_\mu, \mathbbm{D}_\nu\right] = -\mathbbm{F}_{\mu\nu}\,\circ\,,\label{eq:DD}
\end{align}
where the former expression is implied by the latter.
Note that these relations only define the field strengths up to trivial parameters. Under rigid $\mathfrak{e}_9\oleft L_{-1}$ transformations the pair of field strengths behaves as the pairs of gauge parameters and gauge fields, namely is transforms in an indecomposable representation like in \eqref{eq:tpara}.
As is customary in exceptional field theory, covariance of the field strengths under generalised diffeomorphisms requires us to introduce higher (external) forms which enter in their expressions as trivial parameters. Once again, for the purpose of Section~\ref{sec:E9EFT}, we can choose to write
\begin{equation}
\mathbbm{F}_{\mu\nu}=2\,\partial_{[\mu}\mathbbm{A}_{\nu]}-[\mathbbm{A}_{\mu}, \mathbbm{A}_{\nu}]_E+\varpi \mathbf{C}_{\mu\nu}\,,\label{eq:fieldstrengths}
\end{equation}
where $\mathbf{C}_{\mu\nu}$ denotes the set of  two-forms defined in \eqref{Csextuplet}, while $\varpi \mathbf{C}_{\mu\nu}$ is the Dorfman doublet of associated trivial parameter using \eqref{trivialvarpi}. The parts of the field strength components that are independent of the two-forms will be denoted by $\ket{F_{\mu\nu}}$ and $G_{\mu\nu}$. In form notation, their expressions follow from \eqref{eq:Ebracket} and thus read\footnote{We use  the notation $|A^\prime\rangle$ to distinguish the vector field that is not derived by the bra $\langle \partial_A|$, so that $\langle \partial_A | \otimes |A\rangle \otimes |A^\prime\rangle$ means in components $\partial_M A^N\; A^P$.}
\begin{align}
\ket{F}=&\,\dd | A\rangle - \tfrac12 \langle \partial_{A} | A^\prime \rangle | A \rangle + \tfrac12 \eta_{\alpha\beta} \langle \partial_{A} | T^\alpha |A \rangle T^\beta | A^\prime \rangle  + \tfrac12 \langle \partial_{A} | A \rangle | A^\prime \rangle\,,  \nonumber \\[1mm]
G =&\,\dd B - \langle \partial_B | A \rangle B + \eta_{\alpha\beta} \langle \partial_A | T^\alpha | A \rangle T^\beta B+B |A \rangle \langle \partial_A |  + \frac12 \eta_{-1\,\alpha\beta} \Tr(T^\alpha B ) T^\beta B\nn\\
&- \tfrac14 \eta_{1\,\alpha\beta} \langle \partial_{A} | T^\alpha |A  \rangle T^\beta | A^\prime \rangle \langle \partial_{A}|  -\tfrac14 \eta_{1\,\alpha\beta} \langle \partial_{A} | T^\alpha |A ^\prime \rangle T^\beta | A \rangle \langle \partial_{A}|   \,,\label{eq:barefs}
\end{align}
The explicit dependence on  two-forms in \eqref{eq:fieldstrengths} as well as their transformations under generalised diffeomorphisms are given in detail in Appendix~\ref{app:triv}, and are tuned to cancel the non-covariant variations of~\eqref{eq:barefs}. This ultimately ensures that the field strengths transform covariantly as
\begin{equation}
\delta_\mathbbm{\Lambda} \mathbbm{F}=\mathbbm{\Lambda}\circ \mathbbm{F}\,.\label{eq:fsvar}
\end{equation}
We stress that these are not the final transformation properties for the field strengths of $E_9$ exceptional field theory, for several reasons.
In the extended formulation of Section~\ref{sec:extE9exft}, an extended $\circ$ product is necessary to capture the gauge transformations of a larger set of vector fields. 
This is explained in the next subsection.
In the minimal formulation of Section~\ref{sec:E9EFT}, the transformation properties of both the $B$ field and its field strength $\cG$ are modified by extra terms involving the external unimodular metric $\tilde g_{\mu\nu}$. 
Furthermore, both formulations of $E_9$ exceptional field theory imply an Euler--Lagrange equation for $|\cF\rangle$ that appears with a projection. This projected equation is gauge-invariant by construction. If one wants to extend the duality equation to be unprojected and gauge-invariant, the gauge transformation of $\mathbbm{F}$ must be modified by trivial parameters. This applies to both formulations, and we shall see explicitly in the minimal formulation in Section~\ref{sec:E9EFT} that external diffeormorphism invariance requires the inclusion of the unprojected duality equation.

We finish this section by presenting a set of Bianchi identities. Using \eqref{eq:DD} to formally evaluate the action of three antisymmetrised Dorfman covariant derivative on a generic pair $\mathbbm{V}=(\ket{V},V)$ leads to the relation $\mathbbm{D}(\mathbbm{F}\circ\mathbbm{V})=\mathbbm{F}\circ\mathbbm{D}\mathbbm{V}$, which means that 
\begin{equation}
(\mathbbm{D}\mathbbm{F})\,\circ\mathbbm{V}=0\,.\label{eq:Bianchi1}
\end{equation}
On the first component of  $\mathbbm{V}$ this relation reduces to $\mathcal{L}_{\mathbbm{D}\mathbbm{F}}\ket{V}=0$, and implies the following two identities
\begin{equation}
\langle \partial|\mathcal{D}\mathcal{F}\rangle=0\,,\;\;\;\;\;\;\;\;\;\sbm{\mathbbm{D}\mathbbm{F}}_\alpha=0\,,\label{eq:Bianchi2}
\end{equation} 
where the projection in the second equation was defined in \eqref{eq:balpha}.

\subsection{Extended gauge structure}\label{sec:extgauge}

The extension of the group theoretical structure discussed in Section~\ref{sec:extcoset} in the context of two-dimensional supergravity suggests the existence of an extended gauge structure for exceptional field theory. 
In this section, we show that one can indeed consistently enlarge the algebra that is gauged by generalised diffeomorphisms to $\mathfrak{\hat e}_8\oleft \mathfrak{vir}^{-}$.
Many of the identities and results of the previous section still hold in this extended setting and we will allow ourselves to be more schematic.

We start by considering an extension of the generalised Lie derivative \eqref{eq:GL2} in which the rotation term now involves an infinitesimal $\hevirm$ variation. The set of gauge parameters is thus also enlarged, and now consists of 
\begin{equation}
\mathbbm{\Lambda}=(\ket{\Lambda}\,, \Sigma^{\ord{k}})\,\qquad k\in \mathds{N}^+\,.
\end{equation}
On a vector $\ket{V}$ that lies in $R(\Lambda_0)_{-1}$, we have
\begin{align}
\mathcal{L}_{\mathbbm{\Lambda}} |V\rangle &= \langle \partial_V | \Lambda \rangle |V\rangle + \sbm\Lambda_\alpha  \delta^\alpha \ket{V}\nn\\
&=\langle  \partial_V | \Lambda \rangle |V\rangle -\eta_{\alpha\beta}\langle \partial_{\Lambda}|T^\alpha|\Lambda\rangle T^\beta\ket{V}-\langle\partial_\Lambda|\Lambda\rangle \ket{V} -\sum\limits_{k=1}^\infty\eta_{-k\,\alpha\beta}\Tr(T^\alpha \Sigma^{\ord{k}}) T^\beta | V\rangle\,,\label{eq:extGL}
\end{align}
where the parameter $\Sigma^{\ord{1}}$ corresponds to $\Sigma$ in the previous section, and where we have defined $\sbm\Lambda_\alpha$ as the natural extension of the projection \eqref{eq:balpha} onto $\mathfrak{\hat e}_8\oleft \mathfrak{vir}^-$
\begin{equation}
\sbm\Lambda_\alpha\equiv \eta_{\alpha\beta} \langle \partial_\Lambda | T^\beta | \Lambda\rangle+\sum\limits_{k=1}^\infty \eta_{-k\,\alpha\beta} \Tr(T^\beta \Sigma^{\ord{k}})\,.\label{eq:balphaex}
\end{equation}
We recall that the weight term in \eqref{eq:extGL} is due to the fact that $\delta^\alpha$ includes a variation with respect to $\dL$ and not $L_0$. 
Acting on a generic field, we use \eqref{eq:GL2} together with \eqref{eq:balphaex}.

In order for the projection $\sbm\Lambda_\alpha$ to transform in the adjoint, the set of gauge parameters must transform under rigid $\mathfrak{\hat e}_8\oleft \mathfrak{vir}^-$ variations as
\begin{align}
X_\alpha\, \delta^\alpha \ket{\Lambda}&=-X_\alpha T^\alpha\ket{\Lambda}+X_0\ket{\Lambda},\nonumber\\
X_\alpha\, \delta^\alpha \Sigma^{\ord{k}}&= -X_\alpha[T^\alpha, \Sigma^{\ord{k}}]+\!\sum\limits_{0\leq p<k} (2p-k) X_{-p}\Sigma^{(k-p)}+k\,X_{-k}\ket{\Lambda}\bra{\partial_\Lambda}\,,\label{eq:tparaex}
\end{align}
for all $k\in\mathds{N}^+$, where  
\begin{equation}
X_\alpha T^\alpha=\sum\limits_{p=0}^\infty X_{-p}L_{-p}+X_{\dK} \dK+\sum\limits_{n\in\mathds{Z}} X^A_n T^n_A\,.
\end{equation}
The parameters $ \Sigma^{\ord{k}}$ thus carry weight $k$ under $\dL$. Note that, similarly to the unextended case, the gauge parameters $\mathbbm{\Lambda}=(\ket{\Lambda}, \Sigma^{\ord{k}})$ transform in an indecomposable representation under $L_{-p}$ for all $p>0$. Let us also point out that a further extension of generalised diffeomorphisms to include gaugings of the full Virasoro algebra would not be compatible with the section constraints. The set of constraints~\eqref{eq:SC} and~\eqref{eq:SC2} are indeed invariant under $\mathfrak{vir}^-$ transformations, but not under any of the transformations generated $L_{k}$ with $k>0$, for instance an infinitesimal $L_1$ transformation applied to \eqref{eq:SC} maps $\eta_{\alpha\beta}$ to $\eta_{1\,\alpha\beta}$, but the symmetrisation in \eqref{eq:SC2} is not reproduced.

The generalised Lie derivative~\eqref{eq:extGL} still closes as in~\eqref{eq:dfffclos}, but according to an extended E-bracket $[\mathbbm{\Lambda}_1,\mathbbm{\Lambda}_2]_E=(\ket{\Lambda_{12}}, \Sigma_{12}^{\ord{k}})$ which reads explicitly
\begin{align}
\label{eq:Ebracketex}
| \Lambda_{12} \rangle &= \frac12 \left( \mathcal{L}_{(\Lambda_1,0)} |\Lambda_2\rangle-\mathcal{L}_{(\Lambda_2,0)} |\Lambda_1\rangle \right)\,,\nn\\
\Sigma_{12} ^{\ord{k}}&= \mathcal{L}_{(\Lambda_1,0)} \Sigma^{\ord{k}}_2 - \frac12 \sum\limits_{p=1}^\infty \eta_{-p\, \alpha \beta} \Tr(T^\alpha \Sigma^{\ord{p}}_1)T^\beta \Sigma^{\ord{k}}_2-\frac12 \sum\limits_{0<p<k}(2p-k)\Tr(\Sigma_1^{\ord{p}})\Sigma_2^{\ord{k-p}}\\
&\hspace{4.5mm}  
-\frac14 \,\delta_1^k\,\eta_{1\,\alpha\beta}\Bigl(  \langle \partial_{\Lambda_2}| T^\alpha |\Lambda_2\rangle \, T^\beta |\Lambda_1\rangle - \langle \partial_{\Lambda_2}|  T^\alpha |\Lambda_1\rangle \, T^\beta |\Lambda_2\rangle \Bigr) \langle \partial_{\Lambda_2} |
- (1\leftrightarrow 2)\,.\nn
\end{align}
The proof of closure only relies on the use of the section constraints~\eqref{eq:SC} and~\eqref{eq:SC2} which were already necessary for consistency of the unextended gauge algebra, and is presented in Appendix~\ref{app:triv}. Note that $\Sigma_{12}^{\ord{k}}$ is still correctly on section.

The gauge parameters $\Sigma^\ord{k}$ are highly degenerate.
The only new gauge transformations introduced by $\Sigma^\ord{k}$ with $k\ge2$ are those generated by  $L_{-k}$ with $k>1$, associated to $\Tr( \Sigma^{\ord{k}})$ and acting as shifts on the dual potentials $\phi_k$ introduced in Section~\ref{sec:extcoset}.
The $\he8$ transformations generated by $\Sigma^\ord{k}$ can all be reabsorbed into $\Sigma^\ord{1}$ up to trivial parameters.
To see this, we take the $\Sigma$ components of \eqref{eq:balphaex} and apply \eqref{eta and shift ops} to write
\begin{equation}\label{eq:sigmaetashift}
\sum\limits_{k=1}^\infty \eta_{-k\,\alpha\beta} \Tr(T^\alpha  \Sigma^{\ord{k}}) T^\beta 
=
\sum\limits_{k=1}^\infty \left(
   \eta_{\alpha\beta} \Tr(\Sigma^\ord{k} T^\alpha) \cS_{-k}(T^\beta)
 - \Tr(\Sigma^\ord{k} L_{-k}) \dK
 \right)\,.
\end{equation}
By rigid $E_9$ covariance we can assume that the $\Sigma^\ord{k}$ solve the section constraint as in \eqref{eq:secsol}.
Substituting into the above expression and focussing on the gauging of loop generators, one finds that $\Sigma^\ord{1}$ gauges some $\mf e_8$ generators $T^A_0$, a larger set of $T^A_{-1}$ generators, and all $T^A_{-n}$ for $n\ge2$.
Higher $\Sigma^\ord{k}$ then gauge the same algebra shifted by $1-k$ along the negative loop levels.
Hence, each $\Sigma^\ord{k}$ for $k\ge2$ gauges a loop subalgebra of the one gauged by $\Sigma^\ord{1}$.
The only new generators gauged by $\Sigma^\ord{k}$ for $k\ge2$ are then $L_{-k}$, associated with $\Tr(\Sigma^{\ord{k}})$.
Rigid $E_9$ covariance of $\sbm\Lambda_\alpha$ then guarantees that this conclusion holds for any other solution of the section contraint.

In order to simplify the definition and transformation properties of the vector fields and their field strengths, we introduce a $\circ$ product naturally extending \eqref{eq:DorfmanProduct}
\begin{align}
\label{eq:DorfmanProductex} 
\mathbbm{\Lambda}_1\! \circ\! \mathbbm{\Lambda}_2 & = \bigg( \mathcal{L}_{\mathbbm{\Lambda}_1} \Lambda_2 , \,\,
\mathcal{L}_{\mathbbm{\Lambda}_1} \Sigma^{\ord{k}}_2{}
+\delta^{k}_1\,\eta_{1\,\alpha\beta} \langle \partial_{\Lambda_1} |T^\alpha | \Lambda_1\rangle T^\beta |\Lambda_2\rangle \langle \partial_{\Lambda_1}|
+\eta_{\alpha\beta} \Tr( T^\alpha \Sigma^{\ord{k}}_1) T^\beta |\Lambda_2\rangle \langle \partial_{\Sigma_1}| \nn\\
&\hspace{25mm}
-|\Lambda_2\rangle \langle \partial_{\Sigma_1} | \Sigma^{\ord{k}}_1-(k-1)\Big(\Tr(\Sigma_1^{\ord{k}})\ket{\Lambda_2}\bra{\partial_{\Sigma_1}}-\Sigma_1^{\ord{k}}\langle \partial_{\Sigma_1} |\Lambda_2\rangle\Big)
\bigg)\,.
\end{align}
This expression satisfies \eqref{eq:Dorf1}, meaning that it is equivalent to the extended E-bracket \eqref{eq:Ebracketex} up to the addition of trivial parameters.
However, this extended $\circ$ product does not satisfy the Leibniz identity \eqref{eq:Leibniz}.
More precisely, it fails to satisfy it by a trivial parameter.
We can nevertheless use it to define the transformation properties of the vector fields and their field strengths in direct analogy with Section~\ref{sec:gauges}.
We then introduce additional constrained gauge fields $B_\mu^{\ord{k}}$, and their field strengths $\mathcal{G}_{\mu\nu}^{\ord{k}}$, associated to the new generalised diffeomorphisms generated by $L_{-k}$. 
Our notations are thus naturally extended as follows
\begin{equation}
\mathbbm{A}_{\mu}=(\ket{A_\mu}, B_{\mu}^{\ord{k}})\,,\;\;\;\;\;\;\;\mathbbm{F}_{\mu\nu}=(\ket{\mathcal{F}_{\mu\nu}},\mathcal{G}_{\mu\nu}^{\ord{k}})\,, 
\end{equation}
with $k\in \mathds{N}^+$. Under rigid $\mathfrak{\hat e}_9\oleft \mathfrak{vir}^-$ variations these sets of gauge fields and field strengths form indecomposable pairs and transform as the set of parameters $\mathbbm{\Lambda}=(\ket{\Lambda}, \Sigma^{\ord{k}})$ in~\eqref{eq:tparaex}. 
The extended gauge connection still transforms according to \eqref{eq:deltaA} and the field strengths are defined according to the first expression in \eqref{eq:DD} which we reproduce here:
\begin{equation}\label{eq:DDex}
\left[ \mathcal{D}_\mu, \mathcal{D}_\nu\right] = -\mathcal{L}_{\mathbbm{F}_{\mu\nu}}\,.
\end{equation}
The second expression in \eqref{eq:DD} now only holds up to a trivial parameter.
The extended field strengths contain a trivial combination of two-form fields as discussed in the previous section, but now including the novel extended trivial parameters described in Appendix~\ref{app:exttriv}.
Then, combining \eqref{eq:dfffclos} with \eqref{eq:DDex} we find that the transformation property \eqref{eq:fsvar} still holds in the extended setting, with any trivial parameter that may arise on the right-hand side absorbed into the two-form transformation.%
\footnote{As stressed after \eqref{eq:fsvar} and discussed in Section~\ref{sec:E9EFT}, invariance under external diffeomorphisms requires to introduce an unprojected duality equation for $\ket\cF$ which in turn requires to amend \eqref{eq:fsvar} by a trivial parameter.
We will not discuss external diffeomorphisms in the extended formulation, so \eqref{eq:fsvar} will suffice in this setting.}
Finally, the Bianchi identities \eqref{eq:Bianchi2} still hold in the extended setting, as they can be also deduced by formally evaluating the action of three $\cD$ differentials on a covariant field.

\section{\texorpdfstring{Virasoro-extended $E_9$ exceptional field theory}{Virasoro-extended E9 exceptional field theory}}
\label{sec:extE9exft}

We are now equipped to introduce the field content of $E_9$ exceptional field theory and build a pseudo-Lagrangian to determine its dynamics.
The approach we follow in this section is based on the formalism introduced in Section~\ref{sec:extcoset}.
In particular, the twisted self-duality constraint will be the natural covariantisation of \eqref{twsd ungauged, P} and \eqref{twsd ungauged J}.
This makes the connection to two-dimensional supergravity and to the linear system straightforward, and has the further advantage that scalar field currents covariant under generalised diffeomorphisms can be naturally defined in terms of a gauge connection $\mathbbm A$ transforming according to \eqref{eq:deltaA} and \eqref{eq:DorfmanProductex}.
We will however only develop this formalism in the conformal gauge $\tilde g_{\mu\nu}=\eta_{\mu\nu}$.
We refer to this approach as `extended' because of the presence of the $\virm$ scalar fields $\phi_n$, their associated symmetries and gauge fields $B^\ord{n}$.
The `minimal' approach to $E_9$ exceptional field theory and its dynamics is introduced in Section~\ref{sec:E9EFT}.

In this section we always leave wedge products as understood, writing for instance $\fJ\,\fJ$ in place of $\fJ\wedge\fJ$.

\subsection{Covariant currents and twisted self-duality}

We begin by introducing the $(\widehat{E}_{8}\rtimes\Virm )/K(E_9)$ coset representative
\begin{equation}
\cV = \Gamma \hat{V} 
= \,\rho^{-L_0} \, e^{-\phi_1 L_{-1}} \, e^{-\phi_2 L_{-2}} \,  \cdots
  \mathring{V}\,e^{Y_{1A} T^A_{-1}}\,e^{Y_{2A} T^A_{-2}} \, \cdots\,e^{-\sigma\dK}
\label{full coset repr}
\end{equation}
where we work now in a faithful representation of the algebra so that $\dK\neq0$.
We will regard $e^{-\sigma\dK}$ as contained in $\hat{V}$.

These expressions involve products of infinite products. 
However, the coefficients of any generator $T^A_{-m}$ or $L_{-m}$ in the  Maurer--Cartan form are finite expressions that can be computed by truncating the infinite products to the first $m$ factors. 
We then define the generalised metric (of weight 0)%
\footnote{
Because Hermitian conjugation does not map $\Virm$ nor the completed loop group to themselves, expressions like $\cM$ need qualification.
One way is to first identify the $K(E_9)$ singlet element in $\overline{R(\Lambda_0)_0}\otimes_{\mathrm{sym}} \overline{R(\Lambda_0)_0}$, which we denote as $\Delta_{\ul{MN}}$ (the underlined indices transform under the local $K(E_9)$ symmetry), and then regard the generalised metric as a field dependent element of 
$\overline{R(\Lambda_0)_0}\otimes_{\mathrm{sym}} \overline{R(\Lambda_0)_0}$, which we write as 
$\cM_{MN} = \Delta_{\ul{PQ}} \cV^\ul{P}{}_M \cV^\ul{Q}{}_N$.
However, it will turn out to be more convenient to treat $\cM$ as a group element. 
All manipulations we encounter are then justified by switching to the Unendlichbein formulation in terms of the Hermitian and anti-Hermitian projections $\fP$ and $\fQ$ of the Maurer--Cartan form, see~\eqref{MC form and P Q def} and Section~\ref{ssec:unend}.
We provide more details on the definitions of the group $E_9$ and its representations in Appendix~\ref{app:kacmoody}.
}
\begin{equation}
\cM = \cV^\dagger \cV \,,
\label{genmetric extended}
\end{equation}
that transforms under $g\in\hE8\rtimes\Virm$ as
\begin{align}
    \cM \to g^\dagger \cM g \,.
\end{align}
We will always take $\cV$ and $\cM$ in the $R(\Lambda_0)_0$ representation and write explicitly the $\rho$ factors when they act on representations of non-zero conformal weight $h$.

Taking an internal derivative of $\cM$ we define the internal current
\begin{equation}\label{intcurr}
\cJ_{M\,\alpha} T^\alpha = \cM^{-1} \partial_M \cM\,,
\end{equation}
which satisfies the section constraint along the index $M$.
Translating this index to braket notation we shall write 
$\cJ_{M\,\alpha} T^\alpha \bra{e^M} = \bra{\cJ_\alpha} \otimes T^\alpha$ with $\bra{e^M}$ a basis for $\overline{R(\Lambda_0)_{-1}}$.
Under generalised diffeomorphisms $\cM$ transforms covariantly, namely $\delta_{\mathbbm\Lambda}\cM = \cL_{\mathbbm\Lambda} \cM$ with
\begin{align}
\cM^{-1}\cL_{\mathbbm\Lambda}\cM &=
\braket{\cJ_\alpha}{\Lambda} \, T^\alpha
+\bigg[\, 
  \eta_{\alpha\beta}\bra{\partial_\Lambda}T^\beta\ket{\Lambda}
  + \sum_{r=1}^\infty \eta_{-r\,\alpha\beta} \Tr\big[\Sigma^\ord{r} T^\beta \big]
\,\bigg]
\left( T^\alpha + \cM^{-1} (T^\alpha)^\dagger \cM  \right)\,,\CR{}
&=\braket{\cJ_\alpha}{\Lambda}\,T^\alpha  + \sbm{\Lambda}_\alpha \left( T^\alpha + \cM^{-1} (T^\alpha)^\dagger \cM  \right)\,,
\end{align}
Notice that compared to \cite{Bossard:2018utw}, we now include the action of infinitely many Virasoro generators in both the gauge parameters (through $\Tr(\Sigma^\ord{m})\,$) and in the conjugation by $\cM$.
In the second line we have used the shortcut expression $\sbm\Lambda_\alpha$ introduced in \eqref{eq:balphaex}.
We can then define the external current that is covariant under rigid $\widehat{E}_{8}\rtimes\Virm$ transformations as well as generalised diffeomorphisms
\begin{align}
\fJ\ =\ \cM^{-1} \cD \cM\ =\ \cM^{-1}\dd\cM - \braket{\cJ_\alpha}{A} \,T^\alpha
   - \sbm{A}_\alpha \left( T^\alpha + \cM^{-1} (T^\alpha)^\dagger \cM  \right)\,.
\label{covariant J definition}
\end{align}
We expand the algebra components of $\fJ$ as follows:
\begin{equation}\label{fJ components}
\fJ = \fJ_\alpha T^\alpha =
 \sum_{m\in\mathds{Z}}( \, \tJ^m_A\,  T^A_m + \tJ_m\,  L_{m}\,  ) + \tJ_\dK \,\dK\,.
\end{equation}

The twisted self-duality constraint of exceptional field theory will essentially be the covariantisation of the two-dimensional expression \eqref{twsd ungauged J}, but we now need to take into account the $\dK$ component of the current which was trivially represented in Section~\ref{sec:extcoset}.
The shift operators $\cS_m$ and $\cS^\upgamma_m$ introduced in \eqref{eq:shift op def} and at the end of Section~\ref{sec:extcoset} can be interpreted as multiplication by $w^m$ and $\upgamma(w)^m$ respectively, and we can thus relate them by $\Gamma$ conjugation using \eqref{upgamma as series}:
\begin{equation}
\cS^\upgamma_m(X) = 
\cS_0\big(\, \Gamma^{-1} \cS_m(\Gamma X \Gamma^{-1})\,\Gamma \,\big)
\label{shift gamma def}
\end{equation}
for any $X\in\hevir$.
Any $\cS^\upgamma_m$ can be expanded as a series in $\cS_k$ operators, with $k\le m$ and $\rho$ and $\phi_m$ dependent coefficients.
For instance, \eqref{upgamma as series} is reproduced as
\begin{equation}
\cS^\upgamma_1 = 
\frac1\rho \left[
\cS_1 - \phi_1\, \cS_0 - \phi_2\,\cS_{-1} - \phi_3\, \cS_{-2} 
-\Big(\phi_4+\frac12 \phi_2^2\Big) \cS_{-3}
+\ldots
\right] \,.
\label{shift gamma 1 as series}
\end{equation}
The $\cS_0$ operator on the right-hand side of \eqref{shift gamma def} is only necessary to remove any spurious $\dK$ component that might be generated by $\Gamma$ conjugation because of the $\vir$ central charge.
Several properties of these $\Virm$ field dependent shift operators $\cS^\upgamma_m$ are described in Appendix~\ref{app:shift}. 
In particular, notice that because of the factors of $\rho$ generated by $\Gamma$ conjugation all $\cS^\upgamma_m$ operators carry weight 0, whereas $\cS_m$ carry weight $m$.

Consider now the shift $\cS^\upgamma_{m}(\fJ)$ of the covariant current. 
The transformation properties of $\fJ$ and $\Gamma$ imply that under a rigid transformation $X\in\hevirm$  
\begin{equation}
\delta_X \cS_m^\upgamma(\fJ) = 
- [X\,,\cS_m^\upgamma(\fJ)] 
- \omega^{\alpha\beta} X_\alpha \, \big(\cS_m^\upgamma(\fJ)\big)_\beta \, \dK \,,
\label{algebra cocycle from shift gamma}
\end{equation}
where we are varying both $\fJ$ and the $\Gamma$ implicit in the definition of the shift operator.
The expression $\omega^{\alpha\beta}$ is the algebra cocycle corresponding to (minus) the central charge component of the $\hevir$ commutator: given $X,\,Z\in\hevir$ with components
\begin{equation}
X = X_\alpha T^\alpha = \sum_{m\in\mathds{Z}} ( X_A^m T^A_m + X_m L_m ) + X_\dK \dK\,,
\end{equation}
and similarly for $Z$, we define
\begin{equation}
\omega^{\alpha\beta} X_\alpha Z_\beta = -[X,\,Z]\big|_\dK
= -\eta^{AB}\sum_{n\in\mathds{Z}} n \, X_A^n\,Z_B^{-n} 
  -\frac{c_{\vir}}{12} \sum_{n\in\mathds{Z}} (n^3-n) \,X_{n}\,Z_{-n}\,.
\label{algebra cocycle}
\end{equation}
To compensate for this cocycle term in \eqref{algebra cocycle from shift gamma} we follow the same strategy as in \cite{Bossard:2018utw} and define covariant shifted currents
\begin{equation}\label{shifted cov current}
\fJ^\ord{m} = \cS_m^\upgamma(\fJ) + \chi^\upgamma_m \dK\,,
\end{equation}
where the new one-form fields $\chi^\upgamma_m$ satisfy the indecomposable transformation property
\begin{equation}
\delta_X \chi^\upgamma_m = 
\omega^{\alpha\beta} X_\alpha \, \fJ(m)_\beta \,,
\label{rigid trf chigamma}
\end{equation}
where we have used $\omega^{\alpha\dK}=0$ to write $\fJ^\ord{m}$ in place of $\cS_m^\upgamma(\fJ)$.
This transformation property guarantees that the shifted current transforms as an algebra element (of $\hevir$) under \mbox{$X\in\hevirm$}:
\begin{equation}
\delta_X \fJ(m) = -[X,\,\fJ(m)]\,.
\label{rigid trf shifted J}
\end{equation}

More generally, for any $Z\in\hevir$ transforming as an algebra element under $\hevirm$, we may define a $\dK$ completion $\widehat{Z}^\upgamma_m$ of $\cS^\upgamma_m(Z)$ such that
\begin{equation}\label{K completion}
Z^\ord{m} = \cS^\upgamma_m(Z) + \widehat{Z}^\upgamma_m \dK\,,\qquad
\delta_X Z^\ord{m} = -[X,\,Z^\ord{m}]\,.
\end{equation}
When the $\dK$ completion of an object can be expressed entirely in terms of other fields of the theory, we will use the notation introduced above.
If the $\dK$ completion is an independent field, we use a new symbol for it, as we did for $\fJ$ and $\chi^\upgamma_m$.

We define the $\chi^\upgamma_m$ fields to be covariant of weight 0 under generalised diffeomorphisms.
Namely, their transformation descends from \eqref{rigid trf chigamma}:
\begin{equation}
\delta_\mathbbm{\Lambda} \chi^\upgamma_m
= \bra{\partial_\chi}\Lambda\rangle \chi^\upgamma_m
  + \omega^{\alpha\beta} \sbm\Lambda_\alpha \fJ(m)_\beta\,.
\end{equation}
This implies in turn that the $\fJ(m)$ are covariant, their transformation under generalised diffeomorphisms follows from \eqref{eq:GL2} and \eqref{rigid trf shifted J}:
\begin{equation}
\delta_\mathbbm{\Lambda} \, \fJ(m) =
\bra{\partial_{\fJ}}\Lambda\rangle \fJ(m) - \sbm\Lambda_\alpha \, \big[T^\alpha,\,\fJ(m)\big]\,.
\label{shifted J gendiffeo var}
\end{equation}

Notice that \eqref{algebra cocycle} depends implicitly on the  representation in which we write $\fJ$, through the $\vir$ central charge $c_\vir$.
This means that we must specify in which representation we write $\fJ$ in order to define the $\chi^\upgamma_m$ fields.
We have defined $\fJ$ in the $R(\Lambda_0)_0$ representation and hence $c_\vir=8$ as this is the natural choice for $E_9$ exceptional field theory, because its internal derivatives, gauge parameters and generalised metric all sit in (tensor products of) $R(\Lambda_0)_h$ representations and their duals.
Our construction applies to loop groups over any other simple Lie group, and in each case $c_\vir$ will take a different value.
Hence it will be more convenient to leave $c_\vir$ unspecified in \eqref{algebra cocycle} to keep track of where it appears.

The current satisfies $\fJ = \cM^{-1} \fJ^\dagger \cM$. 
This allows us to relate the action of opposite shifts by bringing $\cM$ through a shift operator
\begin{equation}
\cS^\upgamma_m(\fJ) = 
\cM^{-1}\big(\cS^\upgamma_{-m}(\fJ)\big)^\dagger\cM
- \omega^\alpha(\cM) \, \fJ(-m)_\alpha\,\dK\,,
\end{equation}
where we have introduced the (group) cocycle
\begin{align}
\label{eq:omegaalpha}
\omega^\alpha(\cM) 
&= 
\cM^{-1} \cS_0(T^\alpha)^\dagger \cM - \cS_0(\cM^{-1} (T^\alpha)^\dagger \cM)
=
\cM^{-1} (T^\alpha)^\dagger \cM \big|_\dK - \delta^\alpha_\dK\,.
\end{align}
Several properties of such cocycles are collected in Appendix~\ref{app:shift}.
We then see that we can identify $\chi^\upgamma$ fields associated to opposite shifts up to a cocycle, by imposing
\begin{equation}
\cM^{-1}\,(\fJ(m))^\dagger \,\cM = \fJ(-m)
\end{equation}
so that 
\begin{equation}
\chi^\upgamma_{-m} 
= \chi^\upgamma_{m} + \omega^\alpha(\cM) \, \fJ(m)_\alpha
= \chi^\upgamma_{m} - \omega^\alpha(\cM) \, \fJ(-m)_\alpha\,.
\label{opposite chi}
\end{equation}
Finally, we obviously want $\fJ(0)=\fJ$ and hence identify $\chi^\upgamma_0 = \tJ_\dK$.

With the definitions and transformation properties above we can state the covariant twisted self-duality equation of exceptional field theory
\begin{equation}
\star\!\fJ = \fJ(1)\,,
\label{twsd cov J}
\end{equation}
which reduces to \eqref{twsd ungauged J} along the loop and $\vir$ components for $\bra\partial=0$ and $B^\ord{m}=0$.
The $\dK$ component of this equation reads
\begin{equation}
\star\!\tJ_\dK = \chi^\upgamma_1\,,
\end{equation}
which is not a duality equation for $\sigma$ but rather shows that $\chi^\upgamma_1$ is an auxiliary field that does not carry any on-shell degrees of freedom.
While $\star^2\fJ=\fJ$, the shift operator on the right-hand side does not square to $\cS_0$.%
\footnote{We can also rewrite \eqref{twsd cov J} in the equivalent form $\star\fJ=\cM^{-1} (\fJ(1))^\dagger \cM$ so that the operator on the right-hand side is involutive.}
This means that by iterating \eqref{twsd cov J} along the loop and $\vir$ components we recover a cascade of duality relations analogous to \eqref{cascade ungauged, P}.
In order to complete the $\dK$ sector of such relations we then introduce the appropriate $\chi^\upgamma_m$ fields for every shift, obtaining
\begin{equation}\label{cascade cov J}
\star^{|m|}\fJ(n) = \fJ(m+n)\,,\qquad m,\,n\in\mathds{Z}\,.
\end{equation}
We stress that the loop and $\vir$ components of these relations are simply linear combinations of the components of \eqref{twsd cov J} and are therefore redundant.
On the other hand, the $\dK$ component reads $\star^{|m|}\chi^\upgamma_n=\chi^\upgamma_{m+n}$ which relates all $\chi^\upgamma$ fields to each other and to $\tJ_\dK$.
The relations \eqref{twsd cov J} and \eqref{cascade cov J} are manifestly invariant under rigid $\hE8\rtimes\Virm$ and generalised diffeomorphisms, and compatible with \eqref{opposite chi}.

We can also introduce the covariant versions of the Hermitian and anti-Hermitian projections of the Maurer--Cartan one-form
\begin{equation}
\cD\cV\cV^{-1} = \fP + \fQ\,,\qquad
\fP^\dagger=\fP\,,\quad
\fQ^\dagger=-\fQ\,.
\label{MC form and P Q def}
\end{equation}
The relation to the currents $\fJ$ is conjugation by $\cV$:
\begin{equation}
\fJ= 2\cV^{-1}\fP \, \cV \,,
\label{P from J}
\end{equation}
and we can require a similar relation for the shifted currents which, using \eqref{finite cocycle from S to S gamma}, leads to the definitions
\begin{equation}
\fP(m) = \cS_m(\fP) + \tilde\chi_m \dK \,, \qquad
\tilde\chi_m = \tilde\chi_{-m} = 
\frac12 \chi^\upgamma_m - \frac12 \omega^\alpha(\cV) \fJ(m)_\alpha 
\label{shifted P and tilde chi def}
\end{equation}
so that $\fJ(m)= 2\cV^{-1}\fP(m)\cV$.
The one-forms $\tilde\chi_m$ complete the $\dK$ component of $\cS_m(\cP)$ so that $\fP(m)$ transforms with a commutator under $K(\mf{e}_9)$, analogously to the $\dK$ completions of $\cS^\upgamma_m$ introduced in \eqref{K completion}.

With these definitions we can rewrite twisted self-duality \eqref{twsd cov J} and \eqref{cascade cov J} as
\begin{equation}
\star\!\fP = \fP(1)\,,\qquad
\star^{|m|}\fP(n) = \fP(m+n)\,.
\label{twsd covariant P}
\end{equation}
We shall expand $\fP$ and $\cQ$ similarly to \eqref{fJ components}:
\begin{equation}
\fP = \sum_{m\in\mathds{Z}}(\tP_A^m T^A_m + \tP_m L_m)+\tP_\dK\,,\qquad
\cQ = \sum_{m\in\mathds{Z}}(\tQ_A^m T^A_m + \tQ_m L_m)\,.
\label{P and Q components}
\end{equation}
In terms of these components, twisted self-duality implies in particular
\begin{equation}
\tP^m_A = \star^{|m|} \tP^0_A\,,\qquad
\tP_m =  \star^{|m|} \tP_0\,.
\end{equation}

\subsection{Shifted Maurer--Cartan equations}

We will now construct a topological pseudo-Lagrangian for $\virm$ extended $E_9$ exceptional field theory.
The Euler--Lagrange equations obtained by varying this pseudo-Lagrangian must be supplemented with the twisted self-duality constraint \eqref{twsd cov J} (or \eqref{twsd covariant P}) to reproduce the equations of motion.
In particular, in analogy with lower-rank exceptional field theories, we will require that the Euler--Lagrange equations for the $\chi^\upgamma_m$ fields as well as the $B^\ord{m}$ fields vanish upon using twisted self-duality, reflecting the fact that these fields do not encode any physical degrees of freedom.
We will work only in conformal gauge with the two-dimensional unimodular metric fixed to the Minkowski metric: $\tilde g_{\mu\nu}=\eta_{\mu\nu}$.
From the point of view of Kaluza--Klein decomposition of a higher-dimensional metric, it is always possible to partially gauge fix higher-dimensional diffeomorphisms so that a $2\times2$ non-degenerate block on the diagonal of the higher dimensinoal metric is reduced to $\eta_{\mu\nu}$ (or $\delta_{\mu\nu}$).
In other words, even when $\tilde g_{\mu\nu}$ depends on the $Y^M$ coordinates subject to the section constraint, reflecting that our theory is in fact a higher-dimensional supergravity, it is still possible to switch to conformal gauge without loss of generality.
As a result, however, the equations of motion obtained from the pseudo-Lagrangian and twisted self-duality must be supplemented with the covariantised version of the Virasoro constraint \eqref{ungauged Vir constraint}.
We will come back to this in Section~\ref{sec:cov eom}.
We will restore $\tilde g_{\mu\nu}$ in Section~\ref{sec:E9EFT}.

The two-dimensional current \eqref{ungauged J} satisfies the Maurer--Cartan equation $\dd J+\frac12[J,\, J]=0$, 
where the wedge product is understood and the commutators are implicitly graded with the rank of the $p$-forms, namely for $A$ and $B$ any $\hevir$ valued forms, $[A,\,B]=A_\alpha\wedge B_\beta f^{\alpha\beta}{}_\gamma T^\gamma$.
For the covariant current $\fJ$ this expression becomes
\begin{equation}
\cD\fJ + \frac12[\fJ,\, \fJ] + \braket{\cJ_\alpha}{\cF} \, T^\alpha
+\sbm{F}_\alpha \big( T^\alpha+ \cM^{-1} (T^\alpha)^\dagger \cM \big) = 0\,,
\label{cov J maurer cartan eq}
\end{equation}
which is deduced from \eqref{covariant J definition} and the first equation in \eqref{eq:DD}.
We will apply this identity to a shifted current $\fJ(m)$ in order to construct a candidate topological term for exceptional field theory.
The idea is that while all components of $\fJ$ satisfy the above integrability condition, the central charge component of $\fJ(m)$ is $\chi^\upgamma_m$ which is a fundamental field.
Thus, we will be able to write a set of two-forms based on $\fJ(m)$ and its derivatives that transform by conjugation under rigid $\hE8\rtimes\Virm$ and such that, because of the identity \eqref{cov J maurer cartan eq}, only their $\dK$ components are non-vanishing.
As a consequence, these expressions are automatically invariant under rigid $\hE8\rtimes\Virm$.
We will then find that only one such expression is compatible with twisted self-duality, covariant under generalised diffeomorphisms and independent of the two-form fields appearing in the covariant field strengths $\mathbbm F$.
That will be the main part of our pseudo-Lagrangian.

We begin by taking a covariant derivative of $\fJ(m)$:
\begin{align}
\cD\fJ(m) 
&= \cD \cS^\upgamma_m(\fJ) +\cD\chi^\upgamma_m \, \dK  \CR
&= \cS_0\bigg(  \cD\big(\Gamma^{-1}\cS_m(\Gamma\fJ\Gamma^{-1})\Gamma\big)   \bigg)
  + \cD\chi^\upgamma_m \, \dK  \CR
&= \cS_0\bigg( \Gamma^{-1} \big( \cS_m( [\cD\Gamma\Gamma^{-1},\,\Gamma\fJ\Gamma^{-1}] )
               - [\cD\Gamma\Gamma^{-1},\,\cS_m(\Gamma\fJ\Gamma^{-1})] \big) \Gamma   \bigg)
  + \cS^\upgamma_m(\cD\fJ)
  + \cD\chi^\upgamma_m \, \dK  \CR
&= m\sum_{q=0}^\infty (\cD\Gamma\Gamma^{-1})_{-q} \, \cS^\upgamma_{m-q}(\fJ)
   +\cS_m^\upgamma(\cD\fJ) + \cD\chi^\upgamma_m \, \dK\,.
\end{align}
In the last line $(\cD\Gamma\Gamma^{-1})_{-q}$ is the $L_{-q}$ component of the Maurer--Cartan form and we used \eqref{shiftgamma commutator}. 
We now apply \eqref{cov J maurer cartan eq} to the second term and use \eqref{S gamma commutator} to arrive at the identity
\begin{align}
&\cD\fJ(m) 
+ \frac12[\fJ,\,\fJ(m)] 
+ \frac{m}{2}\sum_{q=1}^\infty (\cD\Gamma\Gamma^{-1})_{-q} 
   \left(
      \cS^\upgamma_{m+q}(\fJ) 
      - \cS^\upgamma_{m-q}(\fJ) 
    \right) 
+ \braket{\cJ_\alpha}{\cF}\, \cS^\upgamma_m(T^\alpha)
\CR&
+ \sbm F_\alpha\, \cS^\upgamma_m  
  \big(T^\alpha + \cM^{-1} (T^\alpha)^\dagger\cM  \big)
+ \bigg( 
     \frac12\omega^{\alpha\beta}\, \fJ_\alpha \, \fJ(m)_\beta 
   - \cD\chi^\upgamma_m 
  \bigg)\dK 
\ =\ 0\,.
\label{shifted CM eq J}
\end{align}
This expression contains several shift operators which we now want to complete by adding terms proportional to $\dK$, as described in \eqref{K completion}, so that the completed expression transforms as an algebra-valued object under rigid $\hevirm$.
For each shift of $\fJ$ we will of course add the associated $\chi^\upgamma$ one-form.
The shift of the internal current $\bra{\cJ_\alpha}$ is completed in analogy with the external one and with \cite{Bossard:2018utw} by introducing and \emph{internal} scalar field $\bra{\chi^\upgamma_m}$ in the $\overline{R(\Lambda_0)_{-1}}$ representation, defining
\begin{equation}\label{shifted internal J}
\bra{\cJ^\ord{m}_\alpha} \otimes T^\alpha
=
  \bra{\cJ_\alpha} \otimes \cS^\upgamma_m(T^\alpha)
+ \bra{\chi^\upgamma_m} \otimes \dK\,,
\end{equation} 
and imposing the rigid transformation property
\begin{equation}
\delta_X \bra{\chi^\upgamma_m} = 
  \bra{\chi^\upgamma_m} X
- X_0 \bra{\chi^\upgamma_m}  
+ \omega^{\alpha\beta}X_\alpha \bra{\cJ^\ord{m}_\beta} \,,\qquad 
X\in\hevirm\,.
\end{equation}
The internal index of the internal current (represented by $\bra{\ }$) comes from an internal derivative and satisfies the section constraint \eqref{eq:SC}, therefore $\bra{\chi^\upgamma_m}$ must satisfy the section constraint, too.
Eventually we will find that only one such scalar appears in the pseudo-Lagrangian and that it is related to the $\bra\chi$ field introduced in \cite{Bossard:2018utw}.

Finally, we need to construct the $\dK$ completion of the expression 
$\sbm F_\alpha \cS^\upgamma_m\big(T^\alpha + \cM^{-1} (T^\alpha)^\dagger\cM\big) $ as in \eqref{K completion}.
For reference let us write down
\begin{equation}
\sbm F_\alpha =
\eta_{\alpha\beta}\bra{\partial_\cF} T^\beta \ket{\cF} 
+\sum_{n=1}^\infty \, \eta_{-n\,\alpha\beta} \Tr(\cG^\ord{n} T^\beta)\,.
\label{Falpha}
\end{equation}
This expression transforms covariantly under $\hE8\rtimes\Virm$.
In particular, the rigid indecomposable transformation \eqref{eq:tparaex} of $(\cF,\,\cG^\ord{m})$ guarantees $\Virm$ covariance.
When we apply the shift operator, $\Virm$ covariance of $\sbm F_\alpha \cS^\upgamma_m(T^\alpha)$ still holds because $\cS^\upgamma_m$ commutes with $\Virm$ thanks to the transformation of $\Gamma$ in \eqref{shift gamma def}, and because $\sbm F_\alpha$ does not take values along the positive $\vir$ generators, so that no central charge term is generated. 
The same holds true when we apply $\cM$ conjugation before the shift operator.
Hence, we simply have
\providecommand{\tmp}{\cS^\upgamma_m\big(T^\alpha + \cM^{-1}(T^\alpha)^\dagger\cM\big)}
\begin{equation}
\delta_{L_{-k}} \bigg(\,\sbm F_\alpha \,\tmp \,\bigg) 
= -\sbm F_\alpha \,\big[\,L_{-k}\,,\ \tmp\,\big]\,.
\end{equation}
In order to look at the loop transformation, we first focus on $\sbm F_\alpha \cS^\upgamma_m(T^\alpha)$ and use the fact that $\cS^\upgamma_m$ can be written as a series of constant $\cS_m$ operators such as \eqref{shift gamma 1 as series}, and the coefficients are $\hE8$ invariant:
\begin{equation}\label{shiftgamma generic series}
\cS^\upgamma_m = \sum_{k\le m} a_m^k(\rho,\phi) \, \cS_{k}
\end{equation}
Thus, we look at a single $\cS_k$ operator acting on $\sbm F_\alpha$ and perform a finite $g\in\hE8$ transformation. 
Using \eqref{shifted group cocycle E9} we find
\begin{equation}
\sbm F_\alpha\,\cS_k(g^{-1} T^\alpha g) 
= \sbm F_\alpha\,\left( g^{-1}\cS_k( T^\alpha )g - \omega_{-k}^\alpha(g^{-1}) \dK \right)
\label{loop cocycle from Falpha trf}
\end{equation}
Opening up the cocycle term using
\eqref{cocycle K = 0}, \eqref{move shift from cocycle} and \eqref{eta and shift ops}, it reads
\begin{align}
-\sbm F_\alpha \, \omega^\alpha_{-k}(g^{-1}) &=
-\omega^\alpha(g^{-1}) \bigg( 
 \eta_{k\,\alpha\beta} \bra{\partial_\cF} T^\beta \ket\cF
 +\sum_{n=1}^\infty \, \eta_{k-n\,\alpha\beta} \Tr(\cG^\ord{n} T^\beta)
\bigg) \CR
&=
\bra{\partial_\cF} (g L_k g^{-1} -L_k) \ket\cF 
+ \sum_{n=1}^\infty \Tr\big(\, \cG^\ord{n} (g L_{k-n} g^{-1} -L_{k-n})  \,\big)
\end{align}
Where we used \eqref{cocycle from Vir conjugation} in the second line.
Recalling that this is just the transformation of one term in the series of shifts defining $\cS^\upgamma_m$ and that of course $L_{k-n}=\cS_k(L_{-n})$, we are led to the definition
\begin{equation}
\hbm{F}^\upgamma_m =
-\bra{\partial_\cF} \cS^\upgamma_m(L_0) \ket\cF 
- \sum_{n=1}^\infty \Tr\Big(\,\cG^\ord{n} \cS^\upgamma_m(L_{-n})  \,\Big)\,,
\label{hatF gamma definition}
\end{equation}
such that in the loop variation of the combination
\begin{equation}
\sbm F^\ord{m}_\alpha T^\alpha = \sbm F_\alpha \cS^\upgamma_m(T^\alpha) \,+\,\hbm{F}^\upgamma_m\,\dK
\label{completed shifted F}
\end{equation}
the cocycle terms coming from \eqref{loop cocycle from Falpha trf} cancel out.
Furthermore, one can see that due to the indecomposable $\Virm$ transformation of the field strengths, \eqref{hatF gamma definition} is $\Virm$ invariant.
This guarantees that \eqref{completed shifted F} transforms as an algebra element.
In order to define the $\dK$ completion of the term 
$\sbm F_\alpha \cS^\upgamma_m\big(\cM^{-1} (T^\alpha)^\dagger\cM\big) $,
we observe that the $\cM$ conjugate of \eqref{completed shifted F} also transforms as an algebra element.
Sending $m\to-m$ and writing $\cM= \Gamma^\dagger \widehat{M} \Gamma$ with $\widehat{M}^\dagger=\widehat{M}\in\hE8$, conjugating \eqref{completed shifted F} with $\cM$ and applying Hermitian conjugation gives us
\begin{align}
&\sbm F_\alpha \, \cM^{-1} \big(\cS^\upgamma_{-m}(T^\alpha)\big)^\dagger \cM 
 + \hbm{F}^\upgamma_{-m} \dK \CR
&\qquad=
\sbm F_\alpha \, \bigg(  
  \cM \cS_0\Big(\Gamma^{-1} \cS_{-m} (\Gamma T^\alpha \Gamma^{-1}) \Gamma\Big) \cM^{-1}
\bigg)^\dagger
+ \hbm{F}^\upgamma_{-m} \dK \CR
&\qquad=
\sbm F_\alpha \, \bigg(  
  \cS_0\Big( \Gamma^\dagger \widehat{M} \cS_{-m}(\Gamma T^\alpha \Gamma^{-1}) \widehat{M}^{-1} \Gamma^{-\dagger}\Big)
\bigg)^\dagger 
+\Big( \hbm{F}^\upgamma_{-m} + \omega^\alpha(\cM) \sbm F_\alpha^{\ord{-m}}\Big)\,\dK  \CR
&\qquad=
\sbm F_\alpha \, \cS^\upgamma_m\big(\cM^{-1} (T^\alpha)^\dagger\cM\big)
+\Big( \hbm{F}^\upgamma_{-m} + \omega^\alpha(\cM) \sbm F_\alpha^{\ord{-m}}\Big)\,\dK \CR
&\qquad=
\sbm F_\alpha \, \cS^\upgamma_m\big(\cM^{-1} (T^\alpha)^\dagger\cM\big) \vphantom{\bigg(}
+\mhbm{F}^\upgamma_m\,\dK   \,,
\label{shifted F M conj}
\end{align}
where in the third line we brought $\cM$ through $\cS_0$ generating a cocycle term (written in terms of \eqref{completed shifted F} using $\omega^\dK(\cM)=0$), and in the fourth line we brought $\widehat{M}$ through $\cS_{-m}$, which does not generate cocycles because of the overall $\cS_0$ projection. 
We then propagated Hermitian conjugation through $\cS_0$.
The $\dK$ component of the third and fourth line is the expression we are looking for.
In the last line, for later convenience, we defined
\begin{equation}
\mhbm{F}^\upgamma_m = \hbm{F}^\upgamma_{-m} + \omega^\alpha(\cM) \sbm F_\alpha^{\ord{-m}}\,.
\label{mhbm F def}
\end{equation}

We can now complete each shift operator in \eqref{shifted CM eq J} with the appropriate central charge term arriving at the identity
\begin{align}
& \cD\fJ(m) 
+ \frac12[\fJ,\,\fJ(m)] 
+ m \sum_{n=1}^\infty \tP_n (\fJ(m+n)-\fJ(m-n)) 
\CR&\qquad\ \ \hspace*{.2pt}
+ \braket{\cJ^\ord{m}}{\cF} 
+ \sbm F^\ord{m}_\alpha T^\alpha 
+ \sbm F_\alpha \, \cS^\upgamma_m\big(\cM^{-1} (T^\alpha)^\dagger\cM\big)
+ \mhbm{F}^\upgamma_m \dK
\nonumber\\[1ex]
&\ =\ 
\bigg(\cD\chi^\upgamma_m -\frac12\omega^{\alpha\beta}\fJ_\alpha \fJ(m)_\beta 
+m \sum_{n=1}^\infty \tP_n (\chi^\upgamma_{m+n} - \chi^\upgamma_{m-n})
+\braket{\chi^\upgamma_m}{\cF}
+\hbm{F}^\upgamma_m + \mhbm{F}^\upgamma_m
\bigg)\dK
\nonumber\\[1ex]
&\ \eqqcolon\  \mf{X}^\ord{m} \, \dK\,.
\label{completed shifted CM eq J}
\end{align}
In the first line we have used that the $\virm$ components of the Maurer--Cartan form are proportional to the $\vir$ components of $\cP$: $\tP_{\pm n}=\frac12 (\cD\Gamma\Gamma^{-1})_{-n}$ for $n\neq0$.
The internal shifted current $\bra{\cJ^\ord{m}}$ is defined in \eqref{shifted internal J}.
Because the left-hand side of this equation transforms by conjugation under rigid $\hE8\rtimes\Virm$, so does the right-hand side, which is therefore invariant.

\subsection{Gauge invariant topological term}

We will now show that only $\mf X^\ord{1}$ can be used to construct a pseudo-Lagrangian.
First, we notice that by $\cM$ conjugation $\mf X^\ord{m} = \mf X^{\ord{-m}}$ and hence we focus on $m>0$ ($\mf X^\ord{0}=0$ identically).
Then, we stress that all $\mf X^\ord{m}$ have weight 0 under the generalised Lie derivative:
\begin{equation}
\cL_{\mathbbm\Lambda} \mf X^\ord{m} = \braket{\partial_\mf{X}}{\Lambda} \mf{X}^\ord{m}\,.
\end{equation}
Even assuming a covariant transformation property for $\mf X^\ord{m}$ under generalised diffeomorphisms (so that their variation equals the generalised Lie derivative, which we will prove for $\mf X^\ord{1}$ below), we need objects of weight 1 to be able to integrate out the transport term above.
Our candidate pseudo-Lagrangians are therefore of the form $\rho \mf X^\ord{m}$.
Knowing this, we must check that the Euler--Lagrange equations for the $\chi^\upgamma$ forms and the $B$ fields vanish upon imposing twisted self-duality.
The $B$-field variation is more involved and will in fact require an extra correction to the action, but the $\chi^\upgamma$ variation is straightforward.
Using \eqref{opposite chi} to relate $\chi^\upgamma$ forms associated to opposite shifts and then variying with respect to $\chi^\upgamma_m$, $m>0$, we find
\begin{equation}
\delta(\rho\mf X^\ord{m}) \propto
\tP_0\,\delta\chi^\upgamma_{m}
+m\bigg(\ 
\sum_{p>m}\tP_{p-m}\,\delta\chi^\upgamma_p
-\sum_{p=1}^{m-1}\tP_{m-p}\,\delta\chi^\upgamma_p
-\sum_{p>0}\tP_{m+p}\,\delta\chi^\upgamma_p
\ \bigg)
\end{equation}
and one sees for instance that the $\delta\chi^\upgamma_1$ term is proportional to $\tP_{m-1}+\tP_{m+1}$ for any $m\ge2$, which does not vanish under \eqref{twsd covariant P}.
Instead, for $m=1$ this expression simplifies to 
\begin{equation}\label{chigamma eom}
\sum_{p=1}^\infty(\tP_{p-1}-\tP_{p+1})\delta\chi^\upgamma_p\,,
\end{equation}
which vanishes upon imposing twisted self-duality.
This indicates that the correct topological term should be based on $\mf X^\ord{1}$.%
\footnote{\label{foot:chilagrange}%
Rewriting the Euler--Lagrange equations \eqref{chigamma eom} in terms of $\cD\Gamma\Gamma^{-1}$ one finds an expression analogous to \eqref{dGamma as Vir series}, with covariantised derivatives and $\cD\tilde\rho$ in place of $\star\cD\rho$ (and $\phi_1\equiv2\tilde\rho$).
This shows that the one-forms $\chi^\upgamma_m$ are Lagrange multipliers imposing that $\upgamma(w)$, as defined in \eqref{upgamma as series}, reduces to \eqref{sqrt expansion} and hence essentially to the function used in the Breitenlohner--Maison linear system.
Lagrangian mechanics does not generally allow integrating out one field using the Euler--Lagrange equations of another, hence we cannot naively substitute the relations descending from \eqref{chigamma eom} directly into the pseudo-Lagrangian.%
}

We now prove gauge invariance of $\rho\mf{X}^\ord{1}$ (up to a total internal derivative).
The first three terms in \eqref{completed shifted CM eq J} are manifestly covariant under generalised diffeomorphisms.
Let us focus on $\hbm{F}^\upgamma_1$ and compute its non-covariant variation
\begin{equation}
\Delta_\mathbbm{\Lambda} = \delta_\mathbbm{\Lambda} - \cL_\mathbbm{\Lambda}\,.
\label{noncov var def}
\end{equation}
Because the field strengths transform with the extended $\circ$ product \eqref{eq:DorfmanProductex} as in \eqref{eq:fsvar}, one has
$\Delta_\mathbbm{\Lambda} \mathbbm F = \mathbbm\Lambda \circ \mathbbm F - \cL_\mathbbm{\Lambda}\mathbbm F$.
Substituting into \eqref{Falpha}, \eqref{hatF gamma definition} and \eqref{mhbm F def}, after some work we find
\begin{align}
\Delta_\mathbbm{\Lambda} \sbm F_\alpha &=
-\sbm\Lambda_\alpha \braket{\partial_{\mathbbm\Lambda}}{\cF}\,,
\label{sbmF noncov var}
\\[1ex] 
\Delta_\mathbbm{\Lambda} \hbm{F}^\upgamma_1 &= 
- \, \hbm{\Lambda}^\upgamma_1 \,\braket{\partial_{\mathbbm\Lambda}}{\cF}
+ \rho^{-1} \bra{\partial_\Sigma} \Sigma^\ord{1} \ket\cF 
- \rho^{-1} \Tr(\Sigma^\ord{1}) \braket{\partial_\Sigma}{\cF}\,,
\label{hbmF noncov var}
\\[1ex] 
\Delta_\mathbbm{\Lambda} \mhbm{F}^\upgamma_1 &= 
- \mhbm{\Lambda}^\upgamma_1 \braket{\partial_{\mathbbm\Lambda}}{\cF}
\label{mhbmF noncov var}
\end{align}
where $\hbm{\Lambda}^\upgamma_1 $ and $\mhbm{\Lambda}^\upgamma_1 $  are defined as in \eqref{hatF gamma definition} and \eqref{mhbm F def} but with $(|\Lambda\rangle ,\,\Sigma^\ord{k})$ in place of $(\ket\cF,\,\cG^\ord{k})$.
We prove these results in Appendix~\ref{app:fstrvar}.
Notice that in the last two lines the partial derivatives act on $\mathbbm\Lambda = ( |\Lambda\rangle ,\Sigma^\ord{k})$ contained within $\hbm{\Lambda}^\upgamma_1$ and $\mhbm{\Lambda}^\upgamma_1$, but not on the scalar fields.
We arrive at the result
\begin{equation}
\Delta_\mathbbm{\Lambda} \mf{X}^\ord{1} = 
\Big(\Delta_{\mathbbm\Lambda}\bra{\chi^\upgamma_1}
- \big(\hbm{\Lambda}^\upgamma_1 + \hspace*{-2pt}\mhbm{\Lambda}^\upgamma_1\big) \bra{\partial_{\mathbbm\Lambda}}
+ \rho^{-1} \bra{\partial_\Sigma} \Sigma^\ord{1} 
- \rho^{-1} \Tr(\Sigma^\ord{1}) \bra{\partial_\Sigma}
\Big)\ket\cF\,.
\end{equation}
Notice that the entire expression in the parenthesis satisfies the section constraint.
This means that we can define the generalised diffeomorphism transformation of $\bra{\chi^\upgamma_1}$ to cancel the above variation:
\begin{equation}
\Delta_{\mathbbm\Lambda}\bra{\chi^\upgamma_1} =
\big(\hbm{\Lambda}^\upgamma_1 + \hspace*{-2pt}\mhbm{\Lambda}^\upgamma_1\big) \bra{\partial_{\mathbbm\Lambda}}
- \rho^{-1} \bra{\partial_\Sigma} \Sigma^\ord{1} 
+ \rho^{-1} \Tr(\Sigma^\ord{1}) \bra{\partial_\Sigma}\,.
\label{internal chi gamma noncov var}
\end{equation}
With this definition, we have that $\rho\mf{X}^\ord{1}$ is $\hE8\rtimes\Virm$ invariant and generalised diffeomorphism invariant up to a total internal derivative.
If we follow the same approach for $\rho\mf{X}^\ord{m}$, $m>1$, we find that $\Delta_{\mathbbm\Lambda}\hbm{F}^\upgamma_m$ contains terms that cannot be reabsorbed into the variation of $\bra{\chi^\upgamma_m}$, and therefore $\rho\mf{X}^\ord{m}$ is not invariant under generalised diffeomorphisms.
This is also described in Appendix~\ref{app:fstrvar}.
We again conclude that only $\rho\mf{X}^\ord{1}$ is a suitable candidate for our topological term.

One last cross-check on the suitability of $\rho\mf{X}^\ord{1}$ as a candidate pseudo-Lagrangian is that the two-forms appearing in the covariant field strengths \eqref{eq:fieldstrengths} must not contribute.
These terms can potentially appear in the $\hbm{F}^\upgamma_{\pm1}$ contributions to $\mf{X}^\ord{1}$, but a direct computation shows that they vanish up to a total internal derivative.
To see this, we start by recalling the definition \eqref{hatF gamma definition} and use \eqref{shiftgamma generic series} to write $\cS^\upgamma_m(L_n)$ as a series of $\vir$ generators $L_k$ with $k\le m+n$.
Then, we see that $\hbm{F}^\upgamma_{-1}$ (which appears through \eqref{mhbm F def}) only involves contractions with negative $\vir$ generators and hence no trivial parameters contribute (such traces appear in the $\dK$ component of $\sbm{F}_\alpha$ and therefore anything that contributes to them is by definition not trivial).
A similar argument shows that the two-forms included in $\cG^\ord{k}$, $k\ge2$ do not contribute to $\hbm{F}^\upgamma_{1}$,
and that contributions from the $\ket\cF$ dependent term drop out as total derivatives.
One is then left with evaluating the two-form contributions of $\Tr\big( \cG^\ord{1} L_0 \big)$.
One immediately sees that the two-forms associated to the trivial parameters \eqref{eq:triv1} to \eqref{eq:triv3} appear only through a total derivative and hence are discarded.
The parameters \eqref{eq:triv4} to \eqref{eq:triv6} do not involve any derivative and direct calculation shows their contribution vanishes.
We exemplify this for \eqref{eq:triv4}.
Using the symmetries of the $U$ parameter, its contribution to $\hbm{F}^\upgamma_1$ is proportional to
\begin{align}
& \eta_{1\,\alpha\beta} \big(
    \bra{\pi_1}[L_0,\,T^\alpha]\ket{U_1} 
    \bra{\pi_2} T^\beta \ket{U_2}
  + \bra{\pi_1} T^\alpha \ket{U_1} 
    \bra{\pi_2} [L_0,\,T^\beta] \ket{U_2}
 \big) = 
-\eta_{1\,\alpha\beta} 
  \bra{\pi_1} T^\alpha \ket{U_1} 
  \bra{\pi_2} T^\beta \ket{U_2}
\end{align}
which vanishes by the section constraint since $U$  is symmetric in $\bra{\pi_1}$ and $\bra{\pi_2}$.
The general result can be deduced from equations \eqref{plustwoforms}, \eqref{Ahat triv1} and \eqref{Ahat triv2}.
Again, it is straightforward to see that two-form independence does not hold for $\rho \mf{X}^\ord{m}$ for any $m\ge2$.

The expression $\rho\mf{X}^\ord{1}$ is not the final topological term.
The $\vir$ components of $\cP$ defined in \eqref{P from J} are invariant under rigid $\hE8\rtimes\Virm$ and transform as scalars under generalised diffeomorphisms, hence any bilinear in these components (times $\rho$) defines an invariant action.
This apparent ambiguity is fixed by computing the Euler--Lagrange equations of $\rho$ and the $B$ fields, which we will do in Section~\ref{sec:cov eom}.
For now, we just claim that one such contribution is necessary to reproduce the correct equations of motion, giving
\begin{equation}
\nonumber
\cL_{\text{top}}\ =\ 
  \rho\mf{X}^\ord{1} 
- \rho \frac{c_\vir}{6}\sum_{n=2}^\infty(n^3-n)\tP_n(\tP_{n+1}+\tP_{n-1}) \,,
\end{equation}
{so that explicitly}
\begin{align}
\label{final cov Lagrangian J}
\rho^{-1}\cL_{\text{top}}\ &=\ 
  \cD\chi^\upgamma_1 
- \frac12\omega^{\alpha\beta}\fJ_\alpha \fJ(1)_\beta 
-  \frac{c_\vir}{6}\sum_{n=2}^\infty(n^3-n)\mathsf{P}_n(\tP_{n+1}+\tP_{n-1}) 
\CR&\hphantom{=\ }
+  \sum_{n=1}^\infty \tP_n (\chi^\upgamma_{1+n} - \chi^\upgamma_{1-n})
+ \braket{\chi^\upgamma_1}{\cF}
+ \hbm{F}^\upgamma_1 
+  \mhbm{F}^\upgamma_1\ .
\end{align}

There are several $c_\vir$ dependent couplings in this pseudo-Lagrangian.
Beyond the explicit term we have just added in the first line, the $\omega^{\alpha\beta}$ cocycle contains a $\vir$ component, and a similar cocycle is contained within $\cD\chi^\upgamma_1$:
\begin{equation}\label{Dchigamma}
\cD\chi^\upgamma_1 = 
  (\dd - \braket{\partial_\chi}{A})\chi^\upgamma_1
+ \omega^{\alpha\beta}\sbm A_\alpha \fJ(1)_\beta \,. 
\end{equation}
This reflects the fact that the definition of $\chi^\upgamma_m$ depends on the choice of representation in which we write $\fJ$.
We will see in the next section that all $c_\vir$ dependent couplings cancel out when we rewrite the pseudo-Lagrangian in terms of $\tilde\chi_m$ defined in \eqref{shifted P and tilde chi def}, whose transformation properties are $c_\vir$ independent.

\subsection{Unendlichbein formulation}\label{ssec:unend}

We can reformulate \eqref{final cov Lagrangian J} in terms of the Hermitian projection $\cP$ of the Maurer--Cartan form and the $\tilde\chi_m$ forms introduced in \eqref{shifted P and tilde chi def}.
In order to do so, we first provide more details on how several objects we need are defined and how they transform.
The coset representative \eqref{full coset repr} transforms as follows under infinitesimal rigid $\hE8\rtimes\Virm$ transformations
\begin{equation}
\delta^\alpha \cV = \cV T^\alpha + h^\alpha \cV\,,\qquad
h^\alpha \in K(\mf{e}_9)\,,
\end{equation}
where $h^\alpha$ is a local compensating transformation.
The Maurer--Cartan form and its projections \eqref{MC form and P Q def} are then expanded as%
\footnote{Notice that the local compensating transformation $h^\alpha$ satisfies $(h^\alpha)^\dagger=-h^\alpha$ and does not take values along $\vir$. On the other hand, $\cQ$ also takes values in the anti-Hermitian part of $\vir$.}
\begin{align}
\cD\cV\cV^{-1} &= 
  \dd \cV\cV^{-1} 
- \big(\braket{\partial_\cV}{A} \cV\big)\cV^{-1}
- \sbm A_\alpha \big( \cV T^\alpha \cV^{-1} + h^\alpha \big)
\label{MC def}
\,,\\[1ex]
\cP &= 
  \frac12\big( \dd \cV\cV^{-1}  +{\rm h.c.}\big)
- \braket{\cP_\alpha}{A}\,T^\alpha
- \sbm A_\alpha \frac12 \big( \cV T^\alpha \cV^{-1} +{\rm h.c.}\big)
\label{cP def}
\,,\\[1ex]
\cQ &= 
  \frac12\big( \dd \cV\cV^{-1}  - {\rm h.c.}\big)
- \braket{\cQ_\alpha}{A}\,T^\alpha
- \sbm A_\alpha \frac12 \big( \cV T^\alpha \cV^{-1} -{\rm h.c.}\big)
-\sbm A_\alpha h^\alpha
\,,
\label{cQ def}
\end{align}
where $\bra{\cP_\alpha}$ and $\bra{\cQ_\alpha}$ are the projections of the internal Maurer--Cartan form 
$(\partial_M\cV)\cV^{-1} \bra{e^M}$ which appears in the first line contracted with $\ket A$.
Taking the differential of the above expressions we find the gauged Maurer--Cartan equations%
\footnote{Notice that the rigid $\hevirm$ variation of $h^\alpha$ cannot be entirely specified without reference to the specific gauge choice of $\cV$ and associated Killing vectors.
Only the antisymmetrised part admits a covariant algebraic expression 
$2\delta^{[\alpha}h^{\beta]}=f^{\alpha\beta}{}_\gamma h^\gamma +[h^\alpha,\,h^\beta]$.
This can be used for instance to compute $\delta^\alpha(\cD\cV\cV^{-1}) = \cD h^\alpha + [h^\alpha,\,\cD\cV\cV^{-1}]$ as expected, and hence the second covariant differential acting on $\cD\cV\cV^{-1}$ is well-defined.
On the other hand, to derive \eqref{PQ MC eqs} we simply use $\cD^2=-\cL_{\mathbbm F}$ so that no $\cD h^\alpha$ term appears and we do not need knowledge of $\delta^\alpha h^\beta$.
This is of course standard in computing gauged Maurer--Cartan equations on coset spaces.
}
\begin{align}
\cD\cP -[\cQ,\,\cP] &= 
- \braket{\cP_\alpha}{\cF}\,T^\alpha
- \sbm F_\alpha \frac12 \big( \cV T^\alpha \cV^{-1} +{\rm h.c.}\big)\,,\\[1ex]
\cD\cQ-\frac12[\cQ,\,\cQ]-\frac12[\cP,\,\cP] &=
- \braket{\cQ_\alpha}{\cF}\,T^\alpha
- \sbm F_\alpha \frac12 \big( \cV T^\alpha \cV^{-1} -{\rm h.c.}\big)
-\sbm F_\alpha h^\alpha\,.
\label{PQ MC eqs}
\end{align}

With this information we can in principle repeat the computation of the shifted Maurer--Cartan equation of the previous section in terms of $\cP$ rather than $\fJ$.
More simply, we can take the left-hand side of \eqref{completed shifted CM eq J} (with $m=1$), conjugate the expression with $\cV$ and use \eqref{P from J} and \eqref{shifted P and tilde chi def} to find
\begin{align}
\mf{X}^\ord{1} =\ &
  2\cD\tilde\chi_1
+ 2\omega^{\alpha\beta} \cQ_\alpha \fP(1)_\beta
+ 2\sum_{n=1}^\infty \tP_n (\tilde\chi_{n+1} - \tilde\chi_{n-1})
\CR&
+ 2\braket{\tilde\chi_1}{\cF}
+ \hbm{F}^\upgamma_1
+ \hbm{F}^\upgamma_{-1}
+ \omega^\alpha(\cV) \, 
  \big(
   \sbm F^{(1)}_\alpha + \sbm F^{(-1)}_\alpha 
  \big) \ ,
\label{X1 in terms of P}
\end{align}
where $\bra{\tilde\chi_1}$ is the internal equivalent of the one-form $\tilde\chi_1$ and completes the $\dK$ component of $\bra{\cP_\alpha}\otimes\cS_1(T^\alpha)$.
It is related to $\bra{\chi^\upgamma_1}$ just like its one-form siblings in \eqref{shifted P and tilde chi def}.
The last three terms in the second line are the $\dK$ completions of $\sbm F_\alpha \cS_{\pm1}(\cV T^\alpha \cV^{-1})$.
It is important to notice that in this expression the $h^\alpha$ compensator cancels out. 
To see this, we expand the covariant derivative of $\tilde\chi_1$:
\begin{equation}\label{Dchitilde}
\cD\tilde\chi_1 = 
  \dd\tilde\chi_1 
- \braket{\partial_{\tilde\chi}}{A}\tilde\chi_1 
-\sbm A_\alpha \,(h^\alpha)_\beta\, \omega^{\beta\delta} \fP(1)_\delta
\end{equation}
where $h^\alpha = (h^\alpha)_\beta T^\beta$.
This reflects the fact that $\tilde\chi$ transforms under $K(E_9)$ such that 
\begin{align}
(h^\alpha)_\beta \delta^\beta_{K(\mf{e}_9)}\fP(1)=[h^\alpha,\,\fP(1)]\,.
\end{align}
The cocycle term in $\cD\tilde\chi_1$ then cancels out with the $h^\alpha$ contained in the $\cQ$ connection in the second term.
In fact, the entire first line of \eqref{X1 in terms of P} can be identified with the generalised diffeomorphism and local $K(E_9)$ covariant derivative of $\tilde\chi_1$.
To see this, we can rewrite the last term of the first line as 
$- 2\sum_{n\in\mathds{Z}} \tQ_n \tilde\chi_{1+n}$,
using the relation $\tQ_n = -\mathrm{sgn}(n)\tP_n$, which is valid because the Maurer--Cartan form does not take values along the positive levels of the $\vir$ algebra.%
\footnote{Such sum over $\tQ_n$ appears in the $K(E_9)$ covariant derivative of $\tilde\chi_1$ because constant shift operators $\cS_m$ do not commute with $\vir$ and hence the $\dK$ completions of $\cS_m(\fP)$ must transform equivalently to $w^m$ under the Virasoro components of $\cQ$.}

We anticipated in \eqref{final cov Lagrangian J} that in order to write the full topological Lagrangian we need to subtract a $c_\vir$ dependent term quadratic in the $\vir$ components of $\fP$.
We can now give a better description of such a subtraction.
The algebra cocycle in the first line of \eqref{X1 in terms of P} contains a Virasoro component according to the definition \eqref{algebra cocycle}, and using again that $\tQ_n = -\mathrm{sgn}(n)\tP_n$ we can rewrite it as a term quadratic in $\fP$:
\begin{equation}
-\frac{c_\vir}{12} \sum_{n\in\mathds{Z}} (n^3-n) \,\tQ_n \,\tP_{n+1} =
\frac{c_\vir}{12} \sum_{n=2}^\infty (n^3-n)\, \tP_n \,(\tP_{n+1}+\tP_{n-1})\,.
\label{QP vir cocycle to PP}
\end{equation}
As stated in \eqref{final cov Lagrangian J} and proved below in Section~\ref{sec:cov eom} this is exactly the term that must be removed from the pseudo-Lagrangian in order to reproduce the correct equations of motion.
We can therefore rewrite \eqref{final cov Lagrangian J} as follows%
\begin{align}
\label{final cov Lagrangian P}
\rho^{-1}\cL_{\text{top}}\ =\ &
  2\cD\tilde\chi_1
- 2\eta^{AB}\sum_{n\in\mathds{Z}} n\,\tQ_A^n\,\tP_B^{-n-1}
+ 2\sum_{n=1}^\infty \tP_n (\tilde\chi_{n+1} - \tilde\chi_{n-1})
\\\nonumber&
+\, 2\braket{\tilde\chi_1}{\cF} \,
+\, \hbm{F}^\upgamma_1 \,
+\, \hbm{F}^\upgamma_{\!-1} \,
+\, \omega^\alpha(\cV) \, 
  \big(
   \sbm F^{(1)}_\alpha + \sbm F^{(-1)}_\alpha 
  \big)  \ .
\end{align}

Notice that in this expression there are no $c_\vir$ dependent couplings.
The group cocycle in the second line does not generate $c_\vir$ dependent terms because the Virasoro components of $\sbm F_\alpha$ only run along $\virm$.
Hence, the pseudo-Lagrangian is independent of the representation in which $\fP$ is defined.
This reflects the fact that the $\tilde\chi_m$ only transform under local $K(E_9)$ transformations and so their transformation does not include any $\vir$ cocycle.
This must be contrasted with the $c_\vir$ dependent transformation of $\chi^\upgamma_m$.
In fact, the whole first line of \eqref{final cov Lagrangian P} can be regarded as the generalised diffeomorphisms and $K(E_9)$ covariant derivative of $\tilde\chi_1$ at $c_\vir=0$, a fact that will greatly simplify the computation of the scalar field equations of motion.
The reason $c_\vir$ does not cancel out from \eqref{final cov Lagrangian J} is that the relation \eqref{shifted P and tilde chi def} between $\tilde\chi_1$ and $\chi^\upgamma_1$ is based on a group cocycle that is itself $c_\vir$ dependent.
Such a dependence also disappears if we use generalised diffeomorphisms to gauge fix $\phi_n\to0$ for $n\ge2$. 
This is proved explicitly in Appendix~\ref{sec:gfix details}.

\subsection{\texorpdfstring{$\Virm$ gauge fixing and full pseudo-Lagrangian}{Vir- gauge fixing and full pseudo-Lagrangian}}\label{sec:gfix}

The coset representative \eqref{full coset repr} contains exponentials of all the negative Virasoro generators.
By contrast, the $E_9$ exceptional field theory scalar potential we constructed in \cite{Bossard:2018utw} only includes exponentials of $L_0$ and $L_{-1}$. 
We will not attempt here to generalise the scalar potential of \cite{Bossard:2018utw} to include the other Virasoro fields.
Instead, we will use the gauge freedom associated to $\Tr(B^\ord{n})$ to set 
\begin{equation}
\phi_n \to 0\,,\ \ n\ge2\,.
\label{virm gauge fixing}
\end{equation}
We have stressed in Section~\ref{sec:extcoset} that \eqref{virm gauge fixing} does not solve the two-dimensional twisted self-duality constraint.
In exceptional field theory, this is remedied by the St\"uckelberg coupling of $\phi_n$ to $\Tr(B^\ord{n})$ through the covariant derivative.
This means that we can impose \eqref{virm gauge fixing} at the price of keeping $\Tr(B^\ord{n})\neq0$ in order to respect the $\vir$ components of \eqref{twsd cov J}, as each and all $\vir$ components of $\fJ$ must be dual to $\cD\rho$ and hence generally non-zero.
This has an important consequence.
The gauge-fixed residual generalised diffeomorphisms are generated by $\ket\Lambda$ and $\Sigma^\ord{1}$ (because traceless $\Sigma^{\ord n}$, $n\ge2$ parameters can be reabsorbed into $\Sigma^\ord{1}$ up to trivial parameters), taking the form described in Section~\ref{sec:gendiffeo} and familiar from the previous papers \cite{Bossard:2017aae,Bossard:2018utw}.
However, the $B^\ord{n}$ fields and their field strengths $\cG^\ord{n}$ are non-vanishing for any $n>0$.

We choose to keep $\phi_1\neq0$. 
In fact, we shall substitute $\phi_1\to\tilde\rho$ in the rest of this section, arriving at the coset representative
\begin{equation}
\cV = 
\rho^{-L_0} \, e^{-\tilde\rho L_{-1}} \,
\mathring{V}\,e^{Y_{1A} T^A_{-1}}\,e^{Y_{2A} T^A_{-2}} \, \cdots\,e^{-\sigma\dK}\,,
\label{vir gauge fixed coset repr}
\end{equation}
which matches the one we used in \cite{Bossard:2018utw}.
The duality equation between $\rho$ and $\tilde\rho$ as it descends from \eqref{twsd cov J} reads, expanding the covariant derivatives,
\begin{equation}
2\star\!(\dd-\braket{\partial_\rho}{A}-\braket{\partial_A}{A})\rho 
=
(\dd-\braket{\partial_{\tilde\rho}}{A}-\braket{\partial_A}{A})\tilde\rho 
- \Tr(B^\ord{1})\,.
\end{equation}
The factor of $2$ on the right-hand side does not match the two-dimensional relation \eqref{tilde rho duality ungauged} if we drop internal derivatives and $B$ fields, reflecting the fact that this is really the covariantisation of the duality relation between $\rho$ and $\phi_1$ given in \eqref{vir dualities}, but again this is remedied here by $\Tr(B^\ord{1})$ which trivialises the relation between $\rho$ and $\tilde\rho$ and hence also the distinction between $\tilde\rho$ and $\phi_1$.

The gauge fixing reduces $\upgamma(w)$ to the finite expression
\begin{equation}
\upgamma(w) \to \frac{w-\tilde\rho}{\rho}\,.
\label{gauge fixed upgamma}
\end{equation}
Comparing with the expression \eqref{BMgamma} appearing in the two-dimensional linear system, we see that the gauge-fixed function is missing the square root. 
Reproducing the square root in terms of exponentials of Virasoro generators is what had made it necessary to introduce the $\phi_n$ fields and their duality relations, but again this is now compensated for by the presence of the $B$ fields.
We can use \eqref{gauge fixed upgamma} to expand $\cS^\upgamma_m$ in terms of the constant $\cS_m$ as binomial series.
In particular,
\begin{align}
\cS^\upgamma_1 &\to \rho^{-1} \left( \cS_1 - \tilde\rho\, \cS_0 \right)\,,\qquad
\cS^\upgamma_{-1} \to
\rho \sum_{n=0}^\infty \tilde\rho^n \cS_{-1-n}\,.
\label{gfix shift op expansion}
\end{align}
Notice that the series on the right-hand side appears in the scalar potential we found in \cite{Bossard:2018utw}, acting on the internal current $\bra{\cJ_\alpha}$.
We thus see that $\cS^\upgamma_{-1}$ is the $\virm$ generalisation of that expression (up to an overall factor of $\rho$).

In order to match the expression of the scalar potential and its field content we also need to relate $\bra{\chi^\upgamma_1}$ to the $\bra\chi$ field appearing in \cite{Bossard:2018utw}.
Up to a factor of $\rho$ (which is just due to different weight assignments), the latter gives the $\dK$ completion of the (gauge fixed) $\cS^\upgamma_{-1}$ shift of the internal current, written in terms of the series \eqref{gfix shift op expansion}.
Indeed we defined there
\begin{equation}\label{Jminus def}
\bra{\cJ^{-}_\alpha}\otimes T^\alpha =
\bra{\cJ_\alpha}\,\otimes\,\sum_{n=0}^\infty \tilde\rho^n \cS_{-1-n}(T^\alpha)
+\bra\chi\otimes\dK\,,
\end{equation}
which equals $\rho^{-1}\bra{\cJ^\ord{-1}_\alpha}$ and transforms as an algebra valued object under rigid $\hat{\mf{e}}_8\oleft \langle L_{-1}\rangle$.
To match this relation we perform the field redefinition
\begin{equation}
\label{eq:chiredef}
\bra{\chi^\upgamma_1} 
= \rho \bra{\chi} -2\bra{P_1} - \rho \, \omega^\alpha(\cM) \bra{\cJ^-_\alpha }\,.
\end{equation}
where $\bra{P_1}= -\frac12\rho^{-1}\bra\partial\tilde\rho$.
The presence of $\bra{P_1}$ has no effect on rigid $\hat{\mf{e}}_8\oleft \langle L_{-1}\rangle$ covariance, instead it is motivated by matching the transformation properties of $\bra\chi$ under generalised diffeomorphisms as found in \cite{Bossard:2018utw}.
The non-covariant variation of $\bra{\chi^\upgamma_1}$ is given in \eqref{internal chi gamma noncov var}.
Those of $\bra{P_1}$ and $\bra\cJ$ were derived in \cite{Bossard:2018utw} and we reproduce them here:
\begin{align}
\Delta_\mathbbm{\Lambda} \bra{\cJ_\alpha}\otimes T^\alpha &=
\sbm\Lambda_\alpha \big(T^\alpha +\cM^{-1} (T^\alpha)^\dagger \cM \big) \bra{\partial_{\mathbbm\Lambda}}\,,\\[1ex]
\Delta_\mathbbm{\Lambda} \bra{P_1}  &= 
- \frac{\tilde\rho}{2\rho} \braket{\partial_\Lambda}{\Lambda}\bra{\partial_\Lambda}
- \frac{1}{2\rho}\Tr(\Sigma^\ord{1})\bra{\partial_\Sigma}\,.
\end{align}
Combining these relations with $\Delta_\mathbbm{\Lambda}\omega^\alpha(\cM)=0$ and using relations and definitions analogous to \eqref{hatF gamma definition}, \eqref{shifted F M conj} and \eqref{mhbm F def} for $\mathbbm\Lambda$, we then find
\begin{equation}
\Delta_\mathbbm{\Lambda} \bra{\chi} = 
\frac1\rho \big(\hbm\Lambda^\upgamma_{-1} + \mhbm{\Lambda}^\upgamma_{-1}\big)
  \bra{\partial_{\mathbbm\Lambda}}
- \frac{1}{\rho^2}\bra{\partial_\Sigma}\Sigma^\ord{1}
- \frac{\tilde\rho}{\rho^2} \braket{\partial_\Lambda}{\Lambda}\bra{\partial_\Lambda}\,.
\label{brachi noncov trf}
\end{equation}
We show in Appendix~\ref{sec:gfix details} that this transformation property matches the one we derived in \cite{Bossard:2018utw}.

The final pseudo-Lagrangian density, in form notation, gauge-fixed to \eqref{virm gauge fixing}, reads
\begin{equation}\label{Ltop-pot}
\cL_{\text{ext}} = \cL_{\text{top}} -\star V
\end{equation}
with the topological term expanded as follows%
\footnote{Alternatively, one can use the more compact Unendlichbein formulation \eqref{final cov Lagrangian P}, together with equation (4.116) of \cite{Bossard:2018utw}.
In this case, notice that $\bra{\tilde\chi}$ as defined there equals $2\bra{\tilde\chi_1}+2\bra{P_1}$ as defined here.
}
\begin{align}
\rho^{-1} \cL_{\text{top}} =\ &
  \dd\chi^\upgamma_1
- \braket{\partial_\chi}{A}\chi^\upgamma_1
+ \frac{1}{2\rho}\eta^{AB} \sum_{n\in\mathds{Z}} n\, 
    (\tJ_A^n + 2 \sbm{A}^n_A) \, 
    (\tJ^{-n-1}_B - \tilde\rho \tJ^{-n}_B) \CR&
- \frac12\rho^{-1} \cD\tilde\rho  \,(\chi^\upgamma_{2} - \tJ_\dK) 
+ \frac12 \sum_{n=2}^\infty \rho^{-n} \sum_{k=0}^{n-2}
    \binom{n-2}{k}\,(-\tilde\rho)^k \, \Tr(B^\ord{n-k}) 
    (\chi^\upgamma_{1+n} - \chi^\upgamma_{1-n})  
\nonumber\\[1ex]&
+ \braket{\,\rho\,\chi-2 P_1-\rho\,\omega^\alpha(\cM) \cJ^-_\alpha\,}{\cF\,}
\, +\, \hbm{F}^\upgamma_1
\, +\, \mhbm{F}^\upgamma_1\,,
\label{final cov Lagrangian J GAUGEFIX}
\end{align}
where $\tJ_A^n$ are the loop components of the covariant current \eqref{fJ components}, 
$\sbm{A}_A^m$ are the loop components of $\sbm A_\alpha$ defined analogously to \eqref{eq:balphaex}, 
$\cD\tilde\rho = \dd\tilde\rho-\bra{\partial}( \tilde\rho\,\ket{A}) - \Tr(B^\ord{1})$,
the cocycle $\omega^\alpha(\cM)$ is defined in \eqref{eq:omegaalpha}, the internal shifted current $\bra{\cJ^-_\alpha}$ is defined in \eqref{Jminus def}, and finally $\hbm{F}^\upgamma_1$ and $\mhbm{F}^\upgamma_1$ are defined in \eqref{hatF gamma definition} and \eqref{mhbm F def}.
The $c_\vir$ dependent couplings in \eqref{final cov Lagrangian J} cancel out with the gauge fixing as argued at the end of the previous section and proved explicitly in Appendix~\ref{sec:gfix details}.
The scalar potential $V$ comes from \cite{Bossard:2018utw}:%
\begin{align}
V =\ & 
   \frac{1}{4\rho} \sum_{n\in\mathds{Z}}\sum_{k=0}^\infty \tilde\rho^k 
      \eta^{AB} \bra{J^{k-n}_A} \cM^{-1} \ket{J^n_B}
- \frac{1}{\rho}\bra{P_0}\cM^{-1}\ket{J_\dK} \CR&
-\frac2\rho \bra{P_1}\cM^{-1}\ket{\rho\,\chi-P_1}
- \omega^{\alpha}(\cM)
      \bra{P_1}\cM^{-1}\ket{ \cJ^-_\alpha}    \vphantom{\sum_n}\CR&
- \frac{1}{2\rho} \bra{\cJ_\alpha} T^\beta \cM^{-1} T^\alpha{}^\dagger \ket{\cJ_\beta}
+ \frac{\rho}{2} \bra{\cJ^-_\alpha} T^\beta \cM^{-1} T^\alpha{}^\dagger \ket{\cJ^-_\beta}   
+ \frac{1}{\rho} \bra{P_0} T^\alpha \cM^{-1} \ket{\cJ_\alpha}\,.
\label{scalpot}
\end{align}
We used $\ket{\cJ_\alpha} = (\bra{\cJ_\alpha})^\dag$ (and similarly for other objects) to simplify the notation, thus regarding $\ket{\cJ_\alpha}$ as an element of $R(\Lambda_0)_{+1}$.
The expression $\omega^{\alpha}(\cM) \bra{ \cJ^-_\alpha}$ read instead $\rho^{-2}\Omega^\alpha(\cM)\bra{\cJ_\alpha}$ in \cite{Bossard:2018utw}.
We define $\Omega^\alpha(\cM)$ and prove equivalence of the two expressions in Appendix~\ref{app:shift}.

Note that the topological term and the potential are both invariant under (internal) generalised diffeomorphisms. The relative coefficient between the two in \eqref{Ltop-pot} is fixed by requiring that, when partially soving the section constraint as in \eqref{eq:secsol}, the field equations for the various scalar fields reproduce those of $E_8$ exceptional field theory. This can for instance be verified by using the field equation~\eqref{chife} obtained by varying with respect to $\langle\chi|$. We expect that the coefficient could be alternatively fixed by imposing invariance under conformal external diffeomorphisms.
In the minimal formulation of Section~\ref{sec:E9EFT}, where the conformal gauge is relaxed, the relative coefficient between the topological term and the potential will be fixed (and confirmed) explicitly by requiring invariance under external diffeomorphisms.  

\subsection{Equations of motion}
\label{sec:cov eom}

We shall now take a look at the equations of motion of Virasoro-extended $E_9$ exceptional field theory.
These are obtained as the Euler--Lagrange equations of the pseudo-Lagrangian constructed in the previous sections, combined with the twisted self-duality condition \eqref{twsd cov J} (or equivalently \eqref{twsd covariant P}) and with the covariantisation of the Virasoro constraint \eqref{ungauged Vir constraint}
\begin{equation}\label{cov Vir constraint}
  \cD_\pm \sigma \cD_\pm \rho 
- \frac12 \cD_\pm\cD_\pm \rho
- \frac12 \rho\, \eta^{AB} (\tP_\pm)^0_A (\tP_\pm)^0_B 
\ =\  0 \ ,
\end{equation}
written here in light-cone coordinates.
Up to twisted self-duality it is invariant under rigid $\hE8\rtimes\Virm$ transformations and, by covariance, under generalised diffeomorphisms. 
This constraint must be imposed because we only wrote the covariant action in conformal gauge.
We will relax conformal gauge and see the Virasoro constraint arise as an Euler--Lagrange equation in Section~\ref{sec:E9EFT}. The equivalence of this Euler--Lagrange equation with \eqref{cov Vir constraint} is shown in Appendix~\ref{Virasoro}. 
In Appendix~\ref{app:scaleom match} we will exemplify how the equations of motion derived here reproduce those of $E_8$ exceptional field theory, by matching the equations of motion for the $E_8$ scalar fields.
A complete matching at the level of the action will be given in the minimal formalism of Section~\ref{sec:E9EFT}.
In computing the equations of motion we will not explicitly vary the scalar potential, which was already shown to match $E_8$ exceptional field theory in \cite{Bossard:2018utw}, and instead focus on the variation of the topological term.
These explicit computations are also meant to clarify some details of the $\Virm$ extended formalism, such as the need for the regularisation of certain series induced by the cocycle term in the action and motivating the addition of the extra invariant term in \eqref{final cov Lagrangian J} compared to $\rho \mf{X}^\ord{1}$.

\subsubsection*{The trivial variations}

We have already addressed the Euler--Lagrange equations for the one-forms $\chi^\upgamma_m$ (or equivalently $\tilde\chi_m$) in \eqref{chigamma eom}, which shows that these fields appear as Lagrange multipliers for a subset of the relations imposed by twisted self-duality, and hence their equations of motion are trivialised once the latter is imposed.
Let us now look at the field variations with respect to the $B^\ord{k}$ fields, which should also trivialise upon imposing twisted self-duality.
Without loss of generality we will also gauge fix 
\begin{equation}\label{full vir gfix}
\phi_n \to 0\,,\quad \forall\ n\ge1\,,
\end{equation}
just like we did in the previous section, but now also setting $\tilde\rho\to0$.
The scalar potential of course does not contribute to the variation and we will use \eqref{final cov Lagrangian P} as a starting point.
We therefore consider the variation
\begin{equation}\label{B var of bbA}
\delta\mathbbm A = (0,\,\delta B^\ord{k})\,.
\end{equation}
(We will consider the variation with respect to $\ket{A}$ below, starting from~\eqref{A var of bbA}.)

We have already stressed that the $K(E_9)$ compensator $h^\alpha$ appearing in \eqref{MC def}--\eqref{cQ def} and other expressions cancels out in the topological term.
To simplify the exposition, in this section we will then use the definitions \eqref{MC def}--\eqref{cQ def} and \eqref{Dchitilde}, but removing by hand the compensator
\begin{equation}\label{remove compensator}
h^\alpha \to 0\,,
\end{equation}
since it is guaranteed to cancel out anyway.
With this in mind, we write the $B$ field variation of the Maurer--Cartan form, of $\cP $ and of $\cQ$ as%
\footnote{Because we are ignoring the compensator, $\cD\cV\cV^{-1}$ is formally no longer valued in a parabolic subalgebra of $\hevirm$ but gains positive loop components proportional to $\sbm A_\alpha$. This is reflected in the following computations.}
\begin{align}
\delta(\cD\cV\cV^{-1}) &=
-\sbm*{\delta\mathbbm A}_\alpha \cV T^\alpha \cV^{-1} 
= 
-2 \underline{\sbm*{\delta\mathbbm A}}{}_\alpha T^\alpha
\,,\\[1ex]
\delta \cP &=
 -\underline{\sbm*{\delta\mathbbm A}}{}_\alpha (T^\alpha + \mathrm{h.c.})
\,,\\[1ex]
\delta \cQ &=
 -\underline{\sbm*{\delta\mathbbm A}}{}_\alpha (T^\alpha - \mathrm{h.c.})
 \,,
\end{align}
where we have used \eqref{remove compensator} and we introduced the underlined notation to denote dressing by $\cV$:
\begin{equation}\label{underlined}
\underline{\sbm \Lambda}{}_\alpha T^\alpha 
= 
\frac12\, \sbm\Lambda_\alpha \cV T^\alpha \cV^{-1} \,.
\end{equation}

To further simplify the exposition, we shall use rigid $E_9$ invariance to assume that the solution of the section contraint is of the form \eqref{eq:secsol}.
This guarantees that $\sbm A_\alpha$ takes values in $\virm$ and the non-positive loop levels only, the ones whose exponentials appear in our choice of coset representative \eqref{full coset repr}, except for an $\sbm A^{+1}_A$ component. 
In particular, the  $B^\ord{k}$ fields do not contribute to the $\dK$ component, namely
\begin{equation}
\Tr(B^\ord{k} L_{-k}) = 0\,.
\end{equation}
Furthermore, the $_A$ index of $\sbm A^{+1}_A$ satisfies the $E_8$ section constraint and reads explicitly
\begin{equation}
\sbm A^{+1}_C = \partial_C \braket{0}{A} = \partial_C w\,,
\end{equation}
where in a reduction from three to two dimensions $w$ corresponds to the Kaluza--Klein vector, see~\eqref{3Dmetric}.

Looking at \eqref{final cov Lagrangian P}, the variation of the last term of the first line clearly vanishes upon imposing \eqref{twsd covariant P}.
The only other contribution from the first line comes from the loop cocycle:
\begin{align}\label{cocycle B var}
-2 \rho \, \eta^{AB}\,
\delta\left(
  \sum_{n\in\mathds{Z}} n\,\tQ_A^n\,\tP_B^{-n-1}
\right)
=
  4 \rho\, \eta^{AB} \underline{\sbm*{\delta\mathbbm{A}}}{}^0_A 
     (\star \tP^0_B +  \underline{\sbm A}{}^{+1\,B} )
+4 \rho\, \eta^{AB} 
   \underline{\sbm*{\delta\mathbbm{A}}}{}^{-2}_A 
   \underline{\sbm A}{}^{+1}_B \ ,
\end{align}
where we have repeatedly used twisted self-duality and also used the fact that 
$\sbm*{\delta\mathbbm{A}}^0_A$ satisfies the $E_8$ section (because of the section constraint on $\delta B^\ord{k}$) to remove a contraction with a term $\underline{\sbm A}{}^{+1}_A $ (the coset dressing reduces to conjugation by $\mathring{V}$ which cancels out in the contraction).
The $\underline{\sbm A}{}^{+1}_B$ components appear when converting $\tQ^{\pm1}_A$ to $\tP^{\pm1}_A$ in order to use twisted self-duality on the latter.
Explicitly,
\begin{align}\label{Q to P}
\tQ^1_A = -\tP^1_A -2\underline{\sbm A}^{1}_A\,,\qquad
\tQ^{-1}_A = -\tP^{-1}_A +2\underline{\sbm A}^{1\,A}\,,\qquad
\tQ_A^n = -\mathrm{sgn}(n)\tP^n_A\,,\qquad |n|\ge2\,,
\end{align}
where the position of the $\mf{e}_8$ index in the middle equation is due to the $\mf{e}_8$ transposition $\underline{\sbm A}^{1\,A}= \eta^{AB}  \underline{\sbm A}^{1}_B$ and we could write more explicitly $\tQ^{-1}_A = -\tP^{-1}_A +2 \delta_{AB} \eta^{BC} \underline{\sbm A}^{1}_C$. We will use this notation throughout this section and in Appendix~\ref{app:virdetails}.

We now look at the variation of the second line of \eqref{final cov Lagrangian P}.
The $\braket{\tilde\chi_1}{\cF}$ term does not contain any $B$ fields so we can ignore it.
The other terms can be rewritten in a more compact form thanks to the gauge fixing \eqref{full vir gfix}, which allows us to reinterpret the $\omega^\alpha(\cV)$ terms as arising from dressing a $\vir$ generator with the coset representative, as discussed in Appendix~\ref{app:shift}.
This gives the convenient expression
\begin{align}
&\rho\,\hbm{F}^\upgamma_1 \,
+\rho\,\hbm{F}^\upgamma_{\!-1} \,
+\rho\,\omega^\alpha(\cV) \, 
  \big(
   \sbm F^{(1)}_\alpha + \sbm F^{(-1)}_\alpha 
  \big) \CR 
&\qquad \overset{\text{\eqref{full vir gfix}}}=\ 
- \rho\,\bra{\partial_\cF}\cV^{-1}(L_1+L_{-1})\cV\ket\cF
- \sum_{k=1}^\infty \rho^{1-k} \, \Tr\Big(
    \cG^\ord{k} \cV^{-1}(L_{1-k}+L_{-1-k})\cV
  \Big)\,.
\label{Ftildes}
\end{align}
In order to vary this expression we use closure of the generalised Lie derivative, \eqref{eq:Dorf1}, and \eqref{eq:DDex} to deduce the natural relation%
\footnote{We still use the definition \eqref{eq:Dorfcovdev} for $\mathbbm D$ even if the extended $\circ$ product does not satisfy the Leibniz identity. The trivial parameters on the right-hand side of \eqref{vect variation of bbf} would vanish in the non-extended case.}
\begin{equation}\label{vect variation of bbf}
\delta \mathbbm F = \mathbbm D \delta \mathbbm A + \text{trivial parameters}\,.
\end{equation}
In our case the right-hand side reduces to $(0,\,\cD \delta B^\ord{k})$ using \eqref{B var of bbA} and noticing that the difference between the extended $\circ$ product and the generalised Lie derivative vanishes when the first component of a doubled object is zero.
The trivial parameters can be ignored as we know they at most contribute by a total derivative (equivalently, they can be set to vanish by combining the $B$ field variations with an ad-hoc two-form gauge transformation).
Using now the fact that the right-hand side of the expression $\cV( \cD\delta B^\ord{k}) \cV^{-1}$ still satisfies the section constraint in the form \eqref{eq:secsol}, hence traces with negative Virasoro generators vanish, we find that the $B$ field variation of \eqref{Ftildes} reduces to
\begin{equation}\label{Ftildes B var}
- \Tr\Big(
   \cV (\cD \delta B^\ord{1}) \cV^{-1} L_0
  \Big)
=
  \Tr\Big(
    \cV \delta B^\ord{1} \cV^{-1} [L_0 ,\, \cD\cV\cV^{-1} ]
  \Big)
\end{equation}
up to total (internal and external) derivatives.
We are again using \eqref{remove compensator} to simplify the notation.
Only the loop components of the Maurer--Cartan form contribute.
Furthermore, because $\cV\in E_9$ by \eqref{full vir gfix} and $\eta_{-1\,\alpha\beta}$ is $E_9$ invariant up to a weight term, one can easily combine \eqref{eq:balphaex}, \eqref{B var of bbA} and \eqref{underlined} to find
\begin{equation}
2\rho\underline{\sbm*{\delta\mathbbm A}}{}^n_A 
=
\eta_{AB}\Tr\Big(
  \cV\delta B^\ord{1} \cV^{-1}T^B_{-n-1}
\Big)\,,\qquad
n\ge-2\,,
\end{equation}
where we are restricting to $n\ge-2$ so that higher $\delta B^\ord{k}$ contributions can be removed by trivial parameters (for smaller $n$, the $\cV$ conjugation implies a contribution from $\Tr(\delta B^\ord{2})$ and higher, which are not trivial).
Opening up the commutator one then quickly deduces that \eqref{Ftildes B var} exactly cancels out \eqref{cocycle B var}.
The $B$ field Euler--Lagrange equations vanish upon imposing twisted self-duality as required.

It is now straightforward to check that the $B$ field variation of the extra term we included in \eqref{final cov Lagrangian J} (compared to $\rho \mf{X}^\ord{1}$) does not vanish, and hence its inclusion in the pseudo-Lagrangian is essential for the above result to hold.

\subsubsection*{The scalar field variations}

In varying the constrained scalar field $\bra\chi$ (equivalently $\bra{\chi^\upgamma_1}$ or $\bra{\tilde\chi_1}$) we include explicitly the contribution from the scalar potential.
From \eqref{final cov Lagrangian J GAUGEFIX} and \eqref{scalpot} we find the equation
\begin{equation}
\mathllap{\boxed{\text{\textbf{$\langle\delta\chi|$:}}}\qquad\qquad}
\bra{\delta\chi}\ \Big(
     \ket\cF 
  +2 \star  \rho^{-1} \cM^{-1} \ket{P_1}
  -  \star\ T^\alpha  \cM^{-1} \ket{\cJ^-_\alpha}
\Big) = 0\,.\label{chife}
\end{equation}
This result does not rely on any specific solution of the section constraint, but is based on the gauge-fixing \eqref{virm gauge fixing} in order to display the variation of the scalar potential explicitly.
Recall that $\bra{\delta\chi}$ satisfies the section constraint, hence the equation above only arises from the pseudo-Lagrangian when contracted with a constrained object.

From now on we no longer rely on a gauge-fixing for the Virasoro scalar fields and we do not commit to a specific solution of the section constraint. 
We also do not vary the scalar potential explicitly. 
Nevertheless, all computations displayed below hold if we impose the gauge-fixing \eqref{virm gauge fixing}, and therefore the complete equations of motion can be derived by varying \eqref{scalpot} and combining it with the results in this section.

The variation of the topological term with respect to $\sigma$ is also straightforward, because it only appears in \eqref{final cov Lagrangian P} through $\tilde\chi_0=\cP_\dK=-\cD\sigma$.
Up to twisted self-duality we then get the equation of motion
\begin{equation}\label{sigmaELeq}
\mathllap{\boxed{\text{\textbf{$\delta\sigma$:}}}\hspace{11em}}
\cD\star\cD\rho - \frac12 \star \frac{\delta V}{\delta\sigma} = 0\,.
\end{equation}

The extended Virasoro fields $\phi_n$ are pure gauge, hence their equations of motion are redundant and we do not need to compute them.
We then focus on the variation of the scalars $\rho$, $\mathring V\in E_8$ and $Y^A_n$, $n\ge1$.
We can encode their variation in terms of an Hermitian algebra element $\pi$:
\begin{equation}
  \pi=\pi^\dag = \delta\cV \cV^{-1}\ \in\ \mf{e}_9 \,.
\end{equation}
This means that the variation above is a combination of a field variation and a $K(E_9)$ transformation:
\begin{equation}\label{scalvar and Ke9}
  \delta\cV = \delta_{\text{fields}}\cV + \zeta \cV\,,
  \qquad 
  \zeta \in K(\mf e_9)\,,
  \qquad
  \zeta^n_A = \mathrm{sgn}(n) \pi^n_A \,.
\end{equation}

Recall now that at the end of Section~\ref{ssec:unend} we noticed that the first line of \eqref{final cov Lagrangian P} corresponds to the generalised diffeomorphism and $K(E_9)$ covariant derivative of $\tilde\chi_1$, with $\cQ$ as composite connection, except for the substitution $c_\vir \to 0$ which removes a piece of the cocycle \eqref{QP vir cocycle to PP}.
It is then convenient to include the composite connection $\cQ$ in $\cD$:
\begin{equation}
\widehat\cD \Phi = \cD\Phi + \cQ_\alpha \delta^\alpha_{K(\hevir)}\Phi \,,
\end{equation}
where the second term denotes the transformation of a field $\Phi$ under the local $K(\mf e_9)$, extended by the anti-Hermitian part of $\vir$, where $\cQ$ takes values.%
\footnote{
The presence of $K(\vir)$ can appear strange since only $\vir^-$ are global symmetries of the theory and they do not require compensating local transformations.
This is due to the fact that the ungauged theory is in fact also invariant under infinitesimal positive $\vir$ transformations and the decomposition of the Maurer--Cartan form into $\cP$ and $\cQ$ reflects this.
Most objects we use have therefore natural transformation properties under $K(\vir)$ obtained by extension of their $K(\mf e_9)$ transformation and standard commutation relations, which simplifies many manipulations in the following.
}
With this definition, for instance, the first of \eqref{PQ MC eqs} reads $\widehat\cD\cP=0$ and the variation of $\cP$ and $\cQ$ then reads 
\begin{equation}\label{deltaP deltaQ}
  \delta\cP 
  = \widehat\cD\pi 
  = \cD\pi - [\cQ,\,\pi]
\,,\qquad
  \delta\cQ 
  = [\pi,\,\cP] \,,
\end{equation}
with 
$\cD\pi = 
 \dd\pi -\braket{\partial\pi}{A} - \sbm A_\alpha \smash{\big[}h^\alpha,\,\pi\smash{\big]}$.
Notice that as usual the compensator $h^\alpha$ cancels out between the covariant derivative and the $\cQ$ commutator.

It turns out to be convenient to combine the scalar field variations with a special choice of variation for the $\tilde\chi_m$ forms.
Since the latter contribute trivially to the equations of motion once twisted self-duality is imposed, this choice will not affect our results.
We then introduce a spurious object $\widetilde\pi_m$ meant to complete the $\dK$ component of $\cS_m(\pi)$, just like $\tilde\chi_m$ does for $\cS_m(\cP)$:
\begin{equation}
\label{eq:piord}
\pi^\ord{m} = \cS_m(\pi) + \widetilde\pi_m \dK\,.
\end{equation}
We then choose
$
\delta\tilde\chi_m =
\widehat\cD\widetilde\pi_m
$
so that the variations of the shifted $\cP$ take the simple form
\begin{equation}
\delta\cP^\ord{m} 
= \widehat\cD \pi^\ord{m} 
= \cD \pi^\ord{m} + \cQ_{\alpha} \delta^\alpha_{K(\hevir)}\pi^\ord{m}
\,,
\end{equation}
where $\delta^\alpha_{K(\hevir)}\pi^\ord{m}$ includes a sum over Virasoro shifts just like the last term of the first line in \eqref{X1 in terms of P}.
Explicitly,
\begin{equation}
\cQ_{\alpha} \delta^\alpha_{K(\hevir)}\pi^\ord{m}
=
- [\cQ,\,\pi^\ord{m}]
+ m\sum_{k\in\mathds{Z}} \cQ_k \pi^\ord{m+k}\,.
\end{equation}
After several manipulations, we will set $\widetilde\pi_m\to0$ without loss of generality.

With this setup, in order to compute the variation of the first line in \eqref{final cov Lagrangian P}, we can equivalently compute
\begin{equation}
\delta \big(2\rho\, \widehat\cD\tilde\chi_1\big) =
\delta \big(2\rho\, \widehat\cD\cP^\ord{1}\big) \big|_\dK 
\end{equation}
and set $c_\vir \to 0$ in the cocycle.
We then expand to find
\begin{equation}\label{scalvar 1st line}
\delta \big(2\rho\, \widehat\cD\cP^\ord{1}\big)  
= 
    2 \delta\rho\, \widehat\cD\cP^\ord{1}
  - 2\rho\, [\delta\cQ,\,\cP^\ord{1}]
  + 2\rho\, \widehat\cD \delta \cP^\ord{1} \,,
\end{equation}
where we have already used twisted self-duality to remove a series over the $\vir$ components of $\cQ$.
Let us look at the middle term. 
Projecting onto $\dK$ and using \eqref{deltaP deltaQ} we find that it vanishes.
For instance, using twisted self-duality the loop components of $\pi$ contribute as follows
(recalling that we are keeping wedge producs as understood)
\begin{align}
f^{ABC}\sum_{n,q} (-1)^n n \, \pi^{q}_A \star^{q+1}\!\tP^0_B  \tP^0_C
-\eta^{AB} \sum_{n,q} (-1)^n n\, q\, \pi^{-q}_A \star^{q+1}\tP_0 \tP^0_B 
\ =\ 0 \,.
\end{align}
A similar computation applies to the $L_0$ component of $\pi$ and since we set $c_\vir\to0$, that is all we need.

Let us then look at the last term in \eqref{scalvar 1st line}, which is proportional to $\widehat\cD^2\pi^\ord{1}$.
The square of the covariant differential has the general form
\begin{equation}\label{Dhat squared}
\widehat\cD^2 
= 
- \cL_{\mathbbm F} 
+ \left( \cD\cQ - \frac12[\cQ,\,\cQ] \right)_\alpha \delta^\alpha_{K(\hevir)} \,,
\end{equation}
and we recognise the last term as the covariant field strength of the composite $\cQ$ connection.
We can use the Maurer--Cartan equations \eqref{PQ MC eqs} to further manipulate this expression.
In order to do so, notice that twisted self-duality implies $[\cP,\,\cP]=0$.
The $\dK$ component is straightforward.
For the rest, we use $\cS_0(\cP) = \cS_{m}(\cS_{-m}(\cP))$ for any $m\neq0$ so that
\begin{equation}
\cS_0\big( [\cP,\cP] \big)
=
  \cS_{-m}\big([\cP,\,\cS_m(\cP)]\big) 
+ m \sum_{p\in\mathds{Z}}^\infty \tP_p \, \cS_{p}(\cP) \,.
\end{equation}
Let us now take $m$ odd.
By twisted self-duality, the first term of the right-hand side vanishes.
The second term flips sign under $m\to-m$ while the left-hand side does not, hence they both vanish.
We can then substitute into \eqref{Dhat squared} the left-hand side of the Maurer--Cartan equation \eqref{PQ MC eqs} for the curvature of $\cQ$, and apply it to $\pi^\ord{1}$.
We find (setting $h^\alpha\to0$ as usual as it cancels out anyway)
\begin{align}
\widehat\cD^2 \pi^\ord{1} 
=
&- \braket{\partial_\pi}{\cF}\,\pi^\ord{1}    
+  \big( \tfrac12\braket{\cQ_\alpha}{\cF} + \underline{\sbm F}_\alpha \big) 
    \big[ T^\alpha-T^\alpha{}^\dagger \,,\, \pi^\ord{1} \big] \CR
&+  \sum_{n=1}^\infty \big( \tfrac12\braket{\cQ_n}{\cF} + \underline{\sbm F}_n \big) 
    \big( \pi^\ord{1+n} - \pi^\ord{1-n}  \big) \,.
\end{align}
Projecting this expression on $\dK$ and setting to zero the central charge component $\widetilde\pi_1$ of $\pi^\ord{1}$, we find that only the middle term contributes and arrive at the result
\begin{equation}\label{1st line loopvar}
2\rho\,\widehat\cD \delta \fP(1) \big|_\dK
=
2 \rho\, \eta^{AB} \sum_{n\in\mathds{Z}} n\,
  \Big(
      \braket{Q_A^n}{\cF} 
     + \underline{\sbm F}^{n}_A
     - \underline{\sbm F}^{-n\,A}
  \Big)
  \ \pi^{-n-1}_B \,.
\end{equation}
Notice that the variation of $\rho$ does not contribute to this expression.

Let us now look at the variation of the second line of \eqref{final cov Lagrangian P}.
Using \eqref{finite cocycle from 0 shift} one finds that the $\rho$ variation of $\omega^\alpha(\cV)$ vanishes.
Furthermore, from the definition \eqref{shift upgamma operator} one easily sees that both $\hbm{F}^\upgamma_m$ and $\sbm F_\alpha^\ord{m}$ depend on $\rho$ only through an overall $\rho^{-m}$ factor.
Therefore, the $\rho$ variation of the second line of \eqref{final cov Lagrangian P} reads
\begin{equation}\label{rho var 2nd line}
2 \delta\rho\,\big(\,
     \braket{\tilde\chi_1}{\cF}
  +  \hbm{F}^\upgamma_{-1} 
  +  \omega^\alpha(\cV) \, \sbm F^\ord{-1}_\alpha
\,\big) \,.
\end{equation}
Looking now at the $\hE8$ scalars variation, we must take into account that $\bra{\tilde\chi_1}$ transforms under the $K(\mf e_9)$ transformation $\zeta$ we introduced in \eqref{scalvar and Ke9} to make $\pi$ Hermitian, so that
\begin{equation}\label{intchi KE9 var}
\delta\braket{\tilde\chi_1}{\cF} 
= 
\eta^{AB} \sum_{n\in\mathbbm Z} |n|  \, \pi^n_A \braket{P^{-n-1}_B}{\cF} \,,
\end{equation}
while the variation of $\omega^\alpha(\cV)$ gives
\begin{equation}\label{omegaV loop var}
\delta \omega^{\alpha}(\cV) = [\pi,\,\cV T^\alpha\cV^{-1}]\big|_\dK \,.
\end{equation}

We can now put these results together.
The $\rho$ variation of the topological term is the sum of the first term in \eqref{scalvar 1st line} and of  \eqref{rho var 2nd line}.
Applying twisted self-duality, we find the equation of motion
\begin{equation}\label{rho EL eq}
\mathllap{\boxed{\text{\textbf{$\delta\rho$:}}}\qquad\qquad}
  \widehat\cD\star\cP_\dK 
+ \braket{\tilde\chi_1}{\cF}
+ \hbm{F}^\upgamma_{-1} 
+ \omega^\alpha(\cV) \, \sbm F^\ord{-1}_\alpha
= \frac12 \star \frac{\delta V}{\delta\rho} 
 \,,
\end{equation}
where we need to keep in mind that $c_\vir\to0$ in the cocycle contained in the first term.
We can in fact further apply twisted self-duality to it, in order to make contact with the second order equation for $\sigma$ for two-dimensional supergravity \eqref{sigma 2nd order ungauged}.
To do so, we shall temporarily assume a solution of section of the form \eqref{eq:secsol} and write 
\begin{align}\label{expand hatD PK}
\widehat\cD\star\cP_\dK 
&=
  \cD\star\cP_\dK    
- \eta^{AB} \sum_{n\in\mathds{Z}} n \, \tQ^n_A\,\star \tP^{-n}_B
\CR
&=
- \cD\star\cD\sigma
+ \eta^{AB} \sum_{n\in\mathds{Z}}|n| \, \tP^n_A \,\star \tP^{-n}_B
+ 4\eta^{AB} \underline{\sbm A}^{+1}_A \,\star\tP^{-1}_B
\CR
&=
- \cD\star\cD\sigma
+ 2\eta^{AB} \tP^0_A \, \star\tP^0_B \Big(\sum_{n=1}^\infty (-1)^n n\Big)
+ 4\eta^{AB} \underline{\sbm A}^{+1}_A \,\tP^{0}_B
\CR
&=
- \cD\star\cD\sigma
- \frac12 \eta^{AB} \tP^0_A \, \star\tP^0_B 
+ 4\eta^{AB} \underline{\sbm A}^{+1}_A \,\tP^{0}_B \,,
\qquad\text{(with $h^\alpha\to0$)}
\end{align}
In the second line we have converted $\tQ^n_A$ to $\tP^n_A$ and we stressed that we set $h^\alpha\to0$ everywhere as usual, because it would cancel out between $\cQ$ and the covariant derivative anyway.
In the third line we have used twisted self--duality and in the last step we have regularised the divergent sum as $\sum_{n=1}^\infty (-1)^n n \to -\frac14$.
This is achieved for instance by taking the $z\to1$ limit of a geometric series $\sum_{n=1}^\infty (-z)^n p(n)$ which converges for $|z|<1$ and $p(n)$ any polynomial in $n$.
We motivate this choice of regularisation in Appendices~\ref{app:seriesreg} and \ref{app:scaleom match}.
We then see that if we reduce to two-dimensional supergravity by setting $\bra\partial=0$ and $B^\ord{k}=0$, this equation of motion reduces to \eqref{sigma 2nd order ungauged}.
Notice how this result hinges on the removal of the $c_\vir$ dependent terms of the cocycle in $\widehat\cD\star\cP_\dK$.
Had we not removed the term \eqref{QP vir cocycle to PP} from the pseudo-Lagrangian, it would have contributed here through a term proportional to $(\cD\rho)^2$ (with a regularisation similar to the one above).

Let us now look at the variation of the loop scalar fields.
Using the Maurer--Cartan equation \eqref{PQ MC eqs} for $\cP$ combined with twisted self-duality \eqref{twsd covariant P}, we find the convenient identity
\begin{equation}\label{DP equal F}
\widehat\cD\star^{|n|}\tP^0_A 
= 
- \braket{P^{n+1}_A}{\cF} 
- \underline{\sbm F}^{n+1}_A
- \underline{\sbm F}^{-n-1\,A}\,,
\qquad
n\in\mathds{Z}\,,
\end{equation}
which implies in particular that all the even components of the right-hand side are equal to each other, and the same holds for the odd ones.
Combining this expression with \eqref{1st line loopvar}, \eqref{intchi KE9 var} and \eqref{omegaV loop var}, the latter contracted with $2\rho(\sbm F^\ord{1}_\alpha +\sbm F_\alpha^\ord{-1})$, writing $\bra{Q^n_A}=-\mathrm{sgn}(n)\bra{P^n_A}$ and using that $\pi$ is Hermitian we then find the equation of motion
\begin{equation}\label{scalar eom}
\mathllap{\boxed{\text{\textbf{$\delta \mathring V$, $\delta Y_n$:}}}\quad}
2\rho\,\pi^{0\,A} \big(
    \widehat\cD\star\tP^0_A 
  + 2\underline{\sbm F}^{+1}_A
\big) 
- 4\rho\, \sum_{k=1}^\infty \pi^{-k\,A} \big(
    \underline{\sbm F}^{k-1}_A 
  - \underline{\sbm F}^{k+1}_A 
\big)
=
\star\,\delta V \,.
\end{equation}
The $\pi^0_A$ component of this equation still contains infinitely many dual fields within the covariant derivative. Using twisted self-duality to express it only in terms of physical fields, we are led to a regularisation analogous to the one used for \eqref{expand hatD PK}.
In analogy to that computation we temporarily choose a section solution of the form \eqref{eq:secsol} to simplify the vector field dependence and write explicitly
\begin{align}\label{expand scalar eom}
\widehat\cD\star\tP^0_A 
&=
  \cD\star\tP^0_A 
- f^{CD}{}_A \sum_{n\in\mathds{Z}} \tQ^n_C \star \tP^{-n}_D 
+ \sum_{n\in\mathds{Z}} n \big(
      \tQ_{-n} \star \tP^n_A
    - \tQ^n_A \star \tP_{-n}
  \big) \\
&=
  \cD\star\tP^0_A 
- f^{CD}{}_A \tQ^0_C \star \tP^{0}_D
- 2\sum_{n=1}^\infty n \big(
      \tQ^n_A \star \tP_{-n}
    + \tQ_{n} \star \tP^{-n}_A
  \big) 
\CR&\qquad
+ 2 f^{CD}{}_A (\underline{\sbm A}^1_C - \underline{\sbm A}^{1\,C}) \tP^{0}_D
- 2 \tP_0 (\underline{\sbm A}^1_C +\underline{\sbm A}^{1\,C})
\CR
&=
  \cD\star\tP^0_A 
- f^{CD}{}_A \tQ^0_C \star \tP^{0}_D
+ 2 \tP_0 \star \tP^0_A \sum_{n=1}^\infty (-1)^n n
\CR&\qquad
+ 2 f^{CD}{}_A (\underline{\sbm A}^1_C - \underline{\sbm A}^{1\,C}) \tP^{0}_D
- 2 \tP_0 (\underline{\sbm A}^1_C +\underline{\sbm A}^{1\,C})
\nonumber\\[1.5ex]
&=
  \rho^{-1}\cD(\rho\star\tP^0_A) 
- f^{CD}{}_A \tQ^0_C \star \tP^{0}_D
+ 2 f^{CD}{}_A (\underline{\sbm A}^1_C - \underline{\sbm A}^{1\,C}) \tP^{0}_D
- 2 \tP_0 (\underline{\sbm A}^1_A +\underline{\sbm A}^{1\,A})
\nonumber\,.
\end{align}

We check in Appendix~\ref{app:scaleom match} that the $\pi^0_A$ equation of motion correctly reproduces the equation of motion of the scalar fields in $E_8$ exceptional field theory.
As a further cross-check, we can reduce to two-dimensional supergravity by setting $\bra\partial\to0$ and check that the equations of motion derived so far are then compatible with setting the $B^\ord{k}$ fields to 0 as follows. 
The vector fields $\ket A$ and $\bra{\chi^\upgamma_1}$ decouple when $\bra\partial=0$, and the field equation for $\rho$ reduces to $\dd\star\dd\rho=0$.
Extending then \eqref{DP equal F} also along the $\vir$ components and using the Maurer--Cartan equations as well as the $\pi^0_A$ equation above, we find that all $\virm$ components of $\sbm F_\alpha$, as well as the coset projection of $\underline{\sbm{F}}_A^{n}$ must vanish.
Plugging these results into the rest of equation \eqref{scalar eom}, we then have that all non-trivial components of the $B^\ord{k}$ field strengths must vanish, hence without loss of generality we can set $B^\ord{k}\to0$ as claimed.
Together with the fact that \eqref{twsd cov J} correctly reduces to two-dimensional twisted self-duality when $\bra\partial=0=B^\ord{k}$, this guarantees that two-dimensional supergravity is reproduced from Virasoro-extended $E_9$ ExFT.

\subsubsection*{The $\ket A$ vector field variation}
The starting point is again the pseudo-Lagrangian \eqref{final cov Lagrangian P}, which we will vary with respect to the vector field $\ket{A}$. We recall that we work in the ungauged-fixed setting where $\phi_m\neq0$ for all $m>0 $. We also re-use the notation introduced in \eqref{B var of bbA}, but this time with
\begin{equation}
 \delta \mathbbm{A}=(\ket{\delta A},0)\,.\label{A var of bbA}
\end{equation}
The variations of $\mathcal{P}$ and $\mathcal{Q}$ follow directly from their expressions \eqref{cP def} and \eqref{cQ def}. They read
\begin{align}
\delta \mathcal{P}&=-\langle\mathcal{P}|\delta A\rangle-\underline{\sbm*{\delta\mathbbm A}}{}_\alpha\big(T^\alpha+\mathrm{h.c.}\big)\,,\label{A var of comp P}\\
\delta\mathcal{Q}&=-\langle\mathcal{Q}|\delta A\rangle-\underline{\sbm*{\delta\mathbbm A}}{}_\alpha\big(T^\alpha-\mathrm{h.c.}\big)\,,\qquad\text{(with $h^\alpha\to0$)}\,,\label{A var of comp Q}
\end{align}
where $\underline{\sbm*{\delta\mathbbm A}}{}=\underline{\sbm*{\delta\mathbbm A}}{}_\alpha T^\alpha$ takes values in $\mathfrak{\hat e_8}\oleft \mathfrak{vir}^-$.
As explained previously, we consistently ignore the $h^\alpha$ contributions as they ultimately cancel out in the pseudo-Lagrangian.

Let us start by focusing on the variation of the first term of the first and second line in \eqref{final cov Lagrangian P}. With \eqref{Dchitilde} and \eqref{vect variation of bbf}, we find
\begin{align}
2\rho\, \delta\left( \cD\tilde\chi_1+ \braket{\tilde\chi_1}{\cF}\right)= 2\rho\,\left(\bra{\partial_{\tilde\chi}}\tilde\chi_1-\mathcal{D}\bra{\tilde\chi_1}+\tP_0\bra{\tilde\chi_1}\right)\ket{\delta A}\,,\label{var Dchi term}
\end{align}
up to internal and external total derivatives. For the cocycle term of the first line, one computes
\begin{align}
-2 \rho \, \eta^{AB}\,\delta\left(\sum_{n\in\mathds{Z}} n\,\tQ_A^n\,\tP_B^{-n-1}\right)
=&\,2\rho\,\eta^{AB}\sum_{n\in\mathds{Z}} n \Big(\tQ^{n}_A\bra{\tP_B^{-n-1}}-\tP_A^{-n-1}\bra{\tQ_B^n}\Big)\label{var of QP cocycle}\\
&+4 \rho \, \eta^{AB}\sum_{n\in\mathds{Z}} (n+1)(\tQ+\tP)^n_A\,\underline{\sbm*{\delta\mathbbm A}}_B^{-n-1}\,,\nonumber
\end{align}
where we used twisted self-duality in writing the last line.
For the third term of the first line in \eqref{final cov Lagrangian P}, the variation of $\tP_n$ does not contribute since $(\tilde\chi_{n+1}-\tilde\chi_{n-1})$ vanishes by twisted self-duality. The only contribution  comes from the variation of $\tilde\chi_0=\tP_K$, and we then simply have
\begin{align}
2\rho \,\delta\left(\sum\limits_{n=1}^\infty\tP_n(\tilde\chi_{n+1}-\tilde\chi_{n-1})\right)=2\rho\,\tP_1(\langle P_\dK|\delta A\rangle+2\underline{\sbm*{\delta\mathbbm A}}{}_{\dK})\,.\label{var Pchi term}
\end{align}

The variation of the remaining terms in the second line of the pseudo-Lagrangian \eqref{final cov Lagrangian P} requires more care. To derive the variation of $\hbm{F}_1^\upgamma$ and $\hbm{F}_{-1}^\upgamma$, which are defined in \eqref{hatF gamma definition}, we use the following property which holds exactly for $n<1$ and up to a total derivative for $n=1$:
\begin{align}
\delta\,\hbm{F}^\upgamma_n=\hbm{(\mathbbm D \delta \mathbbm A)}^\upgamma_n&=(\dd-\delta_\mathbbm{A})\hbm{\delta\mathbbm{A}}^\upgamma_n-n\sum\limits_{q=0}^\infty\big(\mathcal{D}\Gamma\Gamma^{-1})_{-q}\,\hbm{\delta\mathbbm{A}}^\upgamma_{n-q}\nonumber\\
&=\mathcal{D}\,\hbm{\delta\mathbbm{A}}^\upgamma_n-\Delta_\mathbbm{A}\hbm{\delta\mathbbm{A}}^\upgamma_n-n\sum\limits_{q=0}^\infty\big(\mathcal{D}\Gamma\Gamma^{-1})_{-q}\,\hbm{\delta\mathbbm{A}}^\upgamma_{n-q}\,,\;\;\;\;\forall n\leq1\,,\label{trick}
\end{align}
where we used \eqref{vect variation of bbf} in the first step. 
Note that the double object $\delta \mathbbm{A}$ transforms with the extended $\circ$ product under a gauge variation, and is therefore treated on the same footing as $\mathbbm{F}$. This allows, in the second step, to extract the differential operator $(\dd-\delta_{\mathbbm{A}})$ out the combination $\hbm{\delta\mathbbm{A}}{}^\upgamma_n$. In the process we however generate the last term of the first line, which substracts the action of the gauge variation on the Virasoro scalars contained in $\hbm{\delta\mathbbm{A}}{}^\upgamma_n$. This is described in Appendix~\ref{app:fstrvar}. The expression of the latter takes the same form as in \eqref{hatF gamma definition}. The non-covariant part of its gauge variation with respect to $\mathbbm{A}$, defined in \eqref{noncov var def} and denoted by $\Delta_{\mathbbm{A}}$, can be deduced from \eqref{hbmF general noncov var}. We thus find 
\begin{align}
\delta(\rho\,\hbm{F}_{-1}^\upgamma)=\ &\rho\,\mathcal{D}\hbm{\delta\mathbbm{A}}_{-1}^\upgamma+\rho\sum\limits_{q=0}^\infty\big(\mathcal{D}\Gamma\Gamma^{-1})_{-q}\,\hbm{\delta\mathbbm{A}}^\upgamma_{-1-q}+\rho \,\hbm{\mathbbm{A}}^\upgamma_{-1} \braket{\partial_{\mathbbm{A}}}{\delta A}\,,\label{delta rho hatF+1}\\
\delta(\rho\,\hbm{F}_1^\upgamma)=\ &\rho\,\mathcal{D}\hbm{\delta\mathbbm{A}}_1^\upgamma-\rho\sum\limits_{q=0}^\infty\big(\mathcal{D}\Gamma\Gamma^{-1})_{-q}\,\hbm{\delta\mathbbm{A}}^\upgamma_{1-q}+ \rho\,\hbm{\mathbbm{A}}^\upgamma_1 \braket{\partial_{\mathbbm{A}}}{\delta A}\nonumber\\
&-  \big(\bra{\partial_B}B^\ord{1} - \Tr( B^\ord{1}) \bra{\partial_B} \big) \ket{\delta A}\,,\label{delta rho hatF-1}
\end{align}
where $\hbm{\mathbbm{A}}^\upgamma_{n}$ is again of the same form as \eqref{hatF gamma definition}. 
The second equation holds only up to a total derivative.
To treat the last terms in the pseudo-Lagrangian \eqref{final cov Lagrangian P} which involve the $\omega^\alpha(\mathcal{V})$ cocycle, let us first consider $\sbm F^{(n)}_\alpha$ defined in \eqref{completed shifted F}. Up to a $\dK$ component, the latter can be written as $\mathcal{S}_n^\upgamma(\sbm{\mathbbm{F}})_\alpha$, 
where in order to simplify the notation in the next few steps, we will write $\sbm{F}=\sbm{F}_\alpha T^\alpha$ and similarly for other objects such as $\sbm{A}$, their derivatives and their variation. 
Taking into account that $\omega^\dK(\mathcal{V})=0$, we then have for $n\leq 1$,
\begin{align}
\delta \Big(\omega^\alpha(\mathcal{V})\sbm F^{(n)}_\alpha\Big)=\ & \,\omega^\alpha(\mathcal{V})\mathcal{S}_n^\upgamma(\sbm{\mathbbm{D}\delta\mathbbm{A}})_\alpha\\
=\ &\,\omega^\alpha(\mathcal{V})\Big(\mathcal{D}\mathcal{S}_n^\upgamma(\sbm{\delta\mathbbm{A}})_\alpha-\Delta_{\mathbbm{A}}\mathcal{S}_n^\upgamma(\sbm{\delta\mathbbm{A}})_\alpha-n\sum\limits_{q=0}^\infty\big(\mathcal{D}\Gamma\Gamma^{-1})_{-q}\,\mathcal{S}_{n-q}^\upgamma(\sbm{\delta\mathbbm{A}})_\alpha\Big)\,.\nonumber
\end{align}
We used the same arguments as in \eqref{trick} in order to extract the differential operator out of the shift operator. At this point, observe that according to \eqref{finite cocycle from S to S gamma}, we can write
\begin{equation}
\omega^\alpha(\mathcal{V})\mathcal{S}_{n}^\upgamma(\sbm{\delta\mathbbm{A}})_\alpha=\mathcal{V}\mathcal{S}_n^\upgamma(\sbm{\delta\mathbbm{A}})\mathcal{V}^{-1}\big|_\dK=\mathcal{V}\sbm{\delta\mathbbm{A}}^{(n)}\mathcal{V}^{-1}\big|_\dK-\hbm{\delta\mathbbm{A}}_n^\upgamma\label{noname1}\,,
\end{equation}
where $\sbm{\delta\mathbbm{A}}^{(n)}$ is, in analogy with \eqref{completed shifted F}, the $\dK$-completed version of $\mathcal{S}_n^\upgamma(\sbm{\delta\mathbbm{A}})$ and therefore transforms with a commutator under rigid $\mathfrak{\hat e}_8\oleft \mathfrak{vir}^-$. 
Bearing this in mind we get, after an integration by parts,
\begin{align}
\delta \Big(\rho\,\omega^\alpha(\mathcal{V})\sbm F^{(n)}_\alpha\Big)=&\,\mathcal{D}\Big(\rho\,\mathcal{V}\sbm{\delta\mathbbm{A}}^{(n)}\mathcal{V}\Big)\big|_\dK-\rho\mathcal{D}\hbm{\delta\mathbbm{A}}_n^\upgamma+\rho\,\tP_0\Big(\hbm{\delta\mathbbm{A}}_n^\upgamma+\omega^\alpha(\mathcal{V})\mathcal{S}_n^\upgamma(\sbm{\delta\mathbbm{A}}\Big)\nonumber\\
&-n\rho\sum\limits_{q=0}^\infty\big(\mathcal{D}\Gamma\Gamma^{-1})_{-q}\,\,\omega^\alpha(\mathcal{V})\mathcal{S}_{n-q}^\upgamma(\sbm{\delta\mathbbm{A}})_\alpha-\rho\,\omega^\alpha(\mathcal{V})\Delta_{\mathbbm{A}}\mathcal{S}_n^\upgamma(\sbm{\delta\mathbbm{A}})_\alpha\,\nonumber\\
&-\rho\big[(\mathcal{P}+\mathcal{Q}),\mathcal{V}\,\mathcal{S}^\upgamma_n(\sbm{\delta\mathbbm A})\mathcal{V}^{-1}\big]\big|_\dK\,,
\end{align}
where the non-covariant variation of $\sbm{\delta\mathbbm{A}}_\alpha$ takes the same form as in \eqref{sbmF noncov var}. The first term can be dropped as it reduces to total internal and external derivatives. Combining this result with \eqref{delta rho hatF+1} and \eqref{delta rho hatF-1} we find, after rearranging the sums of the $(\mathcal{D}\Gamma\Gamma^{-1})$ terms and using twisted self-duality,
\begin{align}
\delta\Big(\rho\,\hbm{F}^\upgamma_1 \,
+&\rho\,\hbm{F}^\upgamma_{\!-1} \,
+\rho\,\omega^\alpha(\cV) \, 
  \big(
   \sbm F^{(1)}_\alpha + \sbm F^{(-1)}_\alpha 
  \big)\Big)\nonumber\\[1.2mm]
  =&\,\rho\sum\limits_{n=1,-1}\Big(\hbm{\mathbbm{A}}^\upgamma_n+\omega^\alpha(\mathcal{V})\,\mathcal{S}^\upgamma_n(\sbm{\mathbbm{A}})_\alpha\Big)\braket{\partial_{\mathbbm{A}}}{\delta A}-  \big(\bra{\partial_B}B^\ord{1} - \Tr( B^\ord{1}) \bra{\partial_B} \big) \ket{\delta A}\nonumber\\
  &-2\rho\,\tP_1\Big(\hbm{\delta\mathbbm{A}}^\upgamma_0+\omega^\alpha(\mathcal{V})\,\mathcal{S}^\upgamma_0(\sbm{\delta\mathbbm{A}})_\alpha\Big)-2\rho\sum\limits_{n=1,-1}\big[(\mathcal{P}+\mathcal{Q}),\mathcal{S}_{n}(\underline{\sbm*{\delta\mathbbm A}})\big]\big|_\dK\,.\label{deltaA Fpart}
\end{align}
Using \eqref{noname1} and $\hbm{\delta\mathbbm{A}}^\upgamma_0=\sbm{\delta\mathbbm{A}}_\dK$, the parenthesis in the last line can be written as $2\underline{\sbm*{\delta\mathbbm A}}{}_{\dK}$ and will therefore cancel against the last term in \eqref{var Pchi term}. Note also that, in writing the last term of \eqref{deltaA Fpart}, we have pulled the conjugation by $\mathcal{V}$ inside of the shift operator and used the notation \eqref{underlined}. The loop contribution of this commutator will cancel against the last line of \eqref{var of QP cocycle}, while its Virasoro contribution should vanish by itself since the pseudo-Lagrangian we vary does not depend on $c_{\vir}$. This is indeed the case because both the Maurer--Cartan form $(\mathcal{P}+\mathcal{Q})$ and $\underline{\sbm*{\delta\mathbbm A}}$ only take values in $\mathfrak{\hat e_8}\oleft\mathfrak{vir}^-$.

Adding up all the variations, we then obtain the following final form for the field equation
\begin{align}
\label{eq:Avar}
{\boxed{\text{\textbf{$ \ket{\delta A}$:}}}\quad\ \ }&\bra{\partial_{\tilde\chi}}\tilde\chi_1-\mathcal{D}\bra{\tilde\chi_1}+\tP_0\bra{\tilde\chi_1}+\tP_1\langle P_\dK|+\eta^{AB}\sum_{n\in\mathds{Z}} n \Big(\tQ^{n}_A\,\bra{\tP_B^{-n-1}}-\tP_A^{-n-1}\bra{\tQ_B^n}\Big)\nonumber\\
&\,+\frac12\sum\limits_{n=1,-1}\Big(\hbm{\mathbbm{A}}^\upgamma_n+\omega^\alpha(\mathcal{V})\,\sbm{\mathbbm{A}}^\ord{n}_\alpha\Big)\bra{\partial_{\mathbbm{A}}}-  \frac{1}{2\rho}\bra{\partial_B}B^\ord{1} +\frac{1}{2\rho} \Tr( B^\ord{1}) \bra{\partial_B}=0\,,
\end{align}
where we divided by an overall factor of $2\rho$ and removed the projection on $\ket{\delta A}$. 
Note that the terms in the bracket of the second line correspond to the $\dK$ completion of ${\sbm{A}}_\alpha\mathcal{S}_n(\cV T^\alpha\cV^{-1})$, analogously to how $\tilde\chi_m$ completes $\cS_m(\fP)$ in \eqref{shifted P and tilde chi def}.
We also emphasise that $\bra{\partial_{\mathbbm{A}}}$ only acts on the vector fields $\ket{A},\,B^\ord{k}$ and not on the scalar fields. 

While it is not manifest, one can verify that the complete set of field equations is invariant under rigid $E_9$. 
The equation~\eqref{eq:Avar} is separately invariant and one way of checking this is by using the invariance of the $\dK$-completion of a shifted Maurer--Cartan equation,  such as~\eqref{shifted CM eq J}, but now with one internal and one external derivative.

\section{\texorpdfstring{Minimal $E_9$ exceptional field theory}{Minimal E9 exceptional field theory}}
\label{sec:E9EFT}

In this section, we shall show that there is a formulation of $E_9$ exceptional field theory with a finite set of fields, i.e.  $\cM\in E_9 ,\, |A\rangle,\, B,\, \chi,\, \langle \chi|$, even though  they are in infinite-dimensional representations of $E_9$. As explained in Section \ref{sec:gfix}, one can gauge-fix all the Virasoro fields $\phi_n$ to zero. The price to pay is that the truncation to two-dimensional supergravity, with fields not depending on the internal coordinates, must necessarily involve non-zero constrained fields  $B^\ord{k}$. This obscures the relation to the linear system in two dimensions. But it has the advantage that this allows us to eliminate the infinite set of constrained one-forms $B^\ord{k}$ and one-forms $\chi^\upgamma_k$ in favour of a single constrained one-form $B \sim B_\mu{}^M{}_N$ and a single one-form $\chi\sim \chi_\mu$ (that still transforms indecomposably with the current) as we shall demonstrate in this section. Moreover, the twisted self-duality equation for the scalar fields can be written using a $E_9$ invariant bilinear form, and the pseudo-Lagrangian takes a form that is more similar to lower-rank ExFTs. In particular, we relax the conformal gauge and define external diffeomorphisms in this minimal formulation. We will exhibit that the complete pseudo-Lagrangian is determined by internal and external diffeomorphism invariance, similar to $E_n$ ExFT for $n\leq 8$. For the appropriate choice of solution to the section constraint, we will finally show that the Euler--Lagrange equations of the pseudo-Lagrangian (combined with the duality equation), reproduce the known field equations of $E_8$ exceptional field theory. All expressions will be finite and we shall not need to resort to formal geometric series summation as in the Virasoro-extended formulation.

\subsection{Integrating out the auxiliary fields}
\label{FromVirtoMin}

At vanishing Virasoro fields $\phi_n$, i.e. for $\Gamma = \rho^{-L_0}$, one can eliminate the constrained fields $B^\ord{k}$ and the scalar fields $\chi^\upgamma_k$ in favour of a single constrained field $B$ and a single scalar field $\chi$. Notice that we also gauge-fix $\phi_1=\tilde{\rho} = 0$, unlike in  Section~\ref{sec:gfix}. In this case, the coset representative $\cV$ in~\eqref{full coset repr} and the generalised metric $\cM$ belong to $E_9$. In order to obtain the minimal duality equation from \eqref{cascade cov J}, it will be convenient to first rewrite \eqref{twsd cov J}  as
\begin{align}
 \label{DualityInterm}  
 \mathfrak{J} = \rho^{-1} \star \bigl( \cM^{-1} \cS_1(\mathfrak{J})^\dagger \cM +\rho  \chi^\upgamma_1 \dK \bigr) \; . \
\end{align}
To analyse the duality equation~\eqref{DualityInterm} we decompose the current~\eqref{covariant J definition} as
\begin{align}
 \label{SplitJA} \mathfrak{J}_\alpha = \cJ^\flat_\alpha - \sum_{k=1}^\infty \Bigl(  \eta_{-k\,  \alpha\beta} \Tr[ T^\beta B^\ord{k}] + \rho^{-2k}  \eta_{k\,  \alpha\beta} \Tr[ \cM ^{-1} T^{\beta \dagger} \cM B^\ord{k}] \Bigr)\; ,
 \end{align}
where
\begin{align}
 \cJ^\flat = \cM^{-1} \cD^\flat \cM =  \cM^{-1}( \dd  -  \langle \partial_\cM | A \rangle ) \cM   - \eta_{\alpha\beta} \langle \partial_A | ( T^\alpha + \cM^{-1} T^{\alpha \dagger} \cM ) | A\rangle T^\beta  \,,
 \label{DflatM} 
 \end{align}
is expressed using a `bare' covariant derivative 
\begin{align}
\label{eq:flatD}
\cD^\flat = \dd - \cL_{(A,0)}
\end{align}
that does not contain the $B^\ord{k}$ terms.

It will be convenient to similarly redefine the fields $\chi^\upgamma_{k}$ in terms of bare fields  $\chi_{k}^\flat$ by separating out the $B^\ord{k}$ components according to
\begin{align}
\label{SplitchiA} \chi^\upgamma_{k} = \rho^{-k} \chi_{k}^\flat + \rho^{-k}  \sum_{q=0}^\infty\Tr \bigl[\bigl( L_{-1+k-q} + \rho^{-2-2q} \cM^{-1} L_{-k-1-q} \cM   \bigr)   B^\ord{1+q } \bigr]  \; . 
\end{align}
The field $\chi_{k}^\flat$ is the $\dK$-completion of $\cS_k(\cJ^\flat)$ and the terms involving $B$ play a similar role for the $B$-terms in~\eqref{SplitJA}. We also recall from the text under~\eqref{cascade cov J} 
that for $k\ge 1$
\begin{align}
 \chi^\upgamma_{1+k} =  \star^{k} \chi^\upgamma_1\; .\label{chiCollapse} 
 \end{align}

The components of \eqref{DualityInterm} along the Virasoro generators give that 
\begin{align}
 \label{TraceBk}  \Tr[ B^\ord{1+k}] = \rho^k \star^k \Tr[ B^\ord{1} ] \, ,
\end{align}
relating the traces of all higher $B^\ord{k}$ to that of $B^\ord{1}$. Therefore we can rewrite the current $\mathfrak{J}$ as
\begin{align}
 \mathfrak{J} &= \cJ^\flat - \eta_{-1\hspace{0.2mm}\alpha\beta} \Tr\Bigl[ \sum_{k=0}^\infty \cS_{-k}(T^\alpha \hspace{-0.2mm}) B^\ord{1+k}\hspace{-0.5mm} \Bigr] \bigr( T^\beta + \cM^{-1} T^{\beta \dagger} \cM \bigr) \nn\\
 &\quad\quad +  \sum_{k=0}^\infty  \rho^k \star^k \Tr[ B^\ord{1} ] \bigl( L_{-1-k} + \cM^{-1} L_{1+k} \cM \bigr) \; 
\end{align}
in which the constrained fields $B^\ord{1+k}$ only appear through 
\begin{align}
 \Tr\Bigl[ \sum_{k=0}^\infty \cS_{-k}(T^\alpha) B^\ord{1+k} \Bigr]  =  \Tr\Bigl[ T^\alpha  \sum_{k=0}^\infty  \widehat{\cS}_{-k}(B^\ord{1+k})\Bigr]\; . 
\end{align} 
The operation  $ \widehat{\cS}_{-k}(B^\ord{1+k})$ is defined implicitly such that the relation \eqref{formal shift ops} holds for any $T^\alpha \in \hat{\mf{e}}_8 \oleft \mf{vir}^-$ as explained in Appendix~\ref{app:exttriv}.
This implies that one can choose a particular gauge for the one-form gauge transformations defined in Appendix~\ref
 {app:exttriv} such that \eqref{TraceBk} extends to the whole constrained fields and fixes for all $k\ge 0$
\begin{align}
  \label{Bcollapse} B^\ord{1+k} = \rho^k \star^k  B \; . 
\end{align}
Note that, although $B^\ord{1}=B$ according to this equation, $B$ does not transform in the same way as $B^\ord{1}$ under internal diffeomorphisms so they should not be understood as being the same constrained fields. This is because the  fields $B^\ord{k}$ only appear through the combination $ \sum_{k=0}^\infty  \widehat{\cS}_{-k}(B^\ord{1+k})$ above, and the identification \eqref{Bcollapse} only holds for this combination traced with a generator $T^\alpha \in \mf{e}_9 \oleft \langle L_{-1} \rangle$. To define the gauge transformation of the field $B$ one identifies the gauge transformation of this combination 
\begin{align}
 \delta_{\Lambda}  \sum_{k=0}^\infty  \widehat{\cS}_{-k}(B^\ord{1+k}) = \cL_{(\Lambda,0)}  \sum_{k=0}^\infty  \widehat{\cS}_{-k}(B^\ord{1+k})  - \eta_{1\, \alpha\beta} \langle \partial_\Lambda | T^\alpha | A\rangle T^\beta | \Lambda\rangle \langle \partial_\Lambda | \; , 
 \end{align}
and defines the gauge transformation of the constrained field $B$ consistently with it, i.e. such that
\begin{align}
  \sum_{k=0}^\infty  \rho^k \star^k \widehat{\cS}_{-k}( \delta_{\Lambda} B) =     \sum_{k=0}^\infty  \rho^k \star^k \widehat{\cS}_{-k}( \cL_{(\Lambda,0)} B)   - \eta_{1\, \alpha\beta} \langle \partial_\Lambda | T^\alpha | A\rangle T^\beta | \Lambda\rangle \langle \partial_\Lambda | +\dots  
 \end{align}
where the ellipses correspond to terms that vanish upon tracing with $T^\alpha $. 
Inverting the geometric series, one obtains the gauge transformation of the constrained field $B$ up to a trivial parameter
\begin{align}
\label{IntermLambdaB}
\delta_\Lambda B  &= \cL_{(\Lambda,0)} B + \bigl( 1- \rho \star \widehat{\cS}_{-1} \bigr) \Bigl(   \eta_{1\, \alpha\beta} \langle \partial_\Lambda | T^\alpha | \Lambda \rangle T^\beta | A \rangle \langle \partial_\Lambda | \Bigr)+\dots  \CR
&= \cL_{(\Lambda,0)} B + \eta_{1\, \alpha\beta} \langle \partial_\Lambda | T^\alpha | \Lambda \rangle T^\beta | A \rangle \langle \partial_\Lambda | \CR
& \hspace{12mm}- \rho \star  \bigl( \eta_{\alpha\beta} \langle  \partial_\Lambda | T^\alpha |  \Lambda \rangle  T^\beta | A \rangle  + \langle \partial_\Lambda |  \Lambda  \rangle | A \rangle -  \langle \partial_\Lambda  | A \rangle |  \Lambda  \rangle\bigr)  \langle \partial_\Lambda | \; . 
\end{align}
Note that although we inverted the formal geometric series to obtain this formula from the Virasoro-extended gauge transformation, it is well defined and $E_9$-covariant. The final line is the definition of the gauge transformation of $B$.

Writing $\chi^\flat \equiv \chi_1^\flat $ for short and substituting \eqref{Bcollapse} on both sides of the duality equation \eqref{DualityInterm}, one obtains 
\begin{align}
&\hspace{4.5mm} \cJ^\flat -  \sum_{k=0}^\infty  \star^k \Bigl( \rho^k \eta_{-1-k\,  \alpha\beta} \Tr[ T^\alpha B ] + \rho^{-2-k}  \eta_{1+k\,  \alpha\beta} \Tr[ \cM ^{-1} T^{\alpha \dagger} \cM B] \Bigr) T^\beta \\
 &\!\!=  \rho^{-1}\hspace{-1mm} \star \! \bigl( \cM^{-1}\hspace{-0.2mm} \cS_1(\hspace{-0.2mm}\mathcal{J}^\flat)^\dagger \cM + \chi^\flat \dK\bigr)\! -\! \sum_{k=1}^\infty \star^{k}  \Bigl( \rho^{-k} \eta_{k-1 \hspace{0.2mm}\alpha\beta} \Tr[ \cM^{-1} T^{\alpha \dagger}\cM B ] + \rho^{k}  \eta_{-1-k\hspace{0.2mm} \alpha\beta} \Tr[  T^{\alpha }  B] \Bigr) T^\beta\nn .
\end{align}
One finds that most of the terms cancel and the remaining equation is
\begin{align}
 \cJ^\flat - \eta_{-1 \alpha\beta} \Tr[ T^\alpha B ] T^\beta= \rho^{-1} \star \bigl( \cM^{-1} \cS_1(\mathcal{J}^\flat)^\dagger \cM -\eta_{ \alpha\beta} \Tr[ T^\alpha  B ]  \cM^{-1} T^{\beta  \dagger}\cM + \chi^\flat \dK\bigr)\; .  
 \end{align}
This equation can be written as
\begin{align}
  \label{DualityEqZ2} 
  \cJ = \rho^{-1} \star \bigl( \cM^{-1} \cS_1(\cJ)^\dagger \cM + \chi \dK\bigr)  \; , 
\end{align}
for the current $\cJ$ and the field $\chi$ defined as
\begin{align} \label{Jchidef}  
\cJ = \cJ^\flat -  \eta_{-1 \alpha\beta} \Tr[ T^\alpha B ] T^\beta\; , \qquad \chi  = \chi^\flat + \Tr[L_0 B] \; ,
\end{align}
distributing the $B$ dependence. 
This provides an alternative formulation of the theory that we refer to as the minimal formulation and discuss in detail in the remainder of this section.

Because $\cJ \ne \cM^{-1} \cJ^\dagger \cM$, this duality equation does not imply an infinite chain of relations between $\cJ$ and its shifts $\cS_n(\cJ)$ as in equation \eqref{cascade cov J} of the Virasoro-extended formulation.
Rather, \eqref{cascade cov J} translates in the minimal formulation into finite linear combinations of the duality equation \eqref{DualityEqZ2} and its Hermitian conjugate where some of the $\cJ^\flat$ components cancel out.  We will see that \eqref{DualityEqZ2} is invariant under internal diffeomorphisms with the gauge transformation~\eqref{IntermLambdaB}.

Let us now consider similar consequences for the pseudo-Lagrangian. The potential term is not modified so we only need to discuss the topological term at vanishing  Virasoro fields $\phi_n=0$ for $ n \ge 1$. We will then rearrange the terms in \eqref{final cov Lagrangian J GAUGEFIX} in the case $\cM\in E_9$, i.e. for $\tilde\rho=0$. By the definition \eqref{hatF gamma definition}, we have 
\bea 
\rho\,  \hbm{F}^\upgamma_1 &=&  - \langle \partial_{\cF}  |L_1 | \cF \rangle - \sum_{k=0}^\infty \Tr\bigl[  L_{-k}  \cG^\ord{1+k} \bigr] \CR
& =& - \langle \partial_{F}  |L_1 | F \rangle - \sum_{k=0}^\infty \Tr\bigl[  L_{-k}  G^\ord{1+k} \bigr]  \CR
&& \qquad + \langle \partial | \Bigl( \tfrac12 \eta_{1\, \alpha\beta} T^\alpha |C_{[1}\rangle \langle \pi_C | T^\beta | C_{2]}\rangle + |C_1^+\rangle \Tr[ C_2^+] - C_2^+ | C_1\rangle \Bigr)  \; , \label{plustwoforms}
\eea
where the two-forms (which enter the field strengths as discussed in Appendix~\ref{twoForm}) appear through a total derivative that we will neglect in the following. For $\cM\in E_9$, one also computes
\begin{align}
 &\hspace{5mm} \cM^{-1} \big(\hbm{F}^\upgamma_{-1} \dK+ \sbm F_\alpha \cS^\upgamma_{-1}(T^\alpha)\big)^\dagger \cM \CR
&=\rho \Bigl( \eta_{-1\, \alpha\beta} \langle \partial_{\cF} | T^\alpha | \cF \rangle + \sum_{k=1}^\infty \eta_{-1-k\, \alpha\beta} \Tr[ T^\alpha \cG^\ord{k}] \Bigr) \cM^{-1} T^{\beta \dagger} \cM \CR
&= \rho^{-1}  \Bigl( \eta_{1\, \alpha\beta} \langle \partial_{\cF}  |\cM^{-1}  T^{\alpha \dagger} \cM | \cF \rangle + \sum_{k=1}^\infty \rho^{-2k} \eta_{1+k\, \alpha\beta} \Tr[ \cM^{-1} T^{\alpha \dagger} \cM  \cG^\ord{k}] \Bigr) T^{\beta} \,, 
\end{align}
and because $\mhbm{F}^\upgamma_1$ is defined as the $\dK$ component of this expression according to \eqref{shifted F M conj}, one obtains
\begin{align}
\rho \,  \mhbm{F}^\upgamma_1    = - \langle \partial_F | \cM^{-1} L_{-1} \cM | F\rangle -  \sum_{k=1}^\infty \rho^{-2k} \Tr[ \cM^{-1} L_{-1-k} \cM  G^\ord{k}] \; , 
\end{align}
where the fact that $\hbm{F}^\upgamma_{-1} $ does not depend on the two-forms was used to replace $|\cF\rangle $ and $\mathcal{G}^\ord{k}$ by  $|F \rangle $ and $G^\ord{k}$, respectively. 

Moreover, one has $\rho^{-1} \cD \tilde{\rho} = -\Tr [B^\ord{1}]$ at $\tilde{\rho}=0$ and one can therefore rearrange  the following terms in~\eqref{final cov Lagrangian J GAUGEFIX} according to
\begin{align} 
&\quad \quad \rho \, \dd \chi_1^\upgamma  - \rho \braket{\partial_\chi}{A}\chi^\upgamma_1
+ \eta^{AB} \sum_{n\in\mathds{Z}} n \sbm{A}^n_A  \,(  \tJ^{-n-1}_B -\tilde{\rho}  \tJ^{-n}_B)\nn\\
&= \cD^\flat ( \rho  \chi_1^\upgamma) - \sum_{n\in \mathds{Z}} (n+1) \sum_{k=0}^\infty \Tr[ T_{n-k}^A B^\ord{1+k} ]   \tJ^{n}_A  + \tfrac12  \mathsf{J}_0 \rho \chi^\upgamma_1 \; , 
\end{align}
where we used the bare  covariant derivative introduced in~\eqref{eq:flatD}. For $\cM \in E_9$ it will be convenient to introduce the shifted cocycles $\omega_{-k}^\alpha(\cM)$ defined in \eqref{shifted group cocycle E9}, that satisfies 
\be \cM^{-1} L_n \cM = \rho^{-2n} L_n - \rho^{2k-2n} \eta_{n-k\, \alpha\beta} \omega^\alpha_{-k}(\cM) T^\beta \; ,  \ee
such that $\omega_0^\alpha(\cM) = \omega(\cM)$. Using \eqref{move shift from cocycle}, the expression \eqref{opposite chi} for $\chi_{-k}^\upgamma$ can be simplified to
\be \chi_{-k}^\upgamma =\chi_{k}^\upgamma  + \rho^k \omega_{-k}^\alpha(\cM)  \mathfrak{J}_\alpha\; . \ee
Combining all these terms, one finally obtains the following form of  the topological term \eqref{final cov Lagrangian J GAUGEFIX} at $\tilde\rho=0$, up to the total derivative in the two-form fields in \eqref{plustwoforms} that we do not include for brevity, 
\begin{multline}
\label{TopPartRough}
\cL_{\scalebox{0.7}{top}} = \cD^\flat ( \rho  \chi_1^\upgamma) - \sum_{n\in \mathds{Z}} (n+1) \sum_{k=0}^\infty \Tr[ T_{n-k}^A B^\ord{1+k} ]   \tJ^{n}_A  + \tfrac12  \mathsf{J}_\dK \Tr B^\ord{1}+ \tfrac12 \eta^{AB} \sum_n n \mathsf{J}{}^n_A\mathsf{J}{}^{-1-n}_B  \\
+\tfrac12 \mathsf{J}_0\rho  \chi^\upgamma_1 + \tfrac12 \sum_{k=1}^\infty \rho^{1-k} \Tr B^\ord{k} \chi^\upgamma_{1+k}  + \tfrac12 \sum_{k=0}^\infty \bigl( \rho^{-1-k} \chi^\upgamma_{1+k} + \omega_{-1-k}^\alpha (\cM) \mathfrak{J}_\alpha \bigr) \Tr B^\ord{2+k}    \\
+\rho \langle \chi^\upgamma_1 | F\rangle - \langle \partial_F  | \bigl( L_1 + \cM^{-1} L_{-1} \cM | F \rangle - \sum_{k=0}^\infty \Tr\bigl[ \bigl( L_{-k} + \rho^{-2-2k} \cM^{-1} L_{-2-k} \cM \bigr) G^\ord{1+k} \bigr] \; . 
\end{multline}
With the split (\ref{SplitJA}) and (\ref{SplitchiA}), the above topological term can be written as
\begin{align} 
\label{TopPartMassage}
\cL_{\scalebox{0.7}{top}} &= \cD^{\flat} \chi^\flat_1  + \tfrac12 \eta^{AB} \sum_n n \cJ^{\flat}{}^n_A\cJ^{\flat}{}^{-1-n}_B+\tfrac12 \eta_{1\, \alpha\beta} \langle \partial_{A} | T^\alpha | A \rangle \langle \partial_{{A}} | \bigl( L_0 +  \rho^{-2} \cM^{-1} L_{-2} \cM \bigr)  T^\beta | {A}^\prime\rangle \nn\\
&\quad +\rho \langle \chi^\upgamma_1 | F\rangle - \langle \partial_F  | \bigl( L_1 + \cM^{-1} L_{-1} \cM | F \rangle  \nn\\[2ex]
&\quad  +\tfrac12 \cJ^\flat_0 \chi^\flat_1 + \tfrac12 \Tr[ \cJ^\flat B^\ord{1} ] + \tfrac12 \Tr\bigl[ \bigl( \cS_{-1}(\cJ^\flat) + \rho^{-2} \chi^\flat_1 + \omega_{-1}^\alpha(\cM) \cJ_\alpha^\flat \bigr) B^\ord{2}\bigr]   \nn\\[2ex]
&\quad + \tfrac12 \sum_{k\ge 1} \Tr \bigl[ \bigl( \cS_{-1-k}(\cJ^\flat) + \rho^{-2-2k} \chi^\flat_{1+k} + \omega_{-1-k}^\alpha(\cM)  \cJ^\flat_\alpha \bigr) \bigl( B^\ord{k+2} -\rho^2 B^\ord{k}\bigr) \bigr]   \nn\\
&\quad 
+\tfrac12 \sum_{k\ge 1} \sum_{q\ge 1} \rho^{-2k} \eta_{1+k-q \, \alpha\beta} \Tr[ T^\alpha B^\ord{q}] \Tr[ \cM^{-1} T^{\beta \dagger} \cM B^\ord{k} ] \; , 
\end{align}
where we used the explicit expression of $G^\ord{k}$ 
\begin{align}
G^\ord{k} &= \cD^\flat B^\ord{k} - \tfrac12 \delta^k_1 \eta_{1\, \alpha\beta} \langle \partial_A | T^\alpha |A\rangle T^\beta |A^\prime\rangle \langle \partial_A | + \tfrac12 \sum_{q=1}^\infty \eta_{-q\, \alpha\beta} \Tr[ T^\alpha B^\ord{q}] T^\beta B^\ord{k} \nn \\[-3mm] 
&\quad + \tfrac12 \sum_{r=1}^{k} (2r-1-k) \Tr[ B^\ord{r} ] B^\ord{1+k-r}  
\end{align} 
and the fact that the covariant derivative of $\chi^\upgamma_1$ compensates the one of these field strengths using 
\begin{align} 
& \quad \cD^\flat (\rho \chi^\upgamma_1 )- \sum_{k=0}^\infty \Tr\bigl[ \bigl( L_{-k} + \rho^{-2-2k} \cM^{-1} L_{-2-k} \cM \bigr) \cD^\flat B^\ord{1+k} \bigr]  \CR
&= \cD^\flat \chi^\flat_1 + \sum_{k=1}^\infty \Tr\bigl[ \cD^\flat \bigl(  \rho^{-2k} \cM^{-1} L_{-1-k} \cM \bigr)  B^\ord{k} \bigr] \; . 
\end{align}

The Euler--Lagrange equations of motion of the fields $B^\ord{k}$ and $\chi^\upgamma_k$ do not imply the duality equation \eqref{DualityInterm}, so one cannot integrate them out in the usual sense. In fact, integrating them out na\"ively gives rise to inconsistencies due to formal indefinite sums that appear in the topological term once one substitutes the solutions to their Euler--Lagrange equations. However, it turns out to be consistent to set them to the values in \eqref{chiCollapse} and \eqref{Bcollapse}. In general, one cannot substitute a  solution  to a duality equation (like (\ref{chiCollapse}) and (\ref{Bcollapse})) into a pseudo-Lagrangian. In the present case, the Euler--Lagrange equations of motions for the fields $\cM,\, |A\rangle$ and $\langle \chi|$ are automatically preserved, but  the Euler--Lagrange equations for $B^\ord{1}$ and $\chi^\upgamma_1$ are not. We shall see nonetheless that the latter are consistent with the duality equation. Indeed, after the substitution of \eqref{chiCollapse} and (\ref{Bcollapse}) in the pseudo-Lagrangian, the $B$ field and $\chi$ field Euler--Lagrange equations become  projected components  of the duality equation \eqref{DualityEqZ2}, which is itself by construction consistent with the original duality equation \eqref{DualityInterm}.  Therefore we obtain that the two pseudo-Lagrangians together with their respective duality equations define the same set of equations. Note that the use of the pseudo-Lagrangian was always to determine the Euler--Lagrange equations of motions for the fields $\cM,\, |A\rangle$ and $\langle \chi|$ only, whereas the Euler--Lagrange equations of the constrained fields $B^\ord{k}$ and $\chi^\upgamma_k$ were redundant with the duality equations. So it may not be that surprising that this manipulation turns out to be consistent.

Because \eqref{chiCollapse} and (\ref{Bcollapse}) involve the Hodge star operator, their substitution in the topological term \eqref{TopPartMassage} gives the sum $\cL_1+\cL_2$ of a kinetic term 
\begin{align}
 \label{LagrangegzeroKin}  
 \cL_1=  \tfrac12 \rho \,  \Tr\bigl[ \bigl( \cS_{-1}(\cJ^\flat) + \rho^{-2} \chi^\flat + \omega_{-1}^\alpha(\cM) \cJ_\alpha^\flat \bigr) \star B\bigr]  
-\tfrac14 \rho^{-1} \eta_{\alpha\beta} \Tr[ T^\alpha B] \star \Tr[ \cM^{-1} T^{\beta \dagger} \cM B ] \; ,
\end{align}
and a topological term
\begin{align} 
\label{LagrangegzeroTop}  
\cL_2 &= \cD^{\flat} \chi^\flat  + \tfrac12 \eta^{AB} \sum_n n J^{\flat}{}^n_A J^{\flat}{}^{-1-n}_B+\tfrac12 \eta_{1\, \alpha\beta} \langle \partial_{A} | T^\alpha | A \rangle \langle \partial_{{A}} | \bigl( L_0 +  \rho^{-2} \cM^{-1} L_{-2} \cM \bigr)  T^\beta | {A}^\prime\rangle \nn\\
&\quad
+\rho \langle \chi^\upgamma_1 | F\rangle - \langle \partial_F  | \bigl( L_1 + \cM^{-1} L_{-1} \cM | F \rangle +\tfrac12 \cJ^\flat_0 \chi^\flat + \tfrac12 \Tr[ \cJ^\flat B ]  \; .  
\end{align}
To obtain this result we have used in particular 
\begin{align}
&\hspace{4.5mm}  \tfrac12 \sum_{k=1}^\infty \sum_{q=1}^\infty \rho^{-2k} \eta_{1+k-q\,  \alpha\beta} \Tr[ T^\alpha B^\ord{q}] \Tr[ \cM^{-1} T^{\beta \dagger} \cM B^\ord{k} ]\CR
&=  \tfrac14 \sum_{k=1}^\infty \sum_{q=1}^\infty ( \rho^{-2k} \eta_{1+k-q\, \alpha\beta} - \rho^{2-2k} \eta_{k-1-q\,  \alpha\beta} ) \Tr[ T^\alpha B^\ord{q}] \Tr[ \cM^{-1} T^{\beta \dagger} \cM B^\ord{k} ]
\CR
&=  - \tfrac14 \rho^{-2}\eta_{\alpha\beta}\Tr[ T^\alpha B^\ord{1} ] \Tr[ \cM^{-1} T^{\beta \dagger} \cM B^\ord{2} ]  \CR
&\hspace{4.5mm}   -\tfrac14 \sum_{k=1}^\infty  \eta_{-k \, \alpha\beta} \Bigl(  \Tr[ T^\alpha B^\ord{k}] \Tr[ \cM^{-1} T^{\beta \dagger} \cM B^\ord{1} ] +\rho^{-2} \Tr[ T^\alpha B^\ord{k+1}] \Tr[ \cM^{-1} T^{\beta \dagger} \cM B^\ord{2} ] \Bigr) \CR
&\hspace{4.5mm}  + \tfrac14 \sum_{k=1}^\infty \sum_{q=1}^\infty \rho^{-2k} \eta_{1+k-q\,  \alpha\beta}  \Tr[ T^\alpha B^\ord{q}] \Tr[ \cM^{-1} T^{\beta \dagger} \cM ( B^\ord{k}-\rho^{-2} B^\ord{k+2}) ] \; . 
\end{align}
This last equation involves indefinite formal sums that we have regularised using geometric series regularisation as in \eqref{expand hatD PK} such that both the second and the last line cancel after the substitution \eqref{Bcollapse}. 

Now one can easily verify that the Euler--Lagrange equations of $B$ and $\chi^\flat$ derived from the pseudo-Lagrangian $\cL_1+\cL_2$ in (\ref{LagrangegzeroKin}) and (\ref{LagrangegzeroTop}), are proportional to the duality equation \eqref{DualityEqZ2}
\begin{align}
\delta ( \cL_1 +  \cL_2)  = \tfrac12 \Tr \Bigl[ \delta B \Bigl(  \rho^{-1} \star \cM^{-1} \bigl(  \cS_1(\cJ) + \chi \dK \bigr)^\dagger \cM - \cJ  \Bigr) \Bigr] + \tfrac12 \delta \chi
^\flat \bigl( \rho^{-1} \star   \Tr [  B] - J_{0} \bigr)  \; .
\end{align}
This concludes the consistency of the substitution of~\eqref{chiCollapse} and (\ref{Bcollapse}) in \eqref{TopPartMassage}. Note that although the relation between the topological term \eqref{TopPartMassage} and $\cL_1+\cL_2$ 
requires a regularisation, the two terms  \eqref{LagrangegzeroKin} and \eqref{LagrangegzeroTop} are well defined and $E_9$ invariant. We shall use them to define the pseudo-Lagrangian in the next section. In this minimal formulation we will see that the Euler--Lagrange equations are finite and do not require any regularisation.

\subsection{Internal diffeomorphisms invariance and pseudo-Lagrangian}
\label{InternalDifMin}

We introduce now the unimodular metric $\tilde{g}_{\mu\nu}$ such that $\det \tilde{g}=-1$ and the two-dimensional metric is $g_{\mu\nu} = e^{2\sigma} \tilde{g}_{\mu\nu}$.  Because this formulation of the theory is not manifestly covariant under $\Sigma$ gauge transformations, it will be useful to use the bare covariant derivative that was  defined in~\eqref{eq:flatD} 
as $\mathcal{D}^\flat = \dd - \cL_{(A ,0)}$ 
instead of $\mathcal{D} = \dd - \cL_{(A ,B)}$. 
For a non-constant unimodular metric $\tilde{g}_{\mu\nu}$, the definition of the current $\cJ^\flat$ has to be generalised from~\eqref{DflatM} to
\begin{align}
  {\cJ}^\flat_{\mu \alpha} T^\alpha =  \cM^{-1} \cD^\flat_\mu \cM  + \tilde{g}^{\nu\sigma} (  \partial_{\nu} - \langle \partial_{\tilde{g}} | A_\nu \rangle ) \tilde{g}_{\mu\sigma}  \dK\; .   
\end{align}
This additional central component is necessary for the covariance of the duality equation under external diffeomorphisms as we shall see later.
It can also be understood by identifying this central component written in terms of the two-dimensional metric $g_{\mu\nu}$
\begin{align}
   {\cJ}^\flat_{\mu \alpha} T^\alpha =  \widetilde{\cM}^{-1} \cD^\flat_\mu \widetilde{\cM}  + {g}^{\nu\sigma} ( \cD_\nu^\flat {g}_{\mu\sigma}  - \cD_\mu^\flat g_{\nu\sigma}) \dK\; ,   
 \end{align}
as the gravitational flux (see e.g. \cite[Eq. 4.13a]{Bossard:2019ksx}), where $\widetilde{\cM}$ does not include the conformal factor $\sigma$.
According to \eqref{Jchidef}, the current $\cJ = \cJ_\alpha T^\alpha$ involves only the $\cM$-independent $B$ term as
\begin{align}
\label{Jdef} 
 {\cJ}_{\mu \alpha} T^\alpha =  \cM^{-1} \cD^\flat_\mu \cM - \eta_{-1\, \alpha\beta} \Tr[ T^\beta B_\mu ] T^\alpha + \tilde{g}^{\nu\sigma} (  \partial_{\nu} - \langle \partial_{\tilde{g}} | A_\nu \rangle ) \tilde{g}_{\mu\sigma}  \dK\; , 
 \end{align}
and therefore $\cJ$ is not equal to $\cM^{-1} \cJ^\dagger \cM$. It does not transform covariantly under generalised diffeomorphisms either, but the duality equation \eqref{DualityEqZ2}, that we reproduce here for convenience
\begin{align}
 \label{DualityEquation} \cJ_{\mu \alpha} T^\alpha = \rho^{-1} \tilde{g}_{\mu\sigma} \varepsilon^{\sigma\nu}    \cM^{-1}  ( \mathcal{S}_1(  \cJ_{\nu  \alpha} T^\alpha ) + \chi_\nu  \dK )^\dagger  \cM \; , 
 \end{align}
is invariant under internal diffeomorphisms, as we prove below. The Hodge star operator is written out with respect to the unimodular metric $\tilde{g}_{\mu\nu}$ with the convention $\varepsilon_{\mu\nu} = \tilde{g}_{\mu\sigma} \tilde{g}_{\nu\rho} \varepsilon^{\sigma\rho}$ and $\varepsilon^{01}= - \varepsilon_{01} = 1$.

We define the gauge transformations of parameter $|\Lambda\rangle$
\begin{align} 
\label{LinGauE11}
\delta_{\Lambda}  \tilde{g}_{\mu\nu}  &=\langle \partial \tilde{g}_{\mu\nu} | \Lambda\rangle \; ,  \nn\\
\delta_{\Lambda}  \cM  &=  \langle \partial_\cM   |\Lambda\rangle \cM  + \eta_{\alpha\beta} \langle \partial_\Lambda | T^\beta  | \Lambda\rangle \bigl( T^{\alpha \dagger} \cM + \cM T^\alpha \bigr)  \; ,  \nn\\
\delta_\Lambda |A_\mu \rangle &= \partial_\mu | \Lambda \rangle + \langle \partial_A | \Lambda\rangle  |A_\mu\rangle - \eta_{\alpha\beta} \langle \partial_\Lambda | T^\alpha | \Lambda\rangle T^\beta   |A_\mu\rangle- \langle \partial_\Lambda | \Lambda\rangle  |A_\mu\rangle  \; , 
\end{align}
and
\begin{align}
\delta_\Lambda  B_\mu &= \langle \partial_ B  | \Lambda \rangle  B_\mu  - \eta_{\alpha\beta} \langle \partial_\Lambda | T^\alpha | \Lambda\rangle [ T^\beta , B_\mu ] +\langle \partial_\Lambda | \Lambda \rangle B_\mu  + \eta_{1\alpha\beta} \langle \partial_\Lambda | T^\alpha |  \Lambda \rangle  T^\beta | A_\mu  \rangle \langle \partial_\Lambda |  \label{Bgauge} \\
& \quad - \rho  \tilde{g}_{\mu\sigma} \varepsilon^{\sigma\nu} \bigl( \eta_{\alpha\beta} \langle  \partial_\Lambda | T^\alpha |  \Lambda \rangle  T^\beta | A_\nu  \rangle  + \langle \partial_\Lambda |  \Lambda  \rangle | A_\nu \rangle -  \langle \partial_\Lambda  | A_\nu \rangle |  \Lambda \rangle\bigr)  \langle \partial_\Lambda | \nn\\[1mm]
\delta_\Lambda  \chi_\mu &= \langle \partial_ \chi  | \Lambda \rangle  \chi_\mu  + \langle \partial_\Lambda | \Lambda \rangle \chi_\mu + \sum_n n \langle \partial_\Lambda | T_n^A | \Lambda \rangle J_{\mu A}^{n-1} - \eta_{\alpha\beta} \langle \partial_\Lambda | T^\alpha |  \Lambda \rangle   \langle \partial_\Lambda |  \cM^{-1} L_{-1}  \cM T^\beta | A_\mu  \rangle\nn\\[-1.5mm]
&  \hspace{10mm} -  \langle \partial_\Lambda  |  \Lambda \rangle \langle \partial_\Lambda  | \cM^{-1} L_{-1} \cM  |   A_\mu \rangle  +  \langle \partial_\Lambda |  A_\mu \rangle  \langle \partial_\Lambda | \cM^{-1}  L_{-1} \cM  | \Lambda \rangle  \nn \\
& \quad - \rho  \tilde{g}_{\mu\sigma} \varepsilon^{\sigma\nu} \bigl( \eta_{\alpha\beta} \langle \partial_\Lambda | T^\alpha |  \Lambda \rangle   \langle \partial_\Lambda |  L_0 T^\beta | A_\nu \rangle +  \langle \partial_\Lambda |  \Lambda  \rangle  \langle \partial_\Lambda | L_0 | A_\nu  \rangle -  \langle \partial_\Lambda  |  A_\nu  \rangle \langle \partial_\Lambda  |L_0 |   \Lambda  \rangle  \bigr) \; . \nonumber 
\end{align}

Note that the gauge transformation of the gauge field is not $\cD_\mu |\Lambda\rangle$ as usual in exceptional field theory, but it differs by a trivial diffeomorphism as shown in \eqref{eq:deltaA}. In particular, the parts of the gauge fields transformations which are independent of the Hodge dual match with~\eqref{GaugeAused} and~\eqref{GaugeBused}. The Hodge dual terms in the variations of $B$ and $\chi$ follow directly from \eqref{IntermLambdaB}. They can also be derived from the gauge transformations in \cite{Bossard:2019ksx}, in which the Hodge dual terms and the gauge field transformation \eqref{LinGauE11} appear naturally.

To prove that the duality equation \eqref{DualityEquation} is invariant under internal diffeomorphisms one computes the gauge transformation of the current 
\begin{align} 
\delta_\Lambda \cJ &= \langle \partial_{\cJ} | \Lambda \rangle \cJ + \eta_{\alpha\beta} \langle \partial_\Lambda | T^\alpha | \Lambda \rangle [ \cJ , T^\beta]  \\
&\hspace{4.5mm}  + \eta_{\alpha\beta}  \bigl( \eta_{\gamma\delta} \langle  \partial_\Lambda | T^\gamma |  \Lambda  \rangle  \langle \partial_\Lambda | T^\alpha T^\delta | A \rangle  + \langle \partial_\Lambda |  \Lambda  \rangle  \langle \partial_\Lambda | T^\alpha | A \rangle -  \langle \partial_\Lambda  | A \rangle \langle \partial_\Lambda | T^\alpha |  \Lambda \rangle \bigr)  \cM^{-1} T^{\beta \dagger} \cM \nn\\
&\hspace{4.5mm}
+ \eta_{-1\, \alpha\beta}  \rho \star  \bigl( \eta_{\gamma\delta} \langle  \partial_\Lambda | T^\gamma |  \Lambda  \rangle  \langle \partial_\Lambda | T^\alpha T^\delta | A \rangle  + \langle \partial_\Lambda |  \Lambda \rangle \langle \partial_\Lambda | T^\alpha | A \rangle -  \langle \partial_\Lambda  | A \rangle \langle \partial_\Lambda | T^\alpha |  \Lambda  \rangle \bigr) T^\beta \; , \nn
\end{align} 
and we identify the first line with its covariant transformation, while the two last lines are found to project out in the duality equation, up to a central element that is compensated by the gauge transformation of $\chi$.

The duality equation \eqref{DualityEquation} is a twisted self-duality equation that can be obtained from a truncation of the $E_{11}$ twisted self-duality equation defined in \cite{Bossard:2019ksx}. Twisted self-duality in $D=2p$ dimensions can normally be written as an equality between a $p$-form field strength dressed with the scalar matrix $\cM\in G$ and its Hodge star contracted with a $G$-invariant bilinear form. For $E_9$, the one-form field strength combines the current $\cJ$ together with the one-form $\chi$  in the module  $\langle L_1\rangle^* \oleft \mf{e}_9 \oleft \langle L_{-1}\rangle $, where $\langle L_1\rangle^* \oleft \mf{e}_9$ is the module conjugate to $ \mf{e}_9 \oleft \langle L_1\rangle$ that describes the indecomposable representation of the field $\chi$ with the current $\cJ$. The symmetric $E_9$-invariant bilinear form is 
\be\rho  \langle ( \chi_\mu,\cJ_\mu ) , (\chi_\nu, \cJ_\nu )\rangle  = \sum_{n\in \mathds{Z}} \eta^{AB} J_\mu{}^n_A J_\nu{}^{-n-1}_B - J_{\mu\,  -1 } J_{\nu \, \dK} -  J_{\mu\,  \dK } J_{\nu\,  -1}- J_{\mu\,  0} \chi_\nu - \chi_\mu J_{\nu\,  0}  \; . \label{bilinearE9} \ee
This bilinear form, together with the action of $E_9$ on the module $\langle L_1\rangle^* \oleft \mf{e}_9 \oleft \langle L_{-1}\rangle $, define the twisted self-duality equation in  $\langle L_1\rangle^* \oleft \mf{e}_9 \oleft \langle L_{-1}\rangle $. 
  \be ( \chi , \cJ ) = ( \cI[\chi],\cI[\cJ]) \; , \label{TwistedSelf} \ee
 with the involution\footnote{The conjugation by $\cM$ uses the $K(E_{9})$ invariant bilinear form defined at $\cM= \mathds{1}$ as
 \be \bigl( (\chi_\mu , \cJ_\mu) , (\chi_\nu , \cJ_\nu) \bigr)  = \sum_{n\in \mathds{Z}} \delta^{AB} J_\mu{}^n_A J_\nu{}^n_B - J_{\mu\, 0} J_{\nu\, \dK} -J_{\mu\, \dK} J_{\nu\, 0} - J_{\mu\, -1} \chi_\nu -  \chi_{\mu\, } J_{\nu\, -1}\; . \vspace{-2ex}\nn 
 \ee }
\begin{align}
\cI[  \mathcal{J} ] = \rho^{-1} \star \cM^{-1} (  \mathcal{S}_1(  \cJ ) + \chi \dK )^\dagger \cM \; ,  \qquad \cI[ \chi ] = \rho \star J_{\dK} - \rho^2 \omega^\alpha_{-1}(\cM)  \,\cI[\cJ]_\alpha \; .  
\end{align}
The equation $\chi  = \cI[\chi] $ follows from the central component of  \eqref{DualityEquation} 
\begin{align} 
 \label{DualityJK} J_{ \dK} = \rho^{-1} \star \chi  + \rho\,  \omega_{-1}^\alpha(\cM) \star \cJ_\alpha   \; .
 \end{align}
One checks indeed that 
\begin{align}
 \cI[ \cI[ \cJ]]  &=   \rho^{-1} \cM^{-1} \star \Bigl(  \mathcal{S}_1 \bigl(  \rho^{-1}  \cM^{-1} \star ( \mathcal{S}_1(   \cJ ) + \chi  \dK )^\dagger  \cM \bigr) +  \chi  \dK \Big)^\dagger  \cM  + \rho^{-1} \star \cI[\chi ]   \dK \nn\\
&=   \cJ   \label{CheckInvol}\; , 
\end{align}
and $ \cI[ \cI[ \chi]]= \chi$. Although \eqref{TwistedSelf} is manifestly involutive, it is more convenient to write  it as an equation in the Lie algebra $\mf{e}_9 \oleft \langle L_{-1}\rangle$ as  \eqref{DualityEquation}.

The pseudo-Lagrangian of the theory for a non-constant unimodular metric $\tilde{g}_{\mu\nu}$ is defined as
\begin{align}
 \label{PseudoLagrangian} 
 \cL_{\text{min}} = \cL_1 + \cL_2 +\frac14 \rho  \varepsilon^{\mu\nu} \varepsilon^{\sigma\rho}  \tilde{g}^{\kappa\lambda} \cD_\mu \tilde{g}_{\sigma\kappa} \cD_\nu  \tilde{g}_{\rho\lambda} + \frac{\rho^{-1}}{4} \langle \partial \tilde{g}^{\mu\nu} | \cM^{-1} | \partial \tilde{g}_{\mu\nu} \rangle -V\; , 
 \end{align}
where the kinetic term
\begin{align} 
\label{eq:kinmin}
\cL_1= \tfrac12 \rho^{-1} \tilde{g}^{\mu\nu} \Bigl(  \Tr \bigl[ \bigl(  \cM^{-1}  \cS_{1}( \cJ_{\mu})^\dagger \cM  + \chi_\mu  \bigr) B_\nu \bigr] +\tfrac12   \eta_{\alpha\beta} \Tr[ T^\alpha B_\mu ] \Tr[ \cM^{-1} (T^{\beta})^\dagger \cM B_\nu ] \Bigr) 
\end{align}
and the topological term\footnote{In this expression we extracted the top-form factor on the left, with the convention that $A B = \varepsilon^{\mu\nu} A_\mu B_\nu  \dd x^0 \!\! \wedge \! \dd x^1 $ and $A\star B = \tilde{g}^{\mu\nu} A_\mu B_\nu  \dd x^0 \!\! \wedge \! \dd x^1  $ for one-forms in the two-dimensional external space.}
\begin{align}
\label{eq:topmin}
 \cL_2  \,\dd x^0 \!\wedge  \dd x^1   &=   \cD \chi +\tfrac12 J_0 \chi +\tfrac12 J_{-1} J_{\dK} +\tfrac12 \eta^{AB} \sum_n n J_A^n J_B^{-1-n}  + \rho^2 \langle  \chi- \omega_{-1}^\alpha(\cM) \cJ_\alpha | F \rangle  \\ 
&  \hspace{-11mm} - \langle \partial_F | ( L_1 + \cM^{-1} L_{-1} \cM ) |F\rangle  - \Tr[ L_0 G]  +\tfrac12 \rho^{-2} \eta_{1\alpha\beta} \langle \partial_{A} | T^\alpha | A \rangle \langle \partial_{{A}} | \cM^{-1} L_{-2} \cM T^\beta | A^\prime \rangle \; , \nn 
\end{align}
are the direct generalisations of \eqref{LagrangegzeroKin}  and \eqref{LagrangegzeroTop} with a unimodular metric  $\tilde{g}_{\mu\nu}$. Note that we wrote explicitly $\langle \chi^\upgamma_1 | = \rho \langle  \chi - \omega_{-1}^\alpha(\cM) \cJ_\alpha | $ in terms of the field $\langle \chi|$ that appears in the potential term $V$ \cite{Bossard:2018utw}. We recall that the (internal) $\langle \chi |$ transforms as the central component of 
\be \langle \cJ^-_\alpha |\otimes T^\alpha =  \langle \cJ_\alpha |\otimes \cS_{-1}(T^\alpha) + \langle \chi | \otimes \dK\; , \label{InternalJminus} \ee
whereas the (external) vector field $\chi$ instead transforms as the central component of $\cS_1(\cJ) + \chi \dK$. In addition to the dependence in the general unimodular metric $\tilde{g}_{\mu\nu}$, \eqref{eq:kinmin} and \eqref{eq:topmin} differ from 
\eqref{LagrangegzeroKin}  and \eqref{LagrangegzeroTop} in that we have included (most of) the dependence in the field $B_\mu$ in the current \eqref{DualityEquation}, the covariant derivative $\cD \chi$ and the field strength $G$, using similar steps as in passing from \eqref{TopPartRough} to \eqref{TopPartMassage}.

In~\eqref{PseudoLagrangian}, the first two terms $\cL_1{+}\cL_2$ together, the two terms involving the derivative of the unimodular metric and the potential $V$ are individually 
invariant under internal diffeomorphisms up to a total derivative. We shall see that their relative coefficients are determined by external diffeomorphisms.

In order to study the invariances of the proposed pseudo-Lagrangian~\eqref{PseudoLagrangian}, we recall a few definitions. The field strength $F$ and $G$ are defined as in \eqref{eq:barefs}, 
while the covariant derivative 
\begin{equation}
\label{Dchi} 
\cD \chi = \dd \chi - \langle \partial_\chi | A \rangle \chi - \langle \partial_A | A \rangle \chi - \sum_n n \langle \partial_A | T_n^A | A\rangle J_A^{n-1} - \sum_n  (n+1) \Tr [ T_n^A B ] J_A^n - \Tr[B] J_{\dK}  \; , 
\end{equation}
follows from the indecomposable representation of the field $\chi$ under $\mf{e}_9 \oleft \langle L_{-1}\rangle$.\footnote{Note that $\cD \chi$ differs from $\cD \chi_1^\upgamma$ in Section \ref{sec:extE9exft}, because $\chi_1^\upgamma$ is invariant under $L_{0}$ and $L_{-1}$, while $\chi$ transforms respectively into itself and $J_\dK$ under these two generators.}  One derives the Bianchi  identity for $\cJ$
\begin{align} 
&\hspace{4.5mm} \cD \cJ + \cJ^2 + \langle \cJ | F \rangle + \eta_{\alpha\beta} \langle \partial_F | ( T^\alpha + \cM^{-1} T^{\alpha \dagger} \cM ) |F \rangle T^\beta+ \eta_{-1\, \alpha\beta} \Tr[ T^\alpha G ] T^\beta \nn\\
&= \tfrac{1}{2} \eta_{-1\, \alpha\beta} \eta_{1 \, \gamma\delta} \langle \partial_{A} | T^\gamma | A\rangle \langle \partial_{A} | T^\alpha T^\delta | A^\prime \rangle \cM^{-1} T^{\beta\dagger} \cM \; .  
\end{align}
The non-covariance of the right-hand-side follows from the property that $\cJ$ does not include the term in $- \eta_{-1\, \alpha\beta} \Tr[ T^\alpha B] \cM^{-1} T^{\beta \dagger} \cM$ in its definition. One can also derive the rigid $E_9$ invariant topological term using the same construction as in \eqref{completed shifted CM eq J}. One can indeed write a Bianchi identity for the shifted current that transforms under ${\mf e}_9$ by a commutator with the ${\mf e}_9$ generators $T^\alpha$. This Bianchi identity is satisfied up to a central element that is equal to the topological term  $\cL_2$  
\begin{align} 
&\quad \cD \bigl( \cS_1(\cJ) + \chi \dK \bigr) + \tfrac12 J_0 \bigl(  \cS_1(\cJ) + \chi \dK \bigr) + \tfrac12 J_{-1} \cJ + \tfrac12  \cJ \bigl(  \cS_1(\cJ) + \chi \dK \bigr)  + \tfrac12 \bigl( \cS_1(\cJ) + \chi \dK \bigr)  \cJ  \nn\\
&+ \langle \cJ_\alpha |  \cS_1(T^\alpha ) |  F\rangle+ \langle \rho^2 ( \chi - \omega_{-1}^\alpha(\cM) \cJ_\alpha )|  F\rangle + \eta_{1\alpha\beta} \langle \partial_F | \bigl( T^\alpha + \cM^{-1} T^{\alpha \dagger}  \cM\bigr) | F \rangle T^\beta\nn\\
 &+ \eta_{\alpha\beta} \Tr[ T^\alpha G ] T^\beta - \tfrac12 \rho^{-2} \eta_{2\,  \alpha\beta} \eta_{1\, \gamma\delta}\langle \partial_{A} | T^\gamma | A \rangle \langle \partial_{A} | \cM^{-1} T^{\alpha \dagger} \cM  T^\delta | A^\prime \rangle  T^\beta = \cL_2 \dK \; ,   
 \end{align} 
which implies that $\cL_2$ is rigid $E_9$ invariant. The term $\cL_1$ is manifestly invariant under rigid $E_9$ and can be determined such that the Euler--Lagrange equation for $B$ is a projection of the duality equation.

Let us now show that $\cL_1{+}\cL_2$ transforms as a density under internal generalised diffeomorphisms. The non-covariant variation of $\cL_1$ gives
\begin{align}
\Delta_\Lambda \cL_1 &=   
 \tfrac12 \rho \tilde{g}^{\mu\nu} \bigl( \eta_{\gamma\delta} \langle \partial_\Lambda | T^\gamma | \Lambda \rangle \langle \partial_\Lambda | T^\alpha T^\delta | A_\mu \rangle  + \langle \partial_\Lambda |\Lambda  \rangle \langle \partial_\Lambda |  T^\alpha | A_\mu \rangle \nn\\ 
 &\quad - \langle \partial_\Lambda | A_\mu \rangle \langle \partial_\Lambda | T^\alpha |\Lambda  \rangle \bigr) \bigl( \cJ_{\nu \alpha} +\cI[\cJ_{\nu \alpha}]  \bigr)  
\end{align} 
To vary $\cL_2$ we must first derive the transformation of the field strengths. To avoid a cumbersome computation one first observes that the gauge transformation of $G$ is mostly determined from the gauge transformation of the field strength $F$ by  the Dorfman product \eqref{eq:DorfmanProduct} according to \eqref{eq:fsvar}. Therefore writing $\delta_\Lambda |F\rangle$ as $\cL_{(\Lambda,0)} |F\rangle $ plus an explicit trivial parameter allows to determine the gauge transformation of $G$, up to  doubly constrained trivial parameters \eqref{eq:triv4}--\eqref{eq:triv6} that only affects $G$ and the non-covariant piece of the $B$ field gauge transformation (i.e. the second line in \eqref{Bgauge}) that must be computed separately. But because the terms in the doubly constrained parameters leave invariant $\Tr[L_0 G]$ in the pseudo-Lagrangian (see Appendix \ref{TrivialPara}),  we  do not need to compute them. One computes in this way  
  \begin{align} 
  \label{Fvariation}
  \delta_\Lambda |F\rangle &= \cL_{(\Lambda,0) } |F\rangle + \tfrac{1}{2} \bigl( \cL_{\dd |\Lambda\rangle } |A\rangle \rangle {-} \cL_{|A\rangle } \dd |\Lambda \rangle \bigr) + \tfrac12 \eta_{-1\alpha\beta} \eta_{1\gamma\delta} \langle \partial_\Lambda | T^\gamma | A\rangle \langle \partial_\Lambda | T^\alpha T^\delta |\Lambda\rangle T^\beta |A\rangle  \\
\delta_\Lambda G &= \cL_{(\Lambda,0)} G + \eta_{1\, \alpha\beta} \langle \partial_\Lambda | T^\alpha | \Lambda \rangle T^\beta | F \rangle \langle \partial_\Lambda |\nn\\
& \hspace{15mm}   + \tfrac12 \eta_{1\, \alpha\beta} \langle \partial_\Lambda  | T^\alpha | A \rangle T^\beta \dd|\Lambda\rangle \langle \partial_\Lambda | - \tfrac12 \eta_{1\alpha\beta} \langle \partial_A  | T^\alpha | A \rangle T^\beta \dd|\Lambda\rangle \langle \partial_A | \nn\\
& \hspace{8mm} - \tfrac12  \eta_{1\, \gamma\delta} \langle \partial_\Lambda | T^\gamma | A\rangle \bigl(  \eta_{\alpha\beta}  T^\alpha  |A\rangle   \langle \partial_\Lambda | T^\beta T^\delta |\Lambda\rangle \langle \partial | -  |A\rangle  \langle \partial |  T^\delta |\Lambda \rangle \langle  \partial_\Lambda |  +T^\delta |\Lambda \rangle   \langle \partial |  A\rangle  \langle  \partial_\Lambda |    \bigr) \nn \\
& \hspace{12mm} + \cD \Bigl(  \rho  \star  \bigl( \eta_{\alpha\beta} \langle  \partial_\Lambda | T^\alpha |  A \rangle  T^\beta | \Lambda \rangle  + \langle \partial_\Lambda |  A  \rangle | \Lambda \rangle -  \langle \partial_\Lambda  | \Lambda \rangle |  A \rangle\bigr)  \langle \partial_\Lambda | \Bigr) + \dots   \; , \label{FGvariation} 
\end{align}
where the first line in $\delta_\Lambda G$ directly follows from the Dorfman product  \eqref{eq:DorfmanProduct}, the second and third lines keep track on the trivial parameter with one constrained index in the transformation of $|F\rangle$, the last line is computed directly and the ellipses stand for a trivial parameter with two constrained indices.

With this we can now study the gauge transformation of the topological term~\eqref{eq:topmin}. When varying $\langle \partial_F |  L_1  |F\rangle + \Tr[ L_0 G]  $, one obtains that all the contributions cancel up to a total derivative, except for  the one obtained by commuting $\langle \partial_F|$ with $\cL_{|\Lambda\rangle}$ on $\delta_\Lambda |F\rangle$ and the last term (involving the Hodge star) in the variation of $\delta_\Lambda G$. To understand this cancellation, one notes that $\langle \partial_F |  L_1  |F\rangle + \Tr[ L_0 G]  $ is  invariant under all one-form gauge transformations up to a total internal derivative~\eqref{Ahat triv2}, and therefore, the trivial parameter components of the variations \eqref{Fvariation} and \eqref{FGvariation} cancel up to a total derivative in the variation of $\langle \partial_F |  L_1  |F\rangle + \Tr[ L_0 G]  $. Eventually one obtains 
\bea & & \Delta_\Lambda \bigl( - 2\langle \partial_F |  L_1  |F\rangle  -2 \Tr[ L_0 G]  \bigr)\nn\\
 &=&  2 \eta_{\alpha\beta} \langle \partial_\Lambda | T^\alpha | \Lambda \rangle  \langle \partial_\Lambda |  L_1   T^\beta |F\rangle+2 \langle \partial_\Lambda  | \Lambda \rangle  \langle \partial_\Lambda |  L_1   |F\rangle +2 \eta_{1\alpha\beta} \langle \partial_\Lambda | T^\alpha | F \rangle  \langle \partial_\Lambda | L_0  T^\beta | \Lambda \rangle \nn\\
 && \qquad - 2 \cD  \bigl[ \star \rho\bigl(  \langle\partial_\Lambda  |T_\alpha | \Lambda \rangle \langle \partial_\Lambda |  L_0 T^\alpha |A\rangle    +\langle \partial_\Lambda |A\rangle \langle \partial_\Lambda | L_0 |A\rangle - \langle \partial_\Lambda | \Lambda \rangle \langle \partial_\Lambda | L_0 | A\rangle \bigr)  \bigr]\nn\\
 &=&  2 \langle \partial_\Lambda  | L_1   | \Lambda \rangle  \langle \partial_\Lambda |  F\rangle  \nn\\
  && - 2 ( \dd - \langle \partial | A\rangle ) \bigl[ \star \rho\bigl(  \langle\partial_\Lambda  |T_\alpha | \Lambda \rangle \langle \partial_\Lambda |  L_0 T^\alpha |A\rangle    +\langle \partial_\Lambda |A\rangle \langle \partial_\Lambda | L_0 |A\rangle - \langle \partial_\Lambda | \Lambda \rangle \langle \partial_\Lambda | L_0 | A\rangle \bigr)  \bigr]\nn\\
 & & -2 \rho \bigl( \langle \partial_A | [ L_0 , T_\alpha] |A\rangle + \eta_{-1\, \alpha\gamma} \Tr[ [ L_0, T^\gamma] B] \bigr)\nn\\
 && \hspace{10mm}  \star \bigl(  \langle\partial_\Lambda  |T_\beta | \Lambda \rangle \langle \partial_\Lambda | T^\alpha  T^\beta |A\rangle    +\langle \partial_\Lambda |A\rangle \langle \partial_\Lambda | T^\alpha |A\rangle - \langle \partial_\Lambda | \Lambda \rangle \langle \partial_\Lambda | T^\alpha  | A\rangle \bigr)  \; , \eea
 where the second line is a total derivative. We also have after some algebra
%
\begin{align} 
 &\hspace{4.5mm} \Delta_\Lambda  \bigl(- 2\langle \partial_F | \cM^{-1} L_{-1} \cM  |F\rangle  +\rho^{-2} \eta_{1\alpha\beta} \langle \partial_{A_2} | T^\alpha | A_{1} \rangle \langle \partial_{{A_2}} | \cM^{-1} L_{-2} \cM T^\beta | {A_{2}}\rangle \bigr) \\
&=
 2 \langle \partial_\Lambda | T_\alpha | \Lambda \rangle  \langle \partial_\Lambda |  \cM^{-1} L_{-1} \cM   T^\alpha |F\rangle
 {+}2 \langle \partial_\Lambda  | \Lambda \rangle  \langle \partial_\Lambda | \cM^{-1}  L_{-1} \cM   |F\rangle  
 {+}2 \langle \partial_\Lambda | A \rangle \langle \partial_\Lambda | \cM^{-1} L_{-1} \cM | \dd \Lambda \rangle \nn\\
 &\! \quad 
 +2 \langle \partial_\Lambda | \dd \Lambda \rangle \langle  \partial_\Lambda | \cM^{-1} L_{-1} \cM | A \rangle
 +  2   \langle \partial_\Lambda | T_\alpha | A \rangle \langle \partial_\Lambda | \cM^{-1} L_{-1} \cM T^\alpha | \dd\Lambda \rangle 
 \nn \\
 &\! \quad 
+  \langle \partial_\Lambda | T_\alpha | \Lambda\rangle \bigl(  \langle \partial_\Lambda | T_\beta T^\alpha |A_1\rangle \langle \partial_{A_2} | \cM^{-1} L_{-1} \cM T^\beta | A_2\rangle - \langle \partial_{A_1} | T_\beta T^\alpha | A_1 \rangle \langle \partial_\Lambda | \cM^{-1} L_{-1} \cM T^\alpha | A_2 \rangle \bigr)  \nn\\
 &\! \quad 
 - \langle \partial_\Lambda | T^\alpha |\Lambda \rangle \bigl(  \langle \partial_\Lambda | A_1\rangle \langle 2\partial_\Lambda {+} \partial_{A_1} {+} 2\partial_{A_2}| \cM^{-1} L_{-1} \cM T^\beta |A_2\rangle + \langle \partial_{A_2} | A_1\rangle \langle \partial_\Lambda | \cM^{-1} L_{-1} \cM T^\alpha |A_2\rangle  \bigr)  \nn\\
 &\! \quad 
 +  \langle \partial_{A_2} | \Lambda\rangle \langle \partial_\Lambda | T_\alpha | A_1\rangle \langle \partial_\Lambda | \cM^{-1} L_{-1} \cM T^\alpha | A_2\rangle +  \langle \partial_\Lambda | \Lambda \rangle \langle   \partial_{A_{1}}  | A_1\rangle \langle \partial_\Lambda  | \cM^{-1} L_{-1} \cM |A_2 \rangle \nn\\
 &\! \quad 
+\langle \partial_{A_2}  | \Lambda \rangle \langle \partial_{\Lambda }  | A_1\rangle \langle \partial_\Lambda  | \cM^{-1} L_{-1} \cM |A_2 \rangle  - 2 \langle \partial_\Lambda | \Lambda \rangle \langle \partial_\Lambda | A_1\rangle \langle  \partial_\Lambda {+} \partial_{A_1} {+} \partial_{A_2} | \cM^{-1} L_{-1} \cM |A_2 \rangle\nn
\end{align} 
where we have written $T_{\alpha} = \eta_{\alpha\beta} T^\beta$ for the  $\mf{e}_9$ generators and one should not forget that the transport term in the Lie derivative of $|A\rangle$ does contribute to the non-covariant variation of the second term when two derivatives act on the parameter $|\Lambda\rangle$. One also computes 
\begin{align}
 \Delta_\Lambda \bigl( 2\langle \rho^2 (\chi- \omega_{-1}^\alpha(\cM) J_\alpha )| F \rangle\bigr)   = - 2 \langle \partial_\Lambda | ( L_1 + \cM^{-1} L_{-1} \cM ) |\Lambda\rangle \langle \partial_\Lambda | F\rangle \; , 
 \end{align}
and  finally  that 
\begin{align}
&\hspace{4.5mm}  \Delta_\Lambda \bigl(  2 D \chi +J_0 \chi +J_{-1} J_{\dK} +\eta^{AB} \sum_n n J_A^n J_B^{-1-n}  \bigr) \\
 &=  \langle \partial_\Lambda | T_\alpha  | A \rangle \langle \partial_\Lambda | \cM^{-1}\cS_1(J)^\dagger \cM T^\alpha | \Lambda \rangle
 + \langle \partial_\Lambda |A  \rangle \langle \partial_\Lambda |  \cM^{-1}\cS_1(J)^\dagger \cM | \Lambda \rangle
  \CR 
&\!\!\quad 
{+} \langle \partial_\Lambda | \Lambda \rangle \langle \partial_\Lambda | \cM^{-1}\cS_1(J)^\dagger \cM |A \rangle     {+}2   \langle \partial_\Lambda | T_\alpha  | F\rangle \langle \partial_\Lambda | \cM^{-1} L_{-1} \cM T^\alpha | \Lambda \rangle
{+}2 \langle \partial_\Lambda |F  \rangle \langle \partial_\Lambda |  \cM^{-1} L_{-1} \cM | \Lambda \rangle  \CR
&\!\!\quad 
{-}2 \langle \partial_\Lambda | \Lambda \rangle \langle \partial_\Lambda | \cM^{-1}L_{-1}  \cM |F \rangle  
{-}2  \langle \partial_\Lambda | T_\alpha  | A \rangle \langle \partial_\Lambda | \cM^{-1} L_{-1} \cM T^\alpha \dd | \Lambda \rangle  \CR
&\!\!\quad 
{-}2 \langle \partial_\Lambda |A  \rangle \langle \partial_\Lambda |  \cM^{-1} L_{-1} \cM \dd  | \Lambda \rangle
{+}2\langle \partial_\Lambda |\dd \Lambda \rangle \langle \partial_\Lambda | \cM^{-1}L_{-1}  \cM |A \rangle  \bigr) 
\CR
&\!\!\quad 
{-} \langle \partial_\Lambda | T_\alpha | \Lambda\rangle \bigl(  \langle \partial_\Lambda | T_\beta T^\alpha |A_1\rangle \langle \partial_{A_2} | \cM^{-1} L_{-1} \cM T^\beta | A_2\rangle 
{-} \langle \partial_{A_1} | T_\beta T^\alpha | A_1 \rangle \langle \partial_\Lambda | \cM^{-1} L_{-1} \cM T^\beta  | A_2 \rangle \bigr)  \nn\\
&\!\!\quad 
{+} \langle \partial_\Lambda | T_\alpha |\Lambda \rangle \bigl(   \langle \partial_\Lambda | A_1\rangle \langle 2\partial_\Lambda {+} \partial_{A_1} {+} 2\partial_{A_2}| \cM^{-1} L_{-1} \cM T^\alpha |A_2\rangle 
{+} \langle \partial_{A_2} | A_1\rangle \langle \partial_\Lambda | \cM^{-1} L_{-1} \cM T^\alpha |A_2\rangle  \bigr)  \nn\\
&\!\!\quad 
 {-} \langle \partial_{A_2} | \Lambda\rangle \langle \partial_\Lambda | T_\alpha | A_1\rangle \langle \partial_\Lambda | \cM^{-1} L_{-1} \cM T^\alpha | A_2\rangle 
{-}  \langle \partial_\Lambda | \Lambda \rangle \langle   \partial_{A_{1}}  | A_1\rangle \langle \partial_\Lambda  | \cM^{-1} L_{-1} \cM |A_2 \rangle  \nn\\
&\!\!\quad 
 {-}\langle \partial_{A_2}  | \Lambda \rangle \langle \partial_{\Lambda }  | A_1\rangle \langle \partial_\Lambda  | \cM^{-1} L_{-1} \cM |A_2 \rangle  
{+} 2 \langle \partial_\Lambda | \Lambda \rangle \langle \partial_\Lambda | A_1\rangle \langle  \partial_\Lambda {+} \partial_{A_1} {+} \partial_{A_2} | \cM^{-1} L_{-1} \cM |A_2 \rangle \nn  \; . 
 \end{align}
Combining all terms one gets eventually
\be \Delta_\Lambda ( \cL_1+\cL_2) =0 \; , 
\ee
such that $\cL_1{+}\cL_2$ is indeed invariant under internal diffeomorphisms. Moreover, one checks that  in $\cL_1{+}\cL_2$ is invariant under the $\Sigma$ and 1-form gauge transformations defined in Appendix~\ref{Closure}.

The potential $V$ was shown to be invariant under internal diffeomorphisms in \cite{Bossard:2018utw}, and the two additional terms in $\tilde{g}_{\mu\nu}$ are manifestly invariant under internal diffeomorphisms up to a total derivative. Therefore the pseudo-Lagrangian \eqref{PseudoLagrangian} consists of three pieces that are separately invariant under internal diffeomorphisms. 

\medskip

We shall see in the next section, that external diffeomorphism invariance also requires to introduce one additional duality equation in the theory. Just like the $B$ field Euler--Lagrange equation, obtained by varying the pseudo-Lagrangian with respect $B$, gives a projection of the duality equation \eqref{DualityEquation}, the Euler--Lagrange equation for the constrained field $\langle \chi|$ gives a projection of the duality equation\footnote{This equation follows from the duality equations in~\cite{Bossard:2019ksx} upon branching to $E_9$.} 
\begin{align}
\label{Fduality}  | \cF_{\mu\nu} \rangle = - \rho^{-1} \varepsilon_{\mu\nu} T^\alpha \cM^{-1} |\cJ^-_\alpha\rangle \; ,
\end{align}
where 
\begin{align}
 \hspace{-0.2mm} |\cF\rangle = |F \rangle\! +\! \langle \partial_C | T_\alpha |C_{(1}\rangle T^\alpha |C_{2)} \rangle \!+\! \langle \pi_C | T_\alpha | C_{[1} \rangle T^\alpha |C_{2]}\rangle \!+ \! 2 \langle \pi_C | C_{[1} \rangle |C_{2]} \rangle \! + \! \eta_{-1\hspace{0.1mm} \alpha\beta } \Tr[ T^\alpha C^+_2] T^\beta |C^+_1\rangle  \; ,
 \end{align}
where we use the same notation as in \eqref{eq:triv1}, \eqref{eq:triv2} and \eqref{eq:triv3} and the shifted internal current $\langle \cJ_\alpha^-|$ is defined in \eqref{InternalJminus}. These two-forms can be written in components
\begin{align}
 \label{Cnotation}
 |C_{(1}\rangle \otimes |C_{2)}\rangle  \Leftrightarrow C^{(MN)} \; , \quad |C_{[1} \rangle \otimes  |C_{2]}\rangle \otimes \langle \pi_C|  \Leftrightarrow C^{[MN]}{}_P \; , \quad  | C^+_1\rangle\otimes C^+_2  \Leftrightarrow C^{+M;N}{}_P\; ,  
 \end{align}
where $C^{(MN)}$ is symmetric, $C^{[MN]}{}_P$ antisymmetric in $MN$, and $C^{[MN]}{}_P$ and $ C^{+M;N}{}_P$ are both constrained on their index $P$. Note in particular that $ |C_{1}\rangle \otimes |C_{2}\rangle $ represents a unique field that does not factorise, so $\langle \partial_C | \otimes |C_{1}\rangle \otimes |C_{2}\rangle  \Leftrightarrow \partial_P C^{MN} $ is a total derivative.  As trivial parameters, they are defined such that they drop out when $|\cF\rangle$ is contracted with a constrained bra, as in the equation of motion of $\langle \chi|$ for example. One checks using \eqref{Fvariation} that this duality equation \eqref{Fduality}  is indeed invariant under internal diffeomorphisms provided one defines the (inhomogeneous) transformation of the two-form potentials as\footnote{Here the inhomogeneous transformation $ \widetilde{\Delta}_\Lambda$ is defined in \eqref{InhomoDelta}. It is the $\mathbbm{C}$-independent component of $\delta_\Lambda \mathbbm{C}$.}
\begin{align} \label{CgaugeTransformation}
 \widetilde{\Delta}_\Lambda |C_1\rangle \otimes |C_{2\, \mu\nu} \rangle   &= \frac{1}{2} \bigl( \partial_{[\mu} |\Lambda\rangle \otimes |A_{\nu]}\rangle - |A_{[\mu} \rangle \otimes  \partial_{\nu]} |\Lambda\rangle \bigr) \; , \\
 \widetilde{\Delta}_\Lambda |C_{[1} \rangle \otimes |C_{2] \mu\nu} \rangle\otimes \langle \pi_C|    &=  \frac{1}{2} \bigl( \partial_{[\mu} |\Lambda\rangle \otimes |A_{\nu]}\rangle + |A_{[\mu}\rangle \otimes  \partial_{\nu]} |\Lambda\rangle \bigr)\otimes \langle  \partial_\Lambda - \partial_A | \; ,  \nn\\
  \widetilde{\Delta}_\Lambda |C^+_{1\, \mu\nu}  \rangle \otimes C_2^+ &=  - \rho^{-1} \varepsilon_{\mu\nu} \cM^{-1} |\partial_\Lambda\rangle \otimes |\Lambda\rangle \langle \partial_\Lambda | - \eta_{1\alpha\beta} \langle \partial_\Lambda | T^\alpha | A_{[\mu}\rangle |A_{\nu]}\rangle \otimes T^\beta |\Lambda\rangle \langle \partial_\Lambda | \; . \nn
\end{align}
Note that the two-form indices can be placed on any $C_i$ since they are representing the same field according to our notation \eqref{Cnotation}. The first term in the non-covariant variation of $|C^+_{1\mu\nu}  \rangle \otimes C_2^+  $ does not follow from the Dorfman structure and does not appear in \eqref{2formGaugeTransf}, but is defined to compensate the non-covariant transformation of the shifted current \cite[Eq. (4.30)]{Bossard:2018utw}
\begin{align}
\Delta_\Lambda T^\alpha \cM^{-1} |\cJ_\alpha^-\rangle &= \langle \partial_\Lambda | T^\alpha | \Lambda \rangle \bigl( \eta_{-1\, \alpha\beta} T^\beta \cM^{-1} |\partial_\Lambda\rangle + \rho^{-2} \eta_{1\, \alpha\beta}  \cM^{-1} T^{\beta \dagger} |\partial_\Lambda\rangle \bigr) \CR
&= \eta_{-1\, \alpha\beta} \langle \partial_\Lambda | T^\alpha | \Lambda \rangle T^\beta \cM^{-1} |\partial_\Lambda\rangle \; , 
\end{align}
where we used that the second term vanishes according to the section constraint. Therefore the field strength does not transform covariantly as in \eqref{eq:fsvar}, but includes moreover the same non-covariant variation as the internal current. 

We shall see in the next subsection that we need to include (at least a projection of) this duality equation in order to obtain a system of equations invariant under external diffeomorphisms. 

\subsection{External diffeomorphisms invariance}
\label{sec:extdiff}

Similarly as in \cite{Hohm:2014fxa}, we define the external diffeomorphisms as 
\begin{align}
\label{CorrectTransform}  
\delta_\xi  \cM &= \xi^\mu \cD^\flat_\mu \cM - \cD_\mu \xi^\mu \cM \; , \nn\\
\delta_\xi \tilde{g}_{\mu\nu} &= \xi^\sigma \cD_\sigma \tilde{g}_{\mu\nu} + 2 \cD_{(\mu} \xi^\sigma \tilde{g}_{\nu)\sigma}- \tilde{g}_{\mu\nu} \cD_\sigma \xi^\sigma\; , \nn\\
\delta_\xi |A_\mu \rangle &=\rho^{-1} \varepsilon_{\mu\nu} \xi^\nu T^\alpha \cM^{-1} |\cJ^-_{\alpha}\rangle + \rho^{-2} \tilde{g}_{\mu\nu} \cM^{-1} |\partial \xi^\nu\rangle \; , 
\end{align}
where 
\begin{align}
 \cD_\mu \xi^\nu = \partial_\mu \xi^\nu - \langle \partial \xi^\nu  | A_\mu \rangle \; .  
 \end{align}
Note that we do not use the full covariant derivative $\cD_\mu \cM$ because $\Sigma$ gauge invariance is not manifest and we have gauge fixed the $L_{-1}$ gauge symmetry by setting $\tilde{\rho}=0$. Writing the duality equation \eqref{Fduality} as $ | \mathcal{E}_{\mu\nu} \rangle=0$ with 
\begin{align}
 | \mathcal{E}_{\mu\nu} \rangle \equiv | \cF_{\mu\nu} \rangle+\rho^{-1} \varepsilon_{\mu\nu} T^\alpha \cM^{-1} |\cJ^-_\alpha\rangle\; ,  
 \end{align}
one finds that this transformation of the vector potential indeed corresponds to a covariant diffeomorphism modulo the duality equation \eqref{Fduality}, i.e. 
\begin{align}
 \delta_\xi |A_\mu \rangle = \xi^\nu |\cF_{\nu\mu}\rangle + \rho^{-2} \tilde{g}_{\mu\nu} \cM^{-1} |\partial \xi^\nu\rangle + \xi^\nu | \mathcal{E}_{\mu\nu}\rangle  \; . 
 \end{align}
 This is the same argument that is used to define \cite[Eq. (3.40)]{Hohm:2014fxa}.

We will  determine the transformation of the fields $\chi_\mu$ and $B_\mu$ such that the twisted self-duality equation \eqref{DualityEquation} 
\begin{align}  \label{DualityEquationE}  \cE_\mu \equiv \cJ_\mu - \rho^{-1}  \cM^{-1} \star ( \cS_1(   \cJ _\mu ) + \chi_\mu  \dK )^\dagger  \cM \;  
\end{align}
transforms into itself and other equations of motion under external generalised diffeomorphisms. We will show below that these transformations can be defined such that 
\begin{align}
\label{EquationVariation} 
\delta_\xi \mathcal{E}_\mu = \xi^\nu \cD^\flat_\nu \mathcal{E}_\mu + \cD_\mu \xi^\nu \mathcal{E}_\nu  +\xi^\nu\Big( -\cM^{-1} \mathcal{L}_{({\cE}_{\mu\nu},0)} \cM +  \rho^{-1}\tilde{g}_{\mu\sigma} \varepsilon^{\sigma\rho}  \cM^{-1} \cS_1 \big(\cM^{-1} \mathcal{L}_{(\mathcal{E}_{\rho\nu},0)} \cM \big)^\dagger \cM \Big) \; , 
\end{align}
so that we find that external diffeomorphism invariance requires to consider 
\begin{align}
  \label{miniFdual}  \eta_{\alpha\beta} \langle \partial_{\mathcal{E}} | T^\beta  | \mathcal{E}_{\mu\nu} \rangle =0 \end{align}
as an equation of motion. This equation is more constraining than the Euler--Lagrange equation $\langle \delta \chi |\mathcal{E}_{\mu\nu} \rangle =0$.

 One may be used to the property that exceptional field theories can usually be defined without being forced to introduce the duality equations for the non-propagating higher form fields, as is the case e.g. in \cite{Hohm:2013vpa,Hohm:2013uia,Hohm:2014fxa}. For $E_9$ one may expect that the Euler--Lagrange equation for the coset scalar fields would imply an equation 
 of the form
\be  \eta_{\alpha\beta} \langle \partial_{\mathcal{E}} | T^\beta  | \mathcal{E}_{\mu\nu} \rangle+\eta_{-1\, \alpha\beta} \Tr[ T^\beta  \mathcal{E}_{\mu\nu} ] \overset{?}{=} 0  \ee
for a $\mathcal{E}_{\mu\nu}$ involving the two-form field strength $\mathcal{G}_{\mu\nu}$ such that this equation would not depend explicitly in the two-form potential $\mathbbm{C}$. However, the scalar fields Euler--Lagrange equations only give a projection of the $\cS_{1}$ shift of such an equation to $\mf{e}_{9}\ominus K({\mf e}_{9})$, which does not imply that it holds.

To compute the variation of \eqref{DualityEquationE}  under external diffeomorphisms, it will be convenient to define the non-covariant component of an external diffeomorphism as
\begin{align}
\delta_\xi = \mathscr{L}_\xi + \Delta_\xi 
\end{align}
such that 
\begin{align}
 \mathscr{L}_\xi \cJ_\mu = \xi^\nu \cD^\flat_\nu  \cJ_\mu+ \cD_\mu \xi^\nu \cJ_\nu \; , 
 \end{align}
and similarly for any vector field. Using this definition, one computes that 
\begin{align}
\Delta_\xi \cJ^\flat_\mu &= \xi^\nu \eta_{-1\, \alpha\beta} \eta_{1\, \gamma\delta} \langle \partial_A | T^\gamma | A_{[\mu}^\prime \rangle \langle \partial_A |T^\alpha T^\delta | A_{\nu]}\rangle ( T^\beta + \cM^{-1} T^{\beta \dagger } \cM ) \\
& \quad - \xi^\nu \bigl(  \langle \cJ_{\alpha } | E_{\mu\nu}\rangle   +\eta_{\alpha\beta}  \langle \partial_E | ( T^\beta + \cM^{-1}  T^{\beta \dagger} \cM )  | E_{\mu\nu} \rangle  \bigr) T^\alpha - \xi^\nu \tilde{g}^{\sigma\rho} \langle \partial \tilde{g}_{\mu\sigma} | E_{\rho\nu}\rangle \dK  \nn\\
&\hspace{4.5mm}
 -\rho^{-2} \tilde{g}_{\mu\nu} \langle \partial \xi^\nu| \cM^{-1} | \cJ_\alpha \rangle T^\alpha - \rho^{-1} \varepsilon_{\mu\nu} \langle \partial \xi^\nu | T_\alpha T^\beta \cM^{-1} |\cJ^-_\beta \rangle ( T^\alpha + \cM^{-1} T^{\alpha \dagger} \cM )  \nn\\
&\hspace{4.5mm} - \rho^{-2}  \langle \partial \tilde{g}_{\mu\nu} | \cM^{-1} | \partial \xi^\nu\rangle \dK + \Big( \cD_\nu \cD_\mu \xi^\nu - 2 \cD_\mu \cD_\nu \xi^\nu + \tilde{g}_{\mu\rho}\tilde{g}^{\nu\sigma}\cD_\nu \cD_\sigma \xi^\rho \Big) \dK\nn\\
&\hspace{4.5mm} - \rho^{-2}  \eta_{\alpha\beta} \tilde{g}_{\mu\nu}  \langle \partial  | ( T^\beta \cM^{-1}  + \cM^{-1} T^{\beta\dagger} ) |\partial \xi^\nu  \rangle T^\alpha-\rho^{-2}   \tilde{g}_{\mu\nu}  \langle \partial \xi^\nu   | [T_\alpha , T^\beta ]   \cM^{-1}  |\cJ_\beta \rangle T^\alpha\nn\; , 
\end{align}
where $\langle \partial | $ in the last line indicates that this term is a total  internal derivative and 
\begin{align}
 | E_{\mu\nu} \rangle = | F_{\mu\nu} \rangle+\rho^{-1} \varepsilon_{\mu\nu} T^\alpha \cM^{-1} |\cJ^-_\alpha\rangle\; ,  
 \end{align}
is the component of equation \eqref{Fduality} that does not include the two-form gauge fields. The first term $ \langle \cJ_{\alpha } | E_{\mu\nu}\rangle = \langle \cJ_{\alpha } | \cE_{\mu\nu}\rangle $, but to get the two-form dependence in the second term, we need to define $\Delta_\xi B$ and $\Delta_\xi \chi$  to depend explicitly on the two-forms. One computes that \eqref{EquationVariation} is satisfied if one defines the variations
{\allowdisplaybreaks  \begin{align}
\Delta_\xi B_\mu &= \eta_{1\,\alpha\beta} \xi^\nu  \langle \partial_A | T^\alpha | A^\prime_{[\mu}\rangle T^\beta |A_{\nu]} \rangle \langle \partial_A | \nn\\
&\qquad - \rho \tilde{g}_{\mu\rho} \varepsilon^{\rho\sigma} \xi^\nu \bigl( \eta_{\alpha\beta}  \langle \partial_A | T^\alpha | A^\prime_{[\sigma}\rangle T^\beta |A_{\nu]}  \rangle  + \langle \partial_A | A^\prime_{[\rho} \rangle |A_{\nu]}\rangle + \langle \partial_A | A_{[\rho} \rangle |A^\prime_{\nu]}\rangle  \bigr) \langle \partial_A | \nn\\
& \quad +  \rho^{-1} \varepsilon_{\mu\nu} \Big( 2 \xi^\nu \cM^{-1} |\partial_\xi\rangle \langle \partial_\xi | + \tilde{g}^{\nu\sigma} \cM^{-1} \big( |\partial\xi^\rho\rangle \langle \partial \tilde{g}_{\rho\sigma} | + |\partial\tilde{g}_{\rho\sigma} \rangle \langle \partial \xi^\rho|\big) \nn\\
&\hspace{35mm} - T^\alpha \cM^{-1} \big(|\partial\xi^\nu \rangle \langle \cJ_\alpha| + |\cJ_\alpha \rangle\langle \partial\xi^\nu|\big) +\cM^{-1} |\partial\xi^\nu  \rangle  \langle J_0 | \Bigr)  \nn\\
&\hspace{4.5mm} + \tilde{g}_{\mu\nu} \Bigl( \cM^{-1} T^{\alpha \dagger} |\partial \xi^\nu \rangle \langle \cJ^{-}_\alpha | + T^\alpha \cM^{-1} | \cJ^{-}_\alpha \rangle \langle \partial \xi^\nu | \Bigr) + \xi^\nu \delta_{ C_{\sigma\nu}}  B_\mu\; ,  \nn\\
\Delta_\xi \chi_\mu &=- \xi^\nu \Bigl( \eta_{\alpha\beta}  \langle \partial_A | T^\alpha | A^\prime_{[\mu}\rangle  \langle \partial_A  | \cM^{-1} L_{-1} \cM  T^\beta |A_{\nu]}  \rangle  + \langle \partial_A | A^\prime_{[\mu} \rangle  \langle \partial_A | \cM^{-1} L_{-1} \cM  |A_{\nu]}\rangle  \nn\\
& \hspace{90mm} +  \langle \partial_A | A_{[\mu} \rangle \langle \partial_A | \cM^{-1} L_{-1} \cM  |A^\prime_{\nu]}\rangle  \Bigr)  \nn\\
&\hspace{4mm}- \xi^\nu \tilde{g}_{\mu\rho} \varepsilon^{\rho\sigma} \bigl( \eta_{\alpha\beta}  \langle \partial_A | T^\alpha | A^\prime_{[\sigma}\rangle  \langle \partial_A  | L_0 T^\beta |A_{\nu]}  \rangle  \!+ \!\langle \partial_A | A^\prime_{[\sigma} \rangle  \langle \partial_A | L_0 |A_{\nu]}\rangle \!+\! \langle \partial_A | A_{[\sigma} \rangle \langle \partial_A | L_0 |A^\prime_{\nu]}\rangle\!  \bigr)  \nn\\
&\hspace{4mm}  - \rho \varepsilon_{\mu\rho} \xi^\nu \langle \partial  \tilde{g}^{\rho\sigma} | E_{\sigma\nu}\rangle + \rho \varepsilon_{\mu\rho} \tilde{g}^{\sigma\rho}  ( \cD_\nu \cD_\sigma \xi^\nu - 2 \cD_\sigma \cD_\nu \xi^\nu ) + \rho \varepsilon_{\mu\nu} \tilde{g}^{\sigma\rho} \cD_\sigma \cD_\rho \xi^\nu   \nn\\
&\hspace{4.5mm} + \rho^{-2} \tilde{g}_{\mu\nu} \Bigl( \langle \partial | ( L_1 \cM^{-1} + \cM^{-1} L_{-1} ) | \partial \xi^\nu\rangle + \langle \partial \xi^\nu | [ L_1 , T^\alpha ] \cM^{-1} | \cJ_\alpha \rangle \nn\\
&\hspace{30mm} +  \langle \partial \xi^\nu | ( L_0 T^\alpha \cM^{-1} + T^\alpha \cM^{-1} L_0)  | \cJ^{-}_\alpha \rangle +  \rho^2 \langle \partial \xi^\nu | \cM^{-1} \cS_1T^{\alpha \dagger} | \cJ_\alpha\rangle  \Bigr) \nn\\
&\hspace{4.5mm} + \rho^{-1} \tilde{g}_{\mu\rho} \varepsilon^{\rho\sigma} \langle \partial \tilde{g}_{\sigma \nu} | ( L_0 \cM^{-1} + \cM^{-1} L_0 - \cM^{-1} ) |\partial \xi^\nu \rangle \nn\\
&\hspace{4.5mm} + \rho^{-1} \varepsilon_{\mu\nu} \Bigl( \langle \partial_\xi + \partial_\cM | ( \cM^{-1} L_0 + L_0 \cM^{-1} ) |\partial \xi^\nu \rangle + \langle \partial \xi^\nu | \cM^{-1} L_0 |J_0 \rangle +\langle  \partial \xi^\nu |T^\alpha \cM^{-1} |\cJ_\alpha \rangle \nn\\
& \hspace{30mm} + \langle \partial \xi^\nu | ( L_1 + \cM^{-1} L_{-1} \cM ) T^\alpha \cM^{-1}| \cJ^-_\alpha \rangle \Bigr) +\xi^\nu \delta_{ C_{\sigma\nu}} \chi_\mu\; , 
\end{align}
where $\xi^\nu \delta_{C_{\sigma\nu}}$ is the one-form gauge transformation of parameter $C_{\sigma\nu} $ where $\sigma$ is the index of the vector parameter whereas $\nu$ is contracted with $\xi^\nu$, i.e.}\footnote{where $\iota_\xi \dd x^\mu = \xi^\mu$.}
\begin{align}
\xi^\nu \delta_{C_{\sigma\nu}}  B &=- \tfrac12  \iota_\xi \eta_{1\alpha\beta} T^\alpha  | C_{[1}\rangle \bigl( \langle \pi_C | T^\beta | C_{2]} \rangle \langle \partial_C | +   \langle \partial_C | T^\beta | C_{2]} \rangle \langle \pi_C | \bigr)\\
&\qquad  + \iota_\xi  \eta_{\alpha\beta} \Tr[ T^\alpha C^+_2] T^\beta |C^+_1 \rangle \langle \partial_{C^+}\! |  - \iota_\xi  |{C^+_1}\rangle \langle \partial_{C^+} | {C^+_2} +  \iota_\xi \langle \partial_{C^+} \! | {C^+_1} \rangle {C^+_2} \nn\\
& \hspace{-7.8mm} +\rho \star  \iota_\xi \bigl( \eta_{\alpha\beta} T^\alpha |C_{[1} \rangle \langle \pi_C | T^\beta |C_{2]}\rangle \hspace{-0.2mm} +\hspace{-0.2mm} 2 |C_{[1} \rangle \langle \pi_C | C_{2]} \rangle \bigr) \langle \partial_C | - \rho \star  \iota_\xi  \eta_{-1\hspace{0.2mm}\alpha\beta}  \Tr[ T^\alpha {C^+_2} ] T^\beta  |{C^+_1} \rangle \langle \partial_{C^+} \! |  \; , \nn\\
\xi^\nu \delta_{C_{\sigma\nu}} \chi &= - \iota_\xi \eta_{\alpha\beta} \langle \pi_C | T^\alpha | C_{[1} \rangle \langle \partial_C | \cM^{-1} L_{-1} \cM T^\beta |C_{2]} \rangle - 2  \iota_\xi  \langle \pi_C | C_{[1}\rangle \langle \partial_C | \cM^{-1} L_{-1} \cM | C_{2]} \rangle\nn\\
& \quad - \rho \star  \iota_\xi \Bigl(   \eta_{\alpha\beta} \langle \pi_C | T^\alpha | C_{[1} \rangle \langle \partial_C | L_0 T^\beta |C_{2]} \rangle + 2 \langle \pi_C | C_{[1}\rangle \langle \partial_C | L_0 | C_{2]} \rangle\Bigr)  \nn\\
& \hspace{1.9mm}  - \eta_{-1\, \alpha \beta }  \iota_\xi \Tr[ T^\alpha {C^+_2} ] \langle \partial_{C^+} \! | \cM^{-1} L_{-1} \cM T^\beta |{C^+_1}\rangle   - \rho \star   \iota_\xi  \eta_{-1\, \alpha \beta } \Tr[ T^\alpha {C^+_2} ] \langle \partial_{C^+}\!  | L_0 T^\beta  |{C^+_1} \rangle \; . \nn
\end{align} 
Because we consider $|\cE_{\mu\nu}\rangle=0$ as an equation of motion, invariance of the system also requires to compute its own transformation under external diffeomorphisms. On computes that $|\cE_{\mu\nu}\rangle$ varies into 
\begin{align}
 \delta_\xi |\cE_{\mu\nu}\rangle = - \rho^{-2} \varepsilon_{\mu\nu} \tilde{g}_{\sigma\lambda} \varepsilon^{\lambda\rho} \cE_\rho \cM^{-1} |\partial \xi^\sigma \rangle \; , 
 \end{align}
with the definitions 
\begin{align}
\delta_\xi \langle \chi | &= - \langle \partial \xi^\mu | \Bigl( \rho^{-1} \tilde{g}_{\mu\sigma} \varepsilon^{\sigma\nu} J_{\nu 0} + \rho^{-2} \bigl( \chi_\mu + \Tr[ \cM^{-1} L_0 \cM B_\mu ] + B_\mu - \Tr[ B_\mu] \bigr) + \omega_{-1}^\alpha(\cM) \cJ_{\mu \alpha}\Bigr) \nn\\
&\hspace{4.5mm} - \sum_n n \xi^\mu  \langle \partial_A | T_{n+1}^A | A_\mu \rangle \langle J_A^n | +\xi^\mu  \langle \partial_A | ( L_{-1} + \cM^{-1} L_1 \cM ) |A_\mu \rangle \langle \partial_A|  \; , 
\end{align}
and 
\begin{align}
\Delta_\xi  |C\rangle \otimes |C_{\mu\nu} \rangle &= \frac{1}{2} \bigl( \delta_\xi  |A_{[\mu} \rangle \otimes |A_{\nu]}\rangle - |A_{[\mu} \rangle \otimes \delta_\xi  |A_{\nu]} \rangle \bigr) \; , \\
\Delta_\xi |C_{[1} \rangle \otimes |C_{2] \mu\nu} \rangle\otimes \langle \pi_C|    &=  \frac{1}{2} \bigl(  \delta_\xi  |A_{[\mu} \rangle  \otimes |A_{\nu]}\rangle + |A_{[\mu}\rangle \otimes  \delta_\xi  |A_{\nu]} \rangle  \bigr)\otimes \langle  \partial_{\delta_\xi A} - \partial_A | \; ,  \nn\\
\Delta_\xi  |C^+_{1\mu\nu}  \rangle \otimes C^+_2 &=  \rho^{-1} \varepsilon_{\mu\nu} \cM^{-1}  \Bigl( |\partial_A \rangle \otimes \xi^\sigma |A_\sigma \rangle \langle \partial_A | - \rho^{-1} \tilde{g}_{\rho\sigma} \varepsilon^{\sigma\nu} |\partial \xi^\rho \rangle \otimes B_\nu \Bigr) \; . \nn
\end{align}
The duality equations \eqref{DualityEquation} and \eqref{Fduality} therefore transform into themselves under external diffeomorphisms. The minimal set of duality equations transforming into each others under external diffeomorphisms is 
\begin{align}
 \cE_{\mu\hspace{0.2mm}\alpha} = 0 \; , \qquad \eta_{\alpha\beta} \langle \partial_{\cE} | T^\beta | \cE_{\mu\nu} \rangle = 0 \; . 
 \end{align}

One may be worried that the external diffeomorphisms variations of the ancillary fields $B$ and $\chi$  depend on the two-form potentials while the pseudo-Lagrangian does not. The resolution is that the pseudo-Lagrangian is only invariant under external diffeomorphisms up to bilinear terms in the duality equations, including equation \eqref{Fduality} that depends on the two-form potentials. To see this we observe that varying $B$ and $\chi$ only in the pseudo-Lagrangian one obtains that the dependence in the two-form potentials can be absorbed into field strengths terms, i.e. that 
\be \Tr [ \cE \delta_{\iota_\xi C} B ]  + \cE_0 \delta_{\iota_\xi C} \chi^\flat  + \langle \partial_\cF | \Bigl( \cS_1(\cE) - \cE_0 L_1  - \cE_0 \cM^{-1} L_{-1} \cM - \cE \rho  \star \Bigr) \iota_\xi |\cF\rangle \ee
does not depend on the two-form potential and so we expect the non-covariant variation of the pseudo-Lagrangian to give
\be \label{ExternalDiffInv} \Delta_\xi \cL =-  \tfrac12  \langle \partial_{|\cE\rangle }  | \Bigl( \cS_1(\cE) - \cE_0 L_1  - \cE_0 \cM^{-1} L_{-1} \cM - \cE \rho \star \Bigr) \iota_\xi |\cE \rangle + \dd \mathcal{B}_\xi - \langle \partial | \mathcal{B}_\xi \rangle  \ee
for some specific boundary terms $\mathcal{B}_\xi$ and $|\mathcal{B}_\xi\rangle$. We stress here that it is enough for the pseudo-Lagrangian to be invariant under a symmetry up to terms quadratic in the duality equations for its Euler--Lagrange equations to transform into themselves and the duality equations under that symmetry. On the contrary, it would not be sufficient for the pseudo-Lagrangian to be invariant up to terms linear in the duality equations. A symmetry of the equations of motion only require that they transform into themselves under the symmetry variations. If the Euler--Lagrange equations of a pseudo-Lagrangian transform under the symmetry into duality equations, then one finds that the pseudo-Lagrangian is only invariant under that symmetry up to terms quadratic in the duality equations. This is discussed in detail in Appendix~\ref{app:pseudoL}. 

As a consistency check, one can verify that the relation between the quadratic terms in the duality equation in \eqref{ExternalDiffInv} are consistently related to the terms proportional to the duality equations in the external diffeomorphism variation of the $B$ fields according to the general discussion of  Appendix~\ref{app:pseudoL}. This can easily be verified, because the equation of motion $ \cE_B=0$ of the constrained field $B$ is simply the projection of \eqref{DualityEquationE} upon tracing with $\delta B$, so its variation under external diffeomorphisms follows directly from  \eqref{EquationVariation} as \footnote{We vary $\delta B$ at  $\delta \chi^\flat =0$, not $\delta \chi =0$.}
\begin{align}
  \Tr[ \delta B \delta_\xi \cE_B ]  &=  - \tfrac12 \Tr[ \delta B \delta_\xi \cE ]  \\
 &= -\tfrac12 \Tr[ \delta B \cL_\xi \cE ] -\tfrac12 \Tr[ T^\alpha \delta B] \iota_\xi \langle \cJ_\alpha | \cE\rangle +\tfrac12 \Tr[ \cM^{-1} \cS_1T^{\alpha \dagger} \cM \delta B ] \rho^{-1} \star \iota_\xi \langle \cJ_\alpha | \cE\rangle \nn\\
&\hspace{4.5mm}  -\tfrac12  \langle \partial_{|\cE\rangle} | \delta B \iota_\xi | \cE\rangle +\tfrac12 \Tr[ \delta B] \iota_\xi  \langle  \partial_{|\cE\rangle} |  \cE\rangle -\tfrac12 \Tr[ \cM^{-1} L_{-1} \cM \delta B] \rho^{-1} \star \iota_\xi \langle  \partial_{|\cE\rangle} | \cE\rangle  \nn\\
&\hspace{4.5mm} + \tfrac12 \eta_{\alpha\beta} \Tr[ \cM^{-1} T^{\beta \dagger} \cM \delta B ] \langle \partial_{|\cE\rangle} |\Bigl(  \rho^{-1} \! \star\! \bigl( \cS_1(T^\alpha) - \delta_0^\alpha ( L_1 {+} \cM^{-1} L_{-1} \cM ) \bigr) - T^\alpha \Bigr) \iota_\xi |\cE\rangle \nn  
\end{align}
where the two first lines are proportional to Euler--Lagrange equations whereas the last line is proportional to the duality equation \eqref{Fduality}. Assuming \eqref{ExternalDiffInv} we can vary the right-hand side with respect to $B$ to obtain 
\begin{align}
 \delta \Delta_\xi \cL &=  \tfrac12  \langle \partial_{|\cE\rangle} | \delta B \iota_\xi | \cE\rangle -\tfrac12 \Tr[ \delta B] \iota_\xi  \langle  \partial_{|\cE\rangle} |  \cE\rangle +\tfrac12 \Tr[ L_0 \delta B]  \iota_\xi \langle  \partial_{|\cE\rangle} | \cE\rangle  \\
&\hspace{4.5mm} + \tfrac12 \eta_{\alpha\beta} \Tr[ \cM^{-1} T^{\beta \dagger} \cM \delta B ] \langle \partial_{|\cE\rangle} |\Bigl(  \rho^{-1} \! \star\! \bigl( \cS_1(T^\alpha) - \delta_0^\alpha ( L_1 + \cM^{-1} L_{-1} \cM ) \bigr) - T^\alpha \Bigr) \iota_\xi |\cE\rangle \; , \nn 
\end{align}
where again the first line is proportional to Euler--Lagrange equations whereas the second matches precisely the term above in $  \Tr[ \delta B \delta_\xi \cE_B ]  $ proportional to the duality equation \eqref{Fduality}, consistently with \eqref{RelationBdeltaL} in Appendix~\ref{app:pseudoL}.

We have not fully derived the invariance of the pseudo-Lagrangian up to quadratic terms in the duality equations. We shall only check that external diffeomorphism invariance fixes all free coefficients in the pseudo-Lagrangian.

We check this in steps. 
 Let us start with  external diffeomorphisms that do not depend on the internal coordinates and with $|A\rangle =0$ for simplicity. We denote the corresponding transformation by $\delta^\ord{0}_\xi$. To check the invariance of the pseudo-Lagrangian under such diffeomorphisms, we only need to consider the terms involving $J_{\mu {\dK}}$ and $\chi$ in $\cL_1+\cL_2$, as well as the third term in \eqref{PseudoLagrangian}. The terms linear in the field $\Tr[ B_\mu]$ transform together covariantly by construction since they are proportional to the duality equation \eqref{DualityJK}. The term that remains is $D\chi +  \frac12 J_0 \chi$ in $\cL_2$. One computes that 
\begin{align}
 \delta_\xi^\ord{0} ( \cL_1 + \cL_2 ) = \tfrac12 \partial_\mu ( \xi^\mu ( \cL_1 + \cL_2 )) +  \rho \,\varepsilon^{\mu\nu} ( \partial_\mu - \langle \vec{\partial} | A_\mu \rangle ) 
 ( \varepsilon^{\sigma\lambda} \tilde{g}_{\lambda\rho} \partial_\nu \partial_\sigma \xi^\rho ) \; .  
\end{align}
To compute the variation of the third term we found useful to use 
\begin{align}
\label{NonCovReduce}  \frac14  \varepsilon^{\mu\nu} \varepsilon^{\sigma\rho}   \tilde{g}^{\kappa\lambda} \partial_{\mu} \tilde{g}_{\sigma\kappa} \partial_{\nu} \tilde{g}_{\rho\lambda}  = \partial_\mu \bigl(  \varepsilon^{\mu\nu} \tilde{g}_{01} \tilde{g}_{11}^{-1} \partial_\nu \tilde{g}_{11} \bigr) \; ,
\end{align}
and vary instead 
\begin{align}
  \delta_\xi^\ord{0} \tilde{g}_{11} &= \xi^\mu \partial_\mu \tilde{g}_{11} +  \tilde{g}_{11} ( \partial_1 \xi^1 - \partial_0 \xi^0) + 2 \tilde{g}_{01} \partial_1 \xi^0 \nn\\
 \delta_\xi^\ord{0} \frac{\tilde{g}_{01}}{\tilde{g}_{11}} &=   \xi^\mu \partial_\mu  \frac{\tilde{g}_{01}}{\tilde{g}_{11}} + \partial_0 \xi^1 +  \frac{\tilde{g}_{01}}{\tilde{g}_{11}} ( \partial_0 \xi^0 - \partial_1 \xi^1 ) +  \frac{\tilde{g}_{00}}{\tilde{g}_{11}} \partial_1 \xi^0 
 \end{align}
from which one computes that 
\begin{multline}
  \delta_\xi^\ord{0}  \Bigl(  \frac{\tilde{g}_{01}}{\tilde{g}_{11}} \partial_\mu \tilde{g}_{11}\Bigr) = \xi^\nu \partial_\nu   \Bigl(  \frac{\tilde{g}_{01}}{\tilde{g}_{11}} \partial_\mu \tilde{g}_{11} \Bigr)  +   \xi^\nu \partial_\mu \xi^\nu \Bigl(  \frac{\tilde{g}_{01}}{\tilde{g}_{11}} \partial_\nu \tilde{g}_{11} \Bigr) - \varepsilon^{\nu\sigma} \tilde{g}_{\sigma\rho} \partial_\mu \partial_\nu \xi^\rho\\
+ \partial_\mu \Bigl( \tilde{g}_{11} \partial_0 \xi^1 + \Bigl( \tilde{g}_{00} + \frac{2}{\tilde{g}_{11}}\Bigr) \partial_1 \xi^0 \Bigr)  \,,
\end{multline}
where the total derivative in the second line will drop out in the variation of the pseudo-Lagrangian. Using this formula one obtains 
\begin{align}
    \delta_\xi^\ord{0} \Bigl( \frac14 \rho \,\varepsilon^{\mu\nu} \varepsilon^{\sigma\rho}   \tilde{g}^{\kappa\lambda} \partial_{\mu} \tilde{g}_{\sigma\kappa} \partial_{\nu} \tilde{g}_{\rho\lambda} \Bigr) = \partial_{\vartheta} \Bigl( \xi^{\vartheta} \frac14 \rho \,\varepsilon^{\mu\nu} \varepsilon^{\sigma\rho}   \tilde{g}^{\kappa\lambda} \partial_{\mu} \tilde{g}_{\sigma\kappa} \partial_{\nu} \tilde{g}_{\rho\lambda} \Bigr)-\rho \,\varepsilon^{\mu\nu} \partial_\mu ( \varepsilon^{\sigma\lambda} \tilde{g}_{\lambda\rho} \partial_\nu \partial_\sigma \xi^\rho )  \,.
\end{align}
Thus, the pseudo-Lagrangian is invariant under purely external diffeomorphisms, i.e. at $|A\rangle =0$ for $\xi^\mu(x)$. 

Now if we take the second order derivative terms in the unimodular metric and the vector field $\xi^\mu(x,y)$, we obtain from $\cL_1 + \cL_2$
\begin{align}
\delta_\xi ( \cL_1 + \cL_2) &= \dots  + \rho^{-1} \Bigl( \langle \partial_\xi | \cM^{-1} | \partial \xi^\sigma\rangle \tilde{g}^{\mu\nu} \cD_\mu \tilde{g}_{\nu\sigma} - \langle \partial \tilde{g}_{\mu\sigma} |\cM^{-1} |\partial \xi^\sigma \rangle \cD_\nu \tilde{g}^{\mu\nu} \Bigr) \CR
&= \dots   + \langle \partial |\bigl( \rho^{-1}  \cM^{-1} | \partial \xi^\sigma\rangle \tilde{g}^{\mu\nu} \cD_\mu \tilde{g}_{\nu\sigma} \bigr) - \partial_\mu \bigl( \rho^{-1} \langle \partial \tilde{g}_{\nu\sigma} | \cM^{-1} |\partial \xi^\sigma \rangle \tilde{g}^{\mu\nu} \bigr) \nn\\& \qquad + \rho^{-1} \Bigl( \langle \partial g_{\nu\sigma}  | \cM^{-1} \cD_\mu | \partial \xi^\sigma\rangle \tilde{g}^{\mu\nu} -\tfrac12  \langle \partial \tilde{g}_{\mu\nu} |\cM^{-1} |\partial \xi^\sigma \rangle \cD_\sigma \tilde{g}^{\mu\nu} \Bigr)  
\end{align}
which cancels the non-covariant variation of $ \frac{\rho^{-1}}{4} \langle \partial \tilde{g}^{\mu\nu} | \cM^{-1} | \partial \tilde{g}_{\mu\nu} \rangle$. 

The variation of the potential gives the terms
\begin{align}
 \Delta_\xi V &= - \tfrac12 \rho^{-1} \langle \partial \xi^\mu | \bigl( T^\alpha \cJ_\mu^\flat + \cJ_\mu^\flat T^\alpha \bigr) \cM^{-1} |\cJ_\alpha\rangle \nn\\ 
 &\quad\quad+ \tfrac12 \rho^{-1} \langle  \partial \xi^\mu | \bigl( T^\alpha  \cS_1(\cJ_\mu^\flat)   + \cS_1(\cJ_\mu^\flat) T^\alpha \bigr) \cM^{-1} |\cJ^-_\alpha\rangle+\dots 
 \end{align}
which are compensated by the  terms 
\begin{align}
 \Delta_\xi \cL_1&= \tfrac12 \rho^{-1} \langle  \partial \xi^\mu |\bigl(  T^\alpha  \cS_1(\cJ_\mu^\flat)  + \cM^{-1} \cS_1(\cJ_\mu^\flat)^\dagger  \cM T^\alpha \bigr)  \cM^{-1} |\cJ^-_\alpha\rangle + \dots \\
 \Delta_\xi \cL_2 &= - \tfrac12 \rho^{-1} \langle \partial \xi^\mu | \bigl( T^\alpha \cJ_\mu^\flat + \cJ_\mu^\flat T^\alpha \bigr) \cM^{-1} |\cJ_\alpha\rangle  \nn\\
 & \qquad + \tfrac12 \rho^{-1} \langle  \partial \xi^\mu | \bigl( \cS_1(\cJ_\mu^\flat) - \cM^{-1}\cS_1(\cJ_\mu^\flat)^\dagger \cM  \bigr)  T^\alpha \cM^{-1} |\cJ^-_\alpha\rangle + \dots  \nonumber \;  \end{align}
in the variation of $\cL_1$ and $\cL_2$, fixing in this way their relative coefficients in $\cL_{\text{min}}$. 

We conclude therefore that the invariance of the pseudo-Lagrangian \eqref{PseudoLagrangian} under external diffeomorphisms, up to total derivatives and terms quadratic in the duality equations, fixes the relative coefficients of all the terms in  \eqref{PseudoLagrangian}. In this section we have proved that the duality equations \eqref{DualityEquation} and \eqref{Fduality} transforms into each others under external diffeomorphisms, and verified several consistency checks for the invariance of the Euler--Lagrange equations. The complete invariance under external diffeomorphisms will be confirmed in the next section, by showing that upon partially solving the section constrained one obtains the $E_8$ exceptional field theory \cite{Hohm:2014fxa}. Because the latter is invariant under three-dimensional external diffeomorphisms, we conclude that the $E_9$ exceptional field theory pseudo-Lagrangian \eqref{PseudoLagrangian} must indeed satisfy \eqref{ExternalDiffInv}.

\subsection{\texorpdfstring{Embedding of $E_8$ exceptional field theory}{Embedding of E8 exceptional field theory}}
\label{E8reduction}

The most convenient way to prove that the $E_9$ exceptional field theory does describe supergravity solution is to show that it reproduces $E_8$ exceptional field theory upon choosing the partial section solution 
\begin{align}
 \langle \partial | = \langle 0 | ( \partial_\varphi + T_1^A \partial_A ) \; , \label{E8Section} 
 \end{align}
where $\partial_A$ is the internal $E_8$ derivative that still satisfies itself the $E_8$ exceptional field theory section condition~\cite{Berman:2012vc,Hohm:2014fxa}
\begin{align}
\label{eq:E8SC}
 \partial_A \otimes \partial_B + \partial_B \otimes \partial_A  + f_{AE}{}^C f^{DE}{}_B \partial_C \otimes \partial_D  = 0 \; . 
 \end{align}
In~\eqref{E8Section} and below we use $\varphi$ to denote the circle coordinate in the Kaluza--Klein ansatz, which is an internal coordinate \eqref{E8Section} in $E_9$ exceptional field theory and an external $S^1$ coordinate in $E_8$ exceptional field theory with three-dimensional external space-time involving a local $S^1$ fibration, see~\eqref{3Dmetric}.

\subsubsection*{Semi-flat current and Kaluza--Klein ansatz}

 We use the $E_8$ parabolic gauge 
\begin{align}
 \cV =  ( \rho^{-L_0} e^{-\sigma \dK} \mathring{V}  ) \, U \label{E8parabolic}
 \end{align}
with $ \mathring{V} \in E_8$ and $U = \prod_{k=1}^\infty \exp(  Y_{kA} T_{-k}^A)$ the negative mode component in  $E_9$. Then one defines the semi-flat current~\cite{Bossard:2019ksx}
\begin{align}
 U^{-1} \tilde{\cJ}^\flat U = \cJ^\flat \;, \quad U^{-1} \tilde{\cJ} U = \cJ\; , \quad \chi  = \tilde{\chi} - \omega^\alpha_{-1}(U) \cJ_\alpha = \tilde{\chi} + \omega^\alpha_{-1}(U^{-1}) {\tilde{\cJ}}_\alpha\; .   
 \end{align}
This gives rise to well-defined finite expressions, because the action of $U^{-1}$ on a constrained bra is finite 
\begin{align}
 \label{DerU}  \langle \partial | U^{-1}  = \langle 0 | ( \partial_\varphi - \eta^{AB} Y_{1A} \partial_B+  T_1^A \partial_A )\; . 
\end{align}
The duality equation~\eqref{DualityEqZ2} then reduces to 
\begin{align}
 \tilde{\cJ} = \rho^{-1} \star \bigl( \cM_{\scalebox{0.5}{$0$}}^{-1} S_1(\tilde{\cJ})^\dagger \cM_{\scalebox{0.5}{$0$}} + \tilde{\chi}\bigr) 
\end{align}
where we define $\cM_{\scalebox{0.5}{$0$}}  = e^{-2\sigma} \rho^{-2L_0} M \in \mathds{R}_+ \times \mathds{R}_+ \times E_8$, the mode 0 component of $\cM$ (i.e. commuting with $L_0$), not to be confused with $M$ that denotes the $E_8$ matrix (written $M_{AB}$ in the adjoint representation of $E_8$).  Since 
\begin{align}
U \cdot \cM^{-1} \dd \cM \cdot U^{-1}   = \cM_{\scalebox{0.5}{$0$}}^{-1} \dd \cM_{\scalebox{0.5}{$0$}}  +  \dd U \cdot U^{-1} + \cM_{\scalebox{0.5}{$0$}}^{-1} (   \dd U  \cdot U^{-1})^\dagger  \cM_{\scalebox{0.5}{$0$}} 
\end{align}
we have similarly 
\begin{align}
\tilde{J}_\alpha T^\alpha &= \cM_{\scalebox{0.5}{$0$}}^{-1} ( \dd - \langle \partial_{\cM_{\scalebox{0.5}{$0$}}} | A\rangle ) \cM_{\scalebox{0.5}{$0$}}  + ( \dd - \langle \partial_{U} | A\rangle )U \cdot U^{-1} + \cM_{\scalebox{0.5}{$0$}}^{-1} (  ( \dd - \langle \partial_{U} | A\rangle ) U \cdot U^{-1})^\dagger  \cM_{\scalebox{0.5}{$0$}}\CR
&\hspace{4.5mm}  - \eta_{\alpha\beta} \langle \partial_A | U^{-1} T^\beta U | A\rangle ( T^\alpha +\cM_{\scalebox{0.5}{$0$}}^{-1} T^{\alpha \dagger}  \cM_{\scalebox{0.5}{$0$}} ) - \eta_{-1\alpha\beta} \Tr[  T^\beta U B U^{-1} ] T^\alpha \; .
\end{align}
The semi-flat current components are then finite by construction and we have in particular the components along $L_0$ and $\dK$
\begin{align}
 \tilde{J}_0 &= - 2 \rho^{-1} ( \dd \rho - \langle \partial_{{A}} + \partial_{\rho} |  {A}\rangle \rho) = - 2 \rho^{-1} \cD \rho  \; , \CR
 \tilde{J}_{\dK} &= - 2 ( \dd \sigma - \langle \partial \sigma |  {A}\rangle )+2  \langle \partial_{{A}} | U^{-1}  L_0  U  | {A}\rangle + \tilde{g}^{\nu\sigma}(  \partial_{\nu} \tilde{g}_{\mu\sigma}  - \langle \partial \tilde{g}_{\mu\sigma}   | A_\nu \rangle )  \dd x^\mu \nonumber \\
 &= - 2 \cD \sigma + \tilde{g}^{\nu\sigma} \cD_{\nu} \tilde{g}_{\mu\sigma}   \dd x^\mu  \; . 
\end{align}

Because the pseudo-Lagrangian involves infinitely many dual fields,  we will need to eliminate them in order to reproduce the $E_8$ exceptional field theory Lagrangian. For this we shall add terms to the pseudo-Lagrangian quadratic in the duality equations, which by construction do not modify the equations of motion since their variation vanishes upon using the duality equation. The duality equation along the loop algebra generators from~\eqref{DualityEqZ2}  are 
\begin{align}
 \label{TwistHermtian}  
 \tilde{J}_A^{-n} = \rho^{2n-1} H^B{}_A \star \tilde{J}^{n-1}_B \; ,
 \end{align}
where we introduced the $E_8$ matrix $H^B{}_A =  \eta^{BC} M_{CA}$ that defines the conjugation $H^A{}_B T^B_0 = M^{-1} (T_0^A)^\dagger M$ and 
\begin{align}
 \tilde{J}^\flat{}^{-n}_A = \rho^{2n} H^B{}_A \tilde{J}^\flat{}^n_B \; . 
\end{align}

We also need the explicit form of the ancillary and constrained field $B$ in the $E_8$ solution~\eqref{E8Section} of the $E_9$ section constraint. As a constrained derivative decomposes as \eqref{E8Section}, one can write  the general ansatz for $B$
\begin{align}
 B = |b\rangle \langle 0 | + |b_A\rangle \langle 0 | T_1^A \; , \label{E8Bexpansion}
\end{align}
with $|b_A\rangle$ constrained on its $A$ index according to the $E_8$ section constraint~\eqref{eq:E8SC}. Writing out the various ways in which $B$ occurs in the pseudo-Lagrangian, this gives the components  
\begin{align} \label{TrBE8expansion} 
 \Tr[ B] &= \langle 0 | b\rangle + \langle 0 | T_1^A | b_A \rangle \equiv b_{\dK} \CR
\Tr[ L_0 B] &= \langle 0 | T_1^A | b_A \rangle \equiv C^A{}_A \CR
\Tr[ L_{-k} B ] &= 0  \quad \forall k\ge 1\CR
\Tr[ T_{-1-k}^A B ] &= 0 \quad \forall k\ge 1 \CR
\eta_{AB} \Tr[ T_{-1}^B B ] &= \langle 0 | b_A\rangle \equiv -B_A \CR
\eta_{AB} \Tr[ T_{0}^B B ] &=  f_{AB}{}^C \langle 0 | T_1^B| b_C\rangle \equiv f_{AB}{}^C C^B{}_C \CR
\eta_{AB} \Tr[ T_{k}^B B ] &=  \eta_{AB} \langle 0 | T_k^B | b\rangle +\eta_{AB} \langle 0 | T_1^C T_k^B | b_C\rangle \equiv \mathcal{B}_A^{-1-k} \quad \forall k\ge 1 
\end{align}
where $B_A$ and $C^B{}_A$ are constrained on their $A$ indices while $b_{\dK}$ and $\mathcal{B}_A^{-1-k}$ are  arbitrary. We shall see that $B_A$ is the vector field in the Kaluza--Klein ansatz of the contrained field in three dimensions, while $C^B{}_A$ is the component of a two-form potential in three dimensions. For the vector field $|A\rangle$ we use the similar ansatz 
\begin{align}
\label{eq:KKAket}
 |A_\mu \rangle = \bigl( w_\mu +\eta_{AB}  A_\mu^A T_{-1}^B + \eta_{AC} \eta_{BD} C_\mu^{A;B} T_{-1}^C T_{-1}^D +  \dots \bigr) |0\rangle\; ,  
 \end{align}
where $w_\mu$ is the Kaluza--Klein vector of the three-dimensional external metric \eqref{3Dmetric}.  The vector $A_{\mu}^A$ in~\eqref{eq:KKAket} is the Kaluza--Klein component of the three-dimensional vector (in the adjoint of $E_8$), and $C^{A;B}$, valued in ${\bf 3875}\oplus {\bf 1}\oplus {\bf 248}$ with 
\begin{align}
 \langle 0 | T_1^A T_1^B |A\rangle  = 2 C^{(A;B)} + f_{CE}{}^B f^{AE}{}_D  C^{C;D} \; , 
 \end{align}
comes from the two-forms in three dimensions~\cite{Hohm:2014fxa}.\footnote{The antisymmetric ${\bf 248}$ component $C^{[A;B]} = - \frac1{60} f^{AB}{}_E f_{CD}{}^E C^{C;D}$ was not considered in \cite{Hohm:2014fxa} because it could be absorbed in a redefinition of constrained two-form $C^B{}_A$ in the $E_8$ theory.} To describe the Kaluza--Klein ansatz for the vectors we note from \cite{Bossard:2018utw} that $Y_1^A = A_\varphi^A$ , while the constrained field 
\begin{align}
 \langle \chi | = \rho^{-1} \langle \tilde{\chi}| + \omega_1^\alpha(U^{-1}) \langle \tilde{J}_\alpha | =  \rho^{-1} \langle   0 | \bigl( \tilde{\chi}_\varphi + \rho^{-1} B_{\varphi A} \bigr) + \omega_1^\alpha(U^{-1}) \langle \tilde{J}_\alpha | \; . 
\end{align}
The Kaluza--Klein ansatz is defined such that the three-dimensional vector fields take the form
\begin{align}
  A^{\scalebox{0.5}{3D} A}  = A_\varphi^A ( \dd\varphi  + w) + A_\mu^A \dd x^\mu \; , \qquad B_{A}^{\scalebox{0.5}{3D}}  = B_{\varphi A} ( \dd \varphi + w) + B_{\mu A} \dd x^\mu + \rho \tilde{g}_{\mu\sigma} \varepsilon^{\sigma\nu} \partial_A w_\nu \dd x^\mu   \; . 
  \end{align}
The additional term in the Kaluza--Klein Ansatz of the constrained vector $B_{A}^{\scalebox{0.5}{3D}} $ is consistent with the $E_8$ section constraint, and can be ascribed to the non-covariant transformation of the field under external diffeomorphisms  \cite{Hohm:2014fxa}. The three-dimensional covariant derivatives 
\be D^{\scalebox{0.5}{3D} } = \dd x^\mu D_\mu + \dd\varphi D_{\varphi} = \dd^{\scalebox{0.5}{3D} } - \cL^{\scalebox{0.5}{3D}}_{(A^{\scalebox{0.4}{3D} A} ,B_{A}^{\scalebox{0.4}{3D}}  )}  \; ,    \ee 
of the metric components are defined in form notations as
\begin{align}
 D^{\scalebox{0.5}{3D} } \sigma &= \dd^{\scalebox{0.5}{3D} } \sigma -   A^{\scalebox{0.5}{3D} A}  \partial_A \sigma - \partial_A A^{\scalebox{0.5}{3D} A} \; , \qquad  D^{\scalebox{0.5}{3D} } \rho = \dd^{\scalebox{0.5}{3D} } \rho -  \partial_A (  A^{\scalebox{0.5}{3D} A}  \rho) \; , \CR 
D^{\scalebox{0.5}{3D} } \tilde{g}_{\mu\nu} &=  \dd^{\scalebox{0.5}{3D} } \tilde{g}_{\mu\nu}  -   A^{\scalebox{0.5}{3D} A}  \partial_A \tilde{g}_{\mu\nu}\; , \hspace{16mm}  D^{\scalebox{0.5}{3D} } w_\mu  =  \dd^{\scalebox{0.5}{3D} }  w_\mu -   A^{\scalebox{0.5}{3D} A}  \partial_A w_\mu \; , \end{align}
and the $E_9$ covariant derivative decomposes as
\be \cD_\mu = D_\mu - w_\mu D_{\varphi} \ee
on $\sigma$, $\tilde{g}_{\mu\nu}$ and $w_\mu$, whereas there is an additional term for $\rho$
\be \cD_\mu   \rho   =  D_\mu \rho - w_\mu D_{\varphi} \rho -\rho  D_\varphi w_\mu  \; .  \label{cDrho} \ee 

Since the field $B$ appears in the pseudo-Lagrangian through $\Tr [T^\alpha U B U^{-1} ]$, we shall also introduce the shorthand notation $\eta_{AB} \Tr[ T_{k}^B U B U^{-1} ]  = \tilde{\mathcal{B}}_A^{-1-k} $ for the loop components with $k\ge 1$. For example, $\tilde{\mathcal{B}}_A^{-2}$ expands as
 \begin{equation}
  \tilde{\mathcal{B}}_A^{-2} =  {\mathcal{B}}_A^{-2} + f_{AB}{}^E f_{EC}{}^D A_\varphi^B C^C{}_D + \eta_{AB} A_\varphi^B b_{\dK} - \tfrac12 f_{AB}{}^E f_{EC}{}^D A_\varphi^B A_\varphi^C B_D- f_{AB}{}^C Y_2^B B_C \; .
\end{equation}
 We also have for the (mode 0) $E_8$ component
\begin{align}
 \eta_{AB} \Tr[ T_{0}^B U B U^{-1}  ] = f_{AB}{}^C \tilde{C}^B{}_C =  f_{AB}{}^C ( {C}^B{}_C - A_\varphi^B B_C )\; .
 \end{align}
 
The duality equations \eqref{TwistHermtian} for $n\ge 2$ give
\be \tilde{\mathcal{B}}_A^{-n} = \tilde{J}^\flat{}_A^{-n} - \rho^{2n-1}   H^B{}_A \star \tilde{J}^\flat{}_B^{n-1}\; ,  \ee
and since the fields $\tilde{\mathcal{B}}_A^{-n}$ are arbitrary, one can integrate them out in the pseudo-Lagrangian by setting them to this value.

\subsubsection*{Expansion of the pseudo-Lagrangian}

We are now ready to analyse the pseudo-Lagrangian \eqref{PseudoLagrangian}. We will first show that  $\cL_1+\cL_2$ can be written as a finite set of terms plus an infinite sum of terms quadratic in the duality equations  \eqref{TwistHermtian}. For this purpose one computes that $\cL_1$ decomposes as
\begin{align}
\label{eq:L1E8}
 \cL_1&= \frac{\rho^{-1} }{2}\tilde{g}^{\mu\nu} \Bigl(   \Tr \bigl[  \tilde{B}_\mu \cM_{\scalebox{0.5}{$0$}}^{-1} \bigl(  \cS_1(\tilde{J}^\flat_\nu) + \tilde{\chi}_\nu K \bigr)^\dagger \cM_{\scalebox{0.5}{$0$}} \bigr] - \tfrac12 \eta_{\alpha\beta}  \Tr [ T^\alpha \tilde{B}_\mu ] \Tr[ \cM_{\scalebox{0.5}{$0$}}^{-1} T^{\beta \dagger} \cM_{\scalebox{0.5}{$0$}} \tilde{B}_\nu] \Bigr)  \\
& \hspace{-7mm} =  \frac{\rho}{2} M^{AB}\tilde{g}^{\mu\nu} \Bigl( -\tilde{J}_\mu^\flat{}_A^0 {B}_{\nu B} + \rho^{-2} \tilde{J}^\flat_\mu{}^{-1}_A f_{BC}{}^D \tilde{C}_{\nu}^C{}_D + \sum_{n=2}^\infty \rho^{-2n} \tilde{J}^\flat_\mu{}_A^{-n} \tilde{\mathcal{B}}_\nu{}_B^{-n}  \Bigr) + \frac{\rho^{-1}}{2}\tilde{g}^{\mu\nu}  ( \tilde{\chi}_\mu^\flat + {\tilde{C}}_\mu^A{}_A)  \tilde{b}_{\nu {\dK}}   \CR
 &\hspace{4.5mm} - \frac{\rho}{4} M^{AB} \tilde{g}^{\mu\nu} \Bigl( {B}_{\mu A}  B_{\nu B} + \rho^{-2} f_{AC}{}^D \tilde{C}_{\mu}^C{}_D  f_{BE}{}^F \tilde{C}_{\nu}^E{}_F+ \sum_{n=2}^\infty \rho^{-2n} \tilde{\mathcal{B}}_\mu{}_A^{-n} \tilde{\mathcal{B}}_\nu{}_B^{-n}\Bigr) \CR
&=- \frac{\rho}{4} M^{AB} \tilde{g}^{\mu\nu} \sum_{n=2}^\infty \rho^{-2n} \bigl( \tilde{J}_\mu{}^{-n}_A - \rho^{2n-1} H^C{}_A (\star \tilde{J})_\mu{}^{n-1}_C \bigr) \bigl( \tilde{J}_\nu{}^{-n}_B - \rho^{2n-1} H^D{}_B (\star \tilde{J})_\nu{}^{n-1}_D \bigr)\CR
 &\hspace{4.5mm}  +\frac{\rho^{-1}}{2}\tilde{g}^{\mu\nu}  ( \tilde{\chi}_\mu- \rho ( \star \tilde{J} )_{\mu   {\dK}} )   ( b_{\nu {\dK}}- \rho ( \star \tilde{J} )_{\nu 0 }) \CR
 &\hspace{4.5mm} -  \frac{\rho^{-1}}{4} M^{AB} \tilde{g}^{\mu\nu} \bigl( \tilde{J}_\mu{}^{-1}_A - \rho H^C{}_A (\star \tilde{J})_\mu{}^{0}_C \bigr) \bigl( \tilde{J}_\nu{}^{-1}_B - \rho H^D{}_B (\star \tilde{J})_\nu{}^{0}_D \bigr)\CR 
 &\hspace{4.5mm} -  \frac{\rho}{4} \eta_{AB}\tilde{g}^{\mu\nu}  \bigl( \eta^{AC}  \tilde{J}_\mu{}_C^0 +M^{AC} B_{\mu C}  \bigr)   \bigl( \eta^{BD}  \tilde{J}_\nu{}_D^0 +M^{BD} B_{\nu D}  \bigr) 
 +\rho \tilde{g}^{\mu\nu}  \tilde{J}_{\mu   {\dK}}  \tilde{J}_{\nu 0 } \nn\\
 &\hspace{4.5mm} + \tfrac12 \varepsilon^{\mu\nu} \eta^{AB} \sum_{n=0}^\infty \tilde{J}_{\mu}{}_A^{n} \tilde{J}_\nu{}_B^{-1-n}  -\tfrac12  \varepsilon^{\mu\nu} \tilde{J}_{\mu {\dK}} \tilde{J}_{\nu -1} -\tfrac12  \varepsilon^{\mu\nu} \tilde{J}_{\mu 0} \tilde{\chi}_{\nu }\; ,  \nn 
 \end{align}
 where we have used in particular that 
 \begin{align}
  \tilde{g}^{\mu\nu} (\star \tilde{J})_\mu{}^{0}_C (\star \tilde{J})_\nu{}^{0}_D= - \tilde{g}^{\mu\nu}  \tilde{J}_\mu{}^{0}_C \tilde{J}_\nu{}^{0}_D\; , \qquad \eta^{AB} B_{\mu A} B_{\nu B} = 0 \; .  
  \end{align}

 For the topological term $\cL_2$, we note in the first place that a rigid $E_9$ transformation $g$ (not to be confused with the metric $g_{\mu\nu}$) in the centrally extended loop group $\hat{E}_8$, i.e. with $\rho(g)=1$, gives 
 \begin{align}
  \cD\chi \rightarrow \cD\chi  + \omega_{-1}^\alpha(g^{-1}) (\cD \cJ)_\alpha \; . 
  \end{align}
 One  computes that 
\begin{align}
\label{eq:Dchi1}
 \cD \chi &= \dd\tilde{\chi} - \langle \partial | (  |{A}\rangle \tilde{\chi} ) - \omega_{-1}^\alpha(U) ( \cD\cJ)_\alpha +\eta^{AB} \sum_{n=1}^\infty n \tilde{J}_A^{-n} \tilde{J}_B^{n-1} - \Tr[ U {B} U^{-1}
] \tilde{J}_{\dK}\CR
&\hspace{4.5mm} + \langle \partial_A| U^{-1} T_{-1}^A U  | A \rangle \tilde{J}_A^{-2} + \langle \partial_A | U^{-1} \cM_{\scalebox{0.5}{$0$}}^{-1} ( T_1^A)^\dagger \cM_{\scalebox{0.5}{$0$}} U  | A \rangle \tilde{J}_A^0  \; , 
\end{align}
where we used 
\begin{align}
 \dd\omega_{-1}^\alpha(U) = R(U^{-1})^\alpha{}_\beta \delta \omega_{-1}^\beta (  \dd U  \cdot U^{-1}) \; , \qquad  \omega_{-1}^\beta ( \exp(X)) =  \delta \omega_{-1}^\beta ( X)  + \mathcal{O}(X^2) \; , 
 \end{align}
and the property that $\langle \partial_A | U^{-1} T_{-n}^A = 0$ for $n\ge 2$ in the $E_8$ partial section solution. 

The first line of~\eqref{eq:Dchi1} gives together with the other terms of $\cL_2$
\begin{align}
 &\hspace{4.5mm}  - 2\omega_{-1}^\alpha(U) ( \cD\cJ)_\alpha +2\eta^{AB} \sum_{n=1}^\infty n \tilde{J}_A^{-n} \tilde{J}_B^{n-1}  - 2\Tr[ U {B} U^{-1} ]  \tilde{J}_{\dK}\CR
&\hspace{4.5mm}  +J_0 \chi +J_{-1} J_{\dK} +\eta^{AB} \sum_n n J_A^n J_B^{-1-n}  + 2\langle \rho^2 (\chi- \omega_{-1}^\alpha(M) \cJ_\alpha )| F \rangle  \CR
&\hspace{4.5mm}  - 2\langle \partial_F | ( L_1 + \cM^{-1} L_{-1} \cM ) |F\rangle  -2 \Tr[ L_0 G]  +\rho^{-2} \eta_{1\, \alpha\beta} \langle \partial_{A} | T^\alpha | A^\prime  \rangle \langle \partial_{{A}} | \cM^{-1} L_{-2} \cM T^\beta | {A}\rangle \CR
&= \tilde{J}_0 \tilde{\chi} - \Tr[\tilde{B}] \tilde{J}_{\dK} -\eta^{AB} \sum_{n=0}^\infty  \tilde{J}_A^n \tilde{J}_B^{-1-n}  + 2\rho \langle \tilde{\chi} | {F} \rangle   -2 \Tr[ L_0  U {G} U^{-1}] \CR
&\hspace{4.5mm}  - 2\langle \partial_{{F}}   | U^{-1}  ( L_1 + \cM_{\scalebox{0.5}{$0$}}^{-1}  L_{-1} \cM_{\scalebox{0.5}{$0$}})  U  |{F}\rangle \\[2mm]
&\hspace{4.5mm} +\rho^{-2} \eta_{1\, \alpha\beta} \langle \partial_{{A}}   | U^{-1} T^\alpha U | {A}^\prime \rangle \langle \partial_{{{A}}}  | U^{-1} \cM_{\scalebox{0.5}{$0$}}^{-1}  L_{-2} \cM_{\scalebox{0.5}{$0$}} T^\beta U  | {{A}}\rangle \,,
\end{align}
where we used the rigid $E_9$-invariance of $\cL_2$. Note that the last line and the second term of the next-to-last vanish when using the partial solution to the section constraint~\eqref{E8Section} associated to $E_8$.  We therefore have in total 
\begin{align}
 \cL_2&=  \dd\tilde{\chi} - \langle \partial | ( U^{-1} |\tilde{A}\rangle \tilde{\chi} ) +\tfrac12 \tilde{J}_0 \tilde{\chi} -\tfrac12  \Tr[B] \tilde{J}_{\dK} -\tfrac12 \eta^{AB} \sum_{n=0}^\infty  \tilde{J}_A^n \tilde{J}_B^{-1-n}  + \rho \langle \tilde{\chi} |  {F} \rangle   - \Tr[ L_0 U {G} U^{-1}]  \CR
&\hspace{4.5mm}   - \langle \partial_{F}  | U^{-1}   L_1  U  |F\rangle + \langle \partial_A| U^{-1} T_{-1}^A U  | A \rangle \tilde{J}_A^{-2} + \langle \partial_A | U^{-1} \cM_{\scalebox{0.5}{$0$}}^{-1} ( T_1^A)^\dagger \cM_{\scalebox{0.5}{$0$}} U  | A \rangle \tilde{J}_A^0\; .  
\end{align}
Combining this with $\cL_1$ from~\eqref{eq:L1E8}, we have 
\begin{align}
 \cL_1 + \cL_2 &= \mathcal{Z}  -  \frac{\rho}{4} \eta_{AB}\tilde{g}^{\mu\nu} \bigl( \eta^{AC}  \tilde{J}_\mu{}_C^0 +M^{AC} B_{\mu C}  \bigr)   \bigl( \eta^{BD}  \tilde{J}_\nu{}_D^0 +M^{BD} B_{\nu D}  \bigr)  + \tfrac12 \rho \tilde{g}^{\mu\nu}  \tilde{J}_{\mu   {\dK}}  \tilde{J}_{\nu 0 }  \nn\\
&\hspace{4.5mm} + \rho \langle \tilde{\chi} | {F} \rangle   - \Tr[ L_0 U {G} U^{-1}]    +\langle \partial_{U}  | U^{-1}   L_1  U  |F\rangle + \langle \partial_A| U^{-1} T_{-1}^A U  | A \rangle \tilde{J}_A^{-2}   \CR
&\hspace{4.5mm} + \langle \partial_A | U^{-1} \cM_{\scalebox{0.5}{$0$}}^{-1} ( T_1^A)^\dagger \cM_{\scalebox{0.5}{$0$}} U  | A \rangle \tilde{J}_A^0  \; ,  
\end{align}
where
\begin{align}\label{Zduality0}
\mathcal{Z} &= - \frac{\rho}{4} M^{AB}  \sum_{n=2}^\infty \rho^{-2n} \bigl( \tilde{J}{}^{-n}_A - \rho^{2n-1} H^C{}_A \star \tilde{J}{}^{n-1}_C \bigr)\star  \bigl( \tilde{J}{}^{-n}_B - \rho^{2n-1} H^D{}_B \star \tilde{J}{}^{n-1}_D \bigr)\CR
 &\hspace{4.5mm}  +\tfrac12 \rho^{-1} ( \tilde{\chi}- \rho  \star \tilde{J}_{ {\dK}} )  \star  ( b_{{\dK}}- \rho \star \tilde{J}_{ 0 })+ \dd\tilde{\chi} - \langle \partial |\bigl(   U^{-1} |\tilde{A}\rangle \tilde{\chi} + U^{-1}   L_1  U  |F\rangle \bigr) \CR
 &\hspace{4.5mm} -  \frac{\rho^{-1}}{4} M^{AB}  \bigl( \tilde{J}{}^{-1}_A - \rho H^C{}_A \star \tilde{J}{}^{0}_C \bigr) \star \bigl( \tilde{J}{}^{-1}_B - \rho H^D{}_B \star \tilde{J}{}^{0}_D \bigr) 
\end{align}
does not contribute to the equations of motion, as it is composed out of bilinear terms in the components of the duality equations  \eqref{TwistHermtian} and total derivatives.

The before to last term in the pseudo-Lagrangian \eqref{PseudoLagrangian} becomes
\begin{align}
  \frac{\rho^{-1}}{4} \langle \partial \tilde{g}^{\mu\nu} | \cM^{-1} | \partial \tilde{g}_{\mu\nu} \rangle =  \frac{\rho^{-1}}{4} e^{2\sigma} D_\varphi   \tilde{g}^{\mu\nu} D_\varphi \tilde{g}_{\mu\nu} + \frac{1}{4} \rho e^{2\sigma} M^{AB} \partial_A \tilde{g}^{\mu\nu} \partial_B \tilde{g}_{\mu\nu} \; . 
  \end{align}
We recall from \cite{Bossard:2018utw} that the $E_9$ potential decomposes as
 \begin{multline}  - V  
 = - e V^{\scalebox{0.5}{3D}} - \frac{1}{4} \rho e^{2\sigma} M^{AB} \partial_A \tilde{g}^{\mu\nu} \partial_B \tilde{g}_{\mu\nu} + \frac{\rho^3}{2} M^{AB} \tilde{g}^{\mu\nu} \partial_A w_\mu \partial_B w_\nu \\
 - \frac14 \rho^{-1} e^{2\sigma} \eta_{AB} j_\varphi^A j_\varphi^B - 2 \rho^{-1} e^{2\sigma} D_\varphi \sigma D_\varphi \sigma  \\
   -\tfrac12  \rho^{-3} e^{2\sigma} \bigl( \rho \tilde{\chi}_{\varphi} - A_\varphi^A B_{\varphi A}  + \partial_A Y_2^A + \tfrac12 f_{AB}{}^C A_\varphi^A \partial_C A_\varphi^B  \bigr)^2 
    \end{multline}
where the derivative of the three-dimensional metric in the first line does not appear in the $E_9$ potential,\footnote{Note that \cite[Eq. (5.34)]{Bossard:2018utw} is written in the conformal gauge $ \dd s_{\scalebox{0.5}{3D}}^2 = e^{2\sigma} \eta_{\mu\nu} \dd x^\mu \dd x^\nu + \rho^2 \dd \varphi^2$ and the internal derivatives of $\tilde{g}_{\mu\nu}$ and $w_\mu$ do not appear in the $E_9$ potential as defined in \cite{Bossard:2018utw}.} but is part of the $E_8$ potential in \cite{Hohm:2014fxa}, so must be subtracted. We also recall that 
\begin{align}
 \langle  \tilde{\chi} | = \langle 0 | ( \tilde{\chi}_\varphi +  \rho^{-1} B_{\varphi A} T_1^A ) \; . 
 \end{align}
To analyse the pseudo-Lagrangian we recombine the various terms coming from the five components of  \eqref{PseudoLagrangian}  into the new five  components 
\begin{align}
 \cL_{\text{min}} = \mathcal{Z} + \cL^{\scalebox{0.5}{3D}}_{\scalebox{0.6}{kin}} + \cL^{\scalebox{0.5}{3D}}_{\scalebox{0.6}{EH}} + \cL^{\scalebox{0.5}{3D}}_{\scalebox{0.6}{CS}} + \cL^{\scalebox{0.5}{3D}}_{\scalebox{0.6}{KK}}  - e V^{\scalebox{0.5}{3D}} \; , \label{E8decompose} 
 \end{align}
with 
\begin{subequations}
\begin{align} 
 \cL^{\scalebox{0.5}{3D}}_{\scalebox{0.6}{kin}} &= -  \frac{\rho}{4} \eta_{AB}\tilde{g}^{\mu\nu} \bigl( \eta^{AC}  \tilde{J}_\mu{}_C^0 +M^{AC} B_{\mu C}  \bigr)   \bigl( \eta^{BD}  \tilde{J}_\nu{}_D^0 +M^{BD} B_{\nu D}  \bigr) - \frac14 \rho^{-1} e^{2\sigma} \eta_{AB} j_\varphi^A j_\varphi^B    \CR
 \hspace{-2mm} &\hspace{4.5mm} +  \langle \partial_A | U^{-1} \cM_{\scalebox{0.5}{$0$}}^{-1} ( T_1^A)^\dagger \cM_{\scalebox{0.5}{$0$}} U  | A \rangle \tilde{J}_A^0  + \frac{\rho^3}{2} M^{AB} \tilde{g}^{\mu\nu} \partial_A w_\mu \partial_B w_\nu\; ,  \label{kin}\\
  \cL^{\scalebox{0.5}{3D}}_{\scalebox{0.6}{EH}} &= \tfrac12 \rho \tilde{g}^{\mu\nu}  \tilde{J}_{\mu   {\dK}}  \tilde{J}_{\nu 0 } - 2 \rho^{-1} e^{2\sigma} D_\varphi \sigma D_\varphi \sigma - \tfrac14 \rho^3 \sqrt{-g} g^{\mu\sigma} g^{\nu\rho} f_{\mu\nu} f_{\sigma\rho} - \rho \tilde{g}^{\mu\nu} \partial_A w_\mu F_{\varphi  \nu }^{\scalebox{0.5}{3D} A}  \CR
&\hspace{4.5mm}    +\frac14 \rho  \varepsilon^{\mu\nu} \varepsilon^{\sigma\rho}  \tilde{g}^{\kappa\lambda} \cD_\mu \tilde{g}_{\sigma\kappa} \cD_\nu  \tilde{g}_{\rho\lambda}+\frac{\rho^{-1}}{4} e^{2\sigma} D_\varphi  \tilde{g}^{\mu\nu} D_\varphi  \tilde{g}_{\mu\nu} \; , \label{EH} \\
 \cL^{\scalebox{0.5}{3D}}_{\scalebox{0.6}{CS}}&=   \rho \langle \tilde{\chi} | {F} \rangle   - \Tr[ L_0 U {G} U^{-1}]    +\langle \partial_{U}  | U^{-1}   L_1  U  |F\rangle + \langle \partial_A| U^{-1} T_{-1}^A U  | A \rangle \tilde{J}_A^{-2}  \CR
 &\hspace{4.5mm} - \bigl( \rho \tilde{\chi}_{\varphi} - A_\varphi^A B_{\varphi A}   +\partial_A Y_2^A+ \tfrac12 f_{AB}{}^C A_\varphi^A \partial_C A_\varphi^B \bigr)\, f+ \rho \tilde{g}^{\mu\nu} \partial_A w_\mu F_{\varphi  \nu }^{\scalebox{0.5}{3D} A}  \; , \label{CS}  \\
  \cL^{\scalebox{0.5}{3D}}_{\scalebox{0.6}{KK}} &=    -\tfrac12  \rho^{-3} e^{2\sigma} \bigl( \rho \tilde{\chi}_{\varphi} - A_\varphi^A B_{\varphi A}  + \partial_A Y_2^A + \tfrac12 f_{AB}{}^C A_\varphi^A \partial_C A_\varphi^B  \bigr)^2 \CR
&\hspace{4.5mm}+ \bigl( \rho \tilde{\chi}_{\varphi} - A_\varphi^A B_{\varphi A}   +\partial_A Y_2^A+ \tfrac12 f_{AB}{}^C A_\varphi^A \partial_C A_\varphi^B \bigr)\, f_{01}+\tfrac14 \rho^3 \sqrt{-g} g^{\mu\sigma} g^{\nu\rho} f_{\mu\nu} f_{\sigma\rho} \, .\label{KK} 
\end{align}
\end{subequations}
Here, we have added by hand the last lines in $  \cL^{\scalebox{0.5}{3D}}_{\scalebox{0.6}{CS}}$ and $  \cL^{\scalebox{0.5}{3D}}_{\scalebox{0.6}{KK}}$ and subtracted the same terms from the first line in $  \cL^{\scalebox{0.5}{3D}}_{\scalebox{0.6}{EH}}$. We have also introduced the Kaluza--Klein field strength 
 \begin{align}
  f = \dd w - A^A \partial_A w - w \partial_{\varphi} w = \langle 0 | F \rangle = \dd x^\mu D_\mu w - w D_\varphi w  \; . 
  \end{align}

Let us start with $ \cL^{\scalebox{0.5}{3D}}_{\scalebox{0.6}{kin}}$. One checks that the combination 
\begin{align}
  \eta^{AC}  \tilde{J}_\mu{}_C^0 +M^{AC} B_{\mu C} &=  j_\mu^A  - \rho \tilde{g}_{\mu\sigma} \varepsilon^{\sigma\nu} (\eta^{AB} + M^{AB}) \partial_B w_\nu \nn\\
&=  j_\mu^{\scalebox{0.5}{3D} A} - w_\mu  j_\varphi^{A} - \rho \tilde{g}_{\mu\sigma} \varepsilon^{\sigma\nu} (\eta^{AB} + M^{AB}) \partial_B w_\nu  \; .
\end{align}
where 
\begin{align}
 j^{\scalebox{0.5}{3D} A}  = j_\varphi^A ( \dd \varphi + w) + j_{\mu}^A \dd x^\mu\; , 
 \end{align}
is the covariant $E_8$ current~\cite{Hohm:2014fxa}.  One computes moreover that
\begin{align}
 \langle \partial_A | U^{-1} \cM_{\scalebox{0.5}{$0$}}^{-1}   (T_{1}^A)^\dagger \cM_{\scalebox{0.5}{$0$}}  U | A\rangle = \rho^2 M^{AB}  \partial_B w \; 
 \end{align}
so that \eqref{kin} gives indeed the scalar field kinetic term of the $E_8$ exceptional field theory 
\begin{align}
  \cL^{\scalebox{0.5}{3D}}_{\scalebox{0.6}{kin}} = -  \frac{\rho}{4} \eta_{AB}\tilde{g}^{\mu\nu} (j_\mu^{\scalebox{0.5}{3D} A} - w_\mu  j_\varphi^{A} ) (j_\nu^{\scalebox{0.5}{3D} B} - w_\nu  j_\varphi^{B} ) - \frac14 \rho^{-1} e^{2\sigma} \eta_{AB} j_\varphi^A j_\varphi^B \; .
  \end{align}
  For the Einstein--Hilbert term we first use that 
 \be [ D_\varphi , \cD_\mu] w_\nu = -F_{\varphi  \mu }^{\scalebox{0.5}{3D} A} \partial_A w_\nu- D_\varphi w_\mu D_\varphi w_\nu\; ,   \ee
 and then compute that\footnote{We used here the following formula for the Einstein--Hilbert term in $D$ dimensions 
\begin{align*} 
 \sqrt{-g} \widehat{R}  &=  \sqrt{-g} \Bigl( - \tfrac14 g^{\mu\nu} g^{\sigma\rho} g^{\kappa\lambda} D_\mu g_{\sigma\kappa} D_\nu g_{\rho\lambda}  + \tfrac{1}{2} g^{\mu\sigma} g^{\nu\rho} g^{\kappa\lambda} D_\mu g_{\rho\kappa} D_\nu g_{\sigma\lambda} +\tfrac14 g^{\mu\nu} g^{\sigma\rho} g^{\kappa\lambda} D_\mu g_{\sigma\rho} D_\nu g_{\kappa\lambda} \\
 &\quad - \tfrac12 g^{\mu\sigma} g^{\nu\rho} g^{\kappa\lambda} D_\mu g_{\kappa\lambda} D_\nu g_{\sigma\rho} \Bigr) 
 + D_\mu \Bigl( \sqrt{-g}g^{\sigma\rho} g^{\mu\nu}  \bigl( D_\sigma g_{\rho\nu} - D_\nu g_{\sigma\rho} \bigr) \Bigr) 
 \end{align*} 
with 
$ D_\sigma g_{\mu\nu} = \partial_\sigma g_{\mu\nu} - A_\sigma^A \partial_A g_{\mu\nu} - \frac{2}{D-2} g_{\mu\nu} \partial_A A_\sigma^A  $ . 
}
\begin{align}
 \cL^{\scalebox{0.5}{3D}}_{\scalebox{0.6}{EH}} &= \rho \Bigl( - \frac14  \tilde{g}^{\mu\nu} \tilde{g}^{\sigma\rho} \tilde{g}^{\kappa\lambda} \cD_\mu \tilde{g}_{\sigma\kappa} \cD_\nu \tilde{g}_{\rho\lambda} + \frac12 \tilde{g}^{\mu\rho} \tilde{g}^{\nu\sigma} \tilde{g}^{\kappa\lambda} \cD_\mu \tilde{g}_{\sigma\kappa} \cD_\nu \tilde{g}_{\rho\lambda}\Bigr)    \\
&\hspace{4.5mm} +  \tilde{g}^{\mu\nu} \cD_\mu \rho  ( 2 \cD_\nu \sigma - \tilde{g}^{\sigma\rho} \cD_\sigma \tilde{g}_{\nu\rho})  + \rho \tilde{g}^{\mu\nu} [ D_\varphi , \cD_\mu ] w_\nu  + \rho \tilde{g}^{\mu\nu} D_\varphi w_\mu D_\varphi w_\nu \CR
&\hspace{4.5mm}  - 2 \rho^{-1} e^{2\sigma} D_\varphi \sigma D_\varphi \sigma -  \tfrac14 \rho^3 \sqrt{-g} g^{\mu\sigma} g^{\nu\rho} f_{\mu\nu} f_{\sigma\rho}+\frac{\rho^{-1}e^{2\sigma}}{4}  D_\varphi   \tilde{g}^{\mu\nu}  D_\varphi    \tilde{g}_{\mu\nu} \CR
&= D_\varphi ( \rho \tilde{g}^{\mu\nu} \cD_\mu w_\nu) -\cD_\mu ( \rho \tilde{g}^{\mu\nu} D_\varphi w_\nu ) \CR
&\hspace{4.5mm} + \rho \Bigl( - \frac14  \tilde{g}^{\mu\nu} \tilde{g}^{\sigma\rho} \tilde{g}^{\kappa\lambda} \cD_\mu \tilde{g}_{\sigma\kappa} \cD_\nu \tilde{g}_{\rho\lambda} + \frac12 \tilde{g}^{\mu\rho} \tilde{g}^{\nu\sigma} \tilde{g}^{\kappa\lambda} \cD_\mu \tilde{g}_{\sigma\kappa} \cD_\nu \tilde{g}_{\rho\lambda}\Bigr)  -  \tfrac14 \rho^3 \sqrt{-g} g^{\mu\sigma} g^{\nu\rho} f_{\mu\nu} f_{\sigma\rho}   \CR
&\hspace{4.5mm} + ( D_\mu -w_\mu  D_\varphi)  \rho ( 2 \tilde{g}^{\mu\nu}  \cD_\nu \sigma  + \cD_\nu \tilde{g}^{\mu\nu}) - 2 \rho \tilde{g}^{\mu\nu} D_\varphi w_\mu \cD_\nu \sigma - \rho \cD_\mu w_\nu  D_\varphi \tilde{g}^{\mu\nu} \CR
&\hspace{4.5mm} + \tilde{g}^{\mu\nu} ( D_\varphi w_\mu  ( D_\nu - w_\nu D_\varphi) \rho - \cD_\mu w_\nu D_\varphi \rho )  - 2 \rho^{-1} e^{2\sigma} D_\varphi \sigma D_\varphi \sigma +\frac{\rho^{-1}e^{2\sigma}}{4}  D_\varphi   \tilde{g}^{\mu\nu}  D_\varphi    \tilde{g}_{\mu\nu} \CR
&=  \sqrt{-g^{\scalebox{0.5}{3D}}} \widehat{R}^{\scalebox{0.5}{3D}}-  \cD_\mu \Bigl(\rho  \tilde{g}^{\sigma\rho} \tilde{g}^{\mu\nu}  \bigl( \cD_\sigma \tilde{g}_{\rho\nu} - \cD_\nu \tilde{g}_{\sigma\rho} \bigr) \Bigr)    \; , \nonumber
\end{align}
where on all individual terms the $E_9$ covariant derivative $\cD_\mu$ gives  $D_\mu - w_\mu D_{\varphi}$, except when it acts as a total covariant derivative, in which case it is a true total  derivative.\footnote{ For total derivatives $\cD_\mu ( \rho X^\mu) =  \partial_\mu (   \rho X^\mu) -\partial_\varphi ( w_\mu \rho X^\mu ) - \partial_A ( A_\mu^A \rho X^\mu )$ using \eqref{cDrho}.} 

The identification of the Chern--Simons term requires more work and we shall first simplify some of the expressions appearing in \eqref{CS}.  Using \eqref{DerU} and $\langle 0 | L_1 = 0 $ one gets 
\begin{align}
  \langle \partial | U^{-1} L_1 U &= \langle 0 | T_2^A  U\partial_A  =    \langle 0 | ( T_2^A   + f_{BC}{}^A Y_1^B T_1^C+ 2 Y_2^A )  \partial_A \; ,  \CR
\langle \partial | U^{-1} T_{-1}^A U  &= \eta^{AB} \langle 0 | \partial_B\; , 
\end{align}
so that
\be
 \langle \partial_{U}  | U^{-1}   L_1  U  |F\rangle + \langle \partial_A| U^{-1} T_{-1}^A U  | A \rangle \tilde{J}_A^{-2}  = 2\partial_A Y_2^A f + f_{BC}{}^A \partial_A A_\varphi^B \langle 0 | T_1^C | F\rangle + \eta^{AB} \partial_A w \tilde{J}_B^{-2}\; . 
 \ee
One computes moreover
\begin{align} 
\langle 0| T_1^A |F\rangle &=  \dd A^A - \frac12 A^B \partial_B A^A - \frac12 f^{EC}{}_D f_{EB}{}^A \partial_C A^D  A^B - \frac12 \partial_B A^B A^A \CR
& \hspace{-6mm}- w \bigl( \partial_{\varphi} A^A +  \partial_B C^{(A;B)}  +\tfrac12 f_{CE}{}^A f^{BE}{}_D \partial_B C^{C;D}   \bigr) + \partial_B w ( C^{(A;B)} +\tfrac12 f_{CE}{}^B f^{AE}{}_D  C^{C;D}) \; ,   \CR
\eta_{AB} \Tr[ T_{-1}^B G ] &= -\dd B_A + A^B \partial_B B_A + \partial_B A^B B_A  - B_B \partial_A A^B + \frac12 f_{CD}{}^B A^C \partial_A \partial_B A^D \CR
& \hspace{-9mm}  + w \bigl( \partial_{\varphi} B_A - \tfrac{1}{2} f_{CD}{}^B \partial_A \partial_B C^{C;D}\bigr) + ( b_{\dK} - C^B{}_B) \partial_A w + \partial_B w C^B{}_A + \tfrac{1}{2} f_{CD}{}^B C^{C;D} \partial_A \partial_B w\; ,  \CR
\Tr[ L_0 G ] &=  \dd C^A{}_A - \partial_A ( A^A C^B{}_B ) - \partial_{\varphi} ( w C^A{}_A ) - \partial_{\varphi} A^A B_A - \frac12 f_{BC}{}^A A^B \partial_{\varphi} \partial_A A^C \CR
&\hspace{4.5mm}\quad + \eta^{AB} \mathcal{B}^{-2}_A \partial_B w + \tfrac1{2} w f_{CD}{}^A \partial_{\varphi} \partial_A C^{C;D} + \tfrac1{2}\partial_{\varphi} \partial_A w f_{CD}{}^A  C^{C;D} \; , 
\end{align}
where we recognise the components of the three-dimensional field strengths \cite{Hohm:2014fxa}
\begin{align}
 F^{\scalebox{0.5}{3D} A}&= \dd A^{\scalebox{0.5}{3D} A} - \frac12 A^{\scalebox{0.5}{3D} B} \partial_B A^{\scalebox{0.5}{3D}  A} - \frac12 f^{EC}{}_D f_{EB}{}^A \partial_C A^{\scalebox{0.5}{3D} D}  A^{\scalebox{0.5}{3D} B} - \frac12 \partial_B A^{\scalebox{0.5}{3D} B} A^{\scalebox{0.5}{3D} A}\; , \CR
G^{\scalebox{0.5}{3D}} _A &=  \dd B^{\scalebox{0.5}{3D}} _A - A^{{\scalebox{0.5}{3D}} B} \partial_B B^{\scalebox{0.5}{3D}} _A - \partial_B A^{{\scalebox{0.5}{3D}} B} B^{\scalebox{0.5}{3D}} _A  + B^{\scalebox{0.5}{3D}} _B \partial_A A^{{\scalebox{0.5}{3D}} B} - \frac12 f_{CD}{}^B A^{{\scalebox{0.5}{3D}} C} \partial_A \partial_B A^{{\scalebox{0.5}{3D}} D}\; . \quad 
\end{align}
In particular, we have 
\begin{align} 
\langle 0| U^{-1} T_1^A U |F\rangle  &= F^A   - w \partial_B  \bigl(  C^{(A;B)}  +\tfrac12 f_{CE}{}^A f^{BE}{}_D  C^{C;D}   \bigr)\nn \\ 
&\quad + \partial_B w \bigl( C^{(A;B)} + A_\varphi^{(A} A^{B)} +\tfrac12 f_{CE}{}^B f^{AE}{}_D  ( C^{C;D}+A_\varphi^{A} A^{B}) \bigr)  \; , 
\end{align}
where $F^A$ is the Kaluza--Klein component of the three-dimensional field strength 
\begin{align}
 F^{\scalebox{0.5}{3D} A} = \eta^{AB}\tilde{J}^\flat{}_B^{-1} (d\varphi + w)   + F^A \ . 
 \end{align}

We also need the expression 
\bea
&& \eta^{AB} \partial_A w   \tilde{J}_B^{-2} \\
&=& \eta^{AB} \partial_A w   \tilde{J}_B^{\flat -2}  - \eta^{AB}\partial_A w \mathcal{B}_B^{-2}  +A_\varphi^A \bigl( (b_{\dK} - C^B{}_B) \partial_A w  + \partial_B w C^B{}_A \bigr) +A_\varphi^A A_\varphi^B B_A \partial_B w \nn 
\eea
where the bare current expands as
\begin{align}
 \eta^{AB} \tilde{J}_B^{\flat -2} &= \bigl(\dd - w \partial_{\varphi} - A^B \partial_B- 2 \partial_B A^B - 2 \partial_{\varphi}  w  \bigr) Y_2^A    - f_{CE}{}^A f^{BE}{}_D ( Y_2^C  \partial_B A^D + A_\varphi^C Y_2^D  \partial_B w )\CR
& \quad + \tfrac12 f_{BC}{}^A A_\varphi^B (\dd - w \partial_{\varphi} - A^D \partial_D) A_\varphi^C- \tfrac12 f^{FG}{}_D f_{CF}{}^A f_{EG}{}^B A_\varphi^C A_\varphi^D \partial_B A^E \CR
&\hspace{4.5mm}  - f_{BC}{}^A A_\varphi^B  \partial_\varphi  A^C  - \tfrac16   f^{FG}{}_C f_{DF}{}^A f_{EG}{}^B A_\varphi^C A_\varphi^D A_\varphi^E \partial_B w \CR
&\hspace{4.5mm}  - f_{BC}{}^A \partial_{\varphi} C^{B;C} - 2 f_{CD}{}^A A_\varphi^C \partial_B C^{(D;B)} - f^{FG}{}_D f_{CF}{}^A f_{EG}{}^B A_\varphi^C \partial_B C^{D;E} \CR
&\hspace{4.5mm} - \langle 0 | T_1^B T_2^A \partial_B | A\rangle  + f^{AB}{}_C Y_3^C \partial_B w \; . 
\end{align}
The last term with the degree 3 fields simplify to a total derivative when contracted with $\partial_A w $ as
\be
 -\partial_A w  \langle 0 | T_1^B T_2^A \partial_B | A\rangle  =  -\partial_B \bigl( \partial_A w  \langle 0 | T_1^B T_2^A  | A\rangle \bigr)
 \ee
using that the $L_0$ degree 3 component of the basic representation does not contain a generic symmetric representation. 

We can now recombine all these terms in \eqref{CS}. We first observe that the dependence in $\mathcal{B}_A^{-2}$ and $b_{\dK}$ cancels between  $- \Tr[ L_0 U {G} U^{-1}] $ and $\langle \partial_A| U^{-1} T_{-1}^A U  | A \rangle \tilde{J}_A^{-2} $. With more work one obtains that the dependence in the three-dimensional two-form components also combine to a total derivative. We compute (where $A^{\scalebox{0.5}{3D} A}$, $F^{\scalebox{0.5}{3D}  A} $ and $B^{\scalebox{0.5}{3D}}_A$ are the forms in two dimensions)
\begin{align}
\label{CSfromE9}
 &\hspace{5mm}   \cL^{\scalebox{0.5}{3D}}_{\scalebox{0.6}{CS}} \\
&= ( B_{\varphi A} +f_{BC}{}^A \partial_A A_{\varphi }^B ) F^{\scalebox{0.5}{3D}  C} + \partial_\varphi A^{\scalebox{0.5}{3D} A} B^{\scalebox{0.5}{3D}}_A + \tfrac12  f_{BC}{}^A A^{\scalebox{0.5}{3D} B} \partial_\varphi \partial_A A^{\scalebox{0.5}{3D} C} \CR 
&\hspace{4.5mm} + A_\varphi^A \Bigl( \dd B^{\scalebox{0.5}{3D} }_A - A^{\scalebox{0.5}{3D} B} \partial_B B^{\scalebox{0.5}{3D} }_A - \partial_B A^{\scalebox{0.5}{3D}  B} B^{\scalebox{0.5}{3D} }_A + B^{\scalebox{0.5}{3D} }_B \partial_A A^{\scalebox{0.5}{3D} B} - \tfrac12 f_{CD}{}^B A^{\scalebox{0.5}{3D}  C} \partial_A \partial_B A^{\scalebox{0.5}{3D} D}  \Bigr)  \CR
&\hspace{4.5mm} - \tfrac12 f^{FG}{}_{(E} f_{C|F}{}^A f_{D)G}{}^B \partial_A \Bigl( \tfrac13 A_\varphi^C A_\varphi^D A_\varphi^E w \partial_B w + w A_\varphi^C \partial_B ( A_\varphi^D A^E) \Bigr) \CR
&\hspace{4.5mm} + \frac12 f_{CD}{}^A\Bigl(   - \partial_{[A} \bigl( w A_\varphi^C \partial_{B]} A_\varphi^D A^B \bigr) + \partial_{(A} \bigl( A_\varphi^B A_\varphi^C \partial_{B)} ( w A^D) \bigr) + \partial_B \bigl( A_\varphi^B \partial_A A_\varphi^C w A^D \bigr)    \Bigr) \CR
&\hspace{4.5mm} -\dd C^A{}_A +\partial_A ( A^A C^B{}_B ) +\partial_{\varphi} ( w C^A{}_A ) - \tfrac12 \partial_A \bigl( w f_{CD}{}^A \partial_{\varphi} C^{C;D}\bigr) - \tfrac12 \partial_{\varphi} \bigl( \partial_A w f_{CD}{}^A C^{C;D}\bigr)  \CR
&\hspace{4.5mm}  - \tfrac12 \bigl( f^{FG}{}_D f_{CF}{}^{[A} f_{EG}{}^{B]} + 2 f_{CD}{}^{[A} \delta_E^{B]} \bigr) \partial_A \bigl( 2 w A_\varphi^C \partial_B C^{(D;E)} + w \partial_B A_\varphi^C C^{(D;E)} \bigr)  \CR
&\hspace{4.5mm} + \tfrac12 f_{CD}{}^{(A} \partial_A \bigl( A_\varphi^{B)} \partial_B ( w C^{C;D})\bigr) \hspace{2mm} +\hspace{2mm}  (2-1)\partial_A Y_2^A (d w - A^A \partial_A w - w \partial_{\varphi} w )  \CR
&\hspace{4.5mm} +\partial_A w \Bigl( \bigl( \dd - w \partial_{\varphi} - A^B \partial_B - 2 \partial_B A^B - 2 \partial_{\varphi} w \bigr) Y_2^A - f_{CE}{}^A f^{BE}{}_D  ( Y_2^C\partial_B A^D + A_\varphi^C Y_2^D \partial_B w) \Bigr) \CR
&=   B_{\varphi A} F^{\scalebox{0.5}{3D}A} + B^{\scalebox{0.5}{3D}}_A \bigl( \dd A_\varphi^A -\partial_\varphi A^{\scalebox{0.5}{3D} A} + A_\varphi^B \partial_B A^{\scalebox{0.5}{3D} A} - A^{\scalebox{0.5}{3D} B} \partial_B A_\varphi^A \bigr) \CR
&\hspace{4.5mm} - \tfrac12 f_{BC}{}^A \bigl( \dd A^{\scalebox{0.5}{3D} B} \partial_A A_\varphi^C - \dd A_\varphi^B \partial_A A^{\scalebox{0.5}{3D} C} + \partial_\varphi A^{\scalebox{0.5}{3D} B} \partial_A A^{\scalebox{0.5}{3D} C} \bigr) \CR
&\hspace{4.5mm} - \tfrac13 f_{CD}{}^A ( A_\varphi^B A^{\scalebox{0.5}{3D} C} \partial_A \partial_B  A^{\scalebox{0.5}{3D} D}-A^{\scalebox{0.5}{3D} B} A_\varphi^C \partial_A \partial_B A^{\scalebox{0.5}{3D} D} +A^{\scalebox{0.5}{3D} B} A^{\scalebox{0.5}{3D} C} \partial_A \partial_B A_\varphi^D \bigr)\CR
&\hspace{4.5mm} - \tfrac16 f^{FG}{}_C f_{FD}{}^A f_{GE}{}^B \bigl( A_\varphi^C \partial_A A^{\scalebox{0.5}{3D} D} \partial_B  A^{\scalebox{0.5}{3D} E} - 2 A^{\scalebox{0.5}{3D} C} \partial_A A_\varphi^D \partial_B  A^{\scalebox{0.5}{3D} E} \bigr)+ \partial_A (\dots\hspace{-0.1mm}) + \dd (\dots\hspace{-0.1mm})  
 \; , \nonumber 
\end{align}
where we have used repeatedly the section constrained and identity \cite[Eq.~(A.1)]{Hohm:2014fxa} to get for example 
\bea
&&  \frac12 f^{FG}{}_C f_{FD}{}^A f_{GE}{}^B A^C \partial_A A_\varphi^D \partial_B  A^E\CR
&=& - \tfrac16 f^{FG}{}_C f_{FD}{}^A f_{GE}{}^B \bigl( A_\varphi^C \partial_A A^D \partial_B  A^E - 2 A^C \partial_A A_\varphi^D \partial_B  A^E \bigr) \CR
&& + \tfrac16 f_{CD}{}^A ( A_\varphi^B A^C \partial_A \partial_B  A^D+2A^B A_\varphi^C \partial_A \partial_B A^D +A^B A^C \partial_A \partial_B A_\varphi^D \bigr)\CR
& &+ \tfrac16 f_{CD}{}^{(A}  \partial_{A}  \bigl( A^{ C} \partial_B (A_\varphi^D A^{B)} ) -\partial_B A^{ C}  A_\varphi^D A^{B)}\bigr)  
\eea
and to cancel the terms symmetric the derivatives indices involving $C^{(D;E)}$. We recognise therefore $\cL^{\scalebox{0.5}{3D}}_{\scalebox{0.6}{CS}}$ as the Chern--Simons term obtained in \cite{Hohm:2014fxa} up to total derivative terms.\footnote{And up to a sign misprint in the third term in \cite{Hohm:2014fxa}.}

Finally, we compute for the Kaluza--Klein term~\eqref{KK} that 
\begin{align}
 \cL^{\scalebox{0.5}{3D}}_{\scalebox{0.6}{KK}} =   -\tfrac12  \rho^{-3} e^{2\sigma} \bigl( \rho \tilde{\chi}_{\varphi} - A_\varphi^A B_{\varphi  A}  + \partial_A Y_2^A + \tfrac12 f_{AB}{}^C A_\varphi^A \partial_C A_\varphi^B - \rho^3 e^{-2\sigma} f_{01}  \bigr)^2\; . 
 \end{align}
This last term does not appear in \cite{Hohm:2014fxa}, but one can integrate out the auxiliary field $\tilde{\chi}_{\varphi} $ to eliminate it. By construction this term is quadratic in the contraction with $\langle 0 |$ of the duality equation \eqref{Fduality} 
\begin{align}
 U  | \cF_{\mu\nu} \rangle = - \rho^{-1} \varepsilon_{\mu\nu}  T^\alpha \cM_{\scalebox{0.5}{$0$}}^{-1}  U^{-1 \dagger} |\tilde{\cJ}^-_\alpha\rangle \; . 
 \end{align}

We have therefore identified 
\begin{align}
 \cL^{\scalebox{0.5}{3D}} =  \cL^{\scalebox{0.5}{3D}}_{\scalebox{0.6}{kin}} +  \cL^{\scalebox{0.5}{3D}}_{\scalebox{0.6}{EH}}  +  \cL^{\scalebox{0.5}{3D}}_{\scalebox{0.6}{CS}}   - e V^{\scalebox{0.5}{3D}} 
 \end{align}
as the $E_8$ exceptional field theory Lagrangian of \cite{Hohm:2014fxa} up to total derivative terms. The $E_9$ exceptional field theory pseudo-Lagrangian \eqref{PseudoLagrangian} on the partial section \eqref{E8Section} decomposes according to \eqref{E8decompose} as the $E_8$ exceptional field theory Lagrangian plus an infinite sum of terms quadratic in the components of the duality equations \eqref{DualityEquation} and \eqref{Fduality}. 

\subsubsection*{Additional duality equations}

The $E_8$ decomposition of the duality equation \eqref{DualityEquation} does not only give the $E_8$ Euler--Lagrange equations, but also an infinite set of duality equations. The duality equations \eqref{TwistHermtian}  for $n\ge 2$ determine the auxiliary fields $\tilde{\mathcal{B}}_A^{-n}$, and do not give any non-trivial new equation. The $n=1$ component of \eqref{TwistHermtian} gives however the non-trivial equation 
\begin{align}
 \cF^{\scalebox{0.5}{3D} A}_{\mu \varphi}  = \rho \tilde{g}_{\mu\sigma} \varepsilon^{\sigma\nu} \eta^{AB} \bigl( j^{\scalebox{0.5}{3D}}_{\nu B} - B^{\scalebox{0.5}{3D}}_{\nu B} - w_\nu ( j_{\varphi B} - B_{\varphi B})\bigr)  \end{align}
with 
\begin{align}
 \cF^{\scalebox{0.5}{3D} A}_{\mu \varphi}  &= F^{\scalebox{0.5}{3D} A}_{\mu \varphi} - \partial_B \Bigl( 2 C_\mu^{(A;B)} + f_{CE}{}^{A} f^{BE}{}_D C_\mu^{C;D} + A_\mu^{(A} A_\varphi^{B)}  + \tfrac12  f_{CE}{}^{(A} f^{B)E}{}_D A_\mu^{C} A_\varphi^{D}\Bigr) \nn\\
&\hspace{4.5mm}  + f^{AB}{}_C \bigl( C_\mu^C{}_B - \tfrac12 w_\mu f_{EF}{}^C A_{\varphi}^E \partial_B A_\varphi^F \bigr)  \nn\\
 &=  F^{\scalebox{0.5}{3D} A}_{\mu \varphi} +\partial_B \Bigl( 2 C_{\mu \varphi}^{\scalebox{0.5}{3D}  (A;B)} + f_{CE}{}^A f^{BE}{}_D C_{\mu\varphi}^{\scalebox{0.5}{3D} C;D} \Bigr)   + f^{AB}{}_C  C_{\mu\varphi} ^{\scalebox{0.5}{3D}C}{}_B\; . 
 \end{align}
One recognises therefore a component of the three-dimensional duality equation 
\begin{align}
\label{3Dduality}  
\cF^{\scalebox{0.5}{3D} A} +\eta^{AB} \star_{\scalebox{0.5}{3D} } ( j^{\scalebox{0.5}{3D} }_B - B^{\scalebox{0.5}{3D} }_{B}) = 0  \; .
\end{align}
This equation is consistent with the three-dimensional equation Euler--Lagrange equation 
\begin{align}
 \delta B_A \bigl(   \cF^{\scalebox{0.5}{3D} A} +\eta^{AB} \star_{\scalebox{0.5}{3D} } j^{\scalebox{0.5}{3D} }_B \bigr) = 0  \; . 
 \end{align}
since the last term in $B_B$ vanishes using the section constraint. 

Note that in the equivalent equation in the Virasoro-extended formulation~\eqref{twsd cov J}, the component $\mathsf{J}_{A}^{-1}$ of the current $\mf{J}$ includes an additional term in $\eta_{AB} \Tr[ T_{-1}^B B^{\scalebox{0.6}{$(2)$}}]$ with respect to $J_A^{-1}$ in \eqref{Jdef}. The component of $B^{\scalebox{0.6}{$(2)$}}$ can be used to eliminate the term in $ B^{\scalebox{0.5}{3D} }_{B}$ in \eqref{3Dduality}. This term can also be reabsorbed in a redefinition of $C_{\mu\varphi} ^{\scalebox{0.5}{3D}C}{}_B$ in the minimal formulation of the theory, but only if one breaks explicitly the covariance of the Ansatz to e.g. $GL(1)\times E_7$. So we may define instead of \eqref{3Dduality}, the three-dimensional duality equation 
\be   
\cF^{\prime \scalebox{0.5}{3D} A} +\eta^{AB} \star_{\scalebox{0.5}{3D} }  j^{\scalebox{0.5}{3D} }_B =  0  \; .  
\ee

Similarly, one can compute the component of the duality equation \eqref{Fduality} coming from $\langle 0| T_1^A U | \cE\rangle = 0 $ to get 
\begin{align}
 \cF^{\scalebox{0.5}{3D} A}_{\mu \nu} + 2 w_{[\mu} \cF_{\nu]\varphi} = - \rho^{-1} e^{2\sigma} \eta^{AB} ( j_{\varphi  B} - B_{\varphi B}) 
 \end{align}
with 
\begin{align}
  \cF^{\scalebox{0.5}{3D} A}_{\mu\nu} &= F^{\scalebox{0.5}{3D} A}_{\mu \nu} + \partial_B \Bigl( 2 C_{\mu\nu}^{(A;B)} + f_{CE}{}^A f^{BE}{}_D C_{\mu\nu}^{C;D} - w_{[\mu} (2 C_{\nu]}^{(A;B)} + f_{CE}{}^A f^{BE}{}_D C_{\nu]}^{C;D} )\nn\\
 &\hspace{4.5mm}  +  w_{[\mu} \bigl(  A_{\nu]}^{(A} A_\varphi^{B)}  + \tfrac12  f_{CE}{}^{(A} f^{B)E}{}_D A_{\nu]}^{C} A_\varphi^{D} \bigr)\Bigr) \nn\\
&\hspace{4.5mm}   + f^{AB}{}_C \bigl( C_{\mu\nu}^C{}_B + \tfrac14 f_{EF}{}^C A_{\varphi}^E A_{[\mu}^F \partial_B w_{\nu]} + 2 w_{[\mu} C_{\nu]}^C{}_B + \rho^{-1} e^{2\sigma} \varepsilon_{\mu\nu} \partial_B A_\varphi^C \bigr)  \nn\\
&= F^{\scalebox{0.5}{3D} A}_{\mu \nu} +\partial_B \Bigl( 2 C_{\mu \nu}^{\scalebox{0.5}{3D}  (A;B)} + f_{CE}{}^A f^{BE}{}_D C_{\mu\nu}^{\scalebox{0.5}{3D} C;D} \Bigr)   + f^{AB}{}_C  C_{\mu\nu} ^{\scalebox{0.5}{3D}C}{}_B\; ,
\end{align}
 where $C_{\mu\nu}^{A;B}$ and $C_{\mu\nu}^B{}_A$ come from the $E_9$ two-forms.  Since external and internal diffeomorphisms in $E_9$ both contain the $E_8$ external diffeomorphisms, it is to be expected that these two equations transform into each other, so that we need to consider \eqref{Fduality}. This is in fact the only non-trivial component of equation \eqref{miniFdual}, 
 because all other components can be solved trivially by fixing an unconstrained two-form.

 The duality equation \eqref{Fduality} includes one more duality equation that can be interpreted as the three-dimensional three-form field strength equation. It gives  
\begin{align}
 \langle 0 | T_1^A T_1^B U |\cE_{\mu\nu} \rangle =  ( \delta^{(A}_C \delta^{B)}_D  + \tfrac12 f_{CE}{}^A f^{BE}{}_D ) \bigl(- \mathcal{G}_{\varphi\mu\nu}^{C;D}  +2 \rho\,  \varepsilon_{\mu\nu}M^{CF} J_{F;}{}^D\bigr) \; \ ,  \label{Threeformduality}  
\end{align}
for a three-form field strength $ \mathcal{G}^{A;B}$  in $E_8$ exceptional field theory. The antisymmetric component  $\langle 0 | T_2^A  |\cF_{\mu\nu} \rangle  $ includes an unconstrained two-form $C_{\varphi\mu\nu}^A$, so this equation can be solved by fixing $C_{\varphi\mu\nu}^A$. The only non-trivial component of  \eqref{Threeformduality} is therefore in the ${\bf 3875}\oplus {\bf 1}$ of $E_8$, as expected from the tensor hierarchy. The field strength  $\langle 0 | T_1^A  T_1^B T_1^C  |\cF_{\mu\nu} \rangle$ also includes an unconstrained two-form in the same $E_8$ representation, and similarly does $\langle 0 | T_1^{A_1} T_1^{A_2}  \cdots   T_1^{A_n}  |\cF_{\mu\nu} \rangle$ for any number $n$ of generators $T_1^A$ greater than three, so all the other components of \eqref{Fduality} can be solved by fixing the corresponding two-form components. 

We conclude therefore that the duality equations \eqref{DualityEquation} and \eqref{Fduality} on section give all the duality equations expected from the $E_8$ tensor hierarchy, and all other components are tautological equations that can be solved by fixing unconstrained auxiliary fields in function of the other fields.

\section{Conclusion}
\label{sec:conclusion}

In this paper, we have constructed the complete dynamics 
of $E_9$ exceptional field theory in two different formulations. 

The Virasoro-extended formulation in Section~\ref{sec:extE9exft} is based on the $\widehat{E}_{8}\rtimes\Virm$ symmetry, where the Virasoro generators $L_m$ for $m\leq 0$ originate in the extended linear system for $D=2$ supergravity~\cite{Julia:1996nu,Paulot:2004hh} discussed in Section~\ref{sec:extcoset}. 
Indeed, this formulation of $E_9$ exceptional field theory allows us to naturally reproduce $D=2$ supergravity and its linear system by setting to zero the internal derivative $\bra\partial$ as well as all constrained fields $B^\ord{k}$ and $\bra{\chi^\upgamma_1}$.
%
In this formulation, a proper gauge connection for generalised diffeomorphisms can be defined, allowing for instance the construction of covariant external currents for the scalar fields, and the vector fields and their field strengths transform in a manner familiar from lower-rank ExFTs.
This also makes the construction of the topological term more intuitive.
The field content of the Virasoro-extended formulation includes the field $\tilde\rho\sim \phi_1$ dual to the $D=2$ dilaton $\rho$.
This field plays an important role when using Weyl coordinates for studying $D=2$ solutions, and the gauging of its shift symmetry is central to the construction of $D=2$ gauged supergravities in \cite{Samtleben:2007an}.
The formulation that we have presented is only phrased in the conformal gauge $\tilde{g}_{\mu\nu} = \eta_{\mu\nu}$ of the $D=2$ external metric. 
This gauge can be imposed for any higher-dimensional solution that one may want to describe in ExFT, but it would nonetheless be interesting to generalise the Virasoro-extended formulation to arbitrary external metrics. 
Even keeping conformal gauge, one should be able to define conformal external diffeomorphisms
and we expect that they will fix all relative coefficients in the pseudo-Lagrangian.
Furthermore, we have only presented a manifestly $\hE8\rtimes\Virm$ invariant formulation of the topological term, while the expression for the scalar potential we have constructed in previous work~\cite{Bossard:2018utw} is based on internal currents gauge-fixed to $\phi_n=0$ for $n\ge2$.
Because these fields are pure gauge, we do not lose any information on the dynamics by this gauge choice, but it would also be interesting to construct the manifest $\Virm$ extension of the scalar potential.
Since we know the transformation properties of all fields under $\hevirm$ extended generalised diffeomorphisms, one can achieve this goal by appropriately acting with local $L_{-n}$ transformations on the known expression for the potential. 
We have derived all the equations of motion implied by our pseudo-Lagrangian.
One may verify that these are compatible with the $E_8$ ExFT equations and we have shown this explicitly for the scalar equation in Appendix~\ref{app:scaleom match}.

The second, minimal, formulation of $E_9$ exceptional field theory was given in Section~\ref{sec:E9EFT}. It involves finitely many fields (in infinite-dimensional representations of $E_9$) and can be obtained from the Virasoro-extended formulation by gauge-fixing all the negative Virasoro fields to zero, including the field $\tilde{\rho}$ associated to $L_{-1}$.  The pseudo-Lagrangian takes a slightly more conventional form, as the sum of a kinetic, a topological and a potential term and is defined for an arbitrary unimodular metric $\tilde{g}_{\mu\nu}$. We have defined the external diffeomorphisms and have shown that they leave the duality equations \eqref{DualityEquation} and \eqref{Fduality} invariant, and fix all free coefficients in the pseudo-Lagrangian in a way familiar from $E_n$ ExFT for $n\leq 8$. We have verified that $E_8$ exceptional field theory, including the Einstein equation, follows correctly from the minimal formulation. In particular one obtains the $E_8$ exceptional field theory Lagrangian from the minimal pseudo-Lagrangian up to terms quadratic in the duality equations. All the expressions are finite in the minimal formulation, and no infinite series regularisation is involved. It is noteworthy that the field content and the duality equations in the minimal formulation are consistent with $E_{11}$ exceptional field theory \cite{Bossard:2017wxl,Bossard:2019ksx}, and could in principle be derived from it upon branching $E_{11}$ under its $SL(2) \times E_9$ subgroup.\footnote{The $E_{11}$ ExFT duality equations were given in \cite{Bossard:2017wxl,Bossard:2019ksx}, and a pseudo-Lagrangian, sharing some features with the $E_9$ minimal formulation, will be presented in \cite{Bossard:2021ebg}. A concrete conjecture for $E_{11}$ and an associated extended space-time in the context of maximal supergravity was first made in~\cite{West:2001as,West:2003fc} and further developed in~\cite{West:2014eza,Tumanov:2016abm}.}
Even though the field $\tilde\rho$ is gauged-fixed to zero in the minimal $E_9$ ExFT of Section~\ref{sec:E9EFT}, it seems plausible that an intermediate formulation including $\tilde\rho$ exists. First investigations show that one can write the duality equation at $\tilde{\rho}\ne 0$ provided one introduces one additional constrained vector field $B^\ord{2}$ in  such a `next-to-minimal' formulation. The field $\tilde\rho$ will be crucial for supersymmetrising $E_9$ ExFT, see~\cite{Nicolai:1987kz,Nicolai:2004nv,Kleinschmidt:2021agj} for work on supersymmetry in ungauged $D=2$ supergravity and fermionic representations of $K(\mf{e}_9)$.

Both the minimal and the Virasoro-extended formulation are based on pseudo-Lagrangians
 that are supplemented by a set of first-order self-duality equations for the scalar fields.
In order to render the model accessible to canonical tools, it may be useful to further extend this framework into a genuine Lagrangian formulation presumably upon sacrificing manifest
two-dimensional Lorentz invariance, along the lines of~\cite{Henneaux:1988gg}, see also \cite{Siegel:1983es,Floreanini:1987as} for earlier work on chiral scalars.

It should be stressed that the formalisms developed in this paper can be directly applied to construct extended field theories based on duality groups $\widehat G$ that are the affine extensions of any finite-dimensional, simple Lie group $G$.
This is simply achieved by exchanging $E_8$ in this paper by $G$, since we have not used detailed information about the structure constants $f^{AB}{}_C$ in~\eqref{eq:e9} anywhere.
The fact that the structure of generalised diffeomorphisms and the section constraint is the same for any simple Lie group $G$ was proved in 
\cite[Sec.~6]{Bossard:2017aae}.
For appropriate choices of $G$, our results define ``half-maximal'' exceptional field theories along the lines of \cite{Ciceri:2016hup,Hohm:2017wtr}, as well as extended field theories based on any symmetric space in three space-time dimensions that lifts at least to four dimensions, irrespective of supersymmetry~\cite{Breitenlohner:1987dg,Cremmer:1999du}.
For instance, taking $G=SL(2)$ we expect that the matching to $E_8$ ExFT discussed in Section~\ref{sec:E9EFT} would instead provide a matching to the $SL(2)$-covariant theory of \cite{Hohm:2013jma}.
Our results then provide a description of $D=4$ general relativity formally covariant under the Geroch group.

\medskip

There are a number of potential applications of $E_9$ exceptional field theory. As with other ExFTs, one use of $E_9$ exceptional field theory is to study in more detail gauged supergravity in $D=2$. Gaugings of $E_9$ have been investigated in~\cite{Samtleben:2007an},
but the general scalar potential is not known. It will be very interesting to construct this in an $E_9$ covariant form from $E_9$ ExFT using a generalised Scherk--Schwarz reduction similar to~\cite{Berman:2012uy,Musaev:2013rq,Hohm:2014qga}. This would lead for instance to theories with AdS${}_2$ vacua, generalising~\cite{Ortiz:2012ib}, and that could be of interest in the AdS${}_2$/CFT${}_1$ correspondence. 
The gaugings defined in \cite{Samtleben:2007an} always involve, beyond some subalgebra of $\hat{\mf{e}}_8$, the $L_{-1}$ generator,\footnote{This is denoted $L_{+1}$ in \cite{Samtleben:2007an}.} and possible non-Lagrangian gaugings involving the $L_0$ generator were discussed in \cite{Bossard:2017aae}.
The Virasoro-extended formulation of $E_9$ exceptional field theory suggests the existence of $D=2$ supergravities that gauge more general subalgebras of $\virm$ involving arbitrarily negative Virasoro generators.
It would certainly be interesting to study this possibility via a generalised Scherk--Schwarz ansatz. 
It would moreover be interesting to  study  generalised Scherk--Schwarz reductions with mild violations of the section constraint to describe for example massive type IIA consistent truncations. This would require to check the invariance under generalised diffeomorphisms in the presence of such a mild violation, or to define a deformed version of $E_9$ ExFT as in \cite{Ciceri:2016dmd} (see also \cite{Cassani:2016ncu}).  

Supergravity in $D=2$ is the natural habitat of exotic branes  that are characterised by having co-dimension at most two and thus sufficiently many isometries to be describable in two dimensions~\cite{deBoer:2010ud,deBoer:2012ma}. The truly exotic branes (with tension scaling like $g_{\rm s}^{-\alpha}$ for $\alpha>2$ in terms of the string coupling) are related by discrete duality transformations to (smeared versions of) the more conventional D- and NS-branes~\cite{Obers:1998fb,Eyras:1999at,Englert:2007qb,Cook:2008bi,Bergshoeff:2012ex,Bergshoeff:2015cba}. $E_9$ ExFT can then provide a framework for studying uplifts of these exotic objects and a unified description of their duality orbits~\cite{Bakhmatov:2017les,Fernandez-Melgarejo:2018yxq,Berman:2018okd,Fernandez-Melgarejo:2019mgd}.

Another interesting avenue of research might be to explore the fate of the known integrable structure of $D=2$ ungauged maximal supergravity~\cite{Nicolai:1987kz,Nicolai:1998gi} within ExFT. The integrable structure arises when the trivial solution $\langle\partial |=0$ to the section constraint is chosen, which corresponds to the toroidal reduction. Whether or not there are any remnants of this integrable structure for more general backgrounds with a non-trivial dependence in the internal coordinates is an open problem, and one may hope that $E_9$ ExFT could shed some light on this question.

\section*{Acknowledgements}
We would like to thank Ergin Sezgin for discussions and comments on the manuscript.
This project has received funding from the European Union’s Horizon 2020 research and innovation programme under the Marie Sk\l odowska-Curie grant agreement No 842991 and under the ERC Advanced Grant agreement No 740209.

\appendix

\section{On trivial parameters and generalised diffeomorphisms}
\label{app:triv}

In this appendix, we collect some details on trivial parameters for the $E_9$ generalised Lie derivative~\eqref{eq:GL} and its Virasoro extension~\eqref{eq:extGL}. These are by definition the non-vanishing pairs $\mathbbm{\Lambda}=(\ket{\Lambda},\Sigma)$,  respectively the non-vanishing $\mathbbm{\Lambda}=(\ket{\Lambda},\Sigma^\ord{k})$, that satisfy
\begin{align}
\mathcal{L}_{\mathbbm{\Lambda}} |V\rangle = 0 
\hspace{15mm} \text{for any $|V\rangle$.}
\end{align}
The existence of such trivial parameters is possible due to the section constraint~\eqref{eq:SC} as usual in exceptional geometry. We first define the (non-extended) trivial parameters in Appendix~\ref{TrivialPara}, show the closure of the algebra of generalised diffeomorphisms~\eqref{eq:deltaA} on the one-form gauge fields and describe the two-form gauge fields and their gauge transformations in Appendix~\ref{twoForm}. Closure of the Virasoro-extended generalised diffeomorphims~\eqref{eq:extGL} is explicitly checked in Appendix~\eqref{app:closureextdif}. Finally, in Appendix~\ref{app:exttriv}, we discuss the trivial parameters of these extended diffeomorphisms, which include the previous ones and an infinite set of additional trivial parameters.

\subsection{Trivial parameters}
\label{TrivialPara}

The commutator of two generalised Lie derivatives gives a generalised Lie derivative according to \eqref{eq:dfffclos}, but the commutator of two generalised diffeomorphisms acting on the Dorfman pair of gauge fields $\mathbbm{A}$ produces an additional one-form gauge transformation $\delta_{\mathbf{R}} \mathbbm{A}$, which is a trivial parameter for the Lie derivative, i.e.
\begin{align}
\label{eq:deltaone}
    \mathcal{L}_{\delta_{\mathbf{R}}  \mathbbm{A}} |V\rangle = 0 
\end{align}
for any $|V\rangle$. It is standard in gauged supergravity that the gauge algebra only closes on the one-form gauge field up to a one-form gauge transformation of the two-form gauge field in the tensor hierarchy \cite{deWit:2005hv}. The $E_n$ gauged supergravity two-forms are valued in the symmetric tensor product of the one-form gauge field representation that we denote $R(\Lambda_n)$ (in Bourbaki labelling for $n\leq 8$), 
in the orthogonal complement of the generic irreducible representation $R(2 \Lambda_n)$. Continuing this structure to $E_9$, this gives a two-form gauge field $C_{\mu\nu}^{M\hspace{-0.2mm}N}$ valued in\footnote{This also follows from a level decomposition of $E_{11}$ (T. Nutma, unpublished).}
\be  
|C_{(1}\rangle \otimes |C_{2)}\rangle \in  R(\Lambda_0)_{-1} \textrm{\large{$\vee$}} R(\Lambda_0)_{-1} \ominus   R(2\Lambda_0)_{-2}\; , \label{TensorHierarchy2} 
\ee
and the corresponding one-form gauge parameter $R^{MN}_\mu$. In exceptional field theory in dimension $D\le 4$, the gauge invariance of the theory requires also the inclusion of an additional two-form field with one constrained index in $\overline{R(\Lambda_n)}$, and the others in the antisymmetric tensor product $\bigwedge^{4-D} R(\Lambda_n)$ \cite{Hohm:2013uia,Hohm:2014fxa}. For $E_9$, this gives a two-form gauge field $C_{\mu\nu}^{PQ}{}_{\! M}$ valued in 
\be  
|C_{[1}\rangle \otimes |C_{2]}\rangle \langle \pi_C | \in  R(\Lambda_0)_{-1} \textrm{\large{$\wedge$}} R(\Lambda_0)_{-1} \otimes \overline{R(\Lambda_0)_{-1}} \; , 
\ee
and the corresponding one-form gauge parameter  $R_{\mu}^{PQ}{}_{\! M}$. In $E_9$ exceptional field theory, there is moreover an additional one-form gauge parameter with one constrained index and one-form gauge parameters with two constrained indices. We expect that $D$-forms with two constrained indices should similarly appear in the field strength of the constrained $(D-1)$-form in lower rank exceptional field theories.

One computes that the following Dorfman pairs of parameters 
$\delta_{\mathbf{R}} \mathbbm{A} =( \delta_{\mathbf{R}} |A\rangle  , \delta_{\mathbf{R}} B)$ are trivial when acting on any $|V\rangle$ according to~\eqref{eq:deltaone} 
\begin{subequations}
\label{eq:trivpars}
\begin{align}
\label{eq:triv1}
&\Big( \eta_{\alpha\beta} \langle \partial_R | T^\alpha |R_{(1}\rangle T^\beta |R_{2)}\rangle ,0\Big) \; , \\ 
\label{eq:triv2}
&\Big( \eta_{\alpha\beta} \langle \pi_R | T^\alpha |R_{[1}\rangle T^\beta |R_{2]}\rangle + 2 \langle \pi_R | R_{[1}\rangle |R_{2]}\rangle
,\nn\\
&\hspace{10mm} \frac12 \eta_{1\,\alpha\beta} T^\alpha |R_{[1}\rangle\left[ \langle \pi_R | T^\beta | R_{2]} \rangle \langle \partial_R| + \langle \partial_R | T^\beta | R_{2]} \rangle \langle \pi_R| \right]
\Big) %
\,,\\
\label{eq:triv3}
&  \Big( \eta_{-1\,\alpha\beta} \Tr[ T^\alpha R^+_2 ] T^\beta |R^+_1\rangle , \nn\\
&\hspace{10mm}
-\eta_{\alpha\beta} \Tr[ T^\alpha R^+_2 ] \,  T^\beta |R^+_1 \rangle \langle \partial_{R^+}|  + |R^+_1\rangle \langle \partial_{R^+} | R^+_2 - \langle \partial_{R^+} |R^+_1\rangle R^+_2 \Big) 
\,,\\
\label{eq:triv4}
& \Big( 0,\tfrac12
\eta_{1\,\alpha\beta}  \langle \pi_{U_{1}} | T^\beta | U_{(1}\rangle T^\alpha |U_{2)}\rangle \langle \pi_{U_{2}} | +\tfrac12\eta_{1\,\alpha\beta}  \langle \pi_{U_{2}} | T^\beta | U_{(1}\rangle T^\alpha |U_{2)}\rangle \langle \pi_{U_{1}} | \Big)\,,\\
\label{eq:triv5}
&\Big( 0, \eta_{\alpha\beta} \Tr[ T^\alpha X_{(1} ] \, T^\beta X_{2)}  + \Tr[ X_{(1}] \, X_{2)} - X_{(1} X_{2)} \Big)\, ,  \\
&\Big( 0, \eta_{-1\,\alpha\beta} \Tr[ T^\alpha W_{(1} ] \, T^\beta W_{2)} \Big) \, . \label{eq:triv6}
\end{align}
\end{subequations}
The notation here for one-form parameters is such that  
\begin{subequations}
\label{Index form of R}
\begin{align}
|R_{(1}\rangle \otimes |R_{2)}\rangle  & \,\Longleftrightarrow \, R^{PQ}= R^{(PQ)}\,,\\
|R_{[1}\rangle \otimes|R_{2]}\rangle\langle \pi_R|  & \,\Longleftrightarrow \, R^{PQ}{}_N = R^{[PQ]}{}_N \quad\text{constrained in $N$}\,,\\
|R^+_1\rangle \otimes R^+_2  & \,\Longleftrightarrow \,
R_+^{P;Q}{}_N \quad\text{constrained in $N$}\,,\\
|U_1\rangle \langle \pi_{U_1} |\otimes  |U_2\rangle  \langle \pi_{U_2}| & \,\Longleftrightarrow \,
U^{P}{}_{M}{}^{Q}{}_{N} = U^{(P}{}_{(M}{}^{Q)}{}_{N)}  \quad\text{constrained in both $M$ and $N$}\,,\\
X_1 \otimes X_2   & \,\Longleftrightarrow \, X^{P}{}_M{}^Q{}_N = X^{Q}{}_N{}^P{}_M\quad\text{constrained in both $M$ and $N$}\; , \\
W_1 \otimes W_2   & \,\Longleftrightarrow \, W^{P}{}_M{}^Q{}_N = W^{Q}{}_N{}^P{}_M\quad\text{constrained in both $M$ and $N$}
\end{align}
\end{subequations}
and in~\eqref{eq:triv5} the notation $X_1X_2$ is the product of operators, which corresponds to the trace $X^M{}_P{}^P{}_N$. The notation $| R_{(1} \rangle \otimes |R_{2)}\rangle$ denotes a single one-form, so that $|\partial_R \rangle$ acts on all of it, and similarly for $|R_{[1}\rangle\otimes  |R_{2]}\rangle \langle \pi_R|$ and $|R_1^+\rangle \otimes R_2^+$. The semi-colon for $R_+^{P;Q}{}_N$ is used to separate the two tensor factors.  

Because the parameter \eqref{eq:triv1} vanishes when $| R_{(1} \rangle \otimes |R_{2)}\rangle \in R(2\Lambda_0)_{-2}$, we can indeed interpret it as the one-form gauge parameter expected from the tensor hierarchy, consistently with \eqref{TensorHierarchy2}. For simplicity we never write the projection to the orthogonal complement of $R(2\Lambda_0)_{-2}$, but this component will always be projected out in the relevant expressions. 

In order to show that these parameters are trivial one has to use the section constraints as well as the invariance of the bilinear forms $\eta_{n\, \alpha\beta}$ under the action of $\hat{\mf{e}}_8$ and
\begin{align}
\eta_{n\, \alpha\beta} [ L_m , T^\alpha ] \otimes T^\beta +  \eta_{n\, \alpha\beta} T^\alpha \otimes  [ L_m , T^\beta ] &= \,(m{-}n) \eta_{n+m\, \alpha\beta} T^\alpha  \otimes T^\beta\nonumber\\
&\quad - \frac{c_\vir}{6} m (m^2{-}1) \delta_{m,-n} \dK \otimes \dK \; . \label{moveL}
\end{align}
To illustrate this, we look at the example where $\delta_{\mathbf{R}} \mathbbm{A}$ is the second parameter~\eqref{eq:triv2}. The transport and weight terms can be shown to vanish easily using~\eqref{eq:SC}, so that we are left with
\begin{align}
\mathcal{L}_{\delta_{\mathbf{R}} \mathbbm{A}}|V \rangle
 &= -\eta_{\alpha\beta} \eta_{\gamma\delta} \langle \pi_R|T^\gamma|R_{[1}\rangle \langle \partial_R |T^\alpha T^\delta | R_{2]}\rangle  T^\beta |V\rangle - 2 \eta_{\alpha\beta} \langle \pi_R |R_{[1}\rangle \langle \partial_R | T^\alpha | R_{2]} \rangle  T^\beta |V\rangle  \nn\\
&\hspace{10mm} + \tfrac12 \eta_{-1\,\alpha \beta} \eta_{1\,\gamma\delta}\bigl(  \langle \partial_R | T^\gamma|R_{[1} \rangle \langle \pi_R | T^\alpha T^\delta |R_{2]} \rangle   + \langle \pi_R | T^\gamma|R_{[1}\rangle \langle \partial_R | T^\alpha T^\delta |R_{2]} \rangle  \bigr) T^\beta |V\rangle\nn \\[1mm]
 &= -\tfrac12 \eta_{\alpha\beta} \eta_{\gamma\delta} \bigl(  \langle \pi_R|T^\gamma|R_{[1}\rangle \langle \partial_R | [ T^\alpha ,T^\delta ]  | R_{2]} \rangle- \langle \pi_R|[  T^\alpha, T^\gamma ] |R_{[1}\rangle \langle \partial_R | T^\delta | R_{2]} \rangle   \bigr)  T^\beta |V\rangle  \nn\\
&\hspace{20mm} -  \eta_{\alpha\beta} \bigl(  \langle  \pi_R |R_{[1}\rangle \langle \partial_R | T^\alpha | R_{2]}  \rangle   +  \langle \partial_R | R_{[1} \rangle \langle \pi_R |T^\alpha |R_{2]} \rangle \bigr) T^\beta |V\rangle  \nn\\
&\hspace{5mm} + \tfrac12 \eta_{-1\,\alpha \beta} \eta_{1\,\gamma\delta}  \bigl( \langle \partial_R | T^\gamma|R_{[1}\rangle \langle \pi_R | [ T^\alpha , T^\delta ]  |R_{2]} \rangle + \langle \pi_R |T^\gamma  |R_{[1}\rangle \langle \partial_R | [ T^\alpha, T^\delta ]  |R_{2]}  \rangle  \bigr) T^\beta |V\rangle \nn \\
&=0\,,
\end{align}
where in the second step we have used~\eqref{eq:SC} and~\eqref{eq:SC2} to write commutators as well as invariance of $\eta_{\gamma\delta}$ to split the first term into two. In the last step we have used the following identity~\cite[Eq.~(2.26)]{Bossard:2017aae} for any two operators $X$ and $Y$ 
\bea && \eta_{n\, \alpha\beta} \eta_{m-n\, \gamma\delta} \Tr[ T^\gamma X] \Tr\bigl[ [ T^\alpha , T^\delta ] Y \bigr] T^\beta- \eta_{p\, \alpha\beta} \eta_{m-p\, \gamma\delta} \Tr[ T^\gamma X] \Tr\bigl[ [ T^\alpha , T^\delta ] Y \bigr] T^\beta \CR
&=&  (n-p) \eta_{m\, \alpha\beta} \bigl( \Tr[ T^\alpha X] \Tr[  Y]  -  \Tr[  X] \Tr[  T^\alpha Y] \bigr) T^\beta -  (n-p) \eta_{m\, \alpha\beta}  \Tr[ T^\alpha X] \Tr[ T^\beta Y]  \dK \CR
&& \quad + \frac{c_\vir}{12} \bigl( n(n^2-1) - p (p^2-1) \bigr) \delta_{m,0} \Tr[ X] \Tr[ Y] \dK \; , 
\eea
with  $m=n=0$ and $p=-1$.

We will now show that these trivial parameters are all the ones we need to close the algebra of generalised diffeomorphisms and define the field strength. In particular we can use these trivial parameters to verify~\eqref{eq:Dorf1}. For this purpose we define the sextuplet of one-form gauge parameters $\mathbf{R}_\mu$ as
\begin{multline}  \mathbf{R} =  \Bigl(  |R_{(1}\rangle \otimes  |R_{2)}\rangle \; , \ 
|R_{[1}\rangle \otimes |R_{2]}\rangle\langle \pi_R| \; , \ 
|R^+_1\rangle \otimes R^+_2 \; , \\ \ |U_{(1}\rangle \langle \pi_{U_1} |\otimes  |U_{2)}\rangle  \langle \pi_{U_2} | \; , \  X_{(1} \otimes X_{2)}  \; , \ W_{(1} \otimes W_{2)}   \Bigr) \; . \label{Rsextuplet} \end{multline}
The trivial parameters \eqref{eq:trivpars} can be written with the linear map $\varpi$ that maps a sextuplet of parameters $ \mathbf{R}$ to a Dorfman doublet as follows 
\bea \varpi \mathbf{R}  \label{trivialvarpi}
\hspace{-2mm} &=& \hspace{-2mm} \Big( \eta_{\alpha\beta} \langle \partial_R | T^\alpha |R_{(1}\rangle T^\beta |R_{2)}\rangle   +  \eta_{\alpha\beta} \langle \pi_R | T^\alpha |R_{[1}\rangle T^\beta |R_{2]}\rangle + 2 \langle \pi_R | R_{[1}\rangle |R_{2]}\rangle \\
&& \quad  +\eta_{-1\,\alpha\beta} \Tr[ T^\alpha R^+_2 ] T^\beta |R^+_1\rangle \ , \CR
&& \qquad    \tfrac12  \eta_{1\,\alpha\beta} T^\alpha |R_{[1}\rangle\left[ \langle \pi_R | T^\beta | R_{2]} \rangle \langle \partial_R| + \langle \partial_R | T^\beta | R_{2]} \rangle \langle \pi_R| \right] \CR
&& \quad -\eta_{\alpha\beta} \Tr[ T^\alpha R^+_2 ] \,  T^\beta |R^+_1 \rangle \langle \partial_{R^+}|  + |R^+_1\rangle \langle \partial_{R^+} | R^+_2 - \langle \partial_{R^+} |R^+_1\rangle R^+_2  \CR
&& \qquad + \tfrac12
\eta_{1\,\alpha\beta}  \langle \pi_{U_{1}} | T^\beta | U_{(1}\rangle T^\alpha |U_{2)}\rangle \langle \pi_{U_{2}} | +\tfrac12\eta_{1\,\alpha\beta}  \langle \pi_{U_{2}} | T^\beta | U_{(1}\rangle T^\alpha |U_{2)}\rangle \langle \pi_{U_{1}} |  \CR
&& \qquad \quad +\eta_{\alpha\beta} \Tr[ T^\alpha X_{(1} ] \, T^\beta X_{2)}  + \Tr[ X_{(1}] \, X_{2)} - X_{(1} X_{2)} 
+ \eta_{-1\,\alpha\beta} \Tr[ T^\alpha W_{(1} ] \, T^\beta W_{2)} \Bigr) \; . \nn \eea
By definition we have that $\cL_{\varpi \mathbf{R} }$ vanishes on any field, and one checks moreover that 
\be \varpi \mathbf{R}  \circ \mathbbm{\Lambda} = 0 \label{wR0}\ee 
for any parameter  $\mathbbm{\Lambda}$. One also computes that 
\begin{align} & \; \Bigl(  \langle \partial_A | L_n | A\rangle +  \Tr[ L_{n-1} B] \Bigr)\Big|_{\mathbbm{A} = \varpi \mathbf{R} }  \CR
=&\; \frac{n}{2} \eta_{n\,  \alpha\beta} \langle \partial_R | T^\alpha | R_{(1}  \rangle \langle \partial_R | T^\beta | R_{2)}  \rangle +  \frac{n}{2} \eta_{n\,  \alpha\beta} \langle \pi_R | T^\alpha | R_{[1} \rangle \langle \partial_R | T^\beta | R_{2]} \rangle\CR
&\;  + \eta_{n-1\,  \alpha\beta} \langle \partial_{R^+} | T^\alpha | R^+_1 \rangle\Tr[   T^\beta R^+_2 ] + \frac{n-2}{2} \eta_{n \, \alpha\beta} \langle \pi_1 | T^\alpha | U_{(1} \rangle \langle \pi_2 | T^\beta | U_{2)}  \rangle \nn\\
& \; + \frac{n-1}{2} \eta_{n-1\, \alpha\beta} \Tr[ T^\alpha X_{(1}] \Tr[ T^\beta X_{2)}] + \frac{n}{2} \eta_{n-2\, \alpha\beta} \Tr[ T^\alpha W_{(1}] \Tr[ T^\beta W_{2)}] \; , \label{Ahat triv1} \end{align} 
which does not vanish for $n\ge 1$. For $n=1$ this term simplifies to 
\be  \Bigl(  \langle \partial_A | L_1 | A\rangle +  \Tr[ L_{0} B] \Bigr)\Big|_{\mathbbm{A} = \varpi \mathbf{R} }   =  \tfrac{1}{2} \eta_{1\,  \alpha\beta} \langle \pi_R | T^\alpha | R_{[1} \rangle \langle \partial_R | T^\beta | R_{2]} \rangle + \eta_{\alpha\beta} \langle \partial_{R^+} | T^\alpha | R^+_1 \rangle\Tr[   T^\beta R^+_2 ] \label{Ahat triv2} \ee
and is a total derivative. For $n\ge 2$ it is not a total derivative. This implies for example that $\hbm{F}^\upgamma_m$ in \eqref{completed shifted CM eq J} depends exlicitly on the two-form fields for $m\ge 2$, so that only $\mf{X}^\ord{1}$ defines an appropriate topological term.

Since the map~\eqref{trivialvarpi} acts only on the internal indices, it applies regardless of whether $\mathbf{R}$ are space-time $p$-forms or scalars. To prove~\eqref{eq:Dorf1} we now
define the bilinear map $\upiota$ that gives for any two  Dorfman doublets $\mathbbm{\Lambda}_1$ and $\mathbbm{\Lambda}_2$ the sextuplet of  parameters
\begin{multline} \upiota (  \mathbbm{\Lambda}_1,\mathbbm{\Lambda}_2) =  \Bigl( - \tfrac12 |\Lambda_{(1} \rangle \otimes | \Lambda_{2)}\rangle \ , \ - \tfrac12  |\Lambda_{[1}\rangle  \otimes |\Lambda_{2]}\rangle \langle \partial_{\Lambda_1}-\partial_{\Lambda_2} | \ , \ - |\Lambda_2\rangle \otimes \Sigma_1 \ , \\
 \ |\Lambda_{(1}\rangle \langle \partial_{\Lambda_1} | \otimes |\Lambda_{2)}\rangle \langle \partial_{\Lambda_1} |  \ , \ - \Sigma_1 \otimes |\Lambda_2 \rangle \langle \partial_{\Lambda_2} | -  |\Lambda_2 \rangle \langle \partial_{\Lambda_2} |  \otimes  \Sigma_1 \ , \ - \Sigma_{(1} \otimes \Sigma_{2)} \Bigr) \; . \end{multline}
With this definition one checks that 
\be    \mathbbm{\Lambda}_1 \circ \mathbbm{\Lambda}_2  - \lb \mathbbm{\Lambda}_1,\mathbbm{\Lambda}_2\rb_E = \varpi \upiota(   \mathbbm{\Lambda}_1 , \mathbbm{\Lambda}_2 ) \; , \label{Trivial}  \ee
which proves~\eqref{eq:Dorf1} by taking respectively the antisymmetric and the symmetric components in $\mathbbm{\Lambda}_1$ and $\mathbbm{\Lambda}_2$. More generally, one can show 
\begin{align}
&(   \mathbbm{\Lambda}_{(1} \circ \mathbbm{\Lambda}_{2)}  ) \circ  \mathbbm{\Lambda}_3 =0\,,\\
&2\,\mathbbm{\Lambda}_{[1}\circ(\mathbbm{\Lambda}_{2]}\circ\mathbbm{\Lambda}_3)=[\mathbbm{\Lambda}_1,\mathbbm{\Lambda}_2]_D\circ \mathbbm{\Lambda}_3\,,
\end{align}
where the first equation follows in fact directly from~\eqref{wR0} and~\eqref{Trivial}. These relations correspond respectively to the triviality of the symmetric bracket $\{\mathbbm{\Lambda}_1,\mathbbm{\Lambda}_2\}=\mathbbm{\Lambda}_{(1}\circ\mathbbm{\Lambda}_{2)}$ and to the closure of the Dorfman product according to the antisymmetric bracket $[\mathbbm{\Lambda}_1,\mathbbm{\Lambda}_2]_D=\mathbbm{\Lambda}_{[1}\circ\mathbbm{\Lambda}_{2]}$. As argued in Section~\ref{sec:gendiffeo}, they together imply the Leibniz identity~\eqref{eq:Leibniz}.

Using the Leibniz identity and \eqref{Trivial}, one then computes that the gauge transformation 
\be 
\delta_{\mathbbm{\Lambda}} \mathbbm{A} = \dd \mathbbm{\Lambda} + \mathbbm{\Lambda} \circ \mathbbm{A} \ee
closes as
\be 
\delta_{\mathbbm{\Lambda}_2} \delta_{\mathbbm{\Lambda}_1} \mathbbm{A}-  \delta_{\mathbbm{\Lambda}_1} \delta_{\mathbbm{\Lambda}_2} \mathbbm{A}  = \delta_{[\mathbbm{\Lambda}_1,\mathbbm{\Lambda}_2]_E} \mathbbm{A} + \varpi \bigl(  \upiota(\mathbbm{\Lambda}_1,\dd \mathbbm{\Lambda}_2) - \upiota(\mathbbm{\Lambda}_2,\dd \mathbbm{\Lambda}_1)  \bigr) \; .  
\ee
In summary, we write the one-form gauge transformation of the gauge fields as
\be 
\delta_{\mathbf{R}} \mathbbm{A} = \varpi \mathbf{R}\; ,
\ee
such that 
\be 
\delta_{\mathbbm{\Lambda}_2} \delta_{\mathbbm{\Lambda}_1} \mathbbm{A}-  \delta_{\mathbbm{\Lambda}_1} \delta_{\mathbbm{\Lambda}_2} \mathbbm{A}  = \delta_{[\mathbbm{\Lambda}_1,\mathbbm{\Lambda}_2]_E} \mathbbm{A} + \delta_{ \mathbf{R}_{12}} \mathbbm{A}   \; , 
\ee
with 
\be 
\mathbf{R}_{12} =  \upiota(\mathbbm{\Lambda}_1,\dd \mathbbm{\Lambda}_2) - \upiota(\mathbbm{\Lambda}_2,\dd \mathbbm{\Lambda}_1) \; . 
\ee

 Note that the Leibniz property is only satisfied up to a trivial parameter in the Virasoro-extended formulation introduced in Section~\ref{sec:extgauge}. This is nonetheless sufficient for the  algebra of generalised diffeomorphisms to close on the gauge field $\mathbbm{A}$. In this case the same construction applies, but $\mathbf{R}_{12}$ is modified by the corresponding violation of the Leibniz identity.

\subsection{Covariant field strengths and two-forms}
\label{twoForm}

The $E_9$ field strength \eqref{eq:fieldstrengths} depends on a sextuplet of two-form fields $ \mathbf{C} $ defined as in \eqref{Rsextuplet} as
\begin{multline} \mathbf{C} =  \Bigl(  |C_{(1}\rangle \otimes |C_{2)}\rangle \; , \ 
|C_{[1}\rangle \otimes |C_{2]}\rangle\langle \pi_C| \; , \ 
|C^+_1\rangle \otimes C^+_2 \; , \\
\ |C^-_{(1}\rangle \langle \pi_{1} |\otimes  |C^-_{2)}\rangle  \langle \pi_{2} | \; , \  C_{(1} \otimes C_{2)}  \; , \ C^+_{(1} \otimes C^+_{2)}   \Bigr) \; , \label{Csextuplet} \end{multline}
with the same symmetries as for $ \mathbf{R}$ in \eqref{Index form of R}. The two-form $C^{M\hspace{-0.3mm} N}$ is the one expected in the tensor hierarchy according to \eqref{TensorHierarchy2}. The two-forms $C^{M\hspace{-0.3mm} N}{}_{\hspace{-0.5mm} P} $
and $C_+^{M;N}{}_{\hspace{-0.5mm} P}$ carry one constrained index, and are similar to the two-forms introduced in \cite{Hohm:2013uia,Hohm:2014fxa} that extend the tensor hierarchy  in exceptional field theory. The two-forms $C_-^M{}_P{}^N{}_Q$, $C^M{}_P{}^N{}_Q$  and $C_+^M{}_P{}^N{}_Q$ carry two constrained indices $P$ and $Q$, and appear in the field strength $\mathcal{G}$ of the one-form $B$,  but not in $|\mathcal{F}\rangle $. 

Using \eqref{Trivial}, one can write the field strength \eqref{eq:fieldstrengths} as
\be  \mathbbm{F} = \dd  \mathbbm{A}- \tfrac12  \lb  \mathbbm{A},  \mathbbm{A} \rb_E + \varpi  \mathbf{C} =  \dd  \mathbbm{A}- \tfrac12  \mathbbm{A} \circ   \mathbbm{A}  + \varpi  \bigl( \mathbf{C} + \tfrac12 \upiota( \mathbbm{A},  \mathbbm{A} ) \bigr)  \; ,  \ee
which is convenient to prove \eqref{eq:fsvar}. Using Leibniz~\eqref{eq:Leibniz} and \eqref{Trivial} one computes the gauge transformation of the field strength
\begin{multline} \delta_{\mathbbm{\Lambda}} \mathbbm{F}  =\mathbbm{\Lambda} \circ   \mathbbm{F}\\  + \varpi \Bigl( \delta_{\mathbbm{\Lambda}} \mathbf{C} +  \upiota( d \mathbbm{\Lambda}  ,   \mathbbm{A} ) + \tfrac12 \upiota(  \mathbbm{\Lambda} \circ \mathbbm{A}, \mathbbm{A}) + \tfrac12 \upiota( \mathbbm{A}, \mathbbm{\Lambda} \circ \mathbbm{A})-\upiota\bigl( \mathbbm{\Lambda}  , \varpi( \mathbf{C} + \tfrac12 \upiota( \mathbbm{A} ,\mathbbm{A} ))\bigr) -  \upiota\bigl(\varpi( \mathbf{C} + \tfrac12 \upiota( \mathbbm{A} ,\mathbbm{A} )), \mathbbm{\Lambda}  \bigr)  \Bigr) \; . \end{multline}
Therefore there exists a gauge transformation of $\mathbf{C}$ such that \eqref{eq:fsvar} is satisfied. This gauge transformation is only defined up to an element in the kernel of $\varpi$. It is convenient to chose a particular element in the kernel to simplify the gauge transformation, so we define 
\begin{multline} \delta_\mathbbm{\Lambda}  \mathbf{C}  = \upiota\bigl( \mathbbm{\Lambda}  , \varpi( \mathbf{C} + \tfrac12 \upiota( \mathbbm{A} ,\mathbbm{A} ))\bigr) +  \upiota\bigl(\varpi( \mathbf{C} + \tfrac12 \upiota( \mathbbm{A} ,\mathbbm{A} )), \mathbbm{\Lambda}  \bigr) \\ - \upiota( \dd \mathbbm{\Lambda}  ,   \mathbbm{A} ) - \tfrac12 \upiota(  \mathbbm{\Lambda} \circ \mathbbm{A}, \mathbbm{A}) - \tfrac12 \upiota( \mathbbm{A}, \mathbbm{\Lambda} \circ \mathbbm{A}) 
- \tfrac12 \kappa( \mathbbm{\Lambda}, \mathbbm{A} ,\mathbbm{A}  )  \; ,\end{multline}
where $\varpi \kappa( \mathbbm{\Lambda}, \mathbbm{A} ,\mathbbm{A}  ) =0$ and we choose
\bea  
\kappa( \mathbbm{\Lambda}, \mathbbm{A} ,\mathbbm{A}  ) \hspace{-2mm} &=&\hspace{-2mm}  \biggl(\quad  |\Lambda\rangle \otimes \eta_{-1\, \alpha\beta} \Tr[ T^\alpha B] T^\beta |A\rangle + \eta_{-1\, \alpha\beta} \Tr[ T^\alpha B] T^\beta |A\rangle \otimes   |\Lambda\rangle  \; , \CR
&& \quad \Bigl(   |\Lambda\rangle \otimes \eta_{-1\, \alpha\beta} \Tr[ T^\alpha B] T^\beta |A\rangle - \eta_{-1\, \alpha\beta} \Tr[ T^\alpha B] T^\beta |A\rangle \otimes   |\Lambda\rangle   \Bigr) \langle \partial_\Lambda - \partial_A - \partial_B | \; , \CR
&&\quad  - |\Lambda\rangle \otimes \Bigl( \Tr[ T_\alpha B ] T^\alpha |A\rangle \langle \partial_B | + |A\rangle \langle \partial_B | B - \langle \partial_B | A\rangle B \CR
&& \hspace{30mm} + \langle \partial_A | T_\alpha | A\rangle T^\alpha B + B |A\rangle \langle \partial_A | - \Tr[ B] |A\rangle \langle \partial_A | \Bigr) \CR
&&\hspace{10mm} - \cL_{ \mathbbm{\Lambda} } ( |A\rangle \otimes B )  + \eta_{-1\, \alpha\beta} \Tr[ T^\alpha B ] T^\beta |A\rangle \otimes \Sigma - \Tr[\Sigma] |A^\prime \rangle \otimes |A\rangle \langle \partial_A |  \CR
&&\hspace{10mm} -  \tfrac12 |\Lambda \rangle \otimes \eta_{1\, \alpha\beta} \Bigl( \langle \partial_A | T^\alpha | A\rangle T^\beta |A^\prime \rangle \langle \partial_A |- \langle \partial_A | T^\alpha | A^\prime \rangle T^\beta |A \rangle \langle \partial_A | \Bigr)\; ,  \CR
&& \hspace{30mm} \dots\;  ,\  \dots\;  ,\  \dots \  \biggr) \; ,
\eea
where we have not computed the expression of the three doubly constrained parameters because they are not needed anywhere.

The homogeneous part in $\mathbf{C}$ is not the Lie derivative of $\mathbf{C}$, but mixes the various components of $\mathbf{C}$. We define  the inhomogeneous  variation of $\mathbf{C}$
\be 
\widetilde\Delta_{\mathbbm{\Lambda} } \mathbf{C} =  \delta_\mathbbm{\Lambda}  \mathbf{C}  -  \upiota\bigl( \mathbbm{\Lambda}  , \varpi  \mathbf{C} \bigr) -  \upiota\bigl(\varpi \mathbf{C}, \mathbbm{\Lambda}  \bigr) \; \; , \label{InhomoDelta}
\ee
that defines the piece depending explicitly on the vector fields $\mathbbm{A}$. One finds 
\begin{align} \label{2formGaugeTransf} 
  \widetilde\Delta_{\mathbbm{\Lambda} } |C_{(1}\rangle \otimes |C_{2)} \rangle   &= \tfrac{1}{4} \bigl(  \dd |\Lambda\rangle \otimes |A\rangle \rangle - |A  \rangle \otimes  \dd |\Lambda\rangle \bigr) \; , \\
 \widetilde\Delta_{\mathbbm{\Lambda} }  |C_{[1} \rangle \otimes |C_{2]} \rangle\otimes \langle \pi_C|    &=  \tfrac{1}{4} \bigl( \dd |\Lambda\rangle \otimes |A \rangle + |A \rangle \otimes  \dd |\Lambda\rangle \bigr)\otimes \langle  \partial_\Lambda - \partial_A | \; ,  \nn\\
 \widetilde\Delta_{\mathbbm{\Lambda} }   |C^+_{1}  \rangle \otimes C_2^+ &= -\tfrac12  \eta_{1\, \alpha\beta} \langle \partial_\Lambda | T^\alpha | A \rangle |A \rangle \otimes T^\beta |\Lambda\rangle \langle \partial_\Lambda | \nn\\
 & \hspace{5mm} - |A\rangle \otimes \dd  \Sigma - \tfrac12  \eta_{\alpha\beta} \Tr[ T^\alpha \Sigma ] |A\rangle \otimes T^\beta |A\rangle \langle \partial_\Sigma | \CR
 & \hspace{5mm} + \tfrac12 |A\rangle \otimes |A\rangle  \langle \partial_\Sigma | \Sigma  + \tfrac12   \Tr[ \Sigma]  |A^\prime \rangle \otimes |A\rangle  \langle \partial_A |   \; . \nn
\end{align}
Note nonetheless that this is {\it not} the gauge transformation that is relevant in minimal $E_9$ exceptional field theory, because the invariance of the duality equation \eqref{Fduality} requires $|\mathcal{F}\rangle $ to transform as the internal current, and the gauge transformation of $   |C^+_{1}  \rangle \otimes C_2^+ $ in exceptional field theory includes an additional term  \eqref{CgaugeTransformation}.\footnote{We only introduced this equation in the minimal formulation, but invariance under conformal diffeomorphisms would require to include this equation in the Virasoro-extended formulation as well.}


\subsection{Closure of extended generalised diffeomorphisms}
\label{app:closureextdif}

The Virasoro-extended generalised diffeomorphisms \eqref{eq:extGL} close according to the extended E-bracket \eqref{eq:Ebracketex}. In order to show this, we only need to consider the commutators $[\cL_{(\Lambda_1,0)},\cL_{(0,\Sigma^{\,\smash{\ord{k}}}_2)}]$ and $[\cL_{(0,\Sigma^{\,\smash{\ord{p}}}_1)},\cL_{(0,\Sigma^{\,\smash{\ord{k}}}_2)}]$, where $\Sigma^\ord{p}_1$ and $\Sigma^\ord{k}_2$ denote two complete sets of extended gauge parameters. Closure of the $\ket{\Lambda}$-diffeomorphisms onto themselves follows directly from the closure of the unextended generalised diffeomorphisms according to \eqref{eq:Ebracket}. 

Evaluating the first commutator on a generalised vector $\ket{V}$ in $R(\Lambda_0)_0$, we find using \eqref{eq:tparaex} and the section constraints,
\begin{align}
[&\cL_{(\Lambda_1,0)},\cL_{(0,\Sigma^{\,\smash{\ord{k}}}_2)}]\ket{V}\nonumber\\
&=-\sum\limits_{k=1}^\infty\eta_{-k\,\alpha\beta}\Tr\Big(T^\alpha\big(\langle\partial_{\Sigma}|\Lambda_1\rangle\Sigma_2^\ord{k}-\eta_{\gamma\delta}\langle\partial_\Lambda|T^\gamma|\Lambda_1\rangle[T^\delta,\Sigma_2^\ord{k}]+k\langle\partial_\Lambda|\Lambda_1\rangle\Sigma_2^\ord{k}\big)\Big)T^\beta\ket{V}\nonumber\\
&=\,\cL_{(0,\,\cL_{(\Lambda_1,0)}\Sigma^{\,\smash{\ord{k}}}_2)}\ket{V}\,.
\end{align}
This confirms the first term in the expression \eqref{eq:Ebracketex} of $\Sigma_{12}$. Let us now consider the second commutator
\begin{align}
[&\cL_{(0,\Sigma^{\,\smash{\ord{p}}}_1)},\cL_{(0,\Sigma^{\,\smash{\ord{k}}}_2)}]\ket{V}\nonumber\\
&=\sum\limits_{k,q=1}^{\infty}\eta_{-k\,\alpha\beta}\,\eta_{-q\,\gamma\delta}\,\Tr(\Sigma_1^\ord{k}T^\alpha)\Tr(\Sigma_2^\ord{q}T^\gamma)\,[T^\beta,T^\delta]\ket{V}\nonumber\\
&=\sum\limits_{k,q=1}^\infty\Big(-\eta_{-k\,\alpha\beta}\eta_{-q\,\gamma\delta}\,\Tr(\Sigma_1^\ord{k}T^\alpha)\Tr(\Sigma_2^\ord{q}T^\delta T^\beta)+(k-q)\eta_{-k-q\,\alpha\gamma}\Tr(\Sigma_1^\ord{k})\Tr(\Sigma_2^\ord{q}T^\alpha)\Big)T^\gamma\ket{V}\nonumber\\
&=-\sum\limits_{q=1}^\infty\eta_{-q\,\gamma\delta}\Tr(\widetilde\Sigma_{12}^\ord{q}T^\delta)\,T^\gamma\ket{V}=\,\cL_{(0,\widetilde\Sigma^{\smash{\ord{q}}}_{12})}\ket{V}\,,
\end{align}
where in the second line we used \eqref{moveL}, and with
\begin{align}
\widetilde\Sigma_{12}^\ord{q}=-\sum\limits_{k=1}^\infty\eta_{-k\,\alpha\beta}\Tr(\Sigma_{[1}^\ord{k}T^\alpha)\,T^\beta\Sigma_{2]}^\ord{q}-\sum\limits_{0<k<q}(2k-q)\Tr(\Sigma_{[1}^\ord{k})\Sigma_{2]}^\ord{q-k}\,.
\end{align}
In writing the above expression, we used the antisymmetry of the commutator in $\Sigma^{\ord{p}}_1$ and $\Sigma^\ord{k}_2$. This reproduces correctly the $\Sigma\Sigma$ terms of $\Sigma^\ord{k}_{12}$ in \eqref{eq:Ebracketex}.

\subsection{Extended trivial parameters}
\label{app:exttriv}
The Virasoro-extended generalised diffeomorphisms~\eqref{eq:extGL} include by construction the trivial parameters $(\delta_{\mathbf{R}}|A\rangle,\delta_{\mathbf{R}}B^\ord{1})$ defined in \eqref{eq:trivpars}, as well as an infinite sequence of parameters, with $ \delta_{\mathbf{R}}|A\rangle =0$,
\begin{multline} \delta_{\mathbf{R}} B^\ord{1} = \sum_{k=2}^\infty \Bigl( \eta_{2-k\, \alpha\beta} \langle \pi_{(1}  | T^\alpha | U_2^\ord{k} \rangle  T^\beta |U_1^\ord{k} \rangle \langle \pi_{2)} | + \eta_{1-k\, \alpha\beta} \Tr[ T^\alpha X_2^\ord{k} ] T^\beta X_1^\ord{k}  \\
 + \eta_{-k\, \alpha\beta} \Tr[ T^\alpha W_2^\ord{k} ] T^\beta W_1^\ord{k} \Bigr) +\langle \pi_{(1} | U_2^\ord{2}\rangle |U_1^\ord{2} \rangle \langle \pi_{2)} |  -  \langle \pi_{(1} | U_1^\ord{2}\rangle |U_2^\ord{2} \rangle \langle \pi_{2)} | \; ,  \end{multline}
 and for $k\ge 2$
\begin{align} \delta_{\mathbf{R}} B^\ord{k}  &{=}\,  \eta_{1\, \alpha\beta} \langle \pi_{(1} | T^\alpha | U_1^\ord{k} \rangle T^\beta |U^\ord{k}_2\rangle \langle \pi_{2)} | + k \langle \pi_{(1} | U_2^\ord{k+1}\rangle |U_1^\ord{k+1} \rangle \langle \pi_{2)} |  - k \langle \pi_{(1} | U_1^\ord{k+1}\rangle |U_2^\ord{k+1} \rangle \langle \pi_{2)} | \CR
&\quad  + \Tr[ T_\alpha X_1^\ord{k} ] T^\alpha X_2^\ord{k} - (k-2) \Tr[ X_1^\ord{k} ] X_2^\ord{k} + (k-1) \Tr[ X_2^\ord{k} ] X_1^\ord{k}- X_2^\ord{k} X_1^\ord{k} \\
&\quad + (k-2) \Tr[ W_2^\ord{k-1} ] W_1^\ord{k-1} - (k-2) \Tr[ W_1^\ord{k-1} ] W_2^\ord{k-1} + \eta_{-1\, \alpha\beta} \Tr[ T^\alpha W_1^\ord{k} ] T^\beta W_2^\ord{k} \; . \nonumber
\end{align}
As in \eqref{Index form of R}, all bra covectors are constrained, but no additional symmetry is assumed 
\begin{align}
\label{trivialK}
|U^\ord{k}_1\rangle \langle \pi_{(1} |\otimes  |U_2^\ord{k} \rangle  \langle \pi_{2)}| & \,\Longleftrightarrow \,
U^{\ord{k}\, P}{}_{M}{}^{Q}{}_{N} = U^{\ord{k}\, P}{}_{(M}{}^{Q}{}_{N)}  \quad\text{constrained in both $M$ and $N$}\,,\CR
X^\ord{k}_1 \otimes X^\ord{k}_2   & \,\Longleftrightarrow \, X^{\ord{k}\, P}{}_M{}^Q{}_N \quad\text{constrained in both $M$ and $N$}\; , \CR
W^\ord{k}_1 \otimes W^\ord{k}_2   & \,\Longleftrightarrow \, W^{\ord{k}\, P}{}_M{}^Q{}_N \quad\text{constrained in both $M$ and $N$}\; . 
\end{align}
These constrained parameters are very redundant, and it is in principle sufficient to only consider the parameters $W^\ord{k}_1 \otimes W^\ord{k}_2$ and $X_1^\ord{2}\otimes X_2^\ord{2}$ to obtain a complete basis of trivial parameters. One can in particular eliminate the traceless component of all $ \delta_{\mathbf{R}}B^\ord{k}$ with $k\ge 3$ using trivial parameters  $W^\ord{k-1}_1 \otimes W^\ord{k-1}_2 = |0 \rangle \langle 0 | \otimes W^\ord{k-1}$. By construction the traces $\Tr[  \delta_{\mathbf{R}}B^\ord{k} ] $ cannot be eliminated, since they appear independently multiplying the Virasoro generator $L_{-1-k}$ in the Lie derivative~\eqref{eq:extGL}.  For any $k\ge 3$ one can indeed always find $W^\ord{n}$ such that
\be 
\delta_{{\bf R}} B^\ord{k} = - (k-2) W^\ord{k-1} + (k-2) \Tr[W^\ord{k-1}] \, |0\rangle \langle 0 |- L_{-1} W^\ord{k} +  \Tr[ \Sigma^\ord{k}  ]  \, |0\rangle \langle 0 |  \; . 
\ee
One can also eliminate the traceless component of $ \delta_{\mathbf{R}} B^\ord{2}$ using an appropriate trivial parameter $X_1^\ord{2}\otimes X_2^\ord{2}$, but not with an $E_8$ invariant ansatz as above for $k\ge 3$. If one uses the $E_8$ solution to the section constraint~\eqref{eq:secsol}, there is a gauge parameter
\be X_1^\ord{2}\otimes X_2^\ord{2} =  |0 \rangle \langle 0 | \otimes X^\ord{2}  + X^C{}_{B;A}T_{-1 C} |0\rangle \langle 0 | T_1^A  \otimes | 0 \rangle\langle 0 | T_{1}^B \ee
with $X^C{}_{B;A}$ satisfying \eqref{eq:E8SC} for its indices $A$ and $B$, such that\footnote{The only traceless $ \delta_{\mathbf{R}}B^\ord{2}$ that cannot be obtained from $ X^\ord{2} $ are such that $L_0  \big[\delta_{\mathbf{R}}B^\ord{2}\big] = 0 $ and $ \delta_{\mathbf{R}}B^\ord{2} |0\rangle = 0$, which gives  $ \delta_{\mathbf{R}}B^\ord{2} =   \delta_{\mathbf{R}}B^\ord{2}_A |0\rangle \langle 0 | T_1^A $. To prove that all constrained vector $ \delta_{\mathbf{R}}B^\ord{2}_A$ can be written as $X^{B}{}_{B;A}$ for a doubly constrained tensor $X^{C}{}_{B;A}$, one can use the $E_{7}$ solution to the constraint  $\eqref{eq:E8SC}$.}
\be  \delta_{\mathbf{R}}B^\ord{2} = - L_0  X^\ord{2}  - X^\ord{2} |0\rangle \langle 0 | +  \Tr[X^\ord{2}] \, |0\rangle \langle 0 |   - X^{B}{}_{B;A} |0\rangle \langle 0 | T_{1}^A - L_{-1} W^\ord{2}  +  \Tr[ \Sigma^\ord{2}  ]  \, |0\rangle \langle 0 |    \; . \ee
This proves, as noted in Section~\ref{sec:extgauge}, that any loop transformation generated by $\Sigma^\ord{k}$ can be reabsorbed into $\Sigma^\ord{1}$.
To write this systematically, we use \eqref{eta and shift ops} to rewrite \eqref{eq:balphaex} as
\begin{equation}\label{eq:sigmaetashift2}
\sbm\Lambda_\alpha =  \eta_{\alpha\beta}\bra{\partial_\Lambda} T^\beta \ket\Lambda
+ \sum_{k=1}^\infty \left(
    \eta_{-1\,\alpha\beta} \Tr\big[\Sigma^\ord{k} \cS_{1-k}(T^\beta)\big]
  - \delta_\alpha^{L_{-k}} \Tr(\Sigma^\ord{k})
\right)\,.
\end{equation}
The first term only contributes to the gauging of $\he8$ generators.
The fact that any loop transformation generated by $\Sigma^\ord{k}$ can be reabsorbed into $\Sigma^\ord{1}$ then means that we can implicitly define shift operators $\widehat\cS_{-k}$ acting on any $\Sigma$ parameter rather than on algebra-valued objects, with
\begin{equation}\label{formal shift ops}
\Tr\Big( \widehat\cS_{-k}(\Sigma)\, T^\alpha \Big)
=
\Tr\Big( \Sigma\ \cS_{-k}(T^\alpha) \Big)\,,\qquad
\forall\ k\in\mathds{N}\,,\quad
T^\alpha \in \hevirm\,.
\end{equation}
For instance, \eqref{eq:balphaex} may be rewritten as
\begin{equation}\label{balphaex with formal shiftop}
\sbm\Lambda_\alpha 
= 
  \eta_{\alpha\beta}\bra{\partial_\Lambda} T^\beta \ket\Lambda
+ \sum_{k=1}^\infty \left(
    \eta_{-1\,\alpha\beta} \Tr\big[ \widehat\cS_{1-k}(\Sigma^\ord{k}) T^\beta \big]
  - \delta_\alpha^{L_{-k}} \Tr(\Sigma^\ord{k}) 
  \right)\,.
\end{equation}
The operators $\widehat\cS_{-k}$ do not admit an explicit $E_9$ covariant expression.
They are defined only up to the addition of trivial parameters of the types $U$, $X$ and $W$ in \eqref{eq:triv4}--\eqref{eq:triv6} and \eqref{trivialK}  and we have used this freedom to make \eqref{formal shift ops} valid also for $T^\alpha = L_0$.

Using these formal shift operators, a basis for the extra trivial parameters appearing in the Virasoro-extended formalism is implicitly given by traceless $\delta_{\mathbf{R}}B^\ord{k}$ parameters satisfying  for fixed  $p\ge2$
\begin{equation}
\Tr(\delta_{\mathbf{R}}B^\ord{p})=0\,,\ \ 
\delta_{\mathbf{R}}B^\ord{1} = - \widehat\cS_{1-p}(\delta_{\mathbf{R}}B^{\ord p})\,,\quad 
\delta_{\mathbf{R}}B^\ord{n}=0\quad \text{for}\ n\notin \{ p,1\}\,.
\end{equation} 

\subsection{One-form gauge transformations in the minimal formulation}
\label{Closure}

In the minimal formulation the gauge transformations of the field $B$ are not \eqref{GaugeBused} and $\varpi\mathbf{R}$, but must instead be modified for the duality equation \eqref{DualityEquation} to be invariant. The $\Sigma$ gauge transformations of the fields that leave \eqref{DualityEquation} invariant are 
\begin{align} 
 \delta_{(0,\Sigma)}  \tilde{g}_{\mu\nu}  =& \ 0  \; ,  \\
 \delta_{(0,\Sigma)}  \cM  =& \   \eta_{-1\,  \alpha\beta} \Tr[ T^\beta \Sigma ]  \bigl( T^{\alpha \dagger} \cM + \cM T^\alpha \bigr)  \; ,  \nn\\
 \delta_{(0,\Sigma)} |A\rangle =& - \eta_{-1\, \alpha\beta} \Tr[ T^\alpha \Sigma ] T^\beta | A\rangle \nn\\
 \delta_{(0,\Sigma)} B =& \; \dd \Sigma + \eta_{\alpha\beta} \Tr[ T^\alpha \Sigma ] T^\beta | A\rangle \langle \partial_\Sigma | - | A\rangle \langle \partial_\Sigma | \Sigma + \eta_{-1\, \alpha\beta} \Tr[ T^\alpha \Sigma ] T^
\beta B \nn\\
& \qquad - \rho \star \widehat{\cS}_{-1} \bigl( \dd \Sigma + \eta_{\alpha\beta} \Tr[ T^\alpha \Sigma ] T^\beta |A\rangle \langle \partial_\Sigma | - |A\rangle \langle \partial_\Sigma | \Sigma \bigr) \nn\\
 \delta_{(0,\Sigma)} \chi =& \; \sum_n (n+1) \Tr[ T_n^A \Sigma ] J_A^n - 2 \rho \star \eta_{-1\, \alpha\beta} \Tr[ T^\alpha \Sigma] \langle \partial_\Sigma | T^\beta |A\rangle  \nn\\ 
&\hspace{4.5mm} -  \Tr\bigl[ (\rho^{-1}  L_{-1} +\rho^{-2} \cM^{-1} L_{-2} \cM) ( \dd \Sigma + \eta_{\alpha\beta} \Tr[ T^\alpha \Sigma ] T^\beta |A\rangle \langle \partial_\Sigma | - |A\rangle \langle \partial_\Sigma | \Sigma  ) \bigr] \; , \nn
\end{align}
where $\widehat{\cS}_{-1}$ is defined implicitly from \eqref{formal shift ops}. The ambiguity in the definition of $\widehat{\cS}_{-1}$ can be absorbed in a redefinition of the one-form gauge parameters $U,\; X$ and $W$. The one-form gauge transformations of $|A\rangle $, $B$ and $\chi$ are defined as
\begin{align} \delta_{\mathbf{R}} |A \rangle &= \langle \partial_R | T_\alpha | R_{(1} \rangle T^\alpha | R_{2)} \rangle+ \langle \pi_R | T_\alpha | R_{[1} \rangle T^\alpha | R_{2]} \rangle + 2 \langle \pi_R | R_{[1} \rangle | R_{2]}\rangle + \eta_{-1\alpha\beta} \Tr [ T^\alpha R^+_2] T^\beta |R_1 \rangle \nn\\
\delta_{\mathbf{R}} B &=  \tfrac12  \eta_{1\,\alpha\beta} T^\alpha |R_{[1}\rangle\left[ \langle \pi_R | T^\beta | R_{2]} \rangle \langle \partial_R| + \langle \partial_R | T^\beta | R_{2]} \rangle \langle \pi_R| \right] \\
& \quad -\eta_{\alpha\beta} \Tr[ T^\alpha R^+_2 ] \,  T^\beta |R^+_1 \rangle \langle \partial_{R^+}|  + |R^+_1\rangle \langle \partial_{R^+} | R^+_2 - \langle \partial_{R^+} |R^+_1\rangle R^+_2  \CR
& \quad - \rho \star \bigl( T_\alpha |R_{[1} \rangle \langle \pi_R | T^\alpha |R_{2]}\rangle + 2 |R_{[1} \rangle \langle \pi_R | R_{2]} \rangle \bigr) \langle \partial_R | + \rho \star \eta_{-1\, \alpha\beta}  \Tr[ T^\alpha R_2^+ ] T^\beta  |R_1^+ \rangle \langle \partial_{R^+} \! |  \nn\\
&\quad + \tfrac12
\eta_{1\,\alpha\beta}  \langle \pi_{U_{1}} | T^\beta | U_{(1}\rangle T^\alpha |U_{2)}\rangle \langle \pi_{U_{2}} | +\tfrac12\eta_{1\,\alpha\beta}  \langle \pi_{U_{2}} | T^\beta | U_{(1}\rangle T^\alpha |U_{2)}\rangle \langle \pi_{U_{1}} |  \CR
& \qquad \quad +\eta_{\alpha\beta} \Tr[ T^\alpha X_{(1} ] \, T^\beta X_{2)}  + \Tr[ X_{(1}] \, X_{2)} - X_{(1} X_{2)} 
+ \eta_{-1\,\alpha\beta} \Tr[ T^\alpha W_{(1} ] \, T^\beta W_{2)} \CR
\delta_{\mathbf{R}} \chi &= \eta_{\alpha\beta} \langle \pi_R | T^\alpha | R_{[1} \rangle \langle \partial_R | \cM^{-1} L_{-1} \cM T^\beta |R_{2]} \rangle + 2 \langle \pi_R | R_{[1}\rangle \langle \partial_R | \cM^{-1} L_{-1} \cM | R_{2]} \rangle\nn\\
& \quad + \rho \star \Bigl(   \eta_{\alpha\beta} \langle \pi_R | T^\alpha | R_{[1} \rangle \langle \partial_\Lambda | L_0 T^\beta |R_{2]} \rangle + 2 \langle \pi_R | R_{[1}\rangle \langle \partial_R | L_0 | R_{2]} \rangle\Bigr)  \nn\\
& \quad  + \eta_{-1\, \alpha \beta } \Tr[ T^\alpha R_2^+ ] \langle \partial_{R^+}\! | \cM^{-1} L_{-1} \cM T^\beta |R_1\rangle   + \rho \star  \eta_{-1\, \alpha \beta } \Tr[ T^\alpha R_2^+ ] \langle \partial_{R^+}\!  | L_0 T^\beta  |R_1 \rangle \; , \nn
\end{align} 
where the only modification of $\delta_{\mathbf{R}} B$ with respect to $\varpi \mathbf{R}$ are the terms involving $\star$. These additional terms are defined such that the duality equation \eqref{DualityEquation} is invariant under these one-form gauge transformations. One moreover checks that the pseudo-Lagrangian \eqref{PseudoLagrangian} is invariant under $\Sigma$ and $\mathbf{R}$ gauge transformations, up to a total derivative. 

\section{Shift operators and cocycles}
\label{app:shift}

We summarise here some details on the shift operators and cocycles encountered in the text and on their transformation properties.
First, we stress that our definition of $\cS_0$ in this paper projects out $\dK$, contrary to $\mathcal{S}_0$ in \cite{Bossard:2018utw} which acts as the identity:
\begin{equation}
\cS_0(\dK)^{\text{[here]}} = 0\,,\qquad 
\mathcal{S}_0(\dK)^{\text{\cite{Bossard:2018utw}}}=\dK\,,\qquad
\cS_k^{\text{[here]}}=\mathcal{S}_k^{\text{\cite{Bossard:2018utw}}}\ \forall\, k\neq0\,.
\end{equation}

For generic algebra elements $X,\,Z\in\hevir$, the shift operator $\cS_m$ defined in \eqref{eq:shift op def} has the commutation property
\begin{equation}
\cS_m\big([X,\,Z]\big) = 
[X,\,\cS_m(Z)] + m \sum_{n\in\mathds{Z}}X_{L_n}\cS_{m+n}(Z) + \omega^{\alpha\beta} X_\alpha \,\big(\cS_m(Z)\big)_\beta\; \dK\,,
\label{shiftgamma commutator}
\end{equation}
where
\begin{equation}
\omega^{\alpha\beta} X_\alpha Z_\beta = -[X,\,Z]\big|_\dK
= -\eta^{AB}\sum_{n\in\mathds{Z}} n \, X_A^n\,Z_B^{-n} 
  -\frac{c_{\vir}}{12} \sum_{n\in\mathds{Z}} (n^3-n) \,X_{n}\,Z_{{-n}}
\end{equation}
is the Lie algebra cocycle in any highest/lowest weight representation of $\he8$.
Notice that $\omega^{\alpha\beta}=-\omega^{\beta\alpha}$.
For a finite $\hE8\rtimes\Virm$ element $g$ we define
\begin{equation}
\omega^\alpha(g)\,\dK = g\cS_0(T^\alpha)g^{-1} - \cS_0(gT^\alpha g^{-1})\,,\qquad
g\in\hE8\rtimes\Virm\,,
\label{finite cocycle from 0 shift}
\end{equation}
which we may also write as $\omega^\alpha(g)=gT^\alpha g^{-1}\big|_\dK - \delta^\alpha_\dK \dK$.
This identity can also be used to define a cocycle for the generalised metric \eqref{genmetric extended}:
\begin{equation}
\omega^\alpha(\cM)\,\dK = \cM\,\cS_0(T^\alpha)\cM^{-1} - \cS_0(\cM\,T^\alpha \cM^{-1})\,.
\label{M cocycle}
\end{equation}
Using $\dK^\dagger=\dK$, one can easily show 
\begin{equation}
\omega^\alpha(\cM) (\cM^{-1} X^\dagger \cM)_\alpha = -\omega^\alpha(\cM) X_\alpha\,,
\end{equation}
which implies for instance $\omega^\alpha(\cM) \fJ(m)_\alpha = -\omega^\alpha(\cM)\fJ(-m)_\alpha$ and $\omega^\alpha(\cM)\fJ_\alpha=0$.
It is also useful to note that
\begin{equation}
\omega^{\alpha \dK}=0=\omega^\dK(\,\cdot\,)\,.
\label{cocycle K = 0}
\end{equation}

Any finite $\hE8\rtimes\Virm$ element can be decomposed as follows:
\begin{equation}
g = F \ell\,,\quad F\in \mathrm{Vir}^-\,,\ \ell\in \hE8\,.
\label{split g}
\end{equation}
Recall that $\cS_m$ in the spectral parameter representation acts as multiplication by $w^m$.
Then, taking into account $F^{-1}w = f(w)$ as in \eqref{Vir minus element}, it is natural to define a shift operator $\cS^f_m$ that acts as multiplication by $f(w)^m$.
This is obtained by conjugating $\cS_m$ with $F$, then projecting out any $\dK$ component generated by $\vir$ cocycles by means of $\cS_0$:
\begin{equation}
\cS^f_m(X) = \cS_0\big(  F^{-1} \cS_m(F\,X\,F^{-1})F   \big)
= \cS_0\big(  g^{-1} \cS_m(g\,X\,g^{-1})g   \big)\,.
\label{shift F operator}
\end{equation}
Then, we can write the finite transformation properties of the standard shift operator as
\begin{equation}
g^{-1} \, \cS_m(g X g^{-1}) \, g 
= \cS^f_m(X) - \omega^\alpha(g) \Big(\cS^f_m(X)\Big)_\alpha\,\dK\,.
\label{finite cocycle from S to S gamma}
\end{equation}

The $\rho$ and $\phi_m$ dependent shift operator $\cS^\upgamma_m$ is defined as in \eqref{shift F operator}:
\begin{equation}
\cS^\upgamma_m(X) = \cS_0\big(  \Gamma^{-1} \cS_m(\Gamma\,X\,\Gamma^{-1})\Gamma   \big)
= \cS_0\big(  \cV^{-1} \cS_m(\cV\,X\,\cV^{-1})\cV   \big)\,,
\label{shift upgamma operator}
\end{equation}
which reproduces \eqref{shift gamma def}.
We can commute $\cS^\upgamma_m$ with an $\hevir$ element to get
\begin{equation}
\cS^\upgamma_m\big([X,\,Z]\big) = 
[X,\,\cS^\upgamma_m(Z)] + m \sum_{n\in\mathds{Z}}(\Gamma X\Gamma^{-1})_{L_n}\cS^\upgamma_{m+n}(Z) + \omega^{\alpha\beta} X_\alpha \,\big(\cS^\upgamma_m(Z)\big)_\beta\; \dK\,.
\label{S gamma commutator}
\end{equation}
This is used for instance in computing the shifted Maurer--Cartan equation \eqref{shifted CM eq J}.

There are several other useful definitions and properties of the cocycles introduced in \cite{Bossard:2018utw} that are only valid for $g\in E_9$.
In this case the decomposition \eqref{split g} is rewritten as
\begin{equation}
g = \rho(g)^{-2L_0}\,\ell\,,
\end{equation}
where $\rho(g)$ is a constant element of the monoparametric subgroup generated by $L_0$ and is not to be confused with the scalar field $\rho$.
We can generalise \eqref{finite cocycle from 0 shift} to $\cS_k$ and define shifted cocycles
\begin{equation}
\omega_{-k}^\alpha(g) \dK = \rho(g)^{-2k}\,g \cS_k(T^\alpha)g^{-1} -\cS_k(g T^\alpha g^{-1})\,,\qquad
g\in E_9\,,
\label{shifted group cocycle E9}
\end{equation}
and we write $\omega_0^\alpha(g) = \omega^\alpha(g)$ for simplicity.
The shift can be moved from the cocycle to the object it contracts:
\begin{equation}
\omega_{-k}^\alpha(g) X_\alpha 
= 
\rho(g)^{-2k}\omega^\alpha(g)\big(\cS_k(X)\big)_\alpha\,,\quad X\in\hevir\,,\quad  g\in E_9\,.
\label{move shift from cocycle}
\end{equation}
This shows that expressions like $\omega^\alpha(g) \big(\cS^\upgamma_k(X)\big)_\alpha$ can be expanded as series of shifted cocycles if $g\in E_9$.
These shifted cocycles also appear in the conjugation of a Virasoro generator by a loop transformation:
\begin{equation}
g^{-1} L_k g 
= \rho(g)^{-2k}\, L_k - \omega_{-k}^\alpha(g) \eta_{\alpha\beta} T^\beta
= \rho(g)^{-2k}\, ( L_k - \omega^\alpha(g) \eta_{k\,\alpha\beta} T^\beta )
\,,\qquad g\in E_9\,.
\label{cocycle from Vir conjugation}
\end{equation}
In the second equality we moved the shift from the cocycle to $\eta_{\alpha\beta}$ and used \eqref{eta and shift ops}.
All the definitions and properties of the shifted cocycles also apply when we substitute $g\to\cM$ (and $\rho(\cM)=\rho$) provided we gauge-fix $\phi_m=0$ (including $\phi_1=\tilde\rho=0$) so that (formally) $\cM\in E_9$.

In \cite{Bossard:2018utw} we also defined a generalisation of $\omega_1^\alpha(\cM)$ to $\tilde\rho\neq0$, using the fact that in this case $\cM\in\hE8\rtimes SL(2)$ and the $SL(2)$ component acts on $w$ as fractional linear transformations:\footnote{Recall that we are using $\cS_0(\dK)=0$ as in the rest of this paper, which differs from \cite{Bossard:2018utw} where we were defining $\cS_0$ to be the identity. 
This slightly affects how we are writing $\Omega^\alpha(\cM)$ here.}
\begin{equation}
\Omega^\alpha(\cM)\dK =
\cM^{-1}\left( \rho^2 \sum_{n=0}^\infty \tilde\rho^n \, \cS_{1+n}(T^\alpha{}^\dagger)
  +\tilde\rho \cS_0(T^\alpha{}^\dagger) \right) \cM
-\cS_1(\cM^{-1} T^\alpha{}^\dagger \cM)\,.
\end{equation}
We will now prove that
\begin{equation}
\Omega^\alpha(\cM) \, \bra{\cJ_\alpha }
=
\rho^2\,\omega^\alpha(\cM) \, \bra{\cJ^-_\alpha} \ .
\end{equation}
To show this, we use \eqref{M cocycle} on the $\cS_0$ term so that we can use the first of \eqref{gfix shift op expansion}, and bring Hermitian conjugation out of the shift operators in the $\tilde\rho$ series so that we can also use the second of \eqref{gfix shift op expansion}.
This gives us
\begin{align}
  \cS^\upgamma_1(\cM^{-1}(T^\alpha)^\dagger\cM)
+ \frac{\tilde\rho}{\rho}\,\omega^\alpha(\cM) \,\dK
&=
  \cM^{-1} \Big( \cS^\upgamma_{-1}(T^\alpha)  \Big)^\dagger \cM 
- \frac1\rho\,\Omega^\alpha(\cM) \,\dK \,.
\end{align}
Contracting with  $\bra{\cJ_\alpha}$ and using
$
\bra{\cJ_\alpha}\otimes \cM^{-1} (T^\alpha)^\dag \cM 
=
\bra{\cJ_\alpha}\otimes T^\alpha
$, 
we find that the cocycle on the left-hand side does not contribute.
On the other hand, we already know from the computation in \eqref{shifted F M conj} that
\begin{equation}
\cS^\upgamma_1(\cM^{-1}X^\dagger\cM) = 
\cM^{-1} \Big( \cS^\upgamma_{-1}(X)  \Big)^\dagger \cM 
- \omega^\alpha(\cM) \big(\cS^\upgamma_{-1}(X) \big)_\alpha\,\dK\,,
\end{equation}
for any $X\in\hevir$.
Substituting $X_\alpha\to\bra{\cJ_\alpha}$, we find that the $\dK$ components of the last two expressions must coincide and using the definition \eqref{Jminus def} we conclude.

\section{Details on the Virasoro-extended formalism}\label{app:virdetails}

This appendix contains additional details for some of the calculations and aspects of the Virasoro-extended formulation of $E_9$ ExFT presented in Section~\ref{sec:extE9exft}.

\subsection{Field strength variations}
\label{app:fstrvar}

The variation of the field strengths under generalised diffeomorphisms was stated in \eqref{sbmF noncov var}--\eqref{mhbmF noncov var} that we prove here, beginning with the first equation.
The field strength $\mathbbm F=(\cF,\,\cG^\ord{k})$ transforms with the extended $\circ$ product \eqref{eq:DorfmanProductex} under generalised diffeomorphisms.
This means that their non-covariant variation \eqref{noncov var def} reads $\Delta_\mathbbm{\Lambda} \mathbbm F = \mathbbm\Lambda \circ\mathbbm F - \cL_\mathbbm{\Lambda}\mathbbm F$ and in particular $\Delta_\mathbbm{\Lambda}\ket\cF=0$.
In order to substitute into \eqref{Falpha}, we then compute
\begin{align}
\label{DELTA braket F}
\Delta_{\mathbbm\Lambda} \bra{\partial_\cF} T^\lambda \ket{\cF} =\ &
-\braket{\partial_\Lambda}{\Lambda}\bra{\partial_\Lambda}T^\lambda\ket\cF
-\sbm\Lambda_\alpha \bra{\partial_{\Lambda,\Sigma}} T^\lambda T^\alpha \ket\cF\,, \\[1ex]
\Delta_{\mathbbm\Lambda} \Tr(\cG^\ord{k} T^\lambda) =\ &
 \delta^k_1 \eta_{1\,\alpha\beta} \bra{\partial_\Lambda} [T^\lambda,\, T^\alpha] \ket\cF
  \bra{\partial_\Lambda} T^\beta \ket\Lambda 
+ \eta_{\alpha\beta} \Tr(\Sigma^\ord{k} T^\alpha ) \bra{\partial_\Sigma} [T^\lambda,\,T^\beta]\ket\cF  \nonumber\\[1ex]&
- k \Tr(\Sigma^\ord{k}) \bra{\partial_\Sigma}T^\lambda\ket\cF
+ (k-1)  \Tr(\Sigma^\ord{k}T^\lambda) \braket{\partial_\Sigma}\cF\,,
\label{DELTA Tr G}
\end{align}
where in the second equation we used the section constraint \eqref{eq:SC} to introduce the commutators in the first line.
Substituting into \eqref{Falpha}, the first equation is contracted with $\eta_{\delta\lambda} T^\delta$ and the second one with $\eta_{-k\,\delta\lambda}T^\delta$ and summed over $k$.
It is then useful to rewrite such expressions in terms of the (rescaled) level 2 coset generators acting on triple tensor products \cite{Bossard:2017aae}
\begin{equation}
\overset{12}{C}_m = -\eta_{m\,\alpha\beta}\ T^\alpha\!\otimes\! T^\beta\! \otimes \!\dK\,, \quad
\overset{13}{C}_m = -\eta_{m\,\alpha\delta}\ T^\alpha\! \otimes\! \dK \!\otimes\! T^\delta\,,\quad
\overset{23}{C}_m = -\eta_{m\,\beta\delta}\ \dK \! \otimes\! T^\beta \otimes\! T^\delta\,,
\label{C def}
\end{equation}
which satisfy several useful relations given in equations (2.24)--(2.26) of \cite{Bossard:2017aae}.
We can then write, using the notation $\Sigma^\ord{k} = \ket{\Sigma^\ord{k}}\bra\pi$ even if $\Sigma^\ord{k}$ is not a tensor product
\begin{align}
\eta_{-k\,\delta\lambda}\Delta_{\mathbbm\Lambda} \Tr(\cG^\ord{k} T^\lambda)\! \otimes\! T^\delta 
=\ &
  \delta_1^k \bra{\partial_\Lambda}\!\otimes\!\bra{\partial_\Lambda}
   \big[\overset{23}C_{-1}\,,\ \overset{12}C_1\big]\ket\Lambda\!\otimes\!\ket\cF
+ \bra\pi\!\otimes\!\bra{\partial_\Sigma} \big[\overset{23}C_{-k}\,,\ \overset{12}C_0 \big]
    \ket{\Sigma^\ord{k}}\!\otimes\!\ket\cF
\CR&
+ \bra\pi\!\otimes\!\bra{\partial_\Sigma}
    \Big( k\, \overset{23}C_{-k}  + (1-k) \overset{13}C_{-k} \Big)
    \ket{\Sigma^\ord{k}}\!\otimes\!\ket\cF
\end{align}
and we now use equations (2.24) and (2.26) of \cite{Bossard:2017aae} (which can be proved using \eqref{moveL}) to write
\begin{align}
\big[\overset{23}C_{-1}\,,\ \overset{12}C_1\big] 
&= 
  \overset{13}C_0-\overset{12}C_0-\overset{23}C_0 
+ \big[\overset{23}C_0\,,\ \overset{12}C_0\big]
\\
\big[\overset{23}C_{-k}\,,\ \overset{12}C_0\big] 
&= 
- k\,\overset{13}C_0 -k \,\overset{12}C_0 +k \,\overset{23}C_0 
+ \big[\overset{23}C_0\,,\ \overset{12}C_{-k}\big] \,.
\end{align}
It is then straightforward to arrive at \eqref{sbmF noncov var} by expanding back the coset generators, repeatedly using the section constraints and adding up all variations into \eqref{Falpha}.

Let us now look at the non-covariant variation of $\hbm{F}^\upgamma_m$ for generic $m\in\mathds{Z}$.
We will use the fact that $\cS^\upgamma_m$ can be expanded in a $\rho$ and $\phi_n$ dependent series of constant $\cS_{n}$ operators with $n\le m$.
Thus, we compute the non-covariant variation of each term in such a series (using $\cS_n(L_p) = L_{p+n}$)
\begin{equation}
\hbm{F}_n =
- \bra{\partial_\cF} L_{n} \ket\cF 
- \sum_{k=1}^\infty \Tr\big( \cG^\ord{k} L_{n-k}  \big)\,,
\end{equation}
and add them up at the end.
We can still use \eqref{DELTA braket F} and \eqref{DELTA Tr G} by setting $T^\lambda \to L_n$ for the former and $T^\lambda \to L_{n-k}$ for the latter.
Looking at the first term of the latter for $k=1$, we can first focus at the commutator term when $T^\beta$ is along the loop components.
We use the identity 
\begin{equation}
[L_{n-1}\,,\ T^A_p] = [L_{n}\,,\ \cS_{-1}(T^A_p)] - \cS_{n-1}(T^A_p)
\end{equation}
which, combined with \eqref{eta and shift ops}, gives us 
\begin{align}
&\eta_{1\,\alpha\beta} 
  \bra{\partial_\Lambda} T^\beta \ket\Lambda
  \bra{\partial_\Lambda} [L_{n-1},\, T^\alpha] \ket\cF  \CR
&=
  \eta_{\alpha\beta} 
    \bra{\partial_\Lambda} T^\alpha \ket\Lambda
    \bra{\partial_\Lambda} [L_{n},\, T^\beta] \ket\cF
- \eta_{n\,\alpha\beta} 
    \bra{\partial_\Lambda} T^\alpha \ket\Lambda
    \bra{\partial_\Lambda} T^\beta  \ket\cF             \CR
&\quad\,
+ \braket{\partial_\Lambda}\Lambda
  \bra{\partial_\Lambda} L_n  \ket\cF
- \bra{\partial_\Lambda} L_n  \ket\Lambda
  \braket{\partial_\Lambda}\cF
\end{align}
where we added and subtracted the $\dK\otimes\vir$ components of each bilinear form to make the identity hold along all components.
This is the only contribution of order $\big(\bra{\partial_\Lambda}\big)^2$ coming from $\Delta_\mathbbm{\Lambda} \cG^\ord{k}$, as the only other ones with two derivatives acting on $\ket\Lambda$ come from \eqref{DELTA braket F}.
Then, putting together all components of order $\big(\bra{\partial_\Lambda}\big)^2$ we find
\begin{equation}
\Delta_\mathbbm{\Lambda} \hbm{F}_n = 
  \bra{\partial_\Lambda} \cS_n(L_0) \ket\Lambda \braket{\partial_\Lambda}{\cF}
+ \eta_{n\,\alpha\beta}  
    \bra{\partial_\Lambda}  T^\alpha\ket\Lambda 
    \bra{\partial_\Lambda} T^\beta \ket\cF
+ \mathcal O(\bra{\partial_\Sigma})
\label{hbmF ncvar step1}
\end{equation}
where in the first term we have highlighted the shift operator coming from writing $\cS^\upgamma_m$ as a series.
With similar steps this expression is completed with the $\Sigma$ dependent terms
\begin{align}
\mathcal O(\bra{\partial_\Sigma}) =\ &
  \sum_{k=1}^\infty  \Tr\big( \Sigma^\ord{k}\cS_n( L_{-k})  \big) \braket{\partial_\Sigma}\cF
+ n \sum_{k=1}^\infty \eta_{n-k\,\alpha\beta} 
     \Tr\big( \Sigma^\ord{k} T^\alpha \big)
     \bra{\partial_\Sigma} T^\beta \ket\cF 
     \,.
\label{hbmF ncvar step2}
\end{align}
We have again highlighted the shift operator $\cS_n$.
The second terms of the last two expressions vanish by the section constraints~\eqref{eq:SC2} if $n<1$. 
When $n=1$, only terms proportional to $\Sigma^\ord{1}$ contribute, and using the section constraint we simplify them to
\begin{equation}
\eta_{0\,\alpha\beta} 
     \Tr\big( \Sigma^\ord{1} T^\alpha \big)
     \bra{\partial_\Sigma} T^\beta \ket\cF 
= 
  \bra{\partial_\Sigma}\Sigma^\ord{1}\ket\cF 
- \Tr\big( \Sigma^\ord{1}\big) \braket{\partial_\Sigma}\cF\,.
\label{hbm var n1 term}
\end{equation}
Recalling that we are really interested in the non-covariant variation of $\hbm F^\upgamma_m$, which is a series in the objects we just varied, for $n\le m$, we find
\begin{align}
\Delta_{\mathbbm\Lambda} \hbm F^\upgamma_m &=
  \bra{\partial_\Lambda} \cS_m^\upgamma(L_0) \ket\Lambda \braket{\partial_\Lambda}{\cF}
+ \sum_{k=1}^\infty  \Tr\big( \Sigma^\ord{k}\cS^\upgamma_m( L_{-k})  \big) \braket{\partial_\Sigma}\cF\,,
\qquad
m<1\,,
\end{align}
and we recognise the definition \eqref{hatF gamma definition} for $\hbm\Lambda^\upgamma_m$.
When $m=1$, from \eqref{shift gamma 1 as series} we see that \eqref{hbm var n1 term} contributes with an overall $\rho^{-1}$ factor.
In total we have
\begin{equation}
\Delta_{\mathbbm\Lambda} \hbm F^\upgamma_m = 
- \hbm\Lambda^\upgamma_m \braket{\partial_{\Lambda,\Sigma}}{\cF}
+ \delta^m_1 \rho^{-1} \big(
     \bra{\partial_\Sigma}\Sigma^\ord{1} 
   - \Tr( \Sigma^\ord{1}) \bra{\partial_\Sigma}
  \big) \ket\cF 
\,,\qquad 
m\le1\,.
\label{hbmF general noncov var}
\end{equation} 
This reproduces \eqref{hbmF noncov var} in particular. 
Combining this result with \eqref{sbmF noncov var} and \eqref{mhbm F def}, \eqref{mhbmF noncov var} is also readily found.
Finally, notice that for $m>1$ \eqref{hbmF general noncov var} is modified by the second terms in \eqref{hbmF ncvar step1} and \eqref{hbmF ncvar step2} which contribute in forms that cannot be simplified by the section constraints.
As a result, the expression contracting $\ket\cF$ in $\Delta_{\mathbbm\Lambda} \hbm F^\upgamma_m$ is not anymore directly subject to the section constraint.
This means that such non-covariant variations appear in the transformation properties of $\mf{X}^\ord{m}$ for $m>1$ and that they cannot be reabsorbed by (re)defining the variation of the term $\braket{\chi^\upgamma_m}{\cF}$.

Let us now prove \eqref{trick}.
The idea is that $\mathbbm D \delta \mathbbm A = (\dd-\delta_{\mathbbm A}) \delta\mathbbm A$ where $\delta_{\mathbbm A}$ is the transformation under generalised diffeomorphisms of $\delta \mathbbm A$, with $\mathbbm A$ as parameter.
Using the definition \eqref{hatF gamma definition} of the operator $\widehat{(\ )}^\upgamma_m$ we then write
\begin{align}\label{trick step1}
\widehat{(\mathbbm D \delta \mathbbm A)}^\upgamma_m 
=
- \bra{\partial_{\mathbbm A}+\partial_{\delta A}}
    \,\cS^\upgamma_m(L_0)\, 
  \ket{(\dd-\delta_\mathbbm{A})\delta A}
- \sum_{k=1}^\infty \Tr\left[
    \big((\dd-\delta_\mathbbm{A})\delta B^\ord{k}\big)
    \,\cS^\upgamma_m(L_{-k})\, 
  \right]\,.
\end{align}
We have included here the variation of the $B^\ord{k}$ fields to be more general, as it does not complicate the proof.
The operator $\dd-\delta_{\mathbbm A}$ commutes with the partial derivatives in the first term so that, in order to bring it out of $\widehat{(\ )}^\upgamma_m$, we just need to add and remove its action on the shift operators.
Using \eqref{shift gamma def} and covariance under generalised diffeomorphisms of the coset representative we are led to compute
\begin{equation}\label{trick step 2}
- \bra{\partial_{\delta A}} 
    \big(\cD \cS^\upgamma_{m}(L_{0})\big)
  \ket{\delta A}
- \sum_{k=1}^\infty \Tr\left[
    \delta B^\ord{k} \cD \cS^\upgamma_{m}(L_{-k})
  \right]\,,
\end{equation}
which must be subtracted from 
$(\dd-\delta_\mathbbm{A})\widehat{\delta\mathbbm A}^\upgamma_m$.
Each derivative of the shift operators reads (with $k=0$ for the first term)
\begin{align}
&\cD\cS_0\Big(\Gamma^{-1} \cS_m(\Gamma L_{-k}\Gamma^{-1})\Gamma \Big)
\CR
&=
\cS_0\Big(\Gamma^{-1} \cS_m\big(
        [\cD\Gamma\Gamma^{-1},\, \Gamma L_{-k} \Gamma^{-1}]
      \big) \Gamma\Big)
-\cS_0\Big( \Gamma^{-1} 
        [\cD\Gamma\Gamma^{-1},\,\cS_m(\Gamma L_{-k}\Gamma^{-1})]
      \Gamma \Big)
\CR
&=
m \sum_{p=0}^\infty (\cD\Gamma\Gamma^{-1})_{-p}\,
  \cS^\upgamma_{m-p}(L_{-k})\,,
\end{align}
where we used \eqref{shiftgamma commutator} in the last step.
Plugging this back into \eqref{trick step 2} we reproduce \eqref{trick}.

\subsection{\texorpdfstring{Some details on $\virm$ gauge fixing}{Some details on vir- gauge fixing}}
\label{sec:gfix details}

The gauge-fixing of $\virm$ in the Virasoro-extended formulation was considered in Section~\ref{sec:gfix}. We here give some additional details. 

\subsubsection*{\texorpdfstring{$c_\vir$ dependent couplings}{cvir dependent couplings}}

We first reconsider
the $c_\vir$ dependent terms in \eqref{final cov Lagrangian J} and show explicitly that after we gauge-fix $\phi_m\to0$, $m\ge2$ they cancel out.
We begin with the term coming from $\omega^{\alpha\beta}\fJ_\alpha\fJ(1)_\beta$.
This is (minus) the contribution of the $\vir$ components of $\fJ$ to the central charge term in the commutator $[\fJ,\,\fJ(1)]$.
The latter can be rewritten as
\begin{equation}
-\frac12[\fJ,\,\fJ(1)]\big|_\dK = 
-2 [\fP,\,\fP(1)]\big|_\dK 
-2 \omega^\alpha(\cV^{-1}) f^{\beta\delta}{}_\alpha \fP_\beta\fP(1)_\delta\,.
\end{equation}
We are interested in the terms quadratic in the $\vir$ components of $\fP$, in which case the cocycle term does not contribute because conjugation by $\rho^{L_0}$ and $e^{\tilde\rho L_{-1}}$ does not generate central terms.
The former term then reads
\begin{equation}
-2 [\fP,\,\fP(1)]\big|_\dK 
= -\frac{c_\vir}{6}\sum_{n\in\mathds{Z}} (n^3-n) \tP_n \tP_{n+1} + \ldots\,,
\end{equation}
where the dots denote terms dependent on the loop components of $\cP$.
Using \eqref{cP def} and the fact that $\tP_n$ only contributes for $|n|\ge2$, we then arrive at
\begin{align}
-\frac12[\fJ,\,\fJ(1)]\big|_\dK 
&=
-\frac{c_\vir}{24} (n^3-n) \tJ_n  \, \tJ_{-n-1} +\ldots\CR
&= -\frac{c_\vir}{12}\sum_{m=0}^\infty \sbm A_{-m} 
     \big[\Gamma L_{-m} \Gamma^{-1}+\mathrm{h.c.},\,\fP(1) \big]\big|_\dK +\ldots\,.
\end{align}
Adding to this expression also the term proportional to $\sbm{A}_n \tJ_{-n-1}$ coming from the cocycle in \eqref{Dchigamma}, we arrive at the identity
\begin{equation}
\frac{c_\vir}{24}\sum_{n\in\mathds{Z}} (n^3-n) (\tJ_n + 2 \sbm{A}_n) \, \tJ_{-n-1}
= -\frac{c_\vir}{12}\sum_{m=0}^\infty \sbm A_{-m} 
     \big[\Gamma L_{-m} \Gamma^{-1}-\mathrm{h.c.}\,,\ \fP(1) \big]\big|_\dK\,.
\end{equation}
Using \eqref{cQ def} it is then straightforward to show that this expression matches the left-hand side of \eqref{QP vir cocycle to PP} and hence cancels out with the last term in the first line of \eqref{final cov Lagrangian J}, removing all $c_\vir$ dependent couplings from the pseudo-Lagrangian.

\subsubsection*{\texorpdfstring{Matching $\bra\chi$ transformation}{Matching chi transformation}}

We now proceed to proving that \eqref{brachi noncov trf} agrees with the transformation of $\bra\chi$ in \cite{Bossard:2018utw}.
The $\hat{\mf e}_8$ invariant bilinear forms \eqref{eta def} are not invariant under generic $\vir$ transformations.
This means in particular that in an expression like
\begin{equation}
\eta_{m\,\alpha\beta}\ \cM^{-1} (T^\alpha)^\dagger \cM \ \otimes\ T^\beta
\end{equation}
we cannot simply bring the $\cM$ conjugation through to the other factor.
Formally writing
\begin{equation}
\cM = \Gamma^\dagger\Gamma\,g_\cM\,,\qquad g_\cM\in\hE8\,,
\end{equation}
we see that the problem lies in $\Gamma^\dagger\Gamma$ which involves exponentials of all $\vir$ generators.
When we gauge fix $\phi_m\to0$, $m\ge2$, however, only exponentials of $L_0$, $L_{-1}$ and $L_1$ are left, and parameterise a well-defined $SL(2)$ element
\begin{equation}
\Gamma^\dagger\Gamma\ \to\ m\in SL(2)\,,
\end{equation}
which acts on $w$ by fractional linear transformations.
In this case it is possible to bring the $\cM$ conjugation through in \eqref{eta M conj} and this was done in the appendix of \cite{Bossard:2018utw}.
We summarise here some results:
\begin{align}
\label{eta M conj}
\eta_{-1\,\alpha\beta}\ \cM^{-1} T^\alpha{}^\dagger \cM \otimes T^\beta &=
  \frac{1}{\rho^2}\left(
     \eta_{1\,\alpha\beta} 
  -2 \tilde\rho \,  \eta_{\alpha\beta}
  +  \tilde\rho^2\, \eta_{-1 \alpha\beta}
  \right) T^\alpha \otimes \cM^{-1} T^\beta{}^\dagger \cM \,,\\[1ex]
\eta_{\alpha\beta}\ \cM^{-1} T^\alpha{}^\dagger \cM \otimes T^\beta &=
   \frac{1}{\rho^2}\left(
     \tilde\rho \, \eta_{1\,\alpha\beta} 
  +  (\rho^2-2 \tilde\rho^2)         \, \eta_{\alpha\beta}
  +  (\tilde\rho^3-\rho^2\tilde\rho) \, \eta_{-1 \alpha\beta}
   \right) T^\alpha \otimes \cM^{-1} T^\beta{}^\dagger \cM \,,\nonumber\\[1ex]
\eta_{1\,\alpha\beta}\ \cM^{-1} T^\alpha{}^\dagger \cM \otimes T^\beta &=
   \frac{1}{\rho^2}\left(
     \tilde\rho^2 \, \eta_{1\,\alpha\beta} 
  +  2(\rho^2\tilde\rho- \tilde\rho^3)         \, \eta_{\alpha\beta}
  +  (\rho^2-\tilde\rho^2)^2 \, \eta_{-1 \alpha\beta}
   \right) T^\alpha \otimes \cM^{-1} T^\beta{}^\dagger \cM \,.\nonumber
\end{align}

Let us then prove that \eqref{brachi noncov trf} matches the non-covariant variation of $\bra\chi$ derived in \cite{Bossard:2018utw}, when gauge fixed to $\phi_m=0$, $m\ge2$ and $\phi_1=\tilde\rho$.
To do so, we must expand $\hbm\Lambda^\upgamma_{-1}$ and $\mhbm\Lambda^\upgamma_{-1}$.
The former is expanded as in equation \eqref{hatF gamma definition}.
The latter is defined as in \eqref{mhbm F def}, with $\mathbbm F\to\mathbbm\Lambda$:
\begin{equation}
\mhbm\Lambda^\upgamma_{-1} =
  \hbm\Lambda^\upgamma_{1} 
+ \omega^\alpha(\cM) \Big(\sbm\Lambda_\beta\cS^\upgamma_1(T^\beta)\Big)_\alpha\,.
\end{equation}
Again, the expansion of the first term on the right-hand side follows \eqref{hatF gamma definition}.
Let us now look at the cocycle term.
We expand the shift operator as in \eqref{gfix shift op expansion} and write $\sbm\Lambda_\beta$ explicitly using \eqref{eq:balpha} (recall that $\Sigma^\ord{m}=0$ for $m\ge2$ because we are gauge-fixed):
\begin{equation}
\omega^\alpha(\cM) \big(\sbm\Lambda_\beta\cS^\upgamma_1(T^\beta)\big)_\alpha =
\omega^\alpha(\cM) \frac1\rho \left(
    (\eta_{1\,\alpha\beta} - \tilde\rho\, \eta_{\alpha\beta} ) \bra{\partial_\Lambda} T^\beta \ket\Lambda
  + (\eta_{\alpha\beta} - \tilde\rho \,\eta_{-1\,\alpha\beta} )\Tr\big(\Sigma^\ord{1} T^\beta  \big)
  \right)\,,
\label{HLambdahat expansion 1}
\end{equation}
where we also used \eqref{eta and shift ops} and $\omega^\dK(\cM)=0$.
We must now expand the contraction of the cocycle with the invariant bilinears.
First, we open the cocycle and write
\begin{align}
\omega^\alpha(\cM)\,\eta_{k\,\alpha\beta} \dK\otimes T^\beta 
&=
\eta_{k\,\alpha\beta} 
  \Big( \cM^{-1} \cS_0(T^\alpha{}^\dagger) \cM - \cS_0(\cM^{-1} T^\alpha{}^\dagger\cM) \Big) 
  \otimes T^\beta  \CR
&=
\eta_{k\,\alpha\beta} 
  \Big( \cM^{-1} T^\alpha{}^\dagger \cM - \cS_0(\cM^{-1} T^\alpha{}^\dagger\cM) \Big) 
  \otimes T^\beta
+\dK\otimes L_k   
\end{align}
Using then \eqref{eta M conj} on this expression for $k=-1,\,0,\,1$, we see that the first term becomes a sum of terms of the form $\eta_{k\,\dK\beta}$ for $k=-1,\,0,\,1$.
The last term instead stays as it is. 
Adding everything up we then find
\begin{align}
\omega^\alpha(\cM) \big(\sbm\Lambda_\beta\cS^\upgamma_1(T^\beta)\big)_\alpha 
=\ &
  \bra{\partial_\Lambda}\cS^\upgamma_1(L_0) \ket\Lambda
- \frac{1}{\rho} \bra{\partial_\Lambda} 
            \cM^{-1} \big( {\tilde\rho} L_0 + (\rho^2-{\tilde\rho^2})L_1\big) \cM 
        \ket\Lambda
\CR&
+ \Tr\Big( \Sigma^\ord{1}  \cS^\upgamma_1(L_{-1}) \Big)     
-\frac{1}{\rho} \Tr\Big( \Sigma^\ord{1} \cM^{-1} \big(  L_0 -{\tilde\rho} L_1\big) \cM \Big) \,.
\end{align}
Combining this expression with $\hbm\Lambda^\upgamma_{1}$ and $\hbm\Lambda^\upgamma_{-1}$, expanding the shift operators and plugging everything into \eqref{brachi noncov trf} we arrive at
\begin{align}
\Delta_{\mathbbm\Lambda}\bra\chi =\ &
- \bra{\partial_\Lambda} \left(
     \sum_{n=0}^\infty \tilde\rho^n L_{-1-n}
   + \cM^{-1} \bigg( \frac{\tilde\rho}{\rho^2} L_0 + \bigg(1-\frac{\tilde\rho^2}{\rho^2}\bigg)L_1\,\bigg) \cM
   \right)\ket\Lambda\bra{\partial_\Lambda} 
- \frac{\tilde\rho}{\rho^2} \braket{\partial_\Lambda}{\Lambda}\bra{\partial_\Lambda}
\CR&
-\Tr\bigg[ \Sigma^\ord{1} \bigg( 
       \sum_{n=0}^\infty \tilde\rho^n L_{-2-n}
       + \cM^{-1} \bigg(  \frac{1}{\rho^2}L_0 -\frac{\tilde\rho}{\rho^2} L_1 \bigg) \cM 
      \bigg)
 \bigg] \bra{\partial_\Sigma}
- \frac{1}{\rho^2}\bra{\partial_\Sigma}\Sigma^\ord{1}\,,
\end{align} 
which reproduces equations (4.67) and (4.92) of \cite{Bossard:2018utw}.

\subsection{Series regularisation and integrability conditions}
\label{app:seriesreg}

In computing the equations of motion in the extended formalism of Section~\ref{sec:cov eom}, we have found that for some of them, using twisted self-duality to eliminate all dual fields leads to divergent sums that need to be regularised.
This necessity is essentially due to the fact that commutators of $\cP$ and $\cQ$ cannot be reduced to a finite expression using only the $E_8$ commutation relations, especially when $\cS_m$ operators are involved, because $\cP$ and $\cQ$ are infinitely extended over both the positive and negative loop and $\vir$ levels.
We give more details on this in Appendix~\ref{app:kacmoody}.
To motivate the regularisation procedure we adopt in this paper, we can look at the integrability condition for twisted self-duality \eqref{twsd ungauged, P} in purely two-dimensional (super)gravity, which we proved in Section~\ref{sec:extcoset}, but formulated now only in terms of $P$ and $Q$ as defined in~\eqref{eq:PQungauged} and their Maurer--Cartan equation.
The matching of integrability conditions of \eqref{twsd ungauged, P} we give in Section~\ref{sec:extcoset} does not involve any regularisation because it is based on the Maurer--Cartan form $\dd\cV\cV^{-1}$ which takes values only along non-positive loop and $\vir$ levels.
Then, the only commutators one encounters are those needed to apply the Maurer--Cartan equation to the linear system \eqref{BMsys} (more precisely, its $w$ expansion with generic $\upgamma(w)$) and they are all finite.
Writing the integrability conditions in terms of $P$ and $Q$ instead, we encounter commutators that are not finite \textit{per se}, and we will see that in order to reproduce the result based on $\dd\cV\cV^{-1}$ the regularisations employed in Section~\ref{sec:extE9exft} become necessary.

The Maurer--Cartan equation for $P$ reads $\dd P - [Q,\,P]=0$.
Applying an $\cS_1$ operator to this expression, using \eqref{shiftgamma commutator} and then \eqref{cascade ungauged, P}, we arrive at
\begin{equation}\label{ungauged intcond P}
\cS_0\big(\,
\dd\star P-[Q,\,\star P]
\,\big) = 0\,.
\end{equation}
The $\mf e_8$ component of this equation was already computed, in ExFT, in Section~\ref{sec:cov eom}.
Let us check that up to a regularisation, \eqref{ungauged intcond P} reproduces \eqref{rho ungauged eq}, \eqref{e8 scalars ungauged equation} and nothing else.
We will need that in the triangular gauge \eqref{Vhat Y expansion} and by twisted self-duality
\begin{align}
P^m_A &= \star ^{m}P^0_A\,,&
Q^m_A &= -\text{sgn}(m) \star ^{m}P^0_A\,,\ m\neq0\,, \\[1ex]
P_m   &= \star ^m (\dd\Gamma\Gamma^{-1})_0\,,&
Q_m   &= -\text{sgn}(m) \star ^m(\dd\Gamma\Gamma^{-1})_0\,,\ \ Q_0=0\,.
\end{align}
Let us take the $L_{-p}$, $p\ge0$ component of this equation which, using twisted self-duality, reads
\begin{align}
& 2 \dd{}{\star ^{p+1}(\dd\Gamma\Gamma^{-1})_0} + [(\dd\Gamma\Gamma^{-1})^\dagger,\,\star (\dd\Gamma\Gamma^{-1})]|_{L_{-p}} \\[1ex]\nn&
= 2 \dd{}{\star ^{p+1}(\dd\Gamma\Gamma^{-1})_0} 
  + \sum_{n\ge0} (2m+p)\, (\dd\Gamma\Gamma^{-1})_{-m} \star (\dd\Gamma\Gamma^{-1})_{-p-m}  \\[1ex]\nn&
= 2 \dd{}{\star ^{p+1}(\dd\Gamma\Gamma^{-1})_0} 
  +(\dd\Gamma\Gamma^{-1})_0   \star ^{p+1} (\dd\Gamma\Gamma^{-1})_0 \Big( 
    2p + 8\sum_{m=1}^\infty (-1)^m m +4p \sum_{m=1}^\infty (-1)^m
  \Big)  \\[1ex]\nn&
= 2 \dd{}{\star ^{p+1}(\dd\Gamma\Gamma^{-1})_0} - 2 (\dd\Gamma\Gamma^{-1})_0   \star ^{p+1} (\dd\Gamma\Gamma^{-1})_0 \\[2ex]\nn&
= -\rho^{-1} \dd\star\dd \rho\ \text{ if $p$ even, 0 otherwise,}
\end{align}
which recovers the free equation of motion for $\rho$.
We have resummed the divergent series $\sum_{m=1}^\infty (-1)^m m\to-1/4$ and $\sum_{m=1}^\infty (-1)^m \to-1/2$ using the $z\to1$ limit of a geometric series  $\sum_{n=1}^\infty (-z)^n p(n)$ for $p(n)$ a polynomial in $n$.
Now for the loop part,
\begin{align}\label{ungauged phys eom loop}
&\dd{}{\star P}|_{T^C_p} - [Q,\,\star P]|_{T^C_p} \\[1ex]\nn&
= \dd{}{\star ^{p+1}P^0_C}
  -f^{AB}{}_C \sum_{n\in\mathds{Z}} Q_A^n  \star P_B^{p-n}
  +\sum_{n\in\mathds{Z}}n\big(  Q_{p-n} \star P_C^n - Q_C^n \star  P_{p-n} \big) \,.
\intertext{%
We redefine $n\to p-n$ in the second-to-last term. Using that, by twisted self-duality, $Q^n_A$ and $Q_n$ are $n$-odd, while the $P$ are $n$-even, we find that several terms cancel out and we are left with%
}
&= \dd{}{\star ^{p+1}P^0_C}
  - f^{AB}{}_C Q^0_A \star ^{p+1}P^0_B
  - 2 \sum_{n=1}^\infty n \big( Q^n_C \star  P_{p-n} + Q_n \star  P_C^{p-n}   \big)\\[1ex]\nn&
= \dd{}{\star ^{p+1}P^0_C} 
  - f^{AB}{}_C Q^0_A \star ^{p+1}P^0_B
  +4 \tfrac{1+(-1)^p}{2} (\dd\Gamma\Gamma^{-1})_0 \star ^{p+1} P^0_C \  \sum_{n=1}^\infty (-1)^n n \\[1ex]\nn&
= \left\{ 
\begin{array}{ll}
\dd P^0_C - f^{AB}{}_C Q^0_A P^0_B = 0  & \text{\ \ $p$ odd,} \\[1ex]
\rho^{-1} \dd(\rho\star P^0_C) 
- f^{AB}{}_C Q^0_A \star P^0_B   &  \text{\ \ $p$ even,}
\end{array}\right.
\end{align}
which reproduces \eqref{e8 scalars ungauged equation} given $P^0_A=\mathring P_A$ and $Q^0_A=\mathring Q_A$, which is valid because of the triangular $K(E_9)$ gauge.
We have again regularised $\sum_{n=1}^\infty (-1)^n n\to-1/4$.

\subsection{\texorpdfstring{Matching of scalar equations of motion to $E_8$ ExFT}{Matching of extended equations of motion to E8 ExFT}}
\label{app:scaleom match}

As a further motivation for the regularisations employed in Section~\ref{sec:cov eom}
and studied in the previous section, we provide here the matching of the $\mathring V$ scalar field equations of motion derived there to those of $E_8$ exceptional field theory.
A full matching of the $E_9$ theory to the $E_8$ one has been given at the level of the pseudo-Lagrangians in the minimal formalism of Section~\ref{sec:E9EFT}.

\def\3D{{\scriptscriptstyle\rm3D}}

We begin with the $\pi^{0\,A}$ component of equation~\eqref{scalar eom}, looking only at the left-hand side since the scalar potential has already been matched to $E_8$ exceptional field theory in \cite{Bossard:2018utw}.
We will rewrite the relevant term as
\begin{equation}\label{scaleom start}
\rho\big(\widehat\cD\star\tP^0_A + 2\underline{\sbm F}^{+1}_A\big) \mathring{V} T^A \mathring{V}^{-1}\,,
\end{equation}
where we have conjugated the expression by the $E_8/Spin(16)$ coset representative $\mathring V$ and rescaled by a factor of 2.
This means that this expression should be traced with $2 \pi^0_B (\mathring V T^B \mathring V^{-1}) = M^{-1} \delta M$, which is the natural expression for the variation of the $E_8$ generalised metric $M_{AB}$.

We will now choose the solution of the section constraint \eqref{eq:secsol}.
With this, we can expand
\begin{align}
\underline{\sbm A}^1_A T^A &= 
\frac12 \rho \, (\partial_A w)\, \mathring{V} T^A \mathring{V}^{-1} \,, \\
\underline{\sbm A}^0_A T^A&= 
\frac12(f_{AB}{}^C\partial_C A^B - B_A + f_{AB}{}^C Y_1^B \partial_C w)\, \mathring{V} T^A \mathring{V}^{-1}\,,
\end{align}
where $T^A\in E_8$ and $w=\braket{0}{A}$ (not to be confused with the spectral parameter).
In analogy with \eqref{eq:KKAket} and \eqref{TrBE8expansion}, we have defined $A^C = \bra{0}T^C_{+1}\ket{A}$ and $-B_A = \Tr(B^\ord{1} T_{-1\,A})$.

Looking at the first two terms in the expansion \eqref{expand scalar eom} of $\widehat\cD\star\tP^0_A$, we write
\begin{align}\label{subs into div}
&\big(\cD(\rho\star\tP^0_A) - \rho\,f^{CD}{}_A \tQ^0_C\star\tP^0_D \big)\mathring{V}^{-1}T^A\mathring{V} \CR
&\ =\ 
\dd(\rho \mathring{V}^{-1}\star\tP^0_A T^A\mathring{V}) 
 -\bra{\partial}\big(\ket{A}\rho \mathring{V}^{-1}\star \tP^0_A T^A \mathring{V}\big)
\CR&\qquad
 +\rho\,\big(f_{AB}{}^C \partial_C A^B -B_A +f_{AB}{}^C Y_1^B \partial_Cw\big) 
   [T^A,\,\mathring{V}^{-1}\star\tP^0_F T^F \mathring{V}]\,.
\end{align}
We now implement the field redefinitions
\begin{equation}\begin{aligned}\label{KK redefs}
A^C\ &=\ A^\3D{}^C - w\,A_\varphi^C\,,\\
B_C\ &=\ B^\3D{}_C - w\,B_{\varphi\,C} -\rho\star\!\partial_C w
\end{aligned}\end{equation}
where $A_\mu^\3D{}^C $ and $B_\mu^\3D{}_C $ are the $x^0$ and $x^1$ components of the $E_8$ vector fields, while $A_\varphi^C$ and $B_{\varphi\,C}$ are their components along the third space-time direction denoted by $\varphi$. 
We do not need a suffix $^\3D$ for the latter, but we identify them with
\begin{equation}
A_\varphi^C = Y_1^C\,,\qquad 
B_{\varphi\,C} = 2\rho\, \eta_{CD} \bra{\tilde\chi_1}T^D_{-1}\ket0\,.
\end{equation}
This was already established in \cite{Bossard:2018utw}.
Using \eqref{cP def}, we then have that 
\begin{equation}
2\,\mathring{V}^{-1}\tP^0_A T^A \mathring{V} = j^\3D-w \,j_\varphi -\rho\star\!\partial_Aw\, (T^A + M^{-1}T^A M)
\end{equation}
where $j^\3D=j^\3D_\mu \dd x^\mu$ is $\mf e_8$ valued and
\begin{equation}
j^\3D_{\widehat\mu} = (j^\3D_\mu,\,j_\varphi) = M^{-1} D_{\widehat\mu} M
\end{equation}
is the $E_{8}$ scalar field current and $D_{\widehat\mu}$ the $E_{8}$ ExFT covariant derivative:
\begin{equation}
M^{-1}D_{\widehat\mu}M = 
  M^{-1}\partial_{\widehat\mu}M 
- A^\3D_{\widehat\mu}{}^C M^{-1}\partial_C M
+ (f_{AB}{}^C\partial_C A^\3D_{\widehat\mu}{}^B -B^\3D_{\widehat\mu\,A}) 
(T^A+M^{-1}(T^A)^\T M)\,.
\end{equation}
We then have that \eqref{subs into div} becomes, after the field redefinitions,
\begin{align}\nn
&
\frac12\big(D-wD_\varphi -(D_\varphi w)\,\big)
\big( \rho\star\!j^\3D -\rho\star\!w\,j_\varphi -\rho^2 \star\!\partial_Aw\,
(T^A+M^{-1}T_A M)\big) \\
&-\frac12\rho^2 \partial_Aw\,\big[T^A,\,j^\3D-w\,j_\varphi-\rho^2\star\!\partial_B w \,(T^B+M^{-1}T_BM)  \big]\,.
\label{E8scal eom piece 1}
\end{align}
The $\partial_Aw*\!\partial_Bw [T^A,\,T^B]$ term vanishes by the section constraint.

We must now add to the above expression the $\underline{\sbm A}^1_A$ dependent terms of \eqref{expand scalar eom} as well as the last one in \eqref{scaleom start}.
For the latter we compute, applying \eqref{KK redefs},
\begin{align}
2 \underline{\sbm{F}}^1_A \mathring{V}^{-1} T^A \mathring{V} 
=
\rho\,\partial_A(\braket{0}{\cF})\, T^A 
=
\rho\,(D-w D_\varphi ) (\partial_A w)\,T^A
  +\rho\,D_\varphi w\,\partial_Aw\,T^A
\label{E8scal eom piece 2}
\end{align}
where $w$ transforms as a scalar (weight 0) under $E_{8}$ generalised diffeomorphisms.
For the last two terms of \eqref{expand scalar eom} we also need $\cD\rho$ after the refinitions \eqref{KK redefs}:
\begin{equation}
\cD\rho = (\,D-wD_\varphi  - (D_\varphi w)\, )\rho  
\end{equation}
so that 
\begin{align}
&2\rho\, f^{CD}{}_A (\underline{\sbm A}^1_C - \underline{\sbm A}^{1\,C}) \tP^{0}_D \mathring{V}^{-1} T^A \mathring{V}
+2\,\cD\rho\,\underline{\sbm A}^1_A \mathring{V}^{-1}(T^A + T_A)\mathring{V} \nn\\[1.5ex]
&\qquad=
\frac12 \rho^2 \partial_A w \big[T^A-M^{-1}T_A M\,,\,\, j^\3D-w j_\varphi  -\rho\star\!\partial_Bw\, (T^B + M^{-1}T_B M) \big]\nn\\[1ex]
&\qquad\quad+\rho (\,D-wD_\varphi  - (D_\varphi w)\, )\rho\,\,  \partial_Aw (T^A + M^{-1}T_A M)\nn\\[1.5ex]
&\qquad\simeq
\rho^2 \partial_A w \big[T^A\,,\,\, j^\3D-w j_\varphi  -\rho\star\!\partial_Bw\, (T^B + M^{-1}T_B M) \big]\nn\\[1ex]
&\qquad\quad+\rho (\,D-wD_\varphi  - (D_\varphi w)\, )\rho\,\,  \partial_Aw (T^A + M^{-1}T_A M)\,,
\label{E8scal eom piece 3}
\end{align}
where in the last step we are using that the whole expression appears traced with $M^{-1} \delta M$. 
Again, the $\partial_Aw\star\!\partial_Bw [T^A,\,T^B]$ term vanishes by the section constraint.
Putting everything together we find
\begin{align}
\frac12 (M^{-1} \delta M)^A \Big[
(\,D-wD_\varphi - (D_\varphi w)\, )(\rho\star\!j^\3D_A -\rho\star\!w\,j_\varphi{}\,_A)
+ \rho^3\,f_{AB}{}^C M^{BD} \partial_Cw\,\star\!\partial_Dw
\Big]+\ldots = 0
\label{E8scal eom all pieces}
\end{align}
where the dots correspond to the contribution from the $E_9$ scalar potential.

Comparing with the $E_8$ ExFT equations of motion, we have that the variation of the kinetic term reads (up to total derivatives, denoting by $e$ the determinant of the 3d vielbein)
\begin{align}
&\quad \delta\left(-\frac14 e g^{\widehat\mu\widehat\nu} \eta^{AB} j^\3D_{\widehat\mu\,A} j^\3D_{\widehat\nu\,B} \right)
=
\frac12 (M^{-1} \delta M)^A D_{\widehat\mu}\big( e g^{\widehat\mu\widehat\nu} j^\3D_{\widehat\mu\,A} \big)\\[1ex]
&=
\frac12(M^{-1} \delta M)^A\big[
  (\,D-wD_\varphi - (D_\varphi w)\, )(\rho\star\!j^\3D_A -\rho\star\!w\,j_{\varphi}{}_A) 
+ D_\varphi(\rho^{-1}e^{2\sigma})j_{\varphi A}
\big] \,, \nn
\end{align}
where the last term is purely internal (only internal derivatives and no 2d vector fields) and hence is reproduced by the $E_9$ scalar potential contributions in the `$\ldots$' above.
Finally, the $E_{8}$ scalar potential contains one term dependent on the Kaluza--Klein vector $w_\mu$:
\begin{align}
-e V_{E_{8}} = 
\ldots + \frac14 e M^{AB} \partial_A g^{\widehat\mu\widehat\nu} \partial_B g_{\widehat\mu\widehat\nu} =
\ldots
-\frac12 \rho^3 \, M^{AB} \partial_A w_\mu\, \partial_B w^\mu\,.
\end{align}
Again the dots represent purely internal terms.
Variation of this term reproduces exactly the last term in \eqref{E8scal eom all pieces}, proving the claim.


\section{\texorpdfstring{The $E_9$ Virasoro constraint}{The E9 Virasoro constraint}}
\label{Virasoro}

In order to match the Virasoro constraint \eqref{cov Vir constraint} in the Virasoro-extended formulation of the theory with the Euler--Lagrange equation for $\tilde{g}_{\mu\nu}$ in its minimal formulation, one must assume that the section constraint is solved consistently with the $E_8$ parabolic gauge  \eqref{E8parabolic} such that $B$ takes the form \eqref{E8Bexpansion}. This is always possible since all solutions to the section constraint are $E_9$ conjugate to this one. Although matching these two equations in this way requires to break $E_9$ symmetry, both equations  are  $E_9$ invariant and they therefore must be equivalent.

The terms in the pseudo-Lagrangian \eqref{PseudoLagrangian} that contribute to the Euler--Lagrange equation for $\tilde{g}_{\mu\nu}$ are 
\be  \cL_1 + \frac{1}{2} \varepsilon^{\mu\nu} J_{\mu \, {-}1} J_{\nu\,  \dK} +\frac12 \rho  \varepsilon^{\mu\nu} \varepsilon^{\sigma\rho}  \tilde{g}^{\kappa\lambda} \cD_\mu \tilde{g}_{\sigma\kappa} \cD_\nu  \tilde{g}_{\rho\lambda} + \frac{\rho^{-1}}{4} \langle \partial \tilde{g}^{\mu\nu} | \cM^{-1} | \partial \tilde{g}_{\mu\nu} \rangle \; . \ee
By construction, the variation of $\cL_1$ involves infinitely many terms, with 
\be
 \delta \cL_1 = \tfrac12 \rho^{-1} \delta \tilde{g}^{\mu\nu} \Bigl(  \Tr \bigl[ \bigl(  \cM^{-1}  \cS_{1}( \cJ^\flat_{\mu})^\dagger \cM  + \chi^\flat_\mu  \bigr) B_\nu \bigr] -\tfrac12   \eta_{\alpha\beta} \Tr[ T^\alpha B_\mu ] \Tr[ \cM^{-1} T^{\beta^\dagger} \cM B_\nu ] \Bigr) \; . 
\ee
But if we assume  \eqref{E8Bexpansion} and therefore \eqref{TrBE8expansion}, one can use the  semi-flat formulation introduced in Section \ref{E8reduction} with the same steps to show that  the variation of the pseudo-Lagrangien \eqref{PseudoLagrangian} with respect to $\tilde{g}_{\mu\nu}$ reduces to the variation of
\begin{multline}
 \mathcal{Z}  -  \frac{\rho}{4} \eta_{AB}\tilde{g}^{\mu\nu} \bigl( \eta^{AC}  \tilde{J}_\mu{}_C^0 +M^{AC} B_{\mu C}  \bigr)   \bigl( \eta^{BD}  \tilde{J}_\nu{}_D^0 +M^{BD} B_{\nu D}  \bigr)  +2  \tilde{g}^{\mu\nu}  \cD_\mu \rho \cD_\nu \sigma + \cD_\mu \rho \cD_\nu \tilde{g}^{\mu\nu}   \\
+\frac12 \rho  \varepsilon^{\mu\nu} \varepsilon^{\sigma\rho}  \tilde{g}^{\kappa\lambda} \cD_\mu \tilde{g}_{\sigma\kappa} \cD_\nu  \tilde{g}_{\rho\lambda} + \frac{\rho^{-1}}{4} \langle \partial \tilde{g}^{\mu\nu} | \cM^{-1} | \partial \tilde{g}_{\mu\nu} \rangle   \; ,\label{tilde g eom}  
\end{multline}
where $ \mathcal{Z}$ is the term \eqref{Zduality0} that vanishes upon using the duality equation \eqref{DualityEquation}.  One can therefore ignore $\mathcal{Z}$ in deriving the Euler--Lagrange equation for $\tilde{g}^{\mu\nu}$ to get a manifestly finite result.  
 
Using the same solution to the section constraint for the fields $B^\ord{k}$ in the extended formulation and setting all the Virasoro fields to zero, one obtains that  
\be  2 \tP_\mu{}^0_A   \mathring{V}^{-1} T_0^A  \mathring{V} = \bigl( \tilde{J}_\mu{}_A^0 +\eta_{AC} M^{BC} B_{\mu B} \bigr) T_0^A\; ,  \ee
where $B_A = -\Tr(B^\ord{1} T_{-1\,A})$.
Substituting this result in the Euler--Lagrange equation~\eqref{tilde g eom} for $\tilde{g}_{\mu\nu}$ evaluated at $\tilde{g}_{\mu\nu}=\eta_{\mu\nu}$ gives precisely  \eqref{cov Vir constraint} at vanishing Virasoro fields.


\section{On symmetries of pseudo-Lagrangians}
\label{app:pseudoL}

In this appendix, we consider in some generality the definition of symmetries of a pseudo-Lagrangian. This discussion is relevant to the invariance of the theory under external diffeomorphisms as discussed in Section~\ref{sec:extdiff}.
Denote the fields of a theory collectively as $\phi^I$, with a pseudo-action $S = \int \cL$, its associated Euler--Lagrange equations $\cE_I = \frac{\delta S}{\delta \phi^I}=0$, and assume a separate set of duality equations $\cE_A = 0$. 

Having a symmetry with parameter $\xi$ means that the equations of motion transform into each other according to
\begin{align}
 \delta_\xi \cE_I = A_I{}^J(\xi,\phi) \cE_J + B_I{}^B(\xi,\phi) \cE_B \; , \qquad   \delta_\xi \cE_A = C_A{}^J(\xi,\phi) \cE_J + D_A{}^B(\xi,\phi) \cE_B\; . \label{symeom}
 \end{align}
This is certainly the case if both the duality equations and the pseudo-action are invariant under that symmetry, in which case $B_I{}^B = C_A{}^J = 0$. But we want to define a minimal requirement. To this purpose consider
\begin{align}
 \frac{\delta\; }{\delta \phi^I} \delta_\xi S  &= \frac{\delta\; }{\delta \phi^I} \int \delta_\xi \phi^J   \frac{\delta S}{\delta \phi^J} \nn\\
&=\int \delta_\xi \phi^J  \frac{\delta\; }{\delta \phi^J}   \frac{\delta S}{\delta \phi^I} + \int  \frac{\delta \delta_\xi \phi^J}{\delta \phi^I}    \frac{\delta S}{\delta \phi^J}  \nn\\ 
&= \delta_\xi \cE_I  + \int  \frac{\delta \delta_\xi \phi^J}{\delta \phi^I}   \cE_J \; .  
\end{align}
One re-obtains that the Euler--Lagrange equations transform into themselves if  $ \delta_\xi S=0$. However, having a symmetry as defined by~\eqref{symeom}, only requires the weaker condition that $\frac{\delta\; }{\delta \phi^I} \delta_\xi S $ must be proportional to the equations of motion $\cE_I$ and the duality equations $\cE_A$. For this to be the case it is sufficient for $ \delta_\xi S $ to be quadratic in the duality equations 
\begin{align}
  \delta_\xi S  = \int \Bigl( \alpha^{IJ}(\xi,\phi) \cE_I \cE_J + \beta^{IB}(\xi,\phi) \cE_I \cE_B + \gamma^{AB}(\xi,\phi) \cE_A \cE_B \Bigr) \; .
  \end{align} 
One can always redefine the symmetry such that the two first terms vanish, with 
\begin{align}
 \delta^\prime_\xi \phi^I = \delta_\xi \phi^I  -  \alpha^{IJ}(\xi,\phi) \cE_J- \beta^{IB}(\xi,\phi)  \cE_B 
 \end{align}
but the last term cannot be eliminated in general. One has then
\begin{align} \label{RelationBdeltaL}
 B_I{}^B(\xi,\phi) = \int \Bigl( \frac{\delta \gamma^{AB}(\xi,\phi)}{\delta \phi^I } \cE_A + 2 \gamma^{AB}(\xi,\phi) \frac{\delta}{\delta \phi^I } \cE_A \Bigr) 
 \end{align}
and we get that $\gamma^{AB}\ne0$ if $B_I{}^B(\xi,\phi) $ does not vanish. So if the Euler--Lagrange equations do not transform into each other under a symmetry, but also mix with the duality equations, then the pseudo-Lagrangian is only invariant up to terms quadratic in the duality equations. However, it is not sufficient that the action is invariant up to terms linear in the duality equations.


\section{On Kac--Moody groups}\label{app:kacmoody}

For a given Kac--Moody Lie algebra there are different definitions of an associated Kac--Moody group and representations, see~\cite{Kumar:2002,Marquis:2018}. In the case of centrally extended loop groups, the two standard notions of a minimal group and maximal (or completed) group can be described as follows.

The minimal group,  denoted $\widehat{E}_8^{\rm m}$ consists of maps from the complex plane into the group $E_8$, where the maps are restricted to Laurent polynomials around infinity (or the origin). In terms of the spectral $w\in \cx$, one starts with elements $g^{\rm m}(w)$ that can be written as 
\begin{align}
g^{\rm m}(w) = \sum_{\ell=\ell_1}^{\ell_2} w^{-\ell} g_{\ell}
\end{align}
with $\ell_1\leq \ell_2$ (finite integers) and such that for each value of $w$ one has $g^{\rm m}(w)\in E_8$ (for a chosen matrix representation). This space of maps forms a group (under pointwise multiplication) and can be centrally extended~\cite{Pressley:1988}. 
The resulting group is $\widehat{E}_8^{\rm m}$.

The completed group $\widehat{E}_8^{{\rm c}-}$ replaces Laurent polynomials by formal Laurent series around infinity. In terms of an expansion in powers of $w$, this means that we are allowing arbitrarily negative powers of $w$
\begin{align}
g^{{\rm c}-}(w) = \sum_{\ell=\ell_1}^{\infty} w^{-\ell} g_{\ell}\,.
\end{align}
The $-$ indicates both that the powers are arbitrarily negative and that this corresponds to a completion in the positive Borel direction, which is consistent with our choice of letting $T_m^A\cong w^m T^A$ correspond to negative (positive) roots for $m>0$ ($m<0$), respectively.
We also require $g^{{\rm c}-}(w)\in E_8$ as a group element in the field of formal Laurent series in $w$. As the conditions for being in the group are algebraic conditions on the matrix entries (such as $\det\!\big( g^{{\rm c}-}(w)\big) =1$), these are also expressible in terms of formal Laurent series and can be imposed without requiring the series to converge.
Since formal Laurent series form a field, elements of this type form a group and can again be centrally extended.\footnote{We note that, when using the Geroch group for generating solutions of the Einstein equations~\cite{Maison:1988zx}, an intermediate version of the loop group is used that is given by meromorphic functions on (covers of) Riemann surfaces.} 
The logarithmic derivative $\dd g^{{\rm c}-}(w)  \left(g^{{\rm c}-}(w) \right)^{-1}$ is an element of the completed Lie algebra $\widehat{\mf{e}}_8^{{\rm c}-}$ that consists of  $\mf{e}_8$-valued Laurent series around infinity with central extension. From the point of view of the root space, we allow for an infinite linear combination of negative root generators in $\widehat{\mf{e}}_8^{{\rm c}-}$ but only a finite linear combination of positive root generators. It possible to define a Lie bracket on $\widehat{\mf{e}}_8^{{\rm c}-}$.

The explicit coset representative that is given in~\eqref{Vhat Y expansion} clearly belongs to this completed group and thus we take for the scalar fields the group $\widehat{E}_8^{{\rm c}-}$. Elements of this group can act on highest weight modules $\overline{R(\Lambda)_h}$ since the exponentials of positivehighest generators all are finite sums (rather than series) due to the existence of a highest weight. When we try to act on a lowest weight modules $R(\Lambda)_h$ the exponentials do not terminate but the computation for any given weight space is a finite sum so that one could consider infinite linear combinations in the space $R(\Lambda)_h$, which is sometimes called the completed module~\cite{Marquis:2018} and which we denote by $R(\Lambda)_h^{{\rm c}-}$. The (algebraic) dual of highest weight module $\overline{R(\Lambda)_h}$ is a completed lowest weight module  $R(\Lambda)_h^{{\rm c}-}$ and the pairing is invariant under $\widehat{E}_8^{{\rm c}-}$.

The Chevalley involution maps the (negatively) completed Lie algebra $\widehat{\mf{e}}_8^{{\rm c}-}$ to the (positively) completed Lie algebra $\widehat{\mf{e}}_8^{{\rm c}+}$ since it interchanges negative and positive roots. If $\widehat{\mf{e}}_8^{{\rm c}-}$  are Laurent series around infinity, then $\widehat{\mf{e}}_8^{{\rm c}+}$ are Laurent series around the origin. Therefore, the projections\footnote{Notice that these are not the same objects as defined in Section~\ref{sec:extcoset} as for simplicity we are not including conjugation by $\Gamma$ in this appendix.} 
\begin{align}
P(w) = \frac12 \left[ \dd V^{{\rm c}-}(w)  \left(V^{{\rm c}-}(w) \right)^{-1}
+ \left(\dd V^{{\rm c}-}(w)  \left(V^{{\rm c}-}(w) \right)^{-1}\right)^\dagger \right]\,,\nn\\
Q(w) = \frac12 \left[ \dd V^{{\rm c}+}(w)  \left(V^{{\rm c}+}(w) \right)^{-1}
- \left(\dd V^{{\rm c}-}(w)  \left(V^{{\rm c}-}(w) \right)^{-1}\right)^\dagger \right]\,,
\end{align}
of the  Mauer--Cartan form lie in the doubly completed space $\widehat{\mf{e}}_8^{{\rm c}+-} \coloneqq \widehat{\mf{e}}_8^{{\rm c}+} + \widehat{\mf{e}}_8^{{\rm c}-}$ for $V^{{\rm c}-}(w)\in \widehat{E}_8^{{\rm c}-}$. There is no Lie bracket on $\widehat{\mf{e}}_8^{{\rm c}+-}$ that can be defined by a finite number of operations from the one on $\mf{e}_8$ since this would require multiplying Laurent series around infinity with Laurent series around zero which is not a well-defined operation. The completed group $\widehat{E}_8^{{\rm c}-}$ does not act on $\widehat{\mf{e}}_8^{{\rm c}+-}$ for the same reason, only $\widehat{E}_8^{{\rm m}}$ does. Therefore, strictly speaking, the current $J(w) =2  \left(V^{{\rm c}-}(w) \right)^{-1} P(w)  V^{{\rm c}-}(w)$ is ill-defined. 
There is a well-defined action of $K(\widehat{E}_8^{\rm m})=K(E_9)$ on the components $P$ and $Q$ of the Maurer--Cartan form. This is sufficient for defining the action of generalised diffeomorphisms since the derivative is a constrained object.
Besides using the Unendlichbein approach to avoid the problem of the ill-defined current, one can consider $E_9$ exceptional field theory in a level decomposition with respect to a finite-dimensional Levi subgroup, such as $E_8$. 
This semi-flat formulation was used in Section~\ref{E8reduction} and yields a current $\tilde{\cJ}$ that is in $\widehat{\mf{e}}_8^{{\rm c}+-}$ since the conjugation of the symmetrised Maurer--Cartan form is reduced to elements of the Levi subgroup and all other objects in the theory are conjugated by the remaining unipotent elements of $\widehat{E}_8^{{\rm c}-}$ appropriately.

Because the coset representative is an element of $\widehat E_8^{\rm c -}$,\footnote{In~\eqref{Vir minus element}, we have also written an infinite exponential of negative Virasoro generators which are therefore in the completed $\Virm=\Vir^{- {\rm c}} $ group as defined there. The associated algebra can act consistently on the completed Lie algebra $\widehat{\mf{e}}_8^{{\rm c}-}$, so that~\eqref{Vhat V relation} is well defined.} the representations it acts on must be of the right type: either completed lowest weight modules (such as $\ket A$) or minimal highest modules (such as $\bra\partial$). 
For instance, typical expressions we encounter are $ \bra\partial V^{-1} \ldots V^{-1 \dagger} |A\rangle$ with $V^{-1 \dagger}\in \widehat{E}_8^{{\rm c}+}$ that can act on completed lowest weight modules (while $V^{-1}$ has finite action on $\bra\partial$).

\bibliographystyle{utphys}
\providecommand{\href}[2]{#2}\begingroup\raggedright\endgroup

\end{document}